\documentclass[11pt,nofrench 
                ]{thesis_IAS}

 \pdfoutput=1
\usepackage{aeguill}

\usepackage{graphicx}

\usepackage[english]{minitoc}
 \usepackage{natbib} 
 \usepackage{wasysym}

\newcommand\upun[1]{\uppercase{\underline{\underline{#1}}}}
\FormatHeadingsWith\upun

\newcommand\itheadings[1]{\textit{#1}}
\FormatHeadingsWith{\itheadings}

\NoChapterNumberInRef
\NoChapterPrefix

\makeglossary
\makeindex

\begin{document}

      \OddHead={{\leftmark\rightmark}{\hfil\slshape\rightmark}}
      \EvenHead={{\leftmark}{{\slshape\leftmark}\hfil}}
      \OddFoot={\hfil\thepage}
      \EvenFoot={\thepage\hfil}
      \pagestyle{ThesisHeadingsII}


\NoNewPageAfterParts


\ResetChaptersAtParts 


\dominitoc


\ThesisTitle{Evolution of intermediate mass galaxies up to  z$\sim$0.7 and studies of  SNe Ia hosts}
\ThesisDate{29 Septembre 2010}
\ThesisAuthor{Myriam ARNAL RODRIGUES}

\ThesisObsParis
\GEPIGalaxyGroup



\NewJuryCategory{family}{\it Directeur de these:}
                        {{\it Directeur de these:}}

\family={Francois HAMMER \\Ana Maria VERGUEIRO MONTEIRO CIDADE MOUR\~{A}O}

\NewJuryCategory{encadrant}{\it Encadrant de these:}
                        {{\it Encadrant de these:}}

\encadrant ={H\'ector FLORES}

\def\blanc{\hspace*{1cm}}

\President    = {Didier PELAT}
\Rapporteurs  = {Jarle BRINCHMANN&\\
                 Daniel SCHAERER \\}
\Examinateurs = {M\'{a}rio Gon\c{c}alo RODRIGUES DOS SANTOS&\\
                 Jorge VENCESLAU COMPRIDO DIAS DE DEUS}

\MakeThesisTitlePage





\begin{ThesisDedication}
To Hypatia, AD 350. 
\end{ThesisDedication}


\DontFrameThisInToc
\begin{ThesisAcknowledgments}

My life is like a Babel tower: it is spread among severals european countries, France, Spain and Portugal. My acknowledgments cannot thus escape to be multi-language. 

Je voudrais remercier avant toute chose mes directeurs de th\`ese officiels et officieux qui m'ont toujours soutenus pendant les quatre ann\'ees de th\`ese. Un grand merci \`a Fran\c{c}ois Hammer pour avoir \'et\'e toujours disponible pour parler de science malgr\'e son agenda de ministre. J'ai \'enorment appris sur mon sujet de th\`ese et sur la science en g\'en\'eral en ta compagnie. Muit\'issima obrigada a Ana Mour\~ao, que a pesar da distancia, sempre esteve presente para me dar apoio e conselhos ao longo do doutoramento. Guardo muito boas lembran\c{c}a das nossas conversa\c{c}\~oes, enfrente a uma boa ch\'avena de ch\'a, tanto em Paris como em Lisboa e no Observat\'orio do Calar Alto.  
Merci \`a Hector pour m'avoir initi\'e avec une grande patience aux observations et traitement de donn\'ees, ainsi que pour ses sages conseils. Je suis grandement redevable \`a Mathieu Puech, qui bien que n'\'etant pas mon encadrant officiel, \`a toujours r\'epondu \`a mes innombrables questions avec une patience infinie.  

Je tiens \`a exprimer mes remerciements aux membres du jury, qui ont accept\'e d'\'evaluer mon travail de th\`ese. Merci \`a Didier Pelat d'avoir accept\'e de pr\'esider le jury de cette th\`ese, et \`a Daniel Schaerer et Jarle Brinchmann  d'avoir accept\'e d'\^etre les rapporteurs de ce manuscrit. Leurs remarques et suggestions lors de la lecture de mon rapport m'ont permis dÕam\'eliorer la qualit\'e de ce dernier.

Un grand merci \`a l'\'equipe de l'\'ecole doctoral, Ana Gomez, Didier Pelat, Daniel Rouan, Jacqueline Plancy et Daniel Michoud pour rendre la partie administrative de la th\`ese tellement plus facile.

Ces quatres ann\'ees auraient \'et\'e bien moins agr\'eables sans l'\'equipe de choc des th\'esards de l'\'equipe extra-galactique du GEPI: Rodney, Loic, Rami, Anand (les PhD pizza turtles), Yan, Sylvain et Benoit, ainsi que les post-docs et stagiaires: Isaura Fuentes-Carrera, Yanbin Yang, Susana Vergani, Paola di Mateo, Leonor Chatain, Karen Disseau. Je garde un chaleureux souvenir de nos discussions anim\'ees durant les pauses caf\'e, de nos longues soir\'ees au labo et de nos sorties nocturnes \`a Paris. 

Je garde de mon aventure meudonnaise un tr\`es agr\'eable souvenir, essentiellement gr\^ace \`a ces personnes qui font de Meudon un lieu sp\'ecial et unique. Un grand merci \`a : L'\'equipe administrative du GEPI, Sabine Kimmel, Laurence Gareau, Pascal Lain\'e. \`A Chantal Balkowski	 pour ces pr\'ecieux conseils et nos conversations artistiques. \`A l'\'equipe Beninoise, Didier Pelat, Jacqueline Plancy, Pascal Gallais, Bernard Talureau, Catherine Boisson, Raphael Galicher, Aude Alapini, Luc di Gallo, Adeline Gicquel et Mathieu Brangier. \`A Bernard Talureau pour les rocks et salsa endiabl\'es sous les coupoles de Meudon!  Enfin, les mots me manquent pour te remercier Jacqueline. Un profond merci pour avoir \'et\'e mon ange gardien et amie depuis mon arriv\'ee \`a Meudon. Longue vie \`a Lulu! 

Por el camino de la vida uno encuentra personas que jamas ha de olvidar. En las frias avenidas de Paris, me tropece contigo Manu. Mi estadia en Paris habr\'ia sido como una pelicula sin banda sonora sin ti. Gracias por haber tra\'ido a Paris tu luz, poesia, una buena dosis de sana locura y tu permanente sonrisa. No te digo cuantos cuentos cuento al escribir estas frases!

Joao, nunca teria imaginado quando nos encontramos no nosso primeiro dia de universidade que viver\'iamos tantas coisas juntos: desde as idas e voltas no comb\'oi para o Porto, jantaradas na resid\^encia de estudantes, passeios por Lisboa, ate as nossas aventuras domesticas em Montrouge e os almo\c{c}os de P\'ascoa longe de casa. Obrigadissima Joao, por ter sido a minha familha durante este quatro anos. 

Merci \`a Raphael Galicher pour toujours avoir \'et\'e l\`a dans les moments difficiles mais aussi pour toutes les aventures v\'ecues ensemble.
 
Obrigada aos meus amigos de Lisboa, Fortunato, David, Marta, Ruben, Hugo, Joana e Mika. A cada vez que volto a minha querida Lisboa, sei que posso contar com voces e que a nossa amizade n\~ao mudo a pesar da distancia. 

Un grand merci, \`a tout ces amis crois\'es au d\'etours de chemins, de conf\'erences et de voyages:  Laure-Anne Nemirouski, Johan Richard, Barry Rothberg, Aurelie Delage. Merci pour vos conseils et votre amiti\'e !

Por fin primos, lo consegui! Ya soy astro .. algo..  "Eso ... astrologa o astronauta ... enfin eso que te gusta astronoma!" Arnal Power y mil abrazos a todos! Finalmente, gracias a mis padres por haberme apoyado sempre, de todos los modos possibles y imaginables!  Gracias por haber aguantado stoicamente el fr\'io haciendo me compa\~n\'ia cuando observaba en el pueblo. 

Et finalement: merci John!

\end{ThesisAcknowledgments}


\NumberAbstractPages
\begin{ThesisAbstract}
  \begin{FrenchAbstract}

La premi\`ere partie de cette th\`ese \'etudie la formation et l'\'evolution des galaxies. Cette \'etude s'est r\'ealis\'ee dans le cadre du relev\'e IMAGES "Intermediate Mass Galaxy Evolution Sequences" qui a pour but de contraindre l'\'evolution des propri\'et\'es globales des galaxies de masses intermediares, de $1.5$ \`a $15\times10^{10}\,M_\odot$, de z=0.9 jusqu' \`a aujourd'hui. Un \'echantillon repr\'esentatif de galaxies distantes du "Chandra Deep Field South" (CDFS) a \'et\'e observ\'e avec le spectrographe int\'egral de champs GIRAFFE au VLT et le spectrographe \`a  fente FORS2 \'egalement au VLT. Je suis responsable au sein de l'\'equipe IMAGES de l'exploitation des donn\'ees FORS2. Je pr\'esente dans ce manuscrit les propri\'et\'es du milieu interstellaire sur un \'echantillon repr\'esentatif de 88 galaxies distantes. En comparant ces observations avec les propri\'et\'es des galaxies locales, je montre que les galaxies ont \'evolu\'e de fa\c{c}on isol\'ees, en convertissant leur gaz en \'etoiles, durant les derniers 8 milliard d'ann\'ees. Aucun apport de gaz ext\'erieur ni d'expulsion de mati\`ere interstellaire n'est n\'ecessaire pour expliquer leur \'evolution r\'ecente. Ces conclusions se basent sur l'\'evolution de la relation fondamental masse- metallicit\'e, de la fraction de gaz, ainsi que de la non d\'etection de vent stellaire intense dans les profiles des raies d'\'emission du gaz ionis\'e. Parall\`element \`a l'\'etude du milieu interstellaire, je me suis \'egalement int\'eress\'ee au contenu en \'etoiles des galaxies. J'ai d\'evelopp\'e une m\'ethode capable d'estimer la masse stellaire de galaxies active en formation stellaire \`a partir de leur distribution spectral d'\'energie. Le principe de cette m\'ethode est de d\'ecoupler la lumi\`ere de la galaxie selon ses diff\'erentes composantes contribuant \`a la masse stellaire: \'etoiles jeunes, d\'age intermediare et vieilles, ainsi que l'effet de la poussi\`ere. Ce probl\`eme est d\'eg\'en\'er\'e mais l'ajout du taux de formation stellaire comme une nouvelle contrainte permet de lever un grand nombre de d\'eg\'en\'erescences. 

 Lors de ma th\'ese, j'ai eu l'opportunit\'e de travailler sur un projet instrumental en phase A pour le Europeen Extremly Large Telescope (E-ELT): le spectrographe multi-objet OPTIMOS-EVE. Mon r\^{o}le dans le consortium OPTIMOS-EVE a \'et\'e d'impl\'ementer une m\'ethode capable d'extraire le bruit du ciel des donn\'ees avec une erreur inf\'erieure \`{a} 1\% au niveau du continu du ciel.  La m\'ethode choisie est bas\'ee sur la reconstruction des variations spatiales du ciel sur tout le champ de vue de l'instrument \`{a} partir d'un \'echantillonnage partiel du ciel avec des fibres d\'edi\'ees. Cette m\'ethode permet d'extraire du ciel des raies tr\`es faibles. Ainsi, une raie d'\'emission avec un flux de $10^{-19} erg/s/cm^2$ sur un continu de magnitude 30 en bande J peut \^etre extraite du ciel  ($M_J$ =18) avec un signal \`{a} bruit sup\'erieur \`{a} 8 apr\`es 40h de temps d'exposition.  

La derni\`ere partie de cette th\`ese est consacr\'ee \`{a} l'\'etude des propri\'et\'es physiques des h\^{o}tes de Supernovae du type Ia (SN Ia). De r\'ecents mod\'eles th\'eoriques d'explosion de SN Ia montrent que les constantes cosmologiques d\'eriv\'ees \`{a} partir des SN Ia peuvent \^{e}tre affect\'ees par des bias issus des propri\'et\'es intrinseques des SN Ia, tels que la metallicit\'e et l'\^{a}ge du prog\'eniteurs lors de l'explosion. L'\'etude directe des progeniteurs de SN Ia n'est pas r\'ealisable actuellement, mais une premi\`ere approximation de leurs caract\'eristiques peut \^etre obtenue \`{a} partir de l'\'etude de leurs galaxies h\^{o}tes. Nous avons initi\'e un relev\'e en spectroscopie int\'egral de champ de galaxies locales, h\^{o}tes de supernova. Je pr\'esente l'\'etude pr\'eliminaire d'une galaxie h\^{o}te. 
 
  \end{FrenchAbstract}
  \begin{EnglishAbstract}
In the first part of this manuscript, I present the results on the properties of the
 interstellar medium and the stellar content of galaxies at z=0.6, from a representative
 sample of distant galaxies observed with the long slit spectrograph 
VLT/FORS2. This study has been realized in the framework of the ESO large program IMAGES 
\textit{"Intermediate MAss Galaxy Evolution Sequences"}, which aims to investigate the evolution of 
the main global properties of galaxies up to z~0.9. I discuss the implications of the observed 
chemical enrichment of the gas on the scenarios of galaxy formation. I also propose a new
 method to estimate reliable stellar masses in starburst galaxies using broadband photometry 
and their total star-formation rate. In a second part, I present a new method to extract, with high accuracy, the sky in spectra acquired with a fiber-fed instrument. I have developed this code in the Framework of the phase A of an instrument proposed for the E-ELT: OPTIMOS-EVE. This is a multi-fiber,able to observe at optical and infrared wavelengths simultaneously. In the third part, I show preliminary results from the CENTRA GEPI- survey at Calar Alto Observatory to study nearby galaxies, hosts of type Ia supernovae, using integral field spectroscopy. I present the first 2D maps of the gas and stellar populations of SNe Ia hosts. The results allow us to directly access the host properties in the immediate vicinity of the SNe Ia. This is a crucial step to  
 investigate eventual correlations between galaxy properties and SNe Ia events and evolution,  
leading to systematic effects on the derivation of the cosmological parameters. 
    
  \end{EnglishAbstract}
  
   \begin{EnglishAbstract}
O manuscrito esta dividido em tr\^es partes. Na primeira parte do manuscrito, exponho os resultados sobre as propriedades do meio interstelar e do conte\'udo estelar das gal\'axias distantes.  O estudo foi realizado numa amostra representativa de gal\'axias a z=0.6 a partir de dados obtidos com o espectrosc\'opio de fenda comprida VLT/FORS2. O estudo  no quadro do  programa observacional da ESO "Intermediate MAss Galaxy Evolution Sequences",  cujo objectivo \'e de investigar a evolu\c{c}\~ao das propriedades globais das gal\'axias de z=0.9 ate hoje.  Em particular, analiso as implicac\~oes da evolu\c{c}\~ao do conte\'udo qu\'imico do g\'as  nos modelos de forma\c{c}\~ao de gal\'axias.  Proponho um novo m\'etodo para estimar a massa estelar em gal\'axias com forma\c{c}\~ao intensa de estrelas a partir de fotometria de banda larga e da taxa de forma\c{c}\~o estelar. Na segunda parte, descrevo um novo m\'etodo para extrair com grande precis\~ao o sinal do seu nos espectros obtidos com espectrografos de fibras.  Desenvolvi este c\'odigo no quadro de uma estudo de fase A de um instrumento proposto para o E-ELT: OPTIMOS-EVE, um espectrografo multi-fibras, com a capacidade de observar v\'arios no domino \'optico e infravermelho. Na ultima parte, descrevo resultados preliminar no estudo de gal\'axias pr\'oximas hospedeiras de SN Ia com objectivo de investigar poss\'iveis sistem\'aticos na deriva\c{c}\~ao dos par\^ametros cosmol\'ogicos devidos as propriedades intr\'insecas das SNe Ia. A equipa come\c{c}o um projecto de observa\c{c}\~ao sistem\'atica das gal\'axias hospedeiras de Sn Ia utilizando espectroscopia integral de campo. Estas observa\c{c}\~oes permitem, pela primeira vez, de mapear as propriedades do g\'as e das popula\c{c}\~oes estelares na regi\~ao onde explodiram as Sn Ia. 
    
  \end{EnglishAbstract}
  
\end{ThesisAbstract}


\DontWriteThisInToc
\begin{TheGlossary} 
\item \textbf{$\Lambda$-CDM} : Cold Dark Matter cosmology
\item \textbf{AGN} : Active Galactic Nuclei
\item \textbf{CDFS} : Chandra Deep Field South
\item \textbf{CFRS} : Canada France Redshift Survey
\item \textbf{CMB} : Cosmological microwave background
\item \textbf{CSP} : Complex Stellar Population
\item \textbf{DTD} : Delay Time Distribution
\item \textbf{E-ELT} : European Extremely Large Telescope 
\item \textbf{EW} : Equivalent width
\item \textbf{FORS2} : Focal Reducer Low Dispersion Spectrograph 
\item \textbf{FoV} : Field of view
\item \textbf{GLAO} : Ground-Layer Adaptive Optics
\item \textbf{GOODS} : Great Observatories Origins Deep Survey
\item \textbf{GRB} : Gamma Ray Burst
\item \textbf{HST} : Hubble Space Telescope 
\item \textbf{IFS} : Integral Field Spectroscopy
\item \textbf{IFU} : Integral Field Unit
\item \textbf{IMAGES} : Intermediate Mass Galaxy Evolution Sequences
\item \textbf{IMF} : Initial Mass Function
\item \textbf{ISM} : Interstellar Medium
\item \textbf{IR} : Infrared
\item \textbf{K-S} : Kennicutt-Scmidt law
\item \textbf{LINER} : Low Ionization Narrow Emission Line 
\item \textbf{LIRG} : Luminous Infrared Galaxy 
\item \textbf{LSB} : Low Surface Brightness 
\item \textbf{MW} : Milky Way 
\item \textbf{M-Z} : Stellar mass metallicity relation 
\item \textbf{$N_e$} : electron density 
\item \textbf{OPTIMOS-EVE} : OPTIMOS Extreme Visual Explorer
\item \textbf{PMAS} : Postdam Multi Aperture Spectrophotometer 
\item \textbf{PCA} : Principal Component Analysis
\item \textbf{PSF} : Point  
\item \textbf{SED} : Spectral Energy Distribution
\item \textbf{SSP} : Single Stellar Population
\item \textbf{SF} : Star forming 
\item \textbf{SFH} : Star Formation History 
\item \textbf{SFR} : Star Formation Rate 
\item \textbf{SN Ia} : Supernova type Ia 
\item \textbf{$T_e$} : electron temperature 
\item \textbf{TP-AGB} : Thermally pulsing asymptotic giant branch
\item \textbf{ULIRG} : Ultra Luminous Infrared Galaxy 
\item \textbf{UV} : Ultraviolet 
\item \textbf{VLT} : Very Large Telescope 
\item \textbf{WD} : White Dwarf 

\end{TheGlossary}


\WritePartLabelInToc
\WriteChapterLabelInToc


\tableofcontents

  
\NumberThisInToc
\mainmatter

\part*{Preface}
 \EmptyNewPage
A fundamental problem in modern cosmology is understanding how galaxies formed. It is widely accepted that this happened within the framework of a Cold Dark Matter ($\Lambda$-CDM) cosmology, whose geometry has now been determined with high precision. Observations of the first emitted photons in the Universe, the cosmological microwave background (CMB), reveal a very isotropic blackbody emission. After 380 000 years, the mass density of the Universe was almost uniform, and galaxies are thought to result from the growth of primordial fluctuations. This cosmological model gives a credible explanation of the formation of structures through the hierarchical assembly of dark matter haloes. In contrast, little is known about the physics of formation and evolution of the baryonic component of gas and stars, because the conversion of baryons into stars is a complex and poorly understood process. 
There are two scenarios to explain the formation of galaxies from the baryonic mass trapped within dark matter halos. In the \textit{Monolithic or secular} scenario, the first galaxies are formed from the gravitational collapse of baryonic matter in high density dark matter haloes. The initial size of the gas cloud, the different initial interactions with the neighboring structures, and the secular accretion of pristine gas explain the variety of morphological types in the Hubble sequence. On the contrary, in the \textit{Hierarchical} scenario the baryonic mass collapses first in the smaller dark matter haloes and then grows by successive merger producing bigger galaxies. In this case, the variety of local galaxy morphologies is explained by the variety of merger histories. Observations and theoretical models converge now on a merger-driven scenario of formation of spheroids, in which elliptical galaxies are the result of successive fusion of disk galaxies in high density regions of the Universe \citep{1992ApJ...389....5T,1994ApJ...437L..47M}. In contrast, the formation of disks still poses major challenges for our current understanding of galaxy evolution.

\subsubsection{Scenario of local disk formation}
Disk galaxies comprise the majority of the galaxy population in the local Universe. They represent 70\% of intermediate-mass galaxies, which in turn contribute to at least two thirds of the present-day stellar mass (e.g \citet{2005A&A...430..115H}). Since the earliest models of disk formation \citep{1962ApJ...136..748E}, mostly motivated by studies of the Milky Way, disk formation is viewed as a gradual process dominated by gas accretion, as opposed to the merger-dominated formation of early-type galaxies. Forming a disk from a collapsing gas cloud requires a relatively high angular momentum. In the 60s/70s, observations of the Milky Way suggested a scenario in which spiral disks acquired their angular momentum at an early epoch by tidal torques induced through interactions with neighboring structures \citep{1969ApJ...155..393P,1984ApJ...286...38W}. In the modern picture of a dark-matter dominated Universe, the angular momentum is inherited from the dark matter halo as the gas collapses \citep{1978MNRAS.183..341W,1980MNRAS.193..189F}. The once-formed disk then grows by further accretion from the halo and other secular processes. 
This theory is faced with at least three major problems. First, this scenario has been implemented assuming that the MW is representative of local spiral galaxies. Unfortunately, \citet{2007ApJ...662..322H} have shown that the Milky Way is an exceptional local spiral.  The Milky Way experienced very few minor mergers and no major merger during the last 10-11 Gyrs \citep{2009IAUS..258...11W,2001ASPC..230....3G}. In contrast, M31 (Andromeda) has a very tumultuous history \citep{2001Natur.412...49I,2004MNRAS.351..117I,2008ApJ...685L.121B}, as testified by the large number of stellar debris in its outskirts. Observations of galaxy outskirts at similar depths are in progress, and structures related to past merger events have been identified surrounding NGC 4013, NGC 5907, M 63 and M 81 \citep{2009ApJ...692..955M,2009AJ....138.1469B}. Second, galaxy simulations demonstrate that such disks can be frequently destroyed by collisions of relatively big satellites \citep{1992ApJ...389....5T}, and such collisions might be too frequent for disks to survive; Third, the disks produced by cosmological simulations in such a way are too small or have a too small angular momentum when compared to the observed ones, and this is so-called the \textit{angular momentum catastrophe}. The monolithic model of disk formation is difficult to incorporate into the hierarchical model, and in contraction to a plethora of observations. Given the high frequency of mergers, the widely accepted assumption that a major merger would unavoidably lead to an elliptical may no longer be tenable: accounting for the large number of major mergers that have apparently occurred since $z\sim3$ would imply that all present-day galaxies should be ellipticals. Motivated by the observational evidence, \citet{2005A&A...430..115H} proposed an alternative scenario for the formation of local disk, the \textit{spiral rebuilding scenario} where disks are formed from gas-rich mergers at intermediate redshifts.

\subsubsection{Diagnosing the ISM and stellar content of distant galaxies}

\begin{figure}[!h]
\centering
\includegraphics[width=14cm]{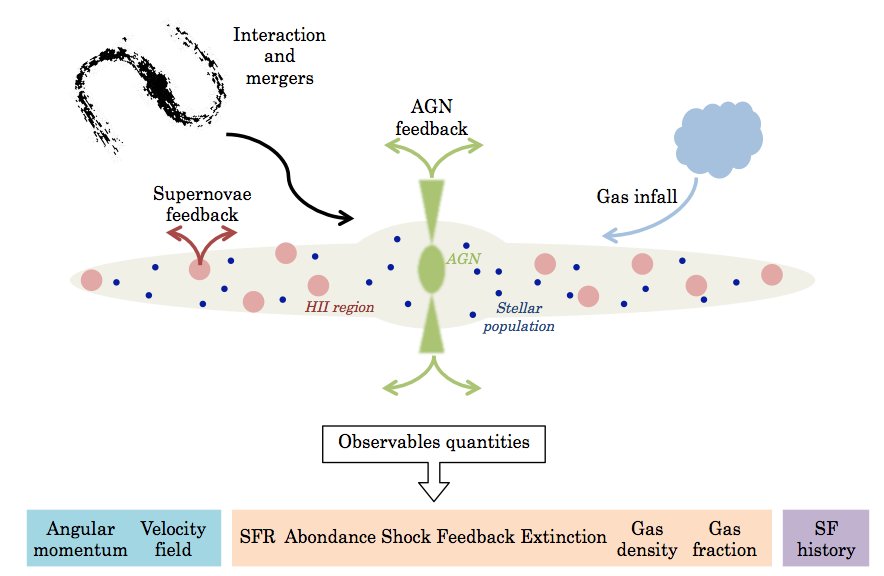}
\caption{{\small The processes taking place in the ISM of a galaxy.}}
\label{Galaxy_physics}
\end{figure}

The study of the interstellar medium (ISM) and the star formation history (SFH) are important tools to shed light on the processes that have led to the formation of present disks. On the first hand, the ISM keeps imprints of the different processes that can affect galaxies, such a the secular star-formation, feedback from supernova or AGN, and any kind of interaction between the galaxy and its environment - secular accretion of gas from cosmic filaments or interaction with other galaxies - see Fig. \ref{Galaxy_physics}. On the other hand, the study of stellar populations in galaxies gives important clues on how galaxies have assembled their stellar mass. Many works have put forth evidence for the fundamental role of stellar mass in galaxy evolution. Indeed, the stellar mass is found to correlate with many galaxy properties, such as luminosity, gas metallicity, color and age of stellar populations, star-formation rate, morphology, and gas fraction, to enumerate a few of them \citep{2000ApJ...536L..77B,2003ApJS..149..289B,2003MNRAS.341...33K,2004ApJ...613..898T,2007ApJ...663..834B}. During my thesis, I have investigated the different methodologies to derive properties of the ISM (\textit{Introduction, Chapter 1}) and stellar populations (\textit{Introduction, Chapter 2}). I dedicate the three first chapters of this report to this topic and I have focused, in particular, on the specific issues inherent to distant galaxy observations (\textit{Introduction, Chapter 3}). 

\subsubsection{The central role of mergers in the formation of local disks}
Understanding the evolution of galaxies as a function of look-back time requires observing galaxies at different epochs in order to constrain the evolution of theirs fundamental quantities. In the framework of the study of disk formation, we have carried out a large observational program, IMAGES, to derive the main properties of the local disk progenitors at $z\sim0.6$. We have investigated the evolution of well-established scaling relations up to z=1, such as those between mass and gas metallicity, the Tully-Fisher relation, and galaxy morphology  (\textit{Part I - Chapter 1}). Within the IMAGES team, I have been responsible for the study of integrated properties from integrated spectroscopy (\textit{Part I - Chapter 2,3}). The central part of this report  deals with the properties of the ISM at z=0.6 and the implications of their evolution on constraining galaxy evolution scenarios (\textit{Part I - Chapter 3}). I have also investigated the issue of stellar mass and star formation history estimations in distant galaxies using integrated spectra and broad-band spectroscopy (\textit{Part I - Chapter 5}). 

\subsubsection{Investigating the physics of first galaxies with the new generation of instruments }

The study of the evolution of galaxies at larger look-back time will require similar observations from large surveys of high-z galaxies. Although galaxies at higher redshift have been already detected, the limitation on the spatial resolution and the low signal-to-noise of the observations and the strong bias on the representativeness of these galaxies prevent us from having a consistent snapshot of galaxy properties at look-back time superior to ten Gyr. During the next decade, such observations will be possible thanks to the advent of a new generation of instruments, such as the European-Extremely Large Telescope (E-ELT). In this framework, I had the opportunity of working on the phase-A of OPTIMOS-EVE a fiber-fed visible to near infrared multi-object spectrograph designed for the E-ELT instrument (\textit{Part III - Chapter 1}). My work in the OPTIMOS-EVE consortium was to define a strategy to resolve the critical issue of sky subtraction for faint object spectroscopy in fiber-fed instruments (\textit{Part III - Chapter 2}).  

\subsubsection{Evaluation of the systematics from host galaxies in cosmology with SN Ia}

In parallel to the topic of the study of distant galaxies, I have applied the method described in \textit{Introduction part - Chapter 1\&2}  to the study of the properties of SN Ia hosts. During the last decade, SNe Ia were used as cosmological standard candles. The observations of SNe Ia led to the discovery of the accelerating expansion of the Universe and dark energy. They are now key tools in the future cosmology experiments to unveil the nature of dark energy. However, subtle systematic uncertainties stemming from our limited physical knowledge of SNe Ia progenitor stars are currently the major obstacles to fully exploit the potential of SNe Ia in cosmology. One approach to get new insight into the properties of SN Ia progenitors is to focus on their host galaxies. 3D spectroscopy allows us to map the physical properties of their hosts. By analyzing in detail the spatial distribution of the kinematics, metallicity, extinction, star formation, electron density, as well as the way they are inter-related, the physical and chemical properties of the gaseous phase in these galaxies can be deduced, and constrains on SN Ia progenitors established (\textit{Part II - Chapter 1}).
In collaboration between GEPI-Observatoire de Paris and the CENTRA-Institute Superior Tecnico of Lisbon -Portugal, I have started a project to study SNe Ia host galaxies in the local Universe. We have observed a sample of nearby SNe Ia hosts with the wide-field PMAS Integral Field Unit to derive the properties of the gas and the stellar populations in the immediate vicinity of the SN. I present here the preliminary results from a pilot study carried out in November 2009 at Calar Alto Observatory (\textit{Part II - Chapter 2}).

\WriteThisInToc
\FrameThisInToc
\NumberThisInToc
\part*{The ISM and stellar population properties from integrated spectroscopy}
 \begin{flushright}
\textit{"Ce ne sont pas seulement des lignes pour moi,\\
chaque nouveau spectre ouvre la porte sur un nouveau monde merveilleux.\\
C'est presque comme si les \'{e}toiles lointaines avaient acquis le don de parler\\
et \'{e}taient capable de raconter leurs conditions physiques et leur constitution"\\
by Annie Jump Cannon}
 \end{flushright}
  \EmptyNewPage


  \FrameThisInToc
\chapter{How to derive the properties of the interstellar medium?}
\minitoc
The emission spectra of star-forming galaxies are dominated by the radiation of the ionized gas emitted in HII regions (Section \ref{Radiation mechanisms in HII regions}). The study of the emission spectra emitted by all the HII regions of a galaxy allows us to characterize the gas of a region with active star-formation: temperature and density (section \ref{Electron density and temperature}), amount of dust (section\ref{Extinction}) and the relative abundance of chemical elements (Section \ref{Metallicity}). Even if stellar radiation is the main source of gas ionization, it is not the only one. Photoionization by active galactic nuclei and ionization produced by radiative shocks in violent star-forming episodes can contribute a large fraction of the observed emission spectrum. These two sources cannot be neglected when characterizing the properties of the ISM (Section \ref{Contamination by active galactic nuclei radiation}). 

I have used the diagnostic described in this chapter to characterize the activity of distant galaxies observed with long-slit spectroscopy (Part I, VLT/FORS2) and integral field spectroscopy observations of local SNe Ia hosts (Part II, Calar Alto/PPAK). 

\section{Radiation mechanisms in HII regions}
\label{Radiation mechanisms in HII regions}
This section briefly describes the physical processes that lead to the characteristic visible HII spectrum. For more details on the radiation mechanism, the reader can consult textbooks such as "\textit{Astrophysics of Gaseous Nebulae and Active Galactic Nuclei}" by \citet{1989agna.book.....O} and "\textit{The Physics and Chemistry of the Interstellar Medium}" by \citet{2005pcim.book.....T}.   

HII regions are induced by massive stars, type O and B, located in their nebulae cocoon. Due to their very high surface temperature, $T_*> 3\times10^4 K$,  massive stars emit a very energetic radiation, mainly in the ultraviolet domain. This energy is transferred to the surrounding gas via photoionisation. The photons with energy higher than the hydrogen potential (E> 13.6 eV) ionize hydrogen atoms and other elements present in the interstellar medium, such as oxygen, nitrogen and carbon. The collisions between free electrons, electrons \& ions distribute the residual energy and heat the gas up to $T\sim 10\,000K$ following a Maxwell-Boltzmann velocity distribution. These collisions excite the ions which decay producing recombination lines. The free electrons eventually recombine with ions. The excited atom created in the recombination process quickly cascades to the ground state emitting several low-energy photons through radiative transition processes.  The HII region fluoresces by converting the stellar ultraviolet light to lower-energy photons, with the bulk of radiation escaping through the Hydrogen recombination lines (e.g. the visible Balmer lines $H\alpha\,\lambda6563$\AA\,, $H\beta\,\lambda4861$\AA\,, $H\gamma\,\lambda4341$\AA\,). 

The elements heavier than hydrogen such as Helium, Neon, Oxygen and Nitrogen are also ionized. However, due to the low density of the interstellar medium,  these elements emit in majority forbidden lines (e.g. [OII]$\lambda 3727$\AA\,, [OIII]$\lambda 5007$\AA\,, [NII]$\lambda 6583$\AA\,). Indeed, the collision between electrons and ions excite low energy levels. As the ISM has a low density ($N_e<10^4\,cm^3$), de-excitation by collision is improbable and de-excitation by radiative transfer of  transitions with low transition probability become possible.  

\begin{figure}[!h]
\centering
\includegraphics[width=16cm]{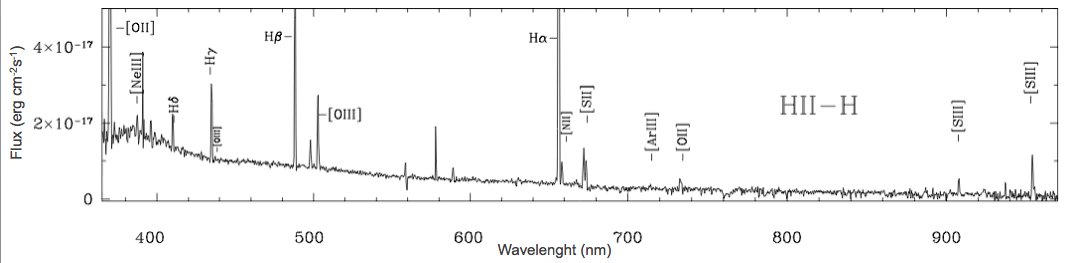}
\caption{VLT spectrum of an HII regions in Sextans B. The typical lines of HII regions are detected : the Balmer serie and [OII] , [OIII], [NII], [SII] forbidden lines.}
\label{HII_spectra}
\end{figure}

A typical optical spectrum of an HII region is composed by the series of hydrogen recombination emission lines and forbidden lines of metals with several degrees of ionization, as shown in Fig. \ref{HII_spectra} \citep{2005A&A...443..115M}. The intensity of emission lines is ruled by the atomic physics and the conditions of the gas: degree of ionization, the shape of the ionizing radiation and the relative abundance of the metals. I will describe in detail the different diagnostics and the related physics in the next sections. 
 
\section{Extinction}
\label{Extinction}
The extinction is a key parameter in the study of the ISM. The dust causing the extinction accounts for 1\% of the total mass of the ISM matter. Dust grains are composed of heavy elements formed during the chemical evolution of galaxies and it is thus an important parameter to study the mechanisms occurring in galaxy evolution. Secondly, dust is the main source of opacity in the ISM and drives the spectral energy distribution of the ISM. Before deriving any physical quantity from the emission lines, it is  necessary to evaluate the amount of extinction due to dust. For details on dust properties, composition and cycle, the reader is refered to "\textit{Astrophysics of Gaseous Nebulae and Active Galactic Nuclei}" by \citet{1989agna.book.....O} and "\textit{The Physics and Chemistry of the Interstellar Medium}" by \citet{2005pcim.book.....T}, \citet{2003ARA&A..41..241D}, \citet{1990ARA&A..28...37M}, \citet{2003pid..book.....K} \textit{"The Physics of Interstellar Dust"}. 
\subsection{The dust}
\begin{figure}[!h]
\centering
\includegraphics[width=0.6\textwidth]{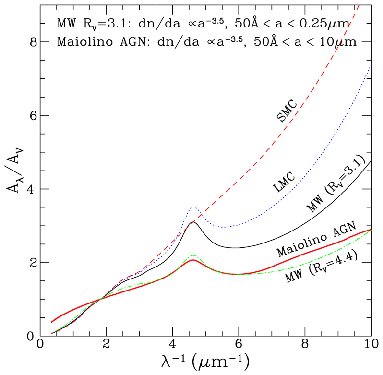}
\caption{{\small Interstellar extinction curves of the Milky Way ($R_V $= 3.1, 4.4), SMC, and LMC. In the Milky Way, the standard value for the diffuse ISM is $R_V$=3.1 while dense clouds, such as Orion nebulae, have higher values ($R_V $= 4.4). Both Magellanic clouds have lower values of $R_V$ because their ISM are diffuse and have low metallicity \citep{1998ApJ...498..735P}. }}
\label{Extinction_law}
\end{figure}

Dust absorbs and scatters part of the light emitted by the astrophysical objects. At a given wavelength, the light coming from an emitting source is dimmed when crossing the interstellar dust: 
\begin{equation}
I_\lambda=I_{\lambda0}e^{-\tau_\lambda}
\end{equation}
where $I_{\lambda0}$ is the intensity of the light emitted and $\tau_\lambda$ is the optical depth at a given wavelength. 
The absorption depends on the wavelength: at a given dust and gas column, more light is scattered in the blue than in the red. This phenomenon, called interstellar reddening, is due to the physical properties of the dust, mainly the dust grain dimension. In a first approximation, we can consider that the properties of the dust are homogeneous in all regions of the ISM. The optical depth can thus be decomposed into two components: 
\begin{equation}
\tau_{\lambda}= cf(\lambda)
\end{equation}
where $c$ is the amount of extinction, or excess of color E(B-V) defined as the difference of total extinction in B and V bands $A(B)-A(V)$, and $f(\lambda)$  is a standard extinction law. 
The extinction law is not universal. It depends on the dust physical conditions in the observed region. However, \citet{1990ARA&A..28...37M} has shown that it can be parametrized by a total-to-selective ratio $R_V$, over 0.1-2$\mu m$, see illustration of the parametrization in Fig.\ref{Extinction_law}. Rv depends on the environment traversed by the line-of-sight \citep{1989ApJ...345..245C}. The standard value in the Milky way is $R_V$=3.1 while in dense clouds $R_V$ have higher values. The extinction can thus be expressed as a function of the color excess E(B-V) and the total-to-selective ratio $R_V$:
\begin{equation}
A_V=\frac{E(B-V)\times R_V}{1.47}
\end{equation}

In the case of extragalactic objects, two extinction sources have to be considered. The intrinsic extinction from the extragalactic object and the extinction of the Milky Way on the line-of-sight. The later source of extinction is easily corrected using maps of galactic dust from \citet{1998ApJ...500..525S} and a galactic extinction law \citep{1999PASP..111...63F}. Nevertheless, distant galaxies are usually observed in cosmological fields located in regions of the sky almost devoid of galactic dust along the line-of-sight and it is therefore useless to apply such a correction.  

\subsection{Balmer decrement}
\label{Balmer decrement}
The determination of the intrinsic extinction of a galaxy can be performed using the method called Balmer decrement. This method is based on the assumption that the ratio of Balmer lines, emitted during recombination processes, depends only and weakly on temperature \citep{1989agna.book.....O}. In the idealized Case B recombination, the HII regions are considered to be optically thick media in which all the Lyman photons are reabsorbed by hydrogen atoms and then reemitted as photons of the Balmer series. Given this assumption, the population of the exited levels of hydrogen depends weakly on the electron density and electron temperature. The ratios of Balmer lines are thus defined by the laws of atomic physic, see Table \ref{Balmer_dec}. Knowing the theoretical ratio between two Balmer lines it is possible to infer the extinction comparing it to the observed ratio of the two lines. 
\begin{equation}
\frac{I(\lambda_1)}{I(\lambda_2)}=\frac{I_0(\lambda_1)}{I_0(\lambda_2)}10^{-c[f(\lambda_1)-f(\lambda_2)]}
\end{equation}
The color excess is thus given by :
\begin{equation}
c=-\frac{1}{[f(\lambda_1)-f(\lambda_2)]}log\left(\frac{R_{obs}}{R_{intr}}\right)
\label{c_extinxtion_eq}
\end{equation}
where $R_{obs}$ and $R_{intr}$ are respectively the observed and the theoretic ratio of lines.
The emission lines can now be corrected for the extinction using the following equation:
\begin{equation}
I_{corr}(\lambda)=I(\lambda)10^{c\times([f(\beta)-f(\inf)]-[f(\beta)-f(\lambda)]}
\end{equation}
where $\lambda$ is the wavelength, $I(\lambda)$ is the observed flux of the line to be corrected and $f(\lambda)$ is the value of the extinction law at $\lambda$. 

\begin{table}[!h]
\begin{center}
\begin{tabular}{c|c|c}
Lines&$I_i/I_{H\beta}$&$f(\lambda_i)-f_\beta$\\
\hline
$H\alpha$&2.87&-0.35\\
$H\gamma$&0.466&0.15\\
\end{tabular}
\caption{{\small Balmer-line intensity relative to $H_\beta$  in the case of He I recombination line Case B, T=10\,000K and the standard interstellar extinction curve, from \citet{1989agna.book.....O} }}
\end{center}
\label{Balmer_dec}
\end{table}

\section{Contamination by active galactic nuclei radiation}
\label{Contamination by active galactic nuclei radiation}
In star-forming galaxies the bulk of the ionizing radiation is dominated by young massive stars located in HII regions. However, some galaxies can host active galactic nuclei and thus produce large amounts of ionizing radiation. A large number of the equations to determine physical quantities, for example the strong line diagnostic for metallicity, are based on the assumption that the emission lines arise only from the ionization of the ISM by massive stars. It is thus important to detect the galaxies that can be contaminated by AGN emission before deriving metallicities, electron temperature or densities. The fraction of AGN is not negligible :  27\% of local galaxies host AGN \citep{2009MNRAS.398.1165G} and the fraction increases in distant galaxies with a peak at z=2-3 \citep{2004MNRAS.349.1397C,1994ApJ...421..412W,2001AJ....121...31F}.  

\begin{figure}[!h]
\centering
\includegraphics[width=0.60\textwidth]{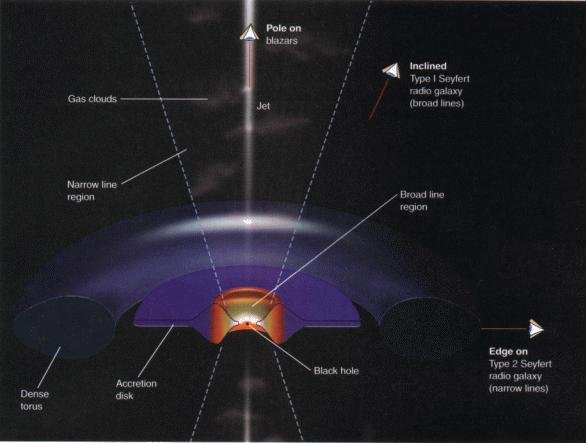}
\caption{{\small Schematic view of an AGN. Active galactic nuclei galaxies host in their center an active black hole, surrounded by an accretion disk in rapid rotation. The matter is intensely heated in the accretion disk and emits an ionizing radiation which photo-ionizes dense fast moving clouds located at proximity to the nuclei (the broad line region). AGN can be Seyfert I or II depending on the line-of-sight from which the nuclei is observed. Seyfert I are AGN saw within a small inclined angle while Seyfert II are seen within a quasi edge-on line-of-sight.}}
\label{AGN_schema}
\end{figure}

Active galactic nuclei galaxies host in their center an active black hole, surrounded by an accretion disk in rapid rotation. The matter is intensely heated in the accretion disk and emit an ionizing radiation which photo-ionizes dense fast moving clouds located at proximity to the nuclei (the broad line region). More diffuse and slower-moving clouds located at larger radii are also ionized (narrow line region). AGN are classified into three categories which depend on the line-of-sight from which the nuclei is observed and the activity of the central black hole, see Figure \ref{AGN_schema}.

\begin{description}
\item[Seyfert I:] The central region is visible and the spectra is composed by large recombination lines emitted by the broad line clouds and  narrow lines from the narrow line region, see Fig.\ref{AGN_schema}. Only the recombination lines have broad emission because the broad line region doesn't emit auroral lines, as [OII] or [OIII]. The medium of broad line clouds is to dense and thus too collisional to allow the emission of forbidden lines. The presence of multiple ionized species as NV, OVI in their spectra reveals the high energy of the ionization source that is incompatible with the energy produced by stellar radiation. 
\item[Seyfert II:] The disk masks the central region and absorbs the lines arising from the broad line region. The spectrum is probably composed by a combination of gas glowing in response to the active nucleus and material ionized by massive stars nearby (narrow line region). The ratio of the strength of lines shows that the ionizing radiation has characteristic electron temperature superior by few hundreds degrees relatively to typical HII region temperature.   
\item[Liners:] Stands for "Low Ionization Narrow Emission Line" galaxy. The collisional lines are very intense, but the hydrogen recombination lines are abnormally faint. The LINERs galaxies have a low degree of ionization.  The mechanism producing the observed ratio of line strength is still open to debate. There are three scenarios:  (1)Low luminosity AGN, powered by accretion (possibly in a radiatively inefficient regime) onto a super-massive black hole; (2) Radiative shock waves; (3) Radiation emitted by old stellar population \citep{2008MNRAS.391L..29S}.
\end{description}

 \begin{figure}[!h]
\centering
\includegraphics[width=0.50\textwidth]{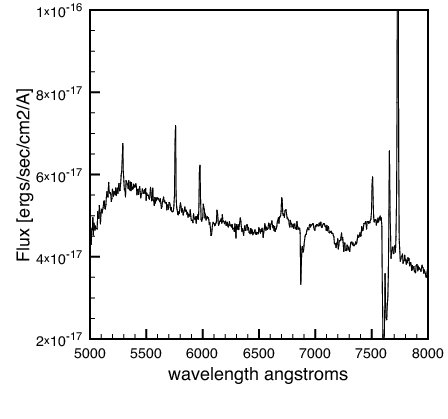}
\includegraphics[width=0.40\textwidth]{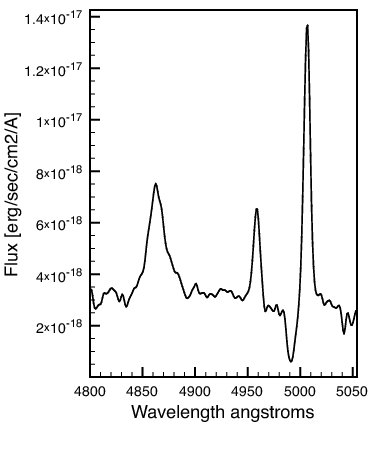}
\caption{{\small Spectra of a Seyfert 1 galaxy from the IMAGES sample, J033208.66-274734.4. Inset zoom of the region $H\beta$-[OIII] in the Seyfert I galaxy J033230.22-274504.6. The broad emission in Balmer recombination lines are visible in both spectra.}}
\label{AGN_Seyfert}
\end{figure}
Seyfert 1 are easily discarded from star-formation dominated galaxies due to their characteristic broad Balmer lines, see Fig.\ref{AGN_Seyfert}. Seyfert 2 and LINERs are isolated from star-formation using diagnostic diagrams based on emission lines ratios. 

\subsection{The BPT diagnostic diagram}
\begin{figure}[!h]
\centering
\includegraphics[width=0.60\textwidth]{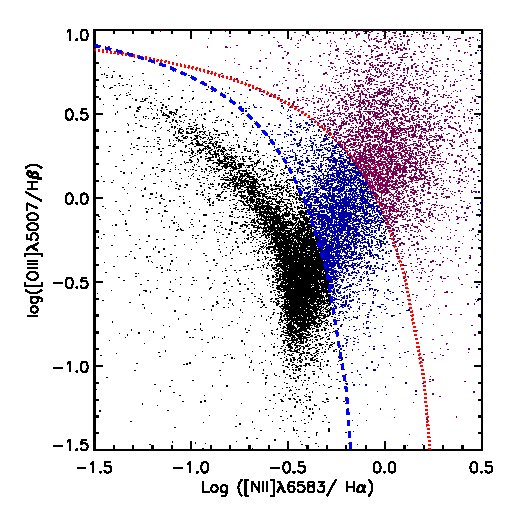}
\caption{{\small BPT diagram for the SDSS galaxies \citep{2003MNRAS.346.1055K}. The black symbols are star-forming galaxies. The red symbols are AGN hosts according to the theoretical delimitation of  \citet{2001ApJS..132...37K}. It corresponds to the theoretical limit of the two ratios of lines when assuming only stellar photo-ionizing radiation. The blue dots, between the theoretical limit of AGN and the empirical limit of star-forming galaxies \citep{2003MNRAS.346.1055K} are composite SF-AGN galaxies. This category includes real AGN hosts and a fraction of no-AGN galaxies whose old stellar population mimic the line ratios produced by AGN radiation \citet{2008MNRAS.391L..29S}.  }}
\label{AGN_K03}
\end{figure}

This diagnostic diagram has been created in a epoch where X-ray observations were not available. It is based on the fact that collisionally excited lines -e.g [NII]- are brighter in nebulae excited by AGN radiation (higher electron temperature) than in HII regions. The \citet{1981PASP...93....5B} BPT diagram is defined as the line ratio of [OIII]$\lambda\lambda4959, 5007/H_{\beta}$ vs. [NII]$\lambda 6584 /H\beta$ and has been widely used \citep{1987ApJS...63..295V, 2003MNRAS.346.1055K}. Galaxies falls into two well-defined branches, see Fig. \ref{AGN_K03}: a left branch, which is populated by star-forming galaxies and a right branch attributed to AGN dominated galaxies (with high $[NII]\lambda 6584 /H\beta$). Several authors have given observational and/or theoretical delimitation lines to classify the galaxies in the BPT diagram. More used ones are those proposed by \citet{2003MNRAS.346.1055K} based on distribution of SDSS galaxies and those of \citet{2006MNRAS.372..961K} derived from photo-ionization modes. It is important to notice, that \citet{2008MNRAS.391L..29S} have recently suggested that part of the galaxies classified as AGN according to the  \citet{2003MNRAS.346.1055K} delimitation can be highly contaminated by galaxies without any AGN source which mimic the LINER emission. Based on stellar population analysis and photoionisation models, they have demonstrated that a large fraction of the systems in the right branch could be galaxies which stopped forming stars, but where the ionizing radiation could be produced by hot post-AGB stars and dwarf. 

\subsection{Other diagnostics}
The BPT diagram is well-suited for local galaxies studies, however at higher redshift the $H\alpha$ and [NII] lines are reshifted to the near-infrared and therefore are difficult to observe. The diagnostic diagram based on [OII]$\lambda 3727 /H\beta$ vs [OIII]$\lambda\lambda4959, 5007/H\beta$ ratio is more adapted to redshifted galaxies. The limitation between Star-formation/Seyfert/LINER can be empirically calibrated from observations or from photoionization models. For instance, the delimitation between SF and AGN in the [OII]/$H\beta$ vs [OIII]/$H\alpha$ diagram of Fig.\ref{AGN_FORS2} (Part II) is the limit ratio predicted by photoionization models for the maximal stellar temperature of 60\,000K, which is the maximum temperature for the most massive stars. The main disadvantage of this diagnostic is its sensitivity to the extinction which affects the ratio [OII]/$H\beta$. It is mandatory to correct the flux of emission lines from dust for this diagnostic due to the presence of the $[OII]/H\beta$ ratio. 

The delimitation between AGN and star-forming galaxies are fuzzy which make the exclusion from SF sample of galaxies falling near the limit lines difficult. Other hints can be investigated for these near-limit objects, as the detection of radio or X-ray emission. In fact, the flow of the hot ionized gas in the accretion disk creates an intense magnetic field which can be strong enough to originate two jets of relativistic plasma. The charged particles are accelerated by the jets and emit a synchrotron radiation and infrared light. The X-ray emission originates in the inner part of the nucleus and/or in the jets.  The presence of lines associated with shock process - as the NeIII$\lambda 3868$ and OI$\lambda$6300 line - and lines from high ionized species are also a good hint for AGN contamination. 

\section{Shocks and gas fall}
\label{Shocks and gas fall}
Radiative shocks are produced by gas infall \citep{1976ApJ...203..361C} and by gas outflow from AGN feedback or violent star formation (supernovae explosion). The interstellar medium is also affected by other source of shocks as the jets and winds associated with stellar objects. But the contribution of these sources are faint and they are not expected to be detectable in integrated spectra where an overwhelming fraction of the light comes from the stellar photo-ionizing radiation.

\begin{figure}[!h]
\centering
\includegraphics[width=0.60\textwidth]{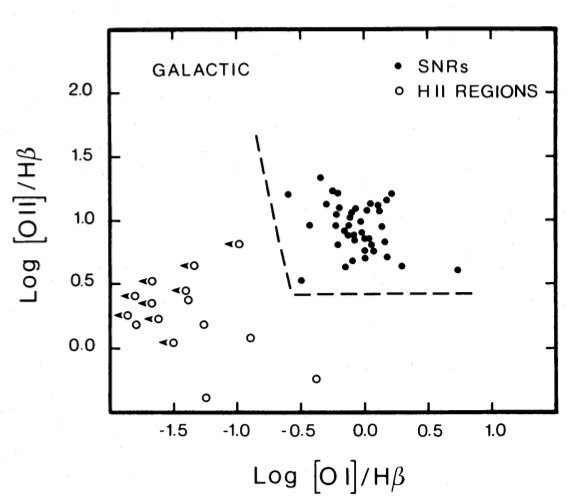}
\caption{{\small The [OII]/$H\beta$ and [OI]/$H\beta$ diagnostic diagram to disentangle between shocked and photo-ionized gas, from \citet{1985ApJ...292...29F}.}}
\label{Diag_shock}
\end{figure}

Shocks compress the medium, increasing the electron density, and thus raise collisional de-excitation transitions. These processes enhance low excitation levels such as [OI]$\lambda$6300\AA\, [OII]$\lambda$3727\AA\,, NI $\lambda$5200\AA\, relatively to hydrogen recombination lines. The ratio between the low excited transition lines and Balmer recombination lines can be used as a diagnostic to discriminate between shocked and photo-ionized gas as illustrated in Figure \ref{Diag_shock}. Some shocks can heat the gas to very high temperature $10^4-5.10^5$K and produce the highly ionized ions \citep{2005pcim.book.....T}. The presence of energetic shocks can thus be inferred from the detection of emission lines such as HeI and NeIII. 

Another way to detect gas flow is by comparing the velocity between the emission and absorption lines ( ionized gas vs. stellar component). The velocity difference gives a good approximation of outflow velocity. It is the case for example of M82, where the gas is expelled perpendicularly to the main axis.

Emission line profiles could also be used to detect gas in-out-fall. Figure \ref{profile_lines} shows a [OIII]$\lambda$5007\AA\, line with a non Gaussian profile which can be due to feedback processes. The line can be fitted by a two Gaussian profile: the first Gaussian profile arises from the bulk of the emission region and it has the same dispersion than the other recombination lines. The second Gaussian function has lower flux and larger velocity dispersion. The center of the line is shifted to lower velocity in this case. This component may arise from gas expelled toward our direction. However, gas outflow are not the only process leading to non Gaussian profile. Indeed, in integrated spectroscopy a line is compounded by the mix of lines arising from different regions of the galaxy which can have different velocities and brightnesses due, for example, to mergers or interactions.

\section{Electron density and temperature}
\label{Electron density and temperature}
The physical state of HII regions can be defined by three observable properties: the electron density $N_{e}$, the electron temperature $T_{e}$ and the ion abundances. These three parameters drive the characteristics of the emission spectrum of HII regions. They can be estimated using lines ratios and equations which combine empirical relations and atomic physics. The diagnostic presented below have been designed for HII region studies. These results have to be interpreted carefully. In integrated spectroscopy of distant galaxies, the emission spectrum arises from several hundreds of HII regions having different physical states. The $N_{e}$ and $T_{e}$ derived are not the average values among all HII regions of a galaxy. The estimation is biased by the brightest HII regions of the galaxy. However, these observables can give some interesting clues on the physical processes taking place in galaxies. For example, in \citet{2006A&A...455..131P} the authors have studied the distribution of the $N_e$ in distant galaxies using VLT/GIRAFFE observations. Combining the $N_e$ maps with high spatial resolution images from HST/ACS, they have identified two mechanisms that can locally increase the $N_e$: (1) intense star-formation episodes in high density clouds; (2) presence of gas inflow/outflow events that produce collisions between molecular clouds of the ISM and the infall/outfall gas. 
 
\subsection{Electron temperature}
The electron temperature does not give fundamental information about the physical mechanisms occurring in galaxies, but it is a key variable for deriving the exact values of physical properties such as the abundance of species in the ISM. According to \citet{1989agna.book.....O}, the typical value for the electron temperature of an HII region with near solar abundance is $\sim$10\,000K.  This value is a good approximation and can be assumed when calculating extinction with the Balmer decrement method or electron density.

\begin{figure}[!h]
\centering
\includegraphics[width=0.6\textwidth]{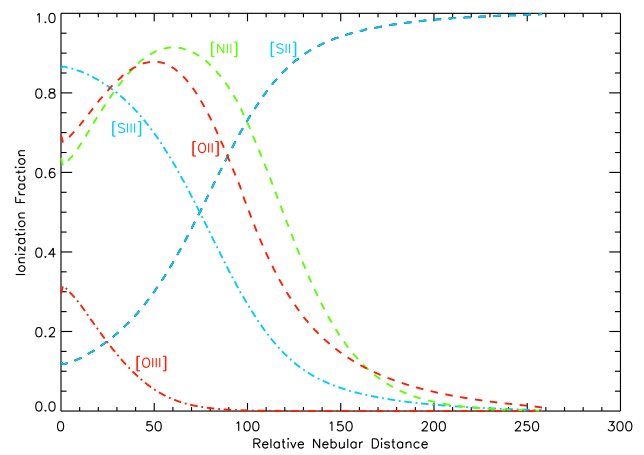}
\caption{{\small Relative ionization fractions for emission lines produced by Mappings III, plotted as a function of relative distance from the inner surface of the nebula. The emission features shown here are [NII] (dashed green line), [OII] (dashed red line), [OIII] (dash-dotted red line), [SII] (dashed blue line), and [SIII] (dash-dotted blue line). From \citet{2010AJ....139..712L}}}
\label{HII_structure}
\end{figure}

For more $T_e$-dependent parameters, like the abundance of ions, a more accurate calculation of the temperature is required. $T_e$ can be determined from the ratio of emission lines from sequential stages of ionization of a single element.  At visible wavelengths the ions having such configuration are [O III] $\lambda\lambda4959,5007/\lambda4363$\AA\, and [N II] $\lambda\lambda6548,6583/\lambda5755$\AA. The energy-level diagrams of these two ions are shown in figure \ref{E_diag_Te}. The temperature is then calculated using the stsdas/temdem IRAF task by giving a $N_e$ and the flux ratio. It is possible to use other ions as temperature diagnostic, like [OII] and [SII], see table \ref{table_Te}. When comparing temperature from different diagnostics, it is important to take into account that the various ionic emissions are not produced in the same regions of the nebulae. The HII regions have an "onions skin" structure where the ionization drops off radially from the center to the outer regions. The electronic and temperature density follow this structures. Low-ionization species, like  $O^0$,$S^+$, $O^+$ are produced in the outer and colder regions of the nebulae while higher-ionization species, e.g.$O^{+2}$, are produced in inner and hotter regions, see Fig.\ref{HII_structure}.  Nevertheless, highly-ionized species can be found in outer regions of nebulae. Indeed, at high distance from the nebulae center, electron temperature increases due to the very low $N_e$ of the ISM. 

\begin{table}[h!]
\begin{center}
\begin{tabular}{cccc}
 Ion&Spectrum&Line-Ratio \AA&Zone\\
 \hline
 $O^0$&[OI]&$\lambda\lambda6300,6363/\lambda5577$&Low\\
 $S^+$&[SII]&$\lambda\lambda6716,6731/\lambda\lambda4068,4076$&Low\\
 $O^+$&[OII]&$\lambda\lambda3726,3729/\lambda\lambda7320,7330$&Low\\
 $N^+$&[NII]&$\lambda\lambda6548,6583/\lambda,5755$&Low\\
 $O^{+2}$&[OIII]&$\lambda\lambda4959,5007/\lambda4363$&Med\\
\end{tabular}
\end{center}
\caption{{\small electron temperature diagnostic from Shaw \& Dufour 1994. }}
\label{table_Te}
\end{table}%

In the case of distant or faint objects, the determination of $T_e$ becomes almost impossible because auroral lines, such as [OIII]$\lambda$4363\AA\, or [NII]$\lambda$5755\AA, are too faint. 
\subsection{Electron density}
\begin{figure}[!h]
\centering
\includegraphics[width=0.6\textwidth]{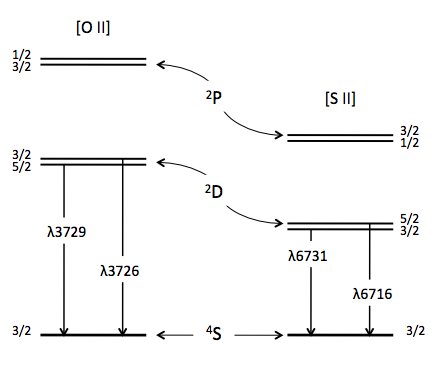}
\caption{{\small Energy-level diagrams of the $2p^3$ ground configuration of [OII] and $3p^3$ ground configuration of [SII], from \citet{1989agna.book.....O}.}}
\label{Ne_diag_E}
\end{figure}

The density can be estimated from two transitions arising from two levels in a same term (almost the same energy level) but with different radiative transition probabilities. The ratio of these levels depends mainly on the electron density, via the collisional de-excitation term, and in a second order on $T_{e}$. There are several ions in the optical region which have this kind of structure. The most used are [SII]$\lambda6716/6731$\AA , [OII]$\lambda3726,3729$\AA\, and [NI]$\lambda5200/5198$\AA. The energy-level diagram of [OII] and [SII] are shown in Fig. \ref{Ne_diag_E}. Other ions can be used as $N_{e}$ diagnostic, e.g. [Cl III], [Ar IV], C III] and [Ne IV] but contrarily to the previous ions whose lines are produced in the inner part of the nebulae, these diagnostic ratios trace the temperature of the outer parts of HII regions. The densities calculated from two diagnostic line ratios formed in different places of the nebulae can give different values. 

\begin{figure}[!h]
\centering
\includegraphics[width=0.8\textwidth]{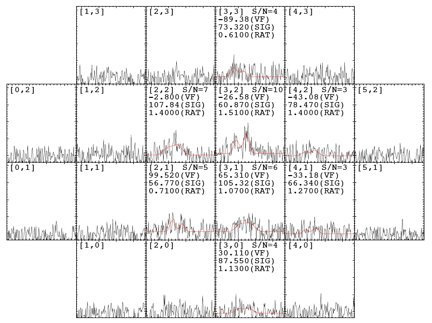}
\caption{{\small \citet{2006A&A...455..131P} have derived $N_{e}$ maps for 6 IMAGES galaxies using GIRAFFE (R=10\,000). }}
\label{Ne_Puech}
\end{figure}

At low redshift, the ratio [SII] is usually used because it is easily detected in the red part of the visible spectra. At higher redshift, the [SII] falls in the near infrared and the [OII] diagnostic is commonly used. The [OII] ratio has the advantage to be formed by two intense lines in star-forming galaxy spectra and to fall in the visible domain for redshifts between $0.4<z<1.0$. However, the lines of the [OII] doublet are very close and high spectral resolution observations are required to resolve the two lines, see Fig. \ref{Ne_Puech}.

The simplest way to estimate the line ratio is measuring the lines fluxes by fitting the two lines of the doublet with a double gaussian profile.  A separation between the two gaussians of 2.783 and equal dispersion values ($\sigma_1=\sigma_2$)  are assumed for the [OII] doublet, see \citet{2006A&A...455..131P}.The $N_e$ can then be computed from the measured doublet ratio using the n-levels atom calculations of the \emph{stsdas/Temdem} IRAF task \citep{1994ASPC...61..327S}. The task computes $N_e$ given $T_e$ and the ratio of a line doublet. The determination of $N_e$ have three main sources of error: the error on the measurement of the ratio; the saturation of the ratio as a function of $Ne$ at low densities ($N_e \leq 10\, cm^{-3}$), as illustrated in Fig.\ref{Ne_Puech}; and the dust extinction which can hide a local density peak. 
  
\section{Metallicity}
\label{Metallicity}
\subsection{Direct method}
Many light elements are observable in the optical spectra of galaxies, including H, He, N, O and Ne. The ionic abundance can be determined from the ratio of ionic lines to a hydrogen recombination line. The abundance of Oxygen relative to Hydrogen, defined by log(O/H), is usually taken to measure the metallicity in the galaxy gas. Oxygen represents about half of the metals of the ISM and have the advantage to emit very intense lines at visible wavelengths.

\begin{figure}[!h]
\centering
\includegraphics[width=0.6\textwidth]{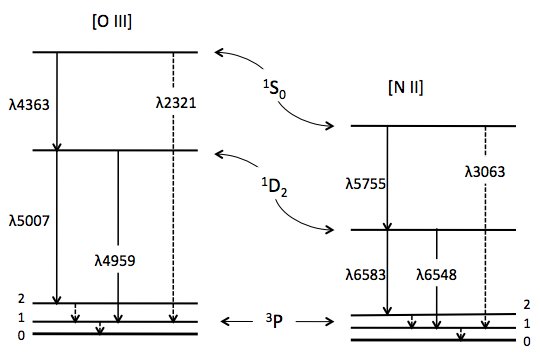}
\caption{{\small Energy-level diagrams of the $2p^2$ ground configuration of [OIII] and [NII], from \citet{1989agna.book.....O}. The strength of the $\lambda$4363\AA line is proportional to the population of the $^1S_0$ level which is proportional to the electron temperature. Therefore the ratio [O III] $\lambda\lambda4959,5007/\lambda4363$\AA\, decreases with increasing $T_e$.}}
\label{E_diag_Te}
\end{figure}

The best method to derive the metallicity is the \textit{$T_e$-directed method}, in which the abundance is directly inferred from the electron temperature \citep{1989agna.book.....O}. The temperature can be calculated from many line ratios as illustrated in table \ref{table_Te}. The most commonly used one is the [O III] $\lambda\lambda4959,5007/\lambda4363$\AA\, ratio. The intensity of [O III] $\lambda\lambda4959,5007$ decreases with $T_e$ while the $\lambda$4363\AA\, line arising from a higher energy level increases (see the energy-level diagram of [OIII] in Fig.\ref{E_diag_Te}). 
This diagnostic allows us  to calculate the abundance of the $O^{++}$ ion and then to infer the O/H metallicities after correcting for unseen stages of ionization \citep{1992MNRAS.255..325P,1992AJ....103.1330G,1993ApJ...411..655S}.The final O/H ratio involves the assumption that:
\begin{equation}
\frac{O}{H}=\frac{O^+}{H}+\frac{O^{++}}{H}. 
\end{equation}
The $T_e$ and derived abundance can be calculated using the stsdas/temdem IRAF task. Unfortunately, it is not possible to apply this diagnostic in all the metallicity range. Indeed, at high metallicities the collisional lines of metals cool the gas in the far infrared and induce a decrease of the $\lambda$4363\AA\, line that becomes immeasurably weak. Other ratios of lines can be used to measure $T_e$ in metal rich galaxies such as the [OII]\,$\lambda\lambda3726,3729-\lambda\lambda7320,7330$\AA\, ratio. In \citet{2007A&A...473..411L}, we have successfully measured direct temperature using this ratio in a sample of metal-rich galaxies from the SDSS, see paper in annex. 

\subsection{Strong line method }
In the majority of cases, the metallicity cannot be measured from the direct method in galaxies because the auroral lines, such as [OIII]$\lambda$4363\AA\, and [NII]$\lambda$5755\AA\,, are too faint. To overcome this difficulty, calibrations between ratios of strong emission lines and the metallicity have been implemented. There is a large number of strong-line parameters in the literature and the selection of the "best" strong-line method is still in debate in the community. All parameters have their advantages and drawbacks: dependence on other variables such as the ionization parameter, the spectral window of the required lines, the use of 'not so strong' lines as $Ar_3O_3$ and $Ne_3O_2$ lines. 

\subsubsection{$R_{23}$ parameter}
The most commonly used strong-line ratio, is the $R_{23}$ defined by \citet{1979MNRAS.189...95P} as:
 \begin{equation}
R_{23}=\frac{[OII]\lambda3727 + [OIII]\lambda\lambda4959,5007}{H\beta}.
\end{equation}
Extensive studies have been performed to calibrate the $R_{23}$ parameter to the gas metallicity. The calibrations have been derived using two independent approaches: an empirical method based on the comparison of the values of $R_{23}$ (or another metallicity parameter) to the metallicity inferred by the $T_e$-direct method \citep{1999ApJ...511..118K, 2001AA...369..594P,2004MNRAS.348L..59P, 2006A&A...459...85N, 2006ApJ...652..257L} and a theoretical method based on photo-ionization models \citep{1994ApJ...426..135M, 2004ApJ...613..898T} also the review from \citet{2002ApJS..142...35K}. The theoretical ratio of the lines at a given metallicity is calculated by photoionization models such as MAPPINGS \citep{1993ApJS...88..253S,2004ApJS..153....9G} or CLOUDY \citep{1998PASP..110..761F}. The choice of the calibration is crucial and depends on the scientific goals and the physical properties of the galaxies studied.  Indeed, abundances determined by different indicators give substantial biases and discrepancies. For example, the difference between calibrations based on electron temperature and photo-ionization model can reach 0.5\,dex \citep{2007A&A...473..411L,2008ApJ...674..172R,2008arXiv0801.1849K}. I will discuss in detail these issues in the next subsection.

\begin{figure}[!h]
\centering
\includegraphics[width=0.45\textwidth]{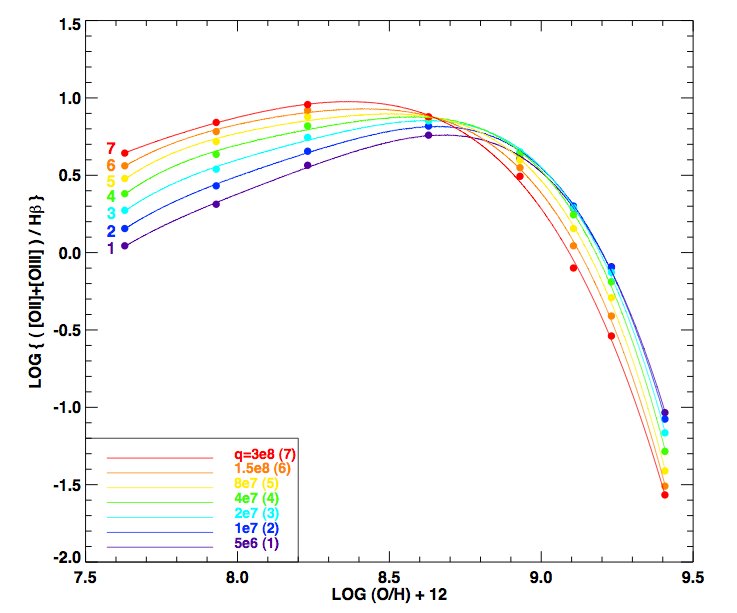}
\includegraphics[width=0.45\textwidth]{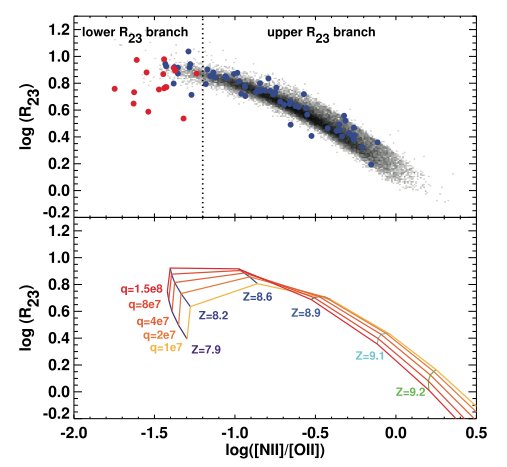}
\caption{{\small Left: The $R_{23}$ diagnostic for abundance versus metallicity from the photo-ionization model MAPPINGS. Curves for each ionization parameter between q = $5\times10^6$ to $3\times10^8$ cm/s are shown. The $R_{23}$ is strongly affected by the ionization parameter in the low metallicity regime. Right: The $N_2O_2$ diagnostic for abundance versus metallicity from the photo-ionization model MAPPINGS. Curves are the same that in the left panel. The $N_2O_2$ varies monotonically with metallicity and it is thus an ideal diagnostic for metallicity or to select the valid branch in the $R_{23}$ calibration. However it can be affected by the extinction and it is difficult to measure in high-z galaxies because the [NII]$\lambda$6584 line falls outside the spectral band. From \citet{2002ApJS..142...35K}. }}
\label{R_23_photoionization}
\end{figure}

Several authors have pointed out that the $R_{23}$ parameter is not an ideal metallicity diagnostic. Indeed, the parameter is affected by three strong drawbacks. Firstly, $R_{23}$ is double valued with $12+log(O/H)$ (Figure \ref{R_23_photoionization}). At low metallicity $R_{23}$ scales with metal abundance, but at higher metallicity ($12+log(O/H)\sim8.3$) gas cooling occurs through metallic lines and $R_{23}$ decreases. However, the degeneracy can be easily broken. The solution branch of the $R_{23}$ vs log(O/H) can be determined by measuring an initial guess of the metallicity with another line ratio diagnostic, such as:
\begin{itemize}
\item The $O_3N_2$ ratio defined as [OIII]$\lambda$5007/[NII]$\lambda$6584 by \citet{1979A&AS...37..361A};
\item The $N_2$ ratio: [NII]$\lambda$6584/$H\alpha$ by \citet{1994ApJ...429..572S};
\item The $N_2O_2$ ratio: [NII]$\lambda$6584/[OII]$\lambda$3727 \citet{2000ApJ...542..224D};
\item The $O_3O_2$ ratio: [OIII]$\lambda$5007/[OII]$\lambda$3727\citet{2006A&A...459...85N}.
\end{itemize}
The majority of the diagnostics use the [NII]$\lambda$6584 line, which has the problem of being very faint in metal poor galaxies and to rapidly fall outside the visible spectral band in high-z galaxy observations. \citet{2006A&A...459...85N} have suggested to use the $O_3O_2$ to break the degeneracy of the $R_{23}$ parameter when the $N_2$ diagnostic cannot be used.

 The second drawback of using the $R_{23}$ parameter is its dependence on the ionization parameter\footnote{The ionization parameter, $U$, is the maximum velocity of an ionization front that can be driven by the local radiation field. }. Figure \ref{R_23_photoionization} illustrates the dependence of the $R_{23}$ parameter on the metallicity and on the ionization parameter. The effect is only substantial in the low metallicity regime  log(O/H)<8.5. Moreover, \citet{1994ApJ...426..135M} and later \citet{2001AA...369..594P} have refined the calibrations to account for the ionization parameter. 
 
 The last but not least inconvenient of the $R_{23}$ parameter is its sensitivity to dust. Due to the large wavelength separation between the [OII] and [OIII] lines, the effect of dust is far from being negligible. The use of this parameter for measuring the metallicity requires a good estimation of the dust extinction.

\subsubsection{$N_2$ parameter}
The $N_2$ ratio have been intensively used in the literature as a metallicity estimator in high redshift studies, 1<z<3 \citep{2004ApJ...612..108S,2005ApJ...635.1006S, 2006ApJ...644..813E, 2007A&A...462..535Y}. This parameter has the advantage to link two lines closely spaced in wavelength, $H\alpha$ and [NII]$\lambda$6584\AA\,, hence it is not affected by extinction. It can be observed in small spectral windows such as the atmospheric bands in the near infrared. 

$N_2$ is not a direct estimator of the oxygen abundance since oxygen lines are not used in the method. The ratio is affected by the oxygen abundance by two factors. Firstly, $N_2$ depends mainly on the ionization parameter. \citet{2006ApJ...647..244D} have shown that the ionization parameter diminishes while metallicity increases and thus that it is indirectly correlated to the metallicity. Secondly, the N/O ratio depends directly on the O/H abundance at high metallicities (log(O/H)+12$\sim$8.3) when the secondary production of nitrogen is effective \citep{2002ApJS..142...35K, 2006ApJ...652..257L}. However, the correlation $O/H$ vs $N/O$ suffers from large scatter due to the variety of star formation history in galaxies. Indeed, as nitrogen is mainly synthesized in intermediate mass stars, its abundance is very sensitive to the star-formation history. The fact that the ratio is an indirect measurement of the oxygen abundance and its dependence on star formation history have led some authors to claim that this diagnostic is not a reliable metallicity estimator \citep{2006ApJ...652..257L,2006A&A...459...85N,2002ApJS..142...35K,2005A&A...434..507S,2007arXiv0704.0348S}. However, it is perfectly suited as a first guess metallicity for selecting the branch of the $R_{23}$ diagnostic to be used. 

\subsubsection{Remarks on the use of strong-line calibration}
\label{Remarks_Z}
The strong-line methods are based on a statistical approach. They are reliable metallicity estimators only if the calibration sample has the same properties than the object to study. From the SDSS data, metallicity calibrations based on the integrated light of star-forming galaxies have been implemented \citep{2006A&A...459...85N, 2006ApJ...652..257L, 2007A&A...462..535Y} and can thus be applied to other extragalactic samples. However, these conversions may not be valid for high-z galaxies. \citet{2005ApJ...635.1006S} and later \citet{2008ApJ...678..758L} have found evidence that HII regions in $z>1$ galaxies have different physical properties than their local counterparts: harder ionizing spectrum, higher ionization parameter, and can be more affected by shocks and AGN. Such differences can be explained by the fact that high-z galaxies have smaller sizes and higher star formation rates than local galaxies. In such a case the metallicity calibrations based on local samples are not reliable any more and thus neither is the conversion between calibrations. Applying them to high-z studies induce biases on the metallicity estimations. Unfortunately, there are no high-z sample with high S/N spectra and multiple spectral features  large enough to calibrate the high-z sample. \citet{2008ApJ...678..758L} have suggested that at z$\sim$1 the metallicities of galaxies are overestimated by $\sim$0.16\,dex. The best tools to probe if a galaxy has different HII region physical conditions than those of the calibration sample are the diagnostic diagrams described in section \ref{Contamination by active galactic nuclei radiation}. Indeed, different locations in the diagnostic diagram reveal differences in the ionization parameter or hardness of the ionizing radiation \citep{1997ApJ...481...49H,1996MNRAS.281..847T}. \citet{2008ApJ...678..758L} have found that $z\sim1$ galaxies are offset from the local star-forming galaxies in the log([OII]]$\,\lambda$3727/$\mathrm{H_{\beta}}$) vs log([OIII] $\lambda\lambda$4959, 5007/$\mathrm{H_{\beta}}$ diagram. They lie in the upper part of the star-forming wing, revealing their high ionization parameters.

\subsection{Is there an absolute metallicity scale?}
\begin{figure}[p]
\centering
\includegraphics[width=0.60\textwidth]{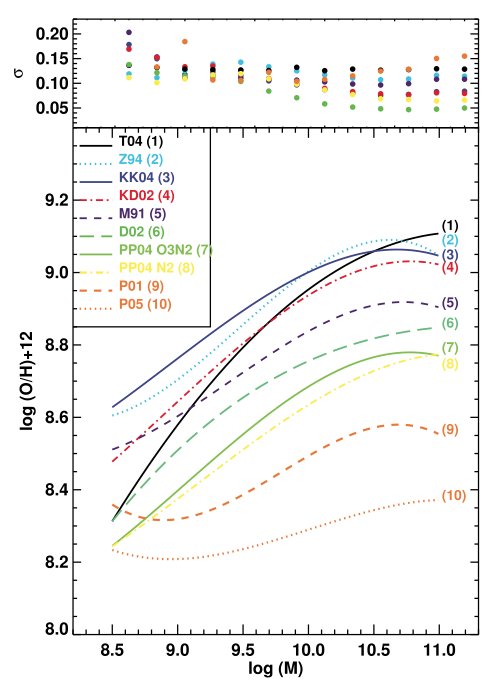}
\caption{{\small Robust best-fit M-Z relations calculated using the different strong-line metallicity calibrations. The top panel shows the rms scatter in metallicity about the best-fit relation for each calibration in 0.1\,dex bins of stellar mass. The y-axis offset, shape, and scatter of the M-Z relation differ substantially, depending on which metallicity calibration is used. From Kewley \& Ellison 2008.}}
\label{Kewley_MZ}
\end{figure}

The determination of the metallicity with different diagnostics, as strong-line and $T_{e}$, give very discrepant results. For example, the calibrations associated with the $R_{23}$ give a difference in metallicities reaching $\sim$0.5 dex, see figure \ref{Kewley_MZ} What is the method giving the best value of the metallicity? Direct methods give nearly the absolute value of the metallicity in a HII region in low and high metallicity environments \footnote{Some authors however argue that the direct method can be affected by systematics in high metallicity regions \citep{2005A&A...434..507S,2002RMxAC..12...62S,2006astro.ph..8410B}.}. The determination of metallicities in metal rich environment (12+log(O/H)>8.6) by direct method is usually not possible in extragalactic studies because the needed lines are too faint. There are few strong-line calibrations calibrated with direct method in metal rich galaxy \citet{2007A&A...473..411L}. For now, it is difficult to point one strong-line calibration as the better approximation of the real mean metallicity in metal rich environments. 

\begin{figure}[!h]
\centering
\includegraphics[width=0.5\textwidth]{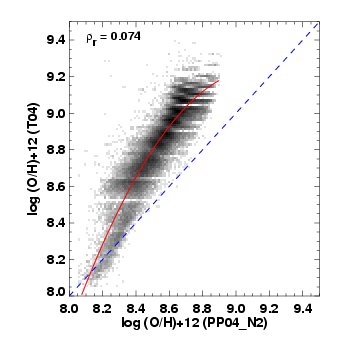}
\caption{{\small Comparison between different metalicities derived by $N_2$ \citet{2004MNRAS.348L..59P} and $R23$ \citet{2004ApJ...613..898T} for the SDSS DR4. The line represents the polynomial fits between the two metallicity-metallicity diagnostics that allows an easy conversion between the two estimators. From \citet{2008arXiv0801.1849K}. }}
\label{N2vsR23}
\end{figure}

Fortunately, despite the lack of a method able to measure an absolute metallicity in extragalactic studies, it is possible to measure with good accuracy relative metallicities. Indeed,\citet{2008arXiv0801.1849K} have shown that the relative metallicities between galaxies can be reliably estimated within $\sim$0.15\,dex when using the same metallicity calibration. In order to compare the metallicities of galaxies measured from different calibrations, several authors have implemented conversions between metallicity calibrations. \citet{2006ApJ...652..257L} and \citet{2008arXiv0801.1849K}  have built conversions between calibrations using the SDSS sample, as the conversion between the $R_{23}$ calibration of \citet{2004ApJ...613..898T} and those of $N_2$ from \citet{2004MNRAS.348L..59P}, illustrated in Fig. \ref{N2vsR23}. See http://www.ifa.hawaii.edu/$\sim$kewley/Metallicity/  and \citet{2008arXiv0801.1849K} for comparison between the metallicity calibrations. 

\section{Star formation rate}

The star formation rate, SFR, represents the mass of stars formed from the gas in a time unit and is thus defined as: 
\begin{equation}
SFR(t)=\frac{dM_{stars}}{dt}=-\frac{dM_{gas}}{dt}.
\end{equation}
By definition, the SFR is directly proportional to the luminosity emitted by the young stars. Hence, the main tracers of the SFR are the UV light, the hydrogen recombination lines produced by the ionizing radiation from very massive stars and the infrared light from the reprocessed UV light by dust grains. The calibration of the luminosity of these tracers to the SFR are done with semi-empirical and photoionization models. The SFR calibrations are thus dependent on the main ingredients of models: the initial mass function (hereafter IMF). The IMF gives the relative distribution of stellar masses in a stellar population formed at the same time from a gas cloud. When comparing SFRs from different calibrations it is crucial to verify if the calibrations used the same IMF. In this study, I have only used the set of calibrations proposed by \citet{1998ARA&A..36..189K} which are all based on the same IMF from \citet{1955ApJ...121..161S}.  

\subsection{SFR from the luminosity of recombination lines}
The Balmer recombination lines are produced by the ionizing radiation produced by the most massive and youngest stars of a galaxy, type O and B stars. Their luminosity is proportional to the number of massive stars and thus to the SFR. Only stars with masses $>10\,M_\odot$ and lifetimes $<20\,Myr$ contribute significantly to the integrated ionizing flux, therefore the SFR estimated by recombination lines gives the instantaneous SFR ($\tau\sim10\,Myr$). The luminosity of $H\alpha$, the most luminous of the hydrogen recombination lines in the optical, is calibrated to the SFR by photoionization models. The calibration of \citet{1998ARA&A..36..189K} gives:
\begin{equation}
SFR_{H\alpha}(M_\odot yr^{-1})=7.9\times10^{-42}L_{H\alpha}(erg\,s^{-1})
\end{equation}
The $H\alpha$ luminosity is calculated from the dust-corrected line flux by: 
\begin{equation}
L_{H \alpha}(erg\,s^{-1})=4\times\pi(3.086\times10^{24}D_L)^2F_{H\alpha}
\end{equation}
where $D_L$ is the luminosity distance in Mpc,  $F_{H\beta}$ is the flux in $H\alpha$ line corrected from extinction in $erg\,s^{-1}\,cm^{-2}$ and the luminosity of the line is $L_{H\beta}$. The $SFR_{H\alpha}$ is very sensitive to uncertainties on the aperture correction and on the extinction correction. 

At high-z, the $H\alpha$ line is redshifted outside the optical window. In such case, the SFR can be estimated from the higher order Balmer lines such as $H\beta$ line or $H\gamma$. The $H\alpha$ flux is estimated from the $H\beta$ flux by applying the theoretical ratio between $H\alpha$ and $H\beta$, see section \ref{Balmer decrement}. 
 
 \subsection{SFR from the ultraviolet luminosity}
In the wavelength interval 912-2500\AA\, the galaxy spectrum is dominated by the powerfull radiation\footnote{The radiation emitted by these stars has high energy but not enough to ionize the surrounding gas. The ionizing radiation, under 912\AA\,, emitted by the most massive stars is reprocess in part in visible light into the Balmer recombination lines, seen by the $SFR_{H\alpha}$} emitted by young stars with typical masses of $M\sim5\,M_\odot$. The SFR(UV) is directly proportional to the UV continuum luminosity and the calibration between the two quantities can be derived using synthesis models. 
The calibration of \citet{1998ARA&A..36..189K} gives:
\begin{equation}
SFR_{UV} (M_\odot yr^{-1})=L_{\nu}(UV)\times1.4\times10^{-28} (ergs\,s^{-1}\, Hz^{-1}).
\label{SFR_UV_kennicutt}
\end{equation}
The SFR(UV) traces the recent SFR, more exactly the integrated SFR over the last 100\,Myr. For example, it is not possible to disentangle between a system which has been forming at constant level of 1$\,M_\odot /yr$ over the last 100\,Myr and one which has formed $4\times10^8M_{\odot}$ in an instantaneous burst 50\,Myr ago \citep{2008ASPC..390..121C}. 

The UV tracer is in particular well-suited for z>0.8 galaxies, for which only the rest-frame UV light is visible in the optical window. Unfortunately, this estimator has a big drawback: it is extremely sensibility to the extinction. A moderate extinction of $A_V$=1\,mag in the V band implies an extinction $A_{1500A}\sim$3\,mag in the UV, and the exact value depends on the geometric distribution of the dust in the galaxy \citep{1998ARA&A..36..189K,2008ASPC..390..121C}. The UV luminosity can be corrected from extinction using the UV slope \citep{1994ApJ...429..582C} or by extrapolating the extinction measured in the optical and infrared domain (see Part II, section 5.3.2). However, the bias due to extinction can still be important after correcting with one of these two diagnostics. The former method can be affected by the age-dust degeneracy \citep{2008ASPC..390..121C,2009A&A...507..693B}. On the other hand the estimation from the optical lines is uncertain due to the extrapolation and the assumption of a selected extinction law (see the large differences between the several extinction laws at UV wavelength in Figure 4.7). The issue of extinction is dramatic in the case of distant galaxies where objects are expected to have high SFR and dust content. A large part of the SFR can be obscured by dust and thus not be taken into account by the UV calibration.

\subsection{SFR from the infrared luminosity}
\begin{figure}[!h]
\centering
\includegraphics[width=0.95\textwidth]{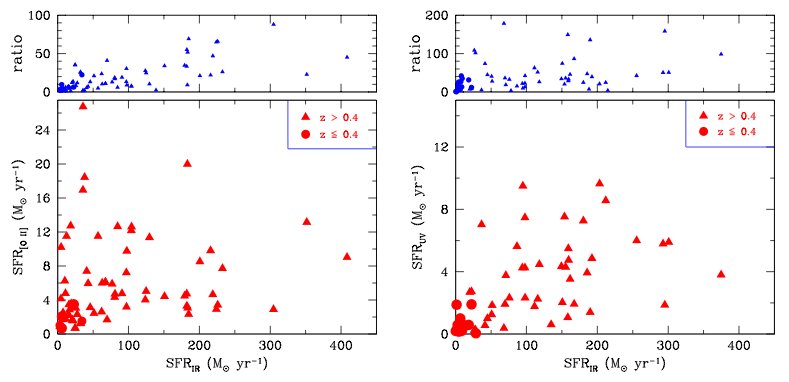}
\caption{{\small Comparison of SFRs calculated using the formalism of \citet{1998ARA&A..36..189K}. (Left panel): [OII]3727 estimates are compared to IR estimates for a sample of 70 galaxies from the CFRS and Marano fields (see Liang et al. 2004b). Upper left panel provides the $SFR_{IR}$/$SFR_{[OII]}$ ratio. Median values of the ratio are 5 and 22 for starbursts (SFR < 20 $M\odot yr^{-1}$) and LIRGs respectively. (Right panel): same comparison for 2800\AA\, UV luminosity estimated for 61 CFRS galaxies. Median values of the SFR(IR)/SFR(UV) ratio are 13 and 36 for starbursts and LIRGs, respectively. From \citet{2005A&A...430..115H}.}}
\label{SFROII_SFRUVvsSFRIR}
\end{figure}

Part of the UV light is absorbed by dust. The absorbed light heats the dust grains which re-emit the light in the mid- and far-infrared domain. As the absorption cross section of the dust peaks in the UV, the IR luminosity is a good tracer of young stars. The SFR(IR) is thus well-suited for dusty and high-SFR system such as LIRGs and starburst galaxies. Like  the IR luminosity, the SFR(UV) is a proxy of the recent star formation ($\sim$100\,Myr). The calibration of \citet{1998ARA&A..36..189K} gives:
\begin{equation}
SFR_{IR}(M_\odot \,yr^{-1})=1.71\times10^{-10}L_{IR}(L_\odot )
\label{SFR_IRkennicutt}
\end{equation}
The SFR(IR) traces the star formation obscured by dust. At intermediate redshift $z\sim0.6$ the fraction of star-formation obscured by dust is not negligible since $\sim$50\% of the integrated optical/UV emission from stars is reprocessed by dust in the infrared \citep{2001ApJ...556..562C}. The right panel of figure \ref{SFROII_SFRUVvsSFRIR} shows the SFR(IR) versus the SFR(UV) for a sample of intermediate galaxies from Hammer et al. (2005). The SFR is underestimated by a factor 2 when using only the UV tracer compared to the IR tracer.   

\begin{figure}[!h]
\centering
\includegraphics[width=0.65\textwidth]{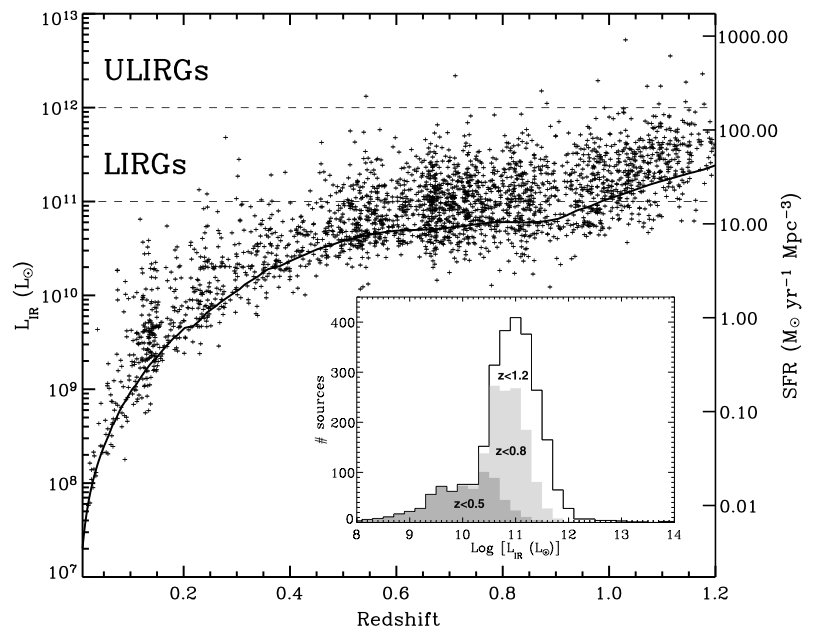}
\caption{{\small Total IR luminosities of the MIPS 24$\mu m$ sources identified with a redshift at 0<z<1.2 ( plus signs). Assuming the calibration from \citet{1998ARA&A..36..189K}, IR luminosities are translated into an 'IR-equivalent SFR' reported on the right vertical axis. The thick solid line indicates as a function of redshift the IR luminosity corresponding to an observed 24$\mu m$ flux of 0.08 mJy (80\% completeness limit of our survey). Inset: Corresponding IR luminosity histogram of the sample. From \cite{2005ApJ...632..169L}}}
\label{LeFloch2005}
\end{figure}

Until recently, the estimation of the dust-obscured star-formation was limited by the low spatial resolution and poor detection limit of the infrared instrument. The high sensibility of the IR spatial telescope Spitzer allows us now to detect small infrared fluxes and thus do not restrict the study to massive ULIRG or nearby galaxies anymore. For example, \citet{2005ApJ...632..169L} have constructed a catalogue of the $24\mu m$ sources of the CDFS field, with a 80\% completeness limit at $\sim$83 $\mu Jy$. They have shown that at z$\sim$0.3 (0.8) objects with SFR $\sim$1$(10)M_\odot\,yr^{-1}$ can be detected,  see Fig.\ref{LeFloch2005}. 

\subsection{Total SFR}
The total recent star formation can be estimated by the sum of $SFR_{IR}$ and $SFR_{UV}$. Indeed, SFR(UV) traces the UV light which has not been obscured by the dust and SFR(IR) trace the UV light thermally reprocessed by the dust. The sum of the two SFR estimators account for all the UV light emitted by the young stars.  

\subsection{SFR from the forbidden OII line}
In high-z galaxy spectra, the Balmer recombination lines fall outside the optical window or the high order Balmer lines ($H\gamma$, $H\beta$) are too faint to derive reliable SFR. \citet{1992ApJ...388..310K} proposed to use the forbidden line [OII]$\lambda$3727\AA\, to derive the SFR. This line has the advantage to be strong in galaxy spectra and to be detected in the optical range up to z$\sim$1.5. Forbidden lines are not directly proportional to the SFR like the recombination lines, their intensity depends on the metallicity and on the ionization parameter among other factors. \citet{1992ApJ...388..310K} claimed to have found a correlation between the EQW([OII]) and EQW($H_\alpha$) in a sample of local galaxies, making possible the implementation of a calibration based on the [OII] luminosity. This tracer has been intensely used in the literature and it is still frequently used to derived SFR in high-z surveys. However, \citet{1997ApJ...481...49H} and later \citet{2001ApJ...551..825J} have shown that the [OII] luminosity is not a good tracer of the SFR. Using a larger survey of intermediate and local galaxies, they found a large dispersion between the EQW([OII]) and EQW($H\alpha$) which invalidates the Kennicutt calibration. The [OII] tracer cumulates all the problems: (1) The [OII] line falls in the blue part of the spectrum and it is thus very sensitive to extinction. (2) The [OII] line intensity depends on the metallicity of the gas. (3) The line is sensitive to the ionization parameter. The left panel of Fig. \ref{SFROII_SFRUVvsSFRIR} shows the dramatic discrepancy between SFR([OII]) and SFR(IR). SFR([OII]) underestimates by several factors the SFR. It is not recommended to estimate SFR using this tracer.

\subsection{Extinction from the $SFR_{H\alpha}$/ $SFR_{IR}$ ratio}
\label{Extinction from the SFR}
Another method to evaluate the dust extinction is from the infrared and optical star formation rates. The SFR indicates the number of stellar masses newly formed and it is thus proportional to the amount of ionizing photons produced by the young massive stars, see subsection 4.3.6.  However, when measured in the optical, using the nebular emission e.g. the intensity of the $H\alpha$, part of the SFR is hidden by the dust. In fact, a large fraction of the ionizing radiation is absorbed by the dust and reprocessed into infrared light. \citet{2004A&A...415..885F} have shown that the optical $SFR_{H\alpha}$ and the $SFR_{IR}$ measured from the IR luminosity give compatible results when $SFR_{H\alpha}$ is correct from extinction and aperture. It is thus possible to estimate the amount of extinction from the ratio of the uncorrected extinction $SFR_{\mathrm{H\alpha}_{nc}}$ (but aperture corrected) and the $SFR_{\mathrm{IR}}$. 
Finally, the ratio has been corrected using the average interstellar extinction law:
\begin{equation}
A_{V}(\mathrm{IR})=\frac{3.1}{1.66}\log{\frac{SFR_{\mathrm{IR}}}{SFR_{\mathrm{H\alpha _{nc}}}}} \, ,
\end{equation}
The factor 1.66 is the value of the term $[f(\beta)-f(\infty)]-[f(\beta)-f(\alpha)]$ from the extinction law and a total-to-selective ratio Rv=3.1 has been assumed.

\section{Gas fraction}
\label{Gas fraction}
The mass of neutral and molecular gas can be measured using HI or CO observations. Unfortunately, the direct measurement of the HI gas mass is not possible above z$\sim$0.3 (Lah et al. 2007). In the future, the interferometric arrays ALMA and SKA will be able to observe the neutral and molecular gas of very distant galaxies. At present, the gas fraction of distant galaxies can only be indirectly estimated using the Kennicutt-Schmidt law \citep{2006ApJ...644..813E}.

\begin{figure}[!h]
\centering
\includegraphics[width=0.55\textwidth]{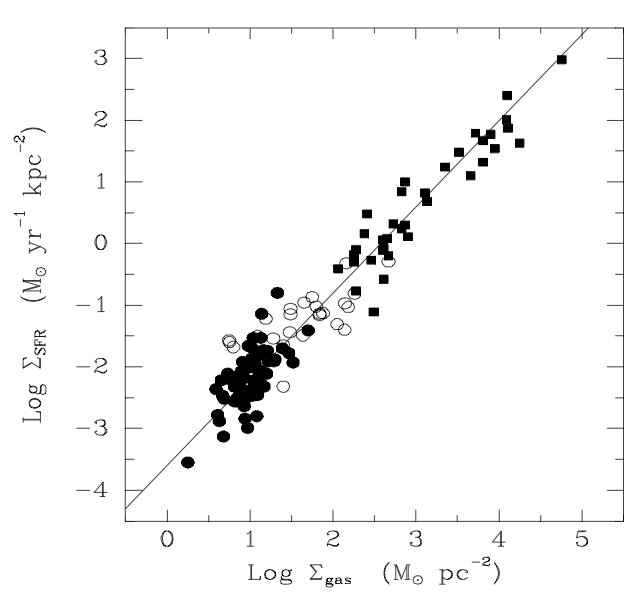}
\caption{{\small The global Kennicutt-Schmidt law in galaxies parameterized in terms of the SFR surface density and the total (atomic plus molecular) gas surface density. The filled circles are normal disks, squares symbols stand for starburst and open circles show the SFRs and gas densities for the center of normal disk galaxies. From \citet{1998ARA&A..36..189K}.}}
\label{KS_law}
\end{figure}

The Kennicutt-Schmidt law (K-S law) links the surface gas density and the star formation density \citep{1998ARA&A..36..189K}: 
\begin{equation}
\Sigma_{SFR} (M_{\odot}\,yr^{-1}\,kpc^{-2})=2.5\times10^{\,-4}\,\Sigma{gas}^{\,1.4}
\end{equation}
The K-S law is valid over a large range of densities, host galaxy properties and over more than 6 orders of magnitude in SFR, as illustrated in Fig. \ref{KS_law}.  The K-S law may not be the result of a unique physical process but produced by a combination of several processes occurring at different scales \citep{1998ARA&A..36..189K}. 

The Kennicutt-Schmidt law can be inverted to estimate the gas mass . The gas mass is proportional to the observed $SFR_{total}$ and the galaxy size $R_{gas}$ in $Kpc$. The method has been used successfully at different redshifts:  local galaxies \citet{2004ApJ...613..898T}, z$\sim$0.6 \citet{2010A&A...510A..68P}, z$\sim$2 \cite{2006ApJ...644..813E}, z$\sim$3 \citet{2009MNRAS.398.1915M}. From the K-S law and the definition of the SFR density: 
\begin{equation}
\Sigma_{SFR} (M_{\odot}\,yr^{-1}\,kpc^{-2})=SFR/(\pi\,R_{gas}^2) 
\end{equation}
The gas density $\Sigma_{gas}$ can be written as a function of the SFR and $R_{gas}$ : 
\begin{equation}
\Sigma_{gas} (M_{\odot}\,yr^{-1}\,kpc^{-2})=SFR^{\,0.714}\times R_{gas}^{\,-1.428}\times1.651\times10^8
\end{equation}
The total mass of gas is computed from $M_{gas}=\Sigma_{gas}\pi\times R_{gas}^{\,2}$ :
\begin{equation}
M_{gas}(M_{\odot})=5.188\,10^8\times SFR^{\,0.714} \times R_{gas}^{\,0.572}
\end{equation}

  \EmptyNewPage
    \FrameThisInToc
\chapter{How to derive the properties of the stellar population?}
\section{Introduction}
The integrated spectrum of a galaxy is a fossil record of its history. Indeed, the age and metal content of young and old stars leave their inprints in the observed colors and absorption features. Scheiner (1899) was probably the first to try to infer the stellar content of a galaxy from integrated spectra. He compared the spectrum of Andromeda to the solar spectrum and found many similarities. He concluded that M31 is composed of solar-type stars and must therefore be a distant stellar system.  Decades after Scheiner, Morgan (1968) and then Spinsad \& Taylor (1972) have re-introduced the problem. From then on, many teams have tried to decompose the stellar component of galaxies into simple stellar populations of different ages and metallicities. 
The study of the stellar content of galaxies is now a key tool to understand galaxy evolution, since it enables to recover the star formation history which provides insight into the crucial issue of stellar mass growth.

\section{Ingredients}
\begin{figure}[!h]
\centering
\includegraphics[width=0.65\textwidth]{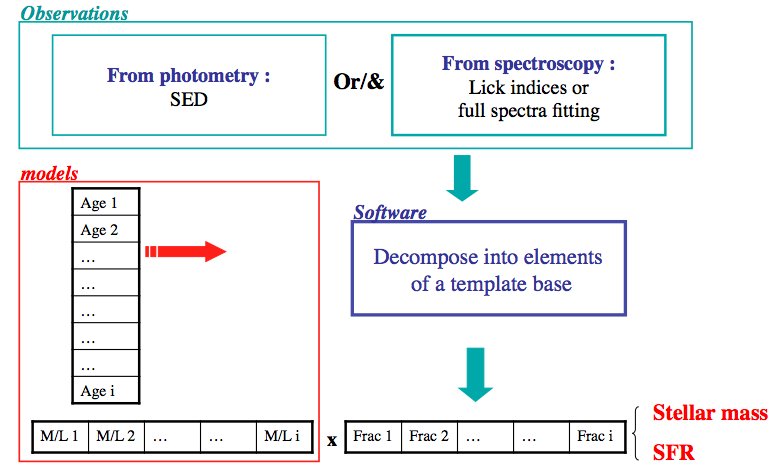}
\caption{{\small General method to retrieve stellar population.}}
\label{Stellar_Pop_problematic}
\end{figure}

The general method consists of fitting stellar features, such as specific absorption lines or all the spectrum, using stellar templates which can be empirical templates or semi-empirical models, see fig.\ref{Stellar_Pop_problematic}. 
\subsection{Base template}
Stellar population decomposition relies on the decomposition of the galaxy SED into a base of simple stellar templates. There are two kinds of templates: empirical templates based on the spectrum of real stellar populations and synthetic templates which are built from the combination of stellar libraries and stellar evolution models.   
\subsubsection{Empirical templates: "Real galaxies consist of real stars"}
In a first attempt, galaxy SEDs can be decomposed into their more fundamental units: stars. A large number of observational campaigns have been carried out to collect panchromatic and high quality star spectra covering the possible range of effective temperature, gravity and metallicity to build stellar librairies \citep{1984ApJS...56..257J,2003A&A...402..433L,2006MNRAS.371..703S}. Nevertheless, spectra of star clusters are usually favored for the study of composite stellar populations \citep{1986A&AS...66..171B,1995A&A...303..753S, 2002MNRAS.335..233P,2005ApJS..160..163S}. According to the scenario of star cluster formation, stars of a star cluster were born simultaneously from a same gas cloud. They are thus an ideal base element for describing a stellar content. The integrated spectra of star clusters are built from Nature's relative proportions of different stars born from a gas cloud at the corresponding metallicity Z. This approach is free of any assumption about the initial mass function (IMF) and details of stellar evolution. Moreover, the main advantage of this method over star libraries is to reduce the number of variables in the grid. Indeed, star cluster libraries are defined by two parameters - age and metallicity - contrarily to the three parameters needed for stellar libraries. The main limitation of empirical templates is that they do not span the all age-Z space homogeneously.


\subsubsection{Evolutionary population synthesis}
Stellar population synthesis computes the spectral energy distribution emitted by specific populations of stars. The first approach of the method has been introduced by \citet{1961MNRAS.122...27C} and then \citet{1978ApJ...222...14T}. The modeling of stellar population synthesis spectra has considerably evolved and it is now possible to reproduce the spectra of globular and star clusters \citep{2008MNRAS.385.1998K,2010MNRAS.403..797G}.

Stellar population synthesis models follow the time evolution of an entire stellar system by combining libraries of evolutionary tracks (isochrones) and stellar spectra with prescriptions for the stellar initial mass function (IMF), star formation and chemical histories. Populations of stars are generated from a stellar initial mass function \citep{1955ApJ...121..161S,2003PASP..115..763C, 2001MNRAS.322..231K}, which gives the distribution of the number of stars per unit of stellar mass formed from the same gas cloud, and a star formation history. The most commonly used parametrization for the star-formation history are: \textit{Single Stellar Population} (SSP), which corresponds to an instantaneous burst; \textit{Complex Stellar Population with exponential decay SFH} (CSP), parametrized by $SFR=exp(-\tau/t)$; \textit{Complex Stellar Population with multiple bursts}, in which the SFH is parametrized by an initial burst with long exponential decay SFR and a younger secondary burst. Then, the evolution of each star of the created stellar population is computed in terms of effective temperature ($T_{eff}$) and luminosity using evolutionary tracks in the Hertzsprung-Russel diagram, called isochrones \citep{1993A&AS...97..851A,1993A&AS..100..647B}. The final step consists in associating to each combination of $T_{eff}$ and luminosity a star spectrum from a spectral library. Stellar libraries can be empirical, i.e. composed by spectra of nearby stars such as STELIB \citep{2003A&A...402..433L} and MILES \citep{2006MNRAS.371..703S}, or theoretical from stellar atmosphere models \citep{2007MNRAS.382..498C}. Since \citet{1993ApJ...405..538B}, who first introduced isochrone synthesis, a large number of teams have created evolutionary population synthesis using different stellar libraries and stellar evolution models, such as: \citet{2003MNRAS.344.1000B,2005MNRAS.362..799M,2009MNRAS.396..462K,2010MNRAS.404.1639V}.
 
Stellar population synthesis models have the advantage of being very versatile. They allow to generate homogenous grids of single stellar populations, with a wide range of age and metallicity, and whose SEDs are defined in a large range of wavelengths (e.g. 3200 to 9500 \AA\, in the case of BC03 models). Moreover, each template is normalized to $1M_\odot$ and it is thus easy to derive stellar masses, SFR and other quantities. However, stellar population synthesis models have important limitations : 
\begin{itemize}
\item Uncertainties on the stellar evolution models, in particular in the advanced stages such as thermally pulsating asymptotic giant branch stars \citep{2005MNRAS.362..799M,2006ApJ...652...85M}.
\item Uncertainties on the stellar libraries. The quality of models depends on the stellar library they use. On the first hand, empirical stellar libraries can be affected by mismatch in the flux calibrations. It is the case of \citet{2003MNRAS.344.1000B} models which have difficulties in reproducing real spectra in the $H\beta$ region due to bad flux calibrations in the stellar library STELIB. On the other hand, the coverage in age, metallicity and surface gravity of the stellar library is also a key parameter of model quality. Stellar population synthesis models have to rely in theoretical stellar atmosphere spectra to fill regions not covered by the empirical library, which add large uncertainties. \citet{2008MNRAS.385.1998K} have shown that \citet{2003MNRAS.344.1000B} have difficulties when recovering properly sub-solar stellar populations due to the limited coverage of its stellar library outside the solar metallicity region. 
\item An assumption of the IMF is needed since stellar population synthesis depends on the selected IMF. 
\end{itemize}
\citet{2009ApJ...699..486C,2010ApJ...708...58C,2010ApJ...712..833C} have made a first attempt to evaluate how these uncertainties propagate through stellar population synthesis modeling. \citet{2010arXiv1002.2013C} have evaluated the model uncertainties on the derivation of SFH by comparing six evolutionary population models.

\subsection{Observables and parameters}
\subsubsection{Absorption features}
\begin{figure}[!h]
\centering
\includegraphics[width=0.9\textwidth]{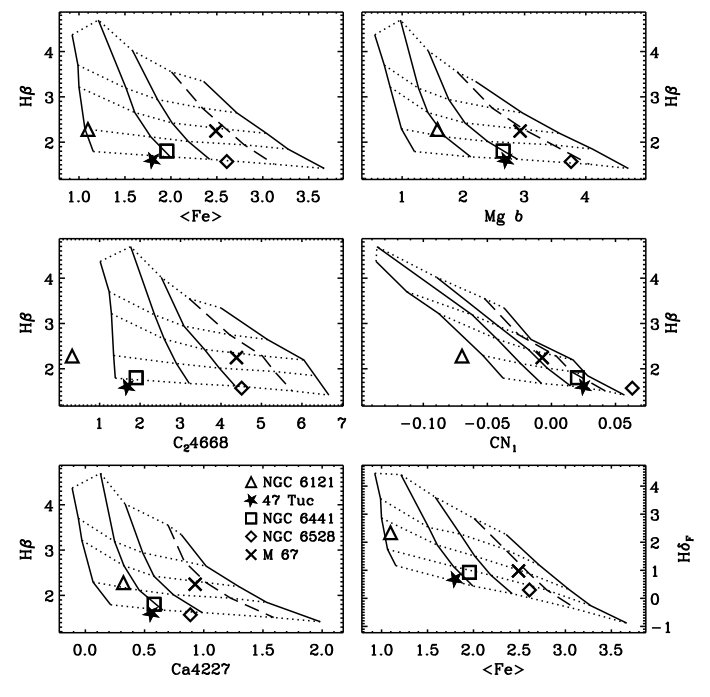}
\caption{{\small Example of estimation of age and metal abundance from Lick indices of cluster data. The location of the object in the grid gives a first approximation of the age and metal content. The grids are produced by the model of \citet{2007ApJS..171..146S}: solid lines show constant [Fe/H] from left to right of -1.3, -0.7, -0.4, 0.0, and +0.2; dotted lines show constant age from top to bottom of 1.2, 2.2, 3.5, 7.0, and 14.1 Gyr. From \citet{2008ApJS..177..446G}.}}
\label{lick_indices}
\end{figure}

The first studies of the stellar content based on spectra have only considered specific absorption features with high information content. These spectral indices are essentially defined by the equivalent width of absorption lines highly sensitive to the age and metallicity. The spectral indices of the Lick Observatory group defined from low resolution spectra \citep{1994ApJS...95..107W}  are still widely used in this kind of studies. The fit of these observables can be done with a grid of models with different star formation histories, e.g. \citet{2003MNRAS.341...33K, 2005MNRAS.362...41G,2004ApJ...613..898T}), or by the linear combination of empirical or synthetic templates, e.g: \citet{1988A&A...195...76B,1997MNRAS.284..365P, 2000A&A...357..850B,2008MNRAS.386..715T,2008ApJS..177..446G} (see figure \ref{lick_indices}) .

\subsubsection{Full-spectra fitting}
The improvement of the base templates in terms of spectral resolution and wavelength coverage now enables to model the full galaxy spectra. This method accounts for all the information contained in the spectrum: absorption features and the continuum shape. Full spectra methods fit flux-calibrated spectra by a linear combination of $N_\star$ simple stellar population templates, usually SSPs. The SED of the synthetic spectrum is defined by:
\begin{equation}
M_\lambda=M_{\lambda 0}\sum^{N_\star}_{j=1}x_j T^j_\lambda\,r_\lambda \otimes \, G(v_\star,\sigma_\star)
\end{equation}
where $M_{\lambda 0}$ is the scaling parameter, $T^j$ are the ($j=1, \ldots,N_\star$) single stellar population templates, $r_\lambda$ is the reddening term and $G(v_\star,\sigma_\star)$ is a gaussian convolution modeling the kinematics of stars within the galaxy ($v_\star$ is the velocity and $\sigma_\star$ is the dispersion of stars).
The contribution of each template $x_j$ ($j=1, \ldots,N_\star$) to the observed spectra is estimated by $\chi^2$ minimization:
\begin{equation}
\chi^2=\sum_\lambda  \frac{(O_\lambda-M_\lambda)^2}{\sigma_{O\lambda}^2} 
\end{equation}
where $O_\lambda$ and $\sigma_{O\lambda}$ are respectively the flux and flux uncertainty on the observed spectrum at wavelength $\lambda$. 

\begin{figure}[!h]
\centering
\includegraphics[width=0.73\textwidth]{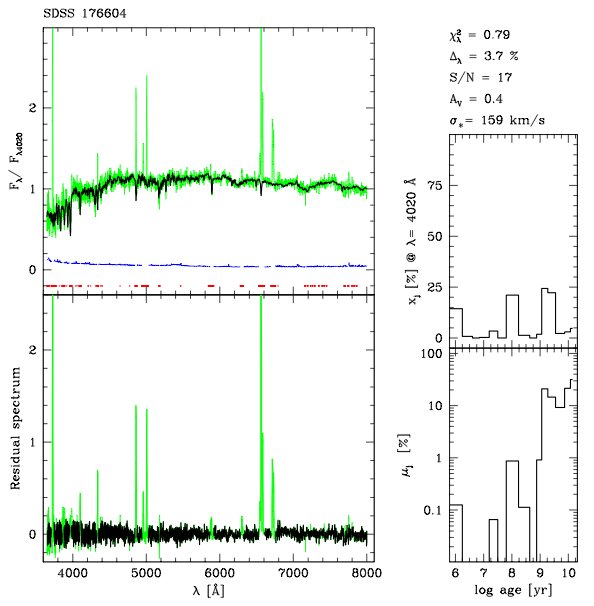}
\caption{{\small Example of full-spectra fitting of a late-type SDSS galaxies using STARLIGHT software. Top-left : the galaxy spectrum (black) has been modeled by a combination of 150 SSP from \citet{2003MNRAS.344.1000B} synthetic models (green). Bottom-left panel: The residual spectrum, i.e. observed - synthetic. Right: Flux (top) and mass (bottom) fractions as a function of the age. From \citet{2005MNRAS.358..363C}. }}
\label{Starlight_pop}
\end{figure}

The method has become popular in the literature as attested by the growing number of softwares performing full spectra fitting of an input spectrum by a vector of templates: \textit{MOPED }\citep{2000MNRAS.317..965H},\textit{PPXF} \citep{2004PASP..116..138C}, \textit{PLATEFIT} \citep{2004ApJ...613..898T,2006A&A...448..893L}, \textit{STARLIGHT} \citep{2005MNRAS.358..363C}, \textit{STECMAP} \citep{2006MNRAS.365...46O}, \textit{ULySS} \citep{2009A&A...501.1269K}, \textit{VESPA} \citep{2007MNRAS.381.1252T}. The method has been widely applied to the SDSS sample, see fig.\ref{Starlight_pop}, among many other:  \citet{2007MNRAS.381..263A,2009MNRAS.396L..71V, 2006MNRAS.370..721M,2009ApJS..185....1T,2009A&A...495..457C, 2010arXiv1004.4336Z}. They are based on different inversion methods and optimization algorithms, but according to the tests reported at IAU Symposium 241 (http://www.astro.rug.nl/~sctrager/challenge/), they all give comparable results (http://www.astro.rug.nl/$\sim$sctrager/challenge/). \citet{2008MNRAS.385.1998K} have validated the full spectrum fitting method and have shown that this approach gives results consistent with Lick indices but three times more precise. 

The use of a discrete SSP-based approach has the advantage not to impose any constraints on the allowed form of the star formation history. However, as full-spectra fitting methods use the information of the stellar continuum, the method is sensitive to flux-calibration mismatches of the observed spectrum and of the stellar templates used to build the synthetic models. 

\subsubsection{Broad-band photometry}
A rough approximation of the stellar mass, extinction, and age of stellar populations can be estimated from broad-band photometry. The galaxy SED is fitted over a large wavelength range to a grid of CSP with different ages, metallicities and SFH parametrization. Due to the small number of observational constrains this method suffers from important degeneracies. Moreover, \cite{2001ApJ...559..620P} have shown that near-IR photometry is mandatory to constrain the fraction of old stars and thus the stellar mass. I will discuss this method in detail in Part I, Chapter 5. 

\section{Limitation of astro-paleontology}
The proper uncovering of a galaxy's star formation history, from spectra or photometry, is limited by the measurement uncertainties, model mismatches and assumptions on the parametrization of the SFH. Moreover, it is also affected by the existence of multiple solutions. The degeneracy arises from intrinsic degeneracies of stellar populations and from algorithms. For instance, the well-known age-metallicity-extinction degeneracy stems from the stellar population itself. The age-Z degeneracy tends to confuse old, metal poor system with young and metal rich galaxies. At old ages, the effect of age reaches a magnitude similar to the effects of dust and metallicity and thus the problem of recovering SFH becomes degenerate. It is thus hazardous to recover SFH beyond 2-4\,Gyr. 

In non-parametric SFH methods, degeneracies can also come from an incorrect definition of the problem. Indeed, the SSPs used in this method can have very similar spectral characteristics but different ages and metallicities. If the elements of the base used in the optimization algorithm are not independent, i.e. that one element can be reproduced by the combination of other elements of the base, the problem turns degenerate. Recently, \citet{2009MNRAS.399.1044R} have investigated this issue and have optimized the elements of the grid, age and metallicity, using a diffusion map algorithm. \citet{2007MNRAS.381.1252T} have shown that degeneracies can be reduced by being less demanding on the detail of SFH. They have implemented a full-spectra fitting algorithm which adjusts the number of elements of the base according to the quality of the input spectrum.

\FrameThisInToc

\chapter{Studying distant galaxies}
\begin{figure}[p]
\centering
\includegraphics[width=0.60\textwidth]{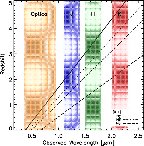}
\caption{{\small Diagram illustrating the redshifts at which H$\alpha$, H$\beta$, H$\gamma$, and [O II]$\lambda$3727 are observable from the ground in the optical (3500-9500\AA\,) and through the JHK near-infrared atmospheric windows. These lines are commonly used to derive physical properties of the ionized ISM in visible spectroscopy. The four lines can be simultaneously observed in the visible domain in galaxies with z<0.3. For galaxies with 0.3<z<0.8, the main properties of the ISM can be derived, as the extinction and metallicity, because the $[OII]$, $H\beta$ and $H\gamma$ still fall in the optical region. At higher redshifts, the lines are redshifted to near infrared wavelength in which only observation in atmospheric windows (J, H, K bands) are possible from ground-based telescopes. From Moustakas, Kennicutt and Tremonti 2006.}}
\label{Lines_red}
\end{figure}

In this chapter I discuss on some aspects of the methodology proper to the study of distant galaxies. Conversely to the local galaxies, distant galaxies are barely resolved, and their physical properties can be only derived for the whole system (section \ref{Interpreting integrated properties}). Moreover, the studies of distant galaxies are limited by strong observational constrains, such as the low signal-to-noise of spectra and the limited number of available lines due to the redshift (see figure \ref{Lines_red}). 

\section{Interpreting integrated properties}
\label{Interpreting integrated properties}

Despite the fact that the number of spatially-resolved observations of distant galaxies has been growing in the last 5 years\footnote{thanks to the new generation of integrated field spectrographs, such as VLT/GIRAFFE and VLT/SINFONI}, the majority of studies relied on integrated spectroscopy. Integrated spectroscopy, using fiber or slit, allow to derive the global properties of large number of distant galaxies thanks to multi-object spectroscopy. 
However, the interpretation of the derived properties is not trivial. Integrated spectra are constituted by a mix of a large number of emitting source hosted by the galaxy, e.g.: 
\begin{itemize}
\item The emission from hundreds of HII regions which can have very different physical states;
\item The combined light from all the stellar populations of the galaxy;
\item The radiation from active galactic nuclei;
\item Radiation produced by shock waves and the thermal emission of the gas. 
\end{itemize}
The role of the spectroscopist is to decompose the integrated spectrum into the individual contribution of its different sources and characterize the properties of each of them (see section \ref{Extracting the ionized gas}). 
\subsection{Robust determination of physical quantities}

The main difficulties in interpreting integrated spectra is due to the fact that different regions of a galaxy do not contribute with the same fraction to the integrated spectrum.  An integrated spectrum is dominated by the regions with higher surface brightness, such as the continuum contribution from the bulge and the emission lines emitted by star-forming regions. It results that the properties derived from global spectra do not represent the mean value of the all galaxy but the properties of these high surface brightness regions. As an example, abundances derived from R23=[O II]$\lambda$3727/H$\beta$ + [O III]$\lambda$4959,5007/H$\beta$ can either sample all the ionized gas spreading over all the disk, or only a single HII giant region in which most of the star formation activity is concentrated.
The gap between the global properties and the mean properties of the all galaxy is almost impossible to constrain since it require prior knowledge on their spatial distribution. To better control the bias on the derivation of properties from global properties, \citet{2005astro.ph..9907H} have proposed to evaluate the physical parameters of faint and distant galaxies with at least 2 independent measurements, see Table \ref{Robust_deter}. Each measurement can be affected by different bias which can be physical, such as the sensitivity of the method to the star formation history, or due to uncertainties on the measurement. The estimation of a quantity using two independent methods allows us to have a better control on the systematics.

\subsubsection{Star-formation}
Star formation rates are usually believed to be very uncertain. First factor of uncertainty is the supposed IMF: most tracers of SFRs are linked with massive, such as IR tracer, or very massive stars, like $H\alpha$ tracer. \citet{1998ARA&A..36..189K} has provided us an useful tool, including formula deriving SFRs from various indicators, all assuming the same IMF. It is also relevant when comparing SFR with stellar mass to consider a common IMF for similar reasons of consistency. For practical reasons, UV or [OII]$\lambda$3727 fluxes have been often used to characterize the SFR, since these emissions are redshifted to the visible window at moderate or high redshifts. Unfortunately, these tracers are very sensitive to metallicity and extinction. Indeed starbursts and LIRGs are so numerous at z> 0.4, that the only viable tracers are those which account for the light reprocessed by dust (IR), or those properly corrected for extinction effect such as H$\alpha$ after accounting for extinction using the $H\alpha$ /$H\beta$ ratio. For a given IMF, mid-IR and extinction corrected H$\alpha$ fluxes can trace the SFR within an accuracy much better than 0.3 dex in logarithmic scale \citep{2004A&A...415..885F}.
\subsubsection{Extinction}
The strong evolution of the Luminous Infra-Red Galaxies (LIRG) number density \citep{2005A&A...430..115H} evidences the need to proper account for extinction when estimating SFR or metal abundance using $R_{23}$. A difficulty is coming from the extinction calculation itself. At moderate redshifts, galaxies host significant populations of intermediate age stars responsible for the strong Balmer absorption lines (see e.g. Hammer et al, 1997), which can easily affect the measurements of Balmer emission lines. One needs a sufficient spectral resolution (R> 1000) and S/N (> 10) to properly correct the underlying absorption, especially for measurements of $H\gamma$ and $H\beta$ lines. Another difficulty is due to the fact that, at z >0.4 the $H\gamma$ line is redshifted to the near-IR, and the $H\alpha$/$H\beta$ ratio has to be estimated using 2 different instruments. A way to circumvent the difficulty (and the cost of near-IR spectroscopy), is to use the ratio $H\gamma$/$H\beta$, although the $H\gamma$ line is often faint. \citet{2004A&A...423..867L} have systematically compared extinction from $H\alpha$/$H\beta$ to that from the ratio of IR to $H\beta$ emission (see Chapter 1 section 1.7.6) and find a good correlation between the two derived extinction.
\subsubsection{Metallicity}
The most accurate method for derived metallicities is direct-$T_{e}$ method (see section )\footnote{ In the case of integrated spectra of galaxies, \citet{1999ApJ...511..118K} have shown that the $T_e$-method does not give the mean metallicity of the galaxy because the contribution of the different HII regions to the spectrum leads to an underestimation of the metallicity.}. However the lines required to derive the metallicities from direct-$T_{e}$ method are usually too faint in distant galaxies and strong-line calibrations are used.  
The strong line ratio $R_{23}$ is sensitive to the extinction through the ratio $[O II]\lambda3727/H\beta$. The metallicities derived with this parameter can be compared to those from the ratio [N II]/$H\alpha$. This ratio is used nitrogen rather than oxygen and can be affected by the production of primary nitrogen (see \citet{2004MNRAS.348L..59P}). However R23 and [N II]/$H\alpha$ seem to correlate well, at least for abundances above 1/4 of the solar values. [N II]/$H\alpha$  measurements can also help to resolve the degeneracy of the O/H-R23 relationship.

\begin{table}[!h]
\begin{center}
\label{Robust_deter}
\begin{tabular}{|l|c|c|c|}
\hline
Quantity&1st measurement&2nd measurement&notes\\
\hline
\hline
Gas extinction & Balmer line ratio & IR/visble ratio& error <0.3$dex$\\
&correct for absorption&aperture corrected& \\
SFR&mid-IR&$H\alpha$ or $H\beta$& error <0.3$dex$\\
&&extinction corrected&\\
Gas abundance& $R23$ extinction correct& [NII]/$H\alpha$&error $\sim$0.1$dex$\\
\hline
\end{tabular}
\caption{{\small Combination of measurements required for a robust determination of physical quantities of distant galaxies. From \citet{2005astro.ph..9907H}.}}
\end{center}
\end{table}

\subsection{Effect of the aperture in integrated spectroscopy}
On the other hand, the properties of the galaxies are not homogenous, e.g:
\begin{itemize}
\item The star formation is concentrated in clumps and filaments (rings, spiral arms, and more complex structures in irregular galaxies). 
\item The properties of the central bulge are very different than those of the disk in spiral galaxies: older stellar population, SFR, metallicities, colors, etc...
\item The gas metallicity has a radial gradient with the central regions having the higher metallicities. For example, late-type spirals have large metallicity gradients (one order of magnitude between the inner and outer disk). On contrary, bared-spirals and early-type galaxies have lower metallicity gradients, see review \citet{1999PASP..111..919H}. 
\item The distribution of the dust is not homogenous and is composed by patch of high density regions and dust lanes. The observed extinction depends thus on the line-of-sight, i.e. if the galaxy is seen edge-on or face-on. According to \citet{1989AJ.....97.1022K}, there is no evidence for a extinction gradient such as metallicity.    
\item The contamination by AGNs only affects the light rising from the nuclear regions.    
\end{itemize}
The radial gradients and variation of the properties of galaxies cause sever bias when comparing observation done with different aperture system which sample only a limited portion of the galaxy. As instance, observing a galaxy with a small projected aperture only probe the properties of the central regions while large aperture will sample the all galaxy. The bias on the derived physical properties depends on the aperture size and on the apparent dimension of the galaxy observed. At higher redshift, the effect of the aperture is less critical because the apparent size of galaxies diminishes and therefore almost all the galaxy is sampled by slit or fiber. It's the reason why it's mandatory to evaluate the bias due to apertures when comparing sample at different redshift.The effect of the aperture size on apparent galaxy properties have been intensively discussed in the literature since decades, e.g the pioneer work of \citet{1961ApJS....5..233D} and \citet{1963AJ.....68..237H}, see \citet{2005PASP..117..227K} for a review. 

\subsubsection{The SDSS 3" aperture fiber}
The case of the SDSS survey is certainly the most documented \citep{2004ApJ...613..898T, 2008arXiv0801.1849K, 2008arXiv0806.2410M, 2009A&A...505..529L,2004MNRAS.351.1151B}.  At z=0.07, the SDSS 3" aperture fiber only probes a central region of 4\,kpc. Depending on the galaxy size, about 60\% to 80\% of the light is lost by the fiber aperture \citep{2005PASP..117..227K}. The SDSS spectra are thus nuclear spectrum and the properties derived are not representative of the all galaxy, but do only characterize the central region. In \citet{2005PASP..117..227K}, the authors have investigated the effect of aperture size on the estimated SFR, metallicity and extinction. They have compared the properties derived from integrated spectra of the Nearby Field Galaxy Survey (NFGS) and those derived from the nuclear spectra of SDSS survey. They have concluded that a minimum covering fraction larger than 20\% is needed to avoid aperture effects. Above this covering fraction, the estimation of the SFR, $A_V$ and metallicity are poorly affected by the aperture bias, see table \ref{table_bias_sdss}. 

\begin{table}[h!]
\begin{center}
\begin{tabular}{|l | c | c|}
  \hline
 Properties & $f_{covering}< 20$\% &$f_{covering}>20$\%\\
 \hline
   \hline

  SFR&$\pm$0.70 &depend on the Hubble type:\\
 &&$\sim$15\% missed in late-type\\
  & & underestimation in early-type\\
  Extinction&$\pm1\,mag$& negligible \\
  Metallicity&$+0.14\,dex$& negligible  \\
  \hline
 \end{tabular}
\end{center}
\caption{{\small Bias and dispersion on the integrated properties derived from SDSS nuclear spectra compared to those derived from global properties. From \citet{2005PASP..117..227K} }}
\label{table_bias_sdss}
\end{table}%
\subsubsection{Effect of slit}
The effect of the aperture size is also valid for slit aperture. The bias due to the aperture does not depend only on the aperture size, but also on the length of the fiber and on the position of the slit on the object. To minimize aperture bias, the slit length has to be orientated along the principal axis of the galaxy. The slit aperture will affect the measured properties not exactly in the same way that fiber aperture. Indeed, a slit centered on a galaxy bulge will also collect light from the disk in it length. At the same aperture size (diameter for fiber and width for slit), the aperture bias in slit is thus expected to be smaller than in fibers.  

The new surveys of nearby galaxies with 3D spectroscopy will allow in the future to evaluate with more precision the effects of aperture (slit and size) on the derivation of the SFR, extinction, etc.. from integrated spectra (see preliminary results of \citet{2010MNRAS.tmp..461R}). It is one of the main goals of the project on SN Ia presented in Part III.

\section{Measuring emission lines} 
\subsection{Extracting the ionized gas component from the galaxy spectra} 
\label{Extracting the ionized gas}
\begin{figure}[!h]
\centering
\includegraphics[width=16cm]{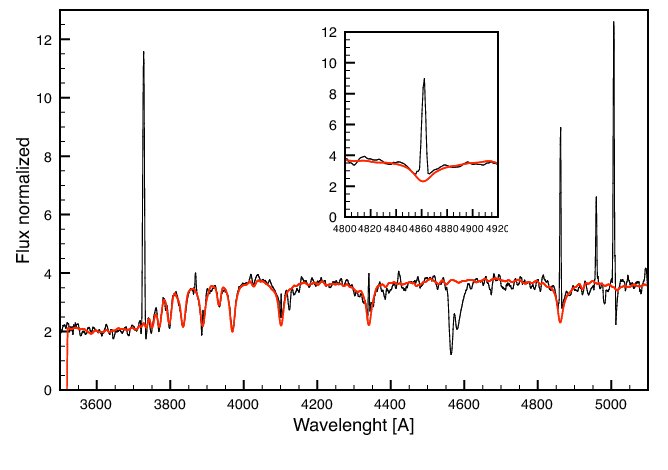}
\caption{ Spectrum of a distant galaxy from the IMAGES sample. The spectrum has been fitted by STARLIGHT with stellar library templates. The contribution from the stellar absorption is plotted in red line. The upper box is a zoom around the $H\beta$ line. The emission line and the underlying absorption are clearly visible in the $H\beta$ line. }
\label{Spectra_with_continum}
\end{figure}

The optical spectrum of a star-forming galaxy has two components: a stellar component - composed by the combination of the spectra from the stellar populations- and an emission spectra from the ionized gas. Before proceeding to the characterization of the gaseous phase from the optical spectra, it is necessary to uncouple the two components. The extraction of the stellar continuum from the gaseous phase is an essential step to measure emission lines. Indeed, the Balmer lines of a star-forming galaxy are composed by an emission lines with an underlying absorption line (See Fig. \ref{Spectra_with_continum}). The emission spectra from the gas is superimposed on the absorption spectra of stars. Not accounting for the underlying absorption when measuring the emission lines leads to important underestimation of their flux and thus induces a bias for all the properties derived \citep{2004A&A...417..905L}. There are two methods to correct the emission lines from the underlying absorption : one based on the equivalent width of absorption lines and another one based on the composition of the stellar population. 

\subsubsection{Equivalent width}
This method consists in only correcting for the underlying absorption of the Balmer lines of Hydrogen. For distant galaxies, the equivalent width of the Balmer absorption lines has often been estimated to be 2 \AA \, \citep[e.g.]{1996MNRAS.281..847T,1999ApJ...511..118K}. In order to have a first approximation of the flux of Balmer lines, a constant equivalent width of  2 \AA \, can be added to the measured uncorrected flux. This method has been widely used in the past because it can be used with low signal-to-noise and in low spectral resolution spectra. However, it can lead to large bias because the absorption in Balmer lines can be larger than 2\AA \,  in severals cases. Indeed, the equivalent width of hydrogen absorption lines depends on the age and metallicity of stellar population of each galaxies. \citet{1997ApJ...481...49H} have found equivalent widths for $H_\beta$ reaching 7-8 \AA \, in some distant galaxies. 

\subsubsection{Full spectra fitting}
A more efficient method consists in fitting a synthetic spectrum from a template library to the spectrum of each galaxy, see previous chapter. The emission lines of the spectrum are masked and the stellar component is fitted by a linear combination of spectra from stellar library or semi-analytic model templates. The main purpose of this method is to derived the composition of stellar population of galaxies from their optical spectra. However, in this stage of the spectrum analysis, we are only concerned to properly subtract the stellar component from the emission lines and the method can be used in a simple level. In the study of distant galaxies, described in Part I Chapter III, the best fit has been computed using the \textit{STARLIGHT} software\footnote{available for download at \textbf{http://www.starlight.ufsc.br/}.}  and a set of 15 stellar spectra from the \citet{1984ApJS...56..257J} stellar library, including B to M stars (e.g B, A, F, G, K and M) with stellar metallicity. A \citet{1989ApJ...345..245C} reddening law has been assumed during the fit. 
 
I give two remarks, one about the choice of the templates and the other one the matching of the spectral resolution. In relation to the choice of the model, the stellar evolutionary population models of \citet{2003MNRAS.344.1000B} are widely used as base templates in the literature. Their use in full-spectra-fitting method are motivated by two reasons: the completeness of the Z-age grid and the wide wavelength coverage, see previous Chapter. Despite these advantages, I have chosen to use stellar templates rather than this stellar evolutionary population models. This is because BC03 models overestimate the absorption in the $\mathrm{H\beta}$ line, which is dramatic for the purpose of correcting $H\beta$ from the absorption component! \citet{2007MNRAS.381..263A,2005MNRAS.358..363C} have suggested that the origin of this bias is due to flux-calibration mismatch of the STELIB library used in the BC03 models. The STELIB spectra have an excess of flux in both sides of the $\mathrm{H\beta}$  when compared to model spectra. From now, as the synthetic spectrum is not used to retrieve physical properties of the stellar population, I have adopted a simple stellar library in order to better control the parameters of the fit.  

My second remarks is about the matching of the spectral resolution between the templates and the observed spectrum. The majority of the software mentioned above matches directly in the best fitting function the spectral resolution of the spectra template and the input spectrum. The matching works reasonably well in high S/N spectra and when the difference between the two spectral resolution is not too large \citep{2005MNRAS.358..363C}. However, for S/N<10 it is highly recommended to adjust previously the template spectra to the observed spectra. Mismatch on the resolution induces bias in the fluxes of Balmer lines. 

\subsection{Measuring emission lines}
\begin{figure}[!h]
\centering
\includegraphics[width=0.80\textwidth]{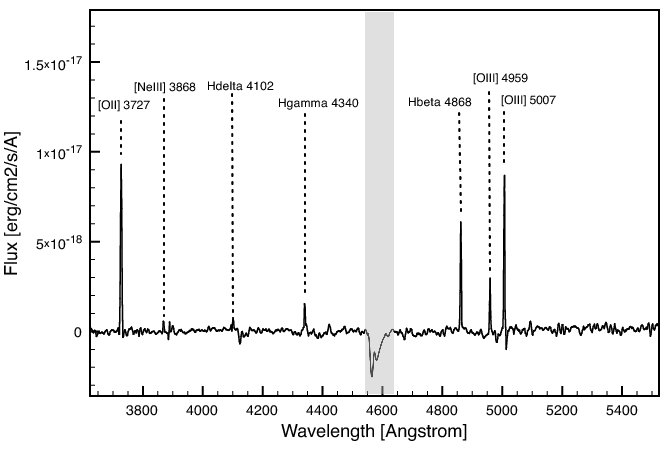}
\caption{{\small The same galaxy as the previous Fig, but only the spectrum from the gaseous phase is plotted. The stellar component has been subtracted using the method of full spectra fitting. The grey region corresponds to the position of sky absorption. The emission lines from gas can be directly measured using for example the IRAF package \textit{splot}}. The physical process which produce this kind of line profile will be discussed in the next section.}
\label{Spectra_without_continum}
\end{figure}

\begin{figure}[!h]
\centering
\includegraphics[width=0.50\textwidth]{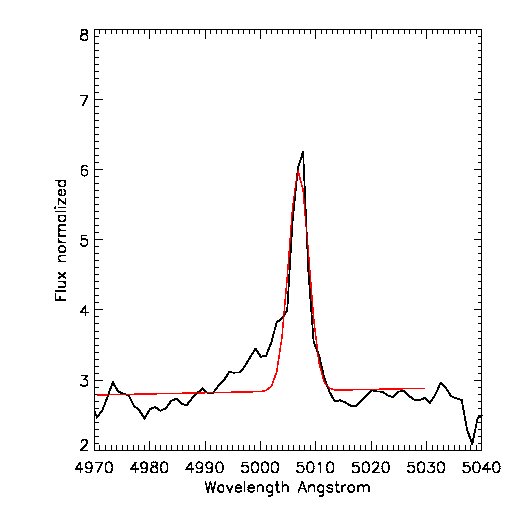}
\caption{{\small [OIII]$\lambda$5007\AA\, lines from a distant galaxy in the FORS2-IMAGES sample. The line cannot be fitted by a simple gaussian. To correctly fit the line, the function must be at least a two gaussians component profile.}}
\label{profile_lines}
\end{figure}

After subtracting the stellar component, the emission lines arising from the gaseous phase of galaxies can be measured using manual or automatic methods fitting. Figures \ref{Spectra_without_continum} shows the spectrum of a galaxy after the subtraction of the continuum. 
Theoretically, emission and absorption lines have gaussian profiles. The width of the line depends on several components: the natural width of the line, the dispersion of velocity of the different emitting regions and the spectral resolution of the instrument. In the case of distant galaxies, the lines can have profiles more complex than a simple gaussian distribution. On one hand, physical processes can lead to none gaussian profile. Indeed, in integrated spectroscopy a line is compounded by the mix of lines arising from different regions of the galaxy which can have different velocities and brightness. Fig.\ref{profile_lines} illustrates an example of lines without gaussian profile: the [OIII]$\lambda$5007\AA\, shows a complex profile which cannot be fitted by a simple gaussian function. In a second hand, the spectra of distant galaxies can be affected by low signal-to-noise. The Poisson noise can apparently deform the gaussian profile of the emission lines. 

I have used the SPLOT routine from the IRAF package to measure the flux of the lines individually. The task evaluates the flux by simply summing the pixels in the portion indicated by the user. A linear continuum is also evaluated from both sides of the line and subtracted in the region of the line. The task returns the line center, continuum at the region center, core intensity, flux above or below the continuum, and the equivalent width. I have chosen this manual method to have a better control on the flux measurement. Automatic methods, like the one used by \citet{2006A&A...448..893L}, can automatically fit all the emission lines and evaluate their flux. The lines are fitted simultaneously with severals gaussians centered in the lines position and with $\sigma$ of the same width. However, an important fraction of the flux can be lost in the case of non-gaussian lines if only the flux in the fitted gaussian is taken. \citet{2006A&A...448..893L} have tested the reliability of their method in a sample of 11 galaxies. They have compared the EQW measured from the automatic software to those measured by hand with IRAF. They found that their automatic software perfectly recovers the flux and EQW. However, this test has been done in a very small sample and a complete analysis with a larger sample may be required to definitively trust the measurements of automatic methods.  

\subsection{The signal-to-noise}
\subsubsection{Detection signal-to-noise}
\begin{figure}[!h]
\centering
\includegraphics[width=1.0\textwidth]{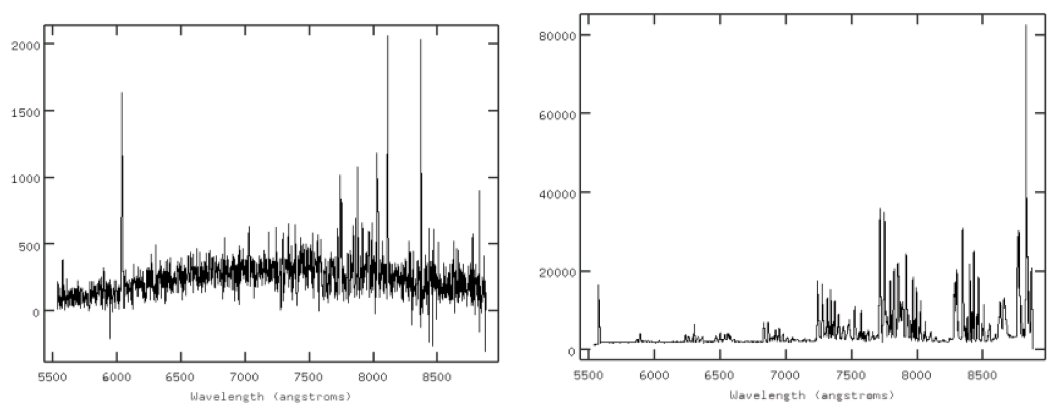}
\caption{{\small \textbf{Left}:1D Spectra of a galaxy after sky subtraction. \textbf{Right} : 1D spectrum of the sky spectrum in counts. The emission lines from the galaxy have the similar intensity than the sky continuum in the region without strong sky lines.}}
\label{SN_spectra}
\end{figure}

The signal-to-noise of galaxy spectrum is given by the ratio at a given wavelength: 
\begin{equation}
S/N=\frac{S_{obj}}{\sqrt{S_{sky}+S_{obj}}}
\end{equation}
In the regions with strong sky lines, the target spectrum can be affected by bad sky extraction, see Fig. \ref{SN_spectra}. In the case of gaussian noise, only regions with S/N> 3 are considered to be detected, since they have a probability superior to 95\% of detection. For other noise distribution, such as the Poisson noise, higher detection limit are commonly used, e.g. S/N>5 . Regions with S/N<3 are not proper to derive physical quantities due to the high uncertainty on their detection. 

\subsubsection{S/N of an emission lines}
The S/N of a emission line can be defined as the ratio between the total flux in the line $F_{line}$ and the noise in the pseudo-continuum $\sigma$ \citep{2006A&A...455..107F}. 
\begin{equation}
<S/N>=\frac{F_{line}}{\sqrt(n)\times\sigma}
\end{equation}
 The S/N is normalized by the number of spectral resolution elements $n$ in which the emission line is spread.

\section{What is the resolution needed to derive a physical quantity?}
\begin{figure}[!ht]
\centering
\includegraphics[width=0.50\textwidth]{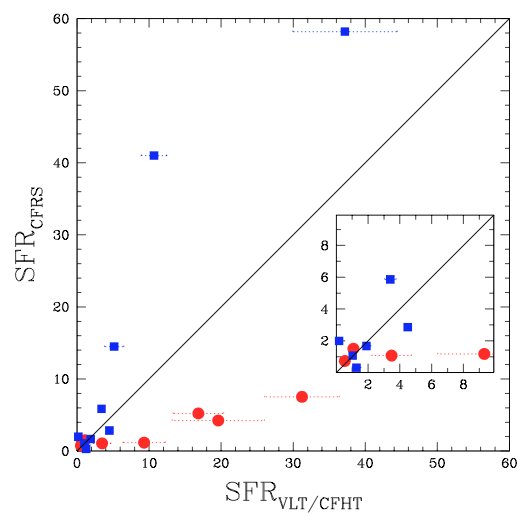}
\caption{{\small Comparison between the SFRs of the galaxies obtained from the low-resolution CFRS spectra (SFR CFRS) and higher quality spectra (SFR VLT/CFHT) studied in this work. The inset zoom the (0,0) region. The red-symbols are galaxies with only $H\alpha$ line visible and the blue symbols represent normal emission galaxies  From \citet{2004A&A...417..905L} }}
\label{Liang_Resolution}
\end{figure}

In several papers, our team has pointed out the necessity of having at least moderate spectral resolution (R>1000) and S/N (>10) to retrieve reliable integrated properties such as extinction, star formation rate and metallicities \citep{2004A&A...417..905L, 2005astro.ph..9907H, 2005A&A...430..115H}.
In \citet{2004A&A...417..905L}, the authors have compared the SFR and extinction derived from low spectral resolution observation from the CFRS ($\lambda\sim$40\AA\,) to moderate spectral resolution observation from the FORS2/VLT ($\lambda\sim$10\AA\,) in a representative sample of low redshift galaxies. They found that low spectral resolution observations induce large biases in the derivation of SFR and extinction, as illustrated in Fig \ref{Liang_Resolution}.

\begin{figure}[!h]
\centering
\includegraphics[width=0.6\textwidth]{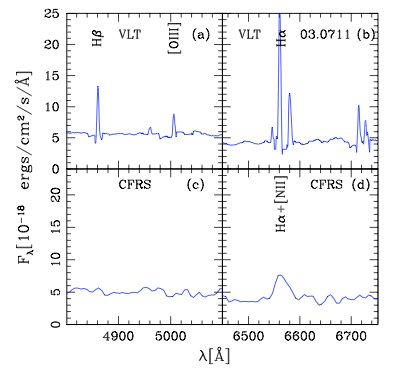}
\caption{{\small a), b) Rest-frame spectra of CFRS03.0711 from the VLT observation around $H\beta$ and $H\alpha$ wavelength positions; c), d) the corresponding low-resolution CFRS spectra. The $H\beta$ and [OIII] emission line lines are completely diluted in the low resolution spectra. At red wavelength the $H\alpha$ line is blended to the  NII $\lambda \lambda$ 6584,6548 lines. From \citet{2004A&A...417..905L}.}}
\label{line_dilu}
\end{figure}

These discrepancies are due to systematic errors in the measurement of the emission line fluxes in low resolution spectra:
\begin{itemize}
\item \textbf{Dilution of emission lines}.
The emission lines are hidden or diluted in the low resolution spectra. For example, faint $H\beta$ lines can be completely diluted by the underlying $H\beta$ absorption line, see Fig.\ref{line_dilu}. In such a case, it is not possible to calculate the extinction from the ratio of Balmer lines, see Chapter 1. In low resolution spectra, the evaluation of the underlying absorption of Balmer lines is difficult to measure. 
\item \textbf{Blended $H\alpha$ and NII emission lines}.
At very low spectral resolution, as in the case of the CFRS spectra, the $H\alpha$ and [NII]$\lambda \lambda$6548,6584 lines are blended, see Fig.\ref{line_dilu}. It is thus not possible to measure the flux of the $H\alpha$ line. Some authors, \citet{1998ApJ...495..691T} have proposed a calibration of the $H\alpha$ flux. Unfortunately, the ratio between $H\alpha$ and [NII]$\lambda \lambda$6584,6 depends on the metallicity and the ionization parameter, see section 4.3.4. The calibration may only work in samples of galaxies with the same properties as the calibration sample. 
\end{itemize}

However, \citet{2006A&A...448..893L} have argued that it is possible to retrieve reliable fluxes in low resolution data. They have downgraded moderate spectral resolution data to lower resolution,  R=250 and R=500 and found almost no dispersion or offset between the fluxes measured in the different spectra. This result is in contradiction with \citet{2004A&A...417..905L}. Below, I investigate the influence of the spectral resolution with a new galaxy sample. 

\subsubsection{New study comparing moderate and low spectral resolution spectroscopy}
\label{New study comparing moderate and low spectral resolution spectroscopy}
\begin{figure}[!h]
\centering
\includegraphics[width=0.70\textwidth]{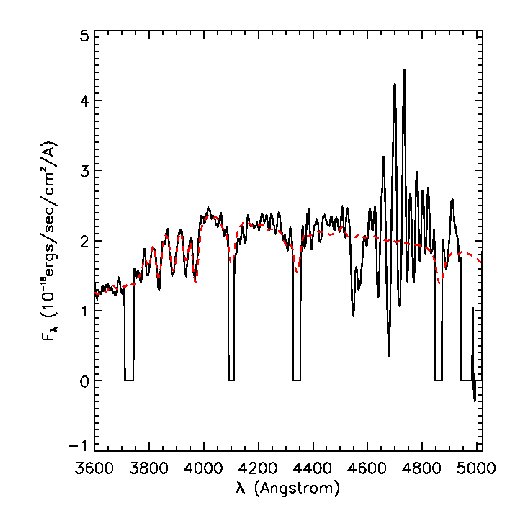}
\caption{{\small Medium R spectrum from GOODS-VIMOS at rest-frame. Emission lines have been masked (regions set to zero) and the continuum has been smoothed to enhance the absorption features. The fitted synthetic spectrum is shown in red-dashed line. The spectrum is very noisy above $\sim$ 4600\AA\, in rest-frame ($\sim$8000\AA\, at observed-frame) due to the emergence of strong sky lines.}}
\label{medium_R_fit}
\end{figure}

I have investigated the issue of the spectral resolution using a similar approach to \citet{2004A&A...417..905L}: extinction and metallicity are derived from medium and low resolution spectra and are compared to those derived from higher spectral resolution spectra. I have searched in the archives low spectral resolution spectra for each of the Sample\,A objects. Table \ref{table_resolution} shows the properties and source of the low resolution (LR), medium resolution (MR) and high resolution (HR) sample. For the medium spectral resolution spectra, the flux of [OII], $H\gamma$, $H\beta$ and the [OIII] lines have been measured following the same methodology as the one applied to the IMAGES-FORS2 sample, see Part I Chapter 3. After masking the emission lines, the continuum has been smoothed and adjusted to a linear combination of stellar population. The emission lines are measured in the continuum subtracted spectra. The adjustment of the continuum is reasonably good in the blue part of the spectra, but the fit quality decreases at redder wavelengths, see Figure \ref{medium_R_fit}. The decrease of the signal-to-noise at wavelengths above $\sim$8000\AA\, is due to the emergence of strong sky lines at this wavelengths. The quality of the sky extraction degrades with decreasing spectral resolution observation. 
\begin{table}[h!]
\begin{center}
\begin{tabular}{| l | c |c|c|}
  \hline
 Sample&Low R&Medium R&High R\\
 \hline
  \hline
 Spectral R&250&500&1200\\
 Number&33&6&33\\
 Survey&LR GOODS& MR GOODS&IMAGES\\
 &VVDS& &\\
 Instrument&VIMOS&VIMOS&FORS2\\
 Reference&Popesso \textit{et al. 2009}& Popesso \textit{et al. 2009}&This work\\
 &Le F\`evre \textit{et al. 2005}&&\\
  \hline
 \end{tabular}
\end{center}
\caption{{\small Description of the low, medium and high spectral resolution samples. }}
\label{table_resolution}
\end{table}%

For the low R spectra it was not possible to fit the continuum and extract reliably the underlying stellar absorption in Balmer lines. Like the medium R spectra, these spectra suffer from very low signal-to-noise levels in the red part of the spectra. The flux of [OIII] and $H\beta$ lines have thus large uncertainties due to the strong sky lines.
 
\begin{figure}[!h]
\centering
\includegraphics[width=0.45\textwidth]{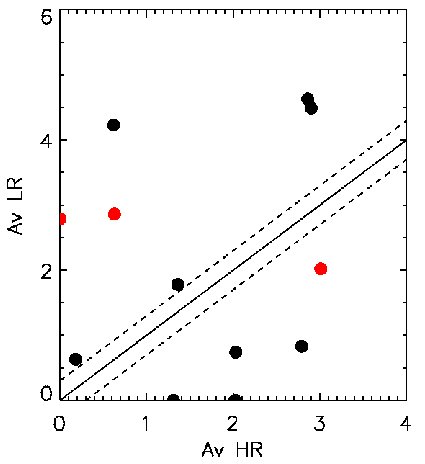}
\includegraphics[width=0.47\textwidth]{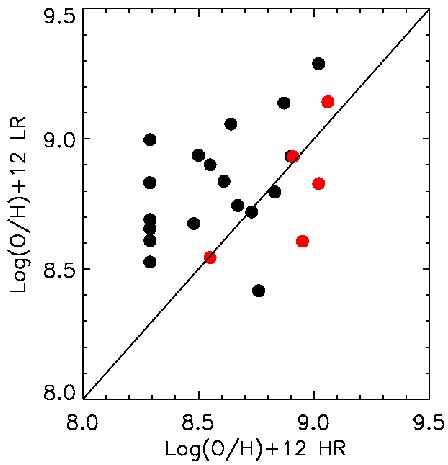}
\caption{{\small \textbf{Left}: Comparison of the extinction derived from FORS2 data ($A_V$ HR high resolution) and those measured in lower resolution data. The red symbol are the observation from medium R, R$\sim$500 and the black symbols stand for low resolution data R$\sim$250. The plain line is the identity.  \textbf{Right}: Metallicity derived from the FORS2 data (HR) versus those derived from lower resolution spectra (LR). The color symbol are the same as the previous figure. The metallicity have been estimated from the $R_{23}$ ratio. The emission lines have been corrected from the extinction. For the low and med-R, as it was not possible to measure directly the extinction from the spectrum, a constant $A_V$=1.57, which corresponds to the medium $A_V$ of the IMAGES sample from the high R spectra has been assumed. The metallicity derived from low resolution data is systematically overestimated compared to those measure on medium or high resolution spectra.}}
\label{Extinction_resolution}
\end{figure}

Figure \ref{Extinction_resolution} left panel compares the extinction estimated from the $H\gamma /H\beta$ ratio measured in the FORS2 spectra (HR R$\sim$1000) and in medium resolution spectra (red symbols) and low resolution spectra (black). There is no correlation between the two estimations. This study shows that it is not possible to measure reliable extinction from spectra with R<1000, as previously found by \citet{2004A&A...417..905L}. 
I have measured the metallicity from the $R_{23}$ parameter, defined as the flux ratio between [OII]$\lambda$3727+[OIII]$\lambda \lambda$4959,5007 over $H\beta$ (see section 4.3.4). The metallicities derived from the R$\sim$500 data are similar to those derived from the FORS2 spectra. The metallicities from the low resolution spectra are systematically overestimated, $\Delta(Log(O/H)+12)=+0.27\,dex$, see Fig.\ref{Extinction_resolution} right panel. 

This study confirms the previous result of \citet{2004A&A...417..905L}: the necessity of having moderate spectral resolution (R>1000) and S/N (>10) to retrieve reliable integrated properties. The discrepancy with \citet{2006A&A...448..893L} may arise from an error of methology in the \citet{2006A&A...448..893L} test. In fact, they have downgraded the resolution of sky subtracted spectra into lower resolution while \citet{2004A&A...417..905L} and this analysis use real low-R data. The downgraded sky subtracted spectra have better signal-to-noise than the real low R observation because they do not account for the issue of the sky subtraction in low resolution spectrum.

  \NumberThisInToc
\FrameThisInToc
\part{Fundamental integrated properties of  distant galaxies}
 \begin{verse}
Caminante son tus huellas  \\
el camino y nada mas; \\
caminante no hay camino\\
se hace camino al andar. 

al andar se hace camino, \\
y al volver la vista atras \\
se ve la senda que nunca\\
se ha de volver a pisar. 
 
Caminante no hay camino \\
sino estelas en la mar. 

\textit{Antonio Machado}
 \end{verse}

 \EmptyNewPage
\FrameThisInToc

\chapter{ IMAGES survey}
The studies on the evolution of the star formation rate \citep{1999ApJ...517..148F, 2001ApJ...556..562C} and of the stellar mass densities \citep[e.g.]{2001MNRAS.326..255C, 2004ApJ...608..742D} have revealed that in present-day galaxies, 30 to 50\% of the baryonic mass locked into stars has condensed at z < 1. This evolution is mostly associated with the evolution of intermediate mass galaxies, with stellar mass between $1.45 -15\times10^{10}\,M_\odot$ \citep{2005A&A...430..115H,2005ApJ...625...23B}. As disk galaxies represent 70\% of intermediate mass galaxies in the local Universe, intermediate mass galaxies at $z\sim0.6$ appear to be the likeliest progenitors of the present day spiral galaxies. 

The 8-10 meter class telescopes allow us to scrutinize the structure of distant galaxies in detail for the first time. The ESO large program IMAGES "Intermediate MAss Galaxy Evolution Sequences" (Principal Investigator F. Hammer) investigates the evolution of the main global properties from z$\sim$0.9 to the local Universe using a representative sample of local spiral progenitors at $z\sim0.6$. The IMAGES strategy aims at reconstituting the puzzle of galaxy evolution, linking by continuity in time the properties of local objects to those of  intermediate redshift ancestors. This is possible because the IMAGES sample has been selected in baryonic mass (paper in progress). In a closed system, mass is conserved as a function of time: even if galaxies are not closed systems in a strict sense, selecting galaxies in baryonic mass allows us to establish a first order causal link between galaxy populations at different epochs. A fully observational scenario of galaxy evolution, and ultimately of their formation, can then be constructed.

\section{The IMAGES survey}
\label{Philosophy_IMAGES}
\subsubsection{Philosophy of the survey}
The objective of IMAGES survey is to gather the maximum of observational constrain of a relatively small, but representative, sample of $z\sim0.6$ intermediate mass galaxies. The IMAGES survey distinguishes from the other intermediate redshift surveys by its selection method. The sample is small when compared with large surveys as GOODS or Combo17. These very large surveys aim to minimize statistical uncertainty (for 1000 galaxies, $\sqrt{N}/N$ is 1\%). However, they are limited by the inevitable use of automatic methods which can induce large bias, at an order of magnitude almost equal to that of the signal to be measured \citep{2006A&A...447..113L, 2008A&A...484..173P}. On contrary, the IMAGES team have chosen to define a smaller number of targets which allow a careful and detailed analysis including spatially-resolved kinematics.  The IMAGES sample certainly faces the small statistic issue (for 100 galaxies, $\sqrt{N}/N$ is 10\%) but this drawback can be compensate by an efficient control of the selection effect. 

\subsubsection{Sample selection}
A sample of $\sim$100 galaxies has been selected on the sole basis of their absolute magnitude in J band, $M_J< -20.3$ corresponding to a $M_{stellar}=1.5\times10^{10}M\odot$ and their redshift $0.4<z<0.9$. To avoid systematic from the cosmic variance the targets have been gathered in three cosmological fields. The representativity of the selected galaxies to the luminosity function at the given redshift has been verified by comparing their luminosity distribution with luminosity function derived by large surveys. 

\subsubsection{Observations}
 In the framework of IMAGES survey and GIRAFFE GTO time, about 100 galaxies have been observed with GIRAFFE an integral field spectrograph at the VLT, see Fig. \ref{GIRAFFE_IFU}. Each galaxy has been observed during 8 to 24 hours. The integral field spectroscopy allows us to derive the spatially resolved kinematic of their ionized gas. In addition, around 200 galaxies have been observed with long slit spectroscopy at moderate spectral resolution from FORS2/VLT, see details of the observation in the next chapter.  These new set of observations offer unprecedent constrains on the kinematics \& velocity field, gas chemistry and stellar population of intermediate redshift galaxies. 
\begin{figure}[!h]
\label{GIRAFFE_IFU}
\centering
\includegraphics[width=0.8\textwidth]{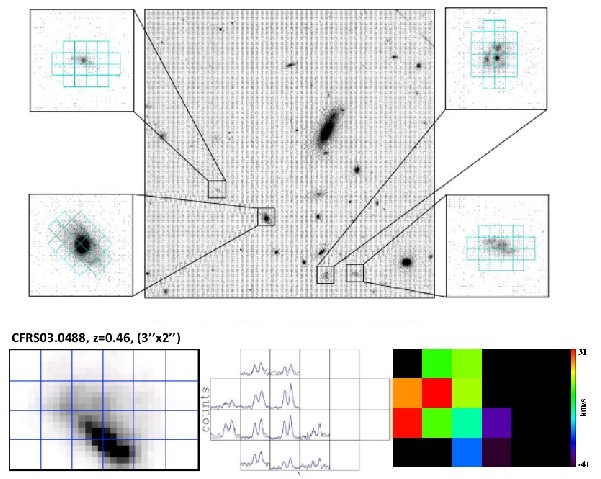}
\caption{{\small GIRAFFE is a multi-object integral field spectrograph able to observe 15 target over a 20 arcmin field-of-view (Upper panel). Each galaxy is observed in up to 20 small apertures or micro-lenses, each of them providing a spectrum at high resolution (R=13000), see bottom-right panel. In the illustration (Bottom-center panel), the OII doublet is well resolved, and each line can be used to trace the velocity and the dispersion of an individual galaxy region.  }}
\end{figure}

 \subsubsection{Complementary data from literature}
 The bulk of the VLT/GIRAFFE observations and all the FORS2 observations have been carried out in the cosmological field Chandra Deep Field South (CDF-S)\footnote{The Chandra Deep Field South (CDF-S) is one of the most depth observation of the X-ray satellite Chandra. It covers a field of view of 0.11 degrees square in the Formaz constellation, around the central point AR=3h32m28.0s, Dec=-27h48m30s (J2000)}.  The CDFS  is a widely surveyed region which has been observed with an unprecedented depth and coverage by ground-based and space telescopes, e.g. large surveys GEMS, COMBO-17, VVDS  and GOODS. The Great Observatories Origins Deep Survey (GOODS) is the principal survey in the field with a covering of 150 arcmin square. It combines imagery and spectroscopy observations, from Radio to X ray, from a wide number of instruments of NASA, ESO and ESA observatories \citep{2003mglh.conf..324D}, see summarize of the GOODS observations in table \ref{CDFS_field}. The ACS/HST data from GOODS provide details of few hundred of parsecs in IMAGES galaxies and a small region has a very deep ACS imagery, exposure time $\sim$100 hours per filter, the Ultra Deep Field. 

\begin{table}[!h]
\begin{center}
\begin{tabular}{|c|c|c|c|}
\hline
Instrument&Spectral domain &Observation& Reference \\
\hline
\hline
ACS@HST& U, B, V, I & Imagery&\citet{2004ApJ...600L..93G} \\
WFI@2.2m-ESO& U, B, V, R, I & Imagery& Arnouts et al. (2001) \\
ISAAC@VLT& J, H, K & Imagery& Retzlaff et al. (2010) \\
SOFI@NTT&J, H & Imagery& Vandame et al. (2001) \\
IRAC@Spitzer&3.6, 4.5, 5.8 and 8.0$\mu m$ & Imagery&Dickinson et al. in prep\\
MIPS@Spitzer&$24\mu m$& Imagery& Chary et al. (2004)\\
PACS@Herchel&100 and 160$\mu m$&Imagery& PI : Elbaz\\
&&& \textit{Observation in progress} \\
VIMOS@VLT&Optical&Spectroscopy& Popesso et al. (2009)\\
& & & Balestra et al. (2010)\\
FORS@VLT&Optical& Spectroscopy&Vanzella et al. (2008)\\
\hline
\end{tabular}
\caption{{\small The GOODS observations}}
\label{CDFS_field}
\end{center}
\end{table}
The GOODS extensive spectroscopic campaign at VLT with FORS2 and VIMOS collected $\sim$2700 secure source redshifts. This catalogue has been used to pre-select the targets, in the IMAGES survey, for the GIRAFFE observations which need secure redshift determination. Nevertheless, GOODs spectra have low spectral resolution and low signal-to-noise and they are thus inappropriate to estimate reliable extinction or metallicity in the CDFS galaxies \citep{2004A&A...417..905L}. It is the reason why IMAGES observed a sample of $\sim$300 galaxies with FORS2 at higher spectral resolution and higher signal-to-noise. 

\subsubsection{IMAGES a unique sample of intermediate redshift galaxies }
\begin{figure}[!h]
\centering
\includegraphics[width=0.9\textwidth]{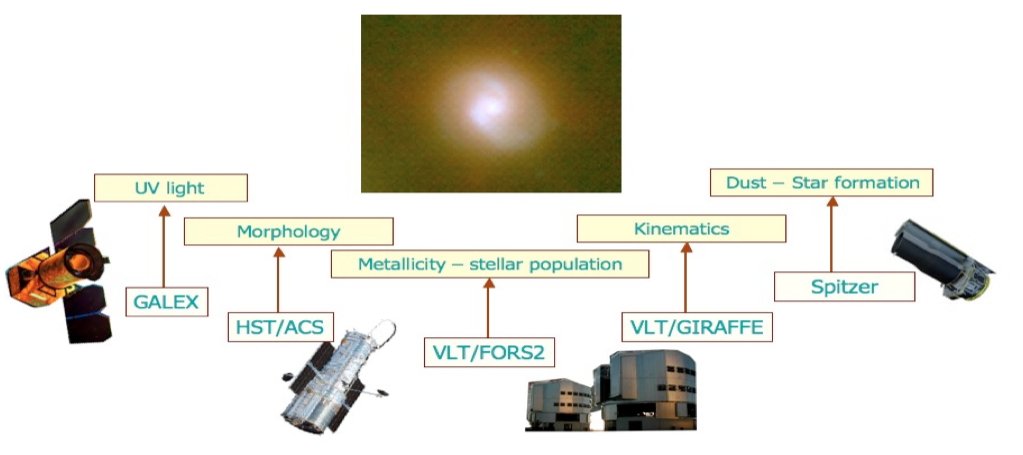}
\caption{{\small The instrumental set-up used by the IMAGES survey and the related observed quantities. The VLT/FORS2 and VLT/GIRAFFE observation are private observation from the IMAGES survey. The other observation arise mainly from the GOODS survey.}}
\label{IMAGES_survey}
\end{figure}

The IMAGEs survey is at the current date the only sample of intermediate redshift galaxies which combine spatially resolved kinematics at 2-3 Kpc scale, deep imagery from the Hubble space telescope giving morphological details at 100\,pc scale and moderate spectral resolution spectroscopy observation allowing the measurement of integrated properties such as metallicity and extinction, see Fig. \ref{IMAGES_survey}. In this framework, we have designed and implemented a set of new methodologies to derive properly the physical parameters from the observation of distant galaxies, see review \citet{2009arXiv0911.3247H} and Chapter 3 of Introduction part.

\section{Observational evidence for the main role of merger}
\begin{figure}[!h]
\centering
\includegraphics[width=0.9\textwidth]{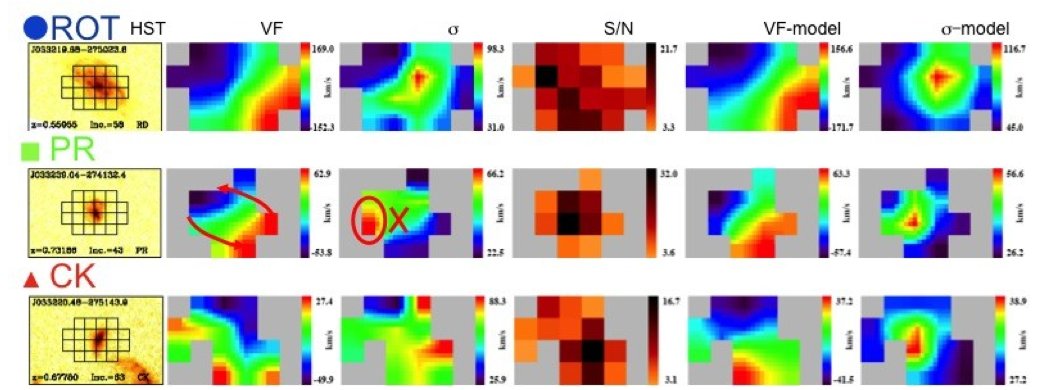}
\caption{{\small It illustrates three different examples (or class) of kinematical properties of distant galaxies. The upper row shows a rotating disk showing a dynamical axis aligned with the optical axis. Because of the low spatial resolution most of the gradient between the two extreme velocities provides a dispersion peak (see 3rd and 6th panels from the left) just at the centre of the velocity field and of the galaxy, i.e. the centre of rotation. Velocity models assume the inclination from the optical image and a velocity gradient from the observations. The middle row shows a galaxy with an almost aligned dynamical axis, thus indicative of a rotation. However the dispersion peak falls far from the centre of rotation, indeed far from any optical emission of the galaxy. This is indicative of a strong disturbance of the gaseous emission, probably due to shocks in gas flows, which perturbs the rotation (called perturbed rotation, PR). In the bottom row, the kinematics is complex (called complex kinematics, CK) from both velocity field and dispersion maps. }}
\label{VF_yang}
\end{figure}

\begin{figure}[p]
\centering
\includegraphics[width=0.8\textwidth]{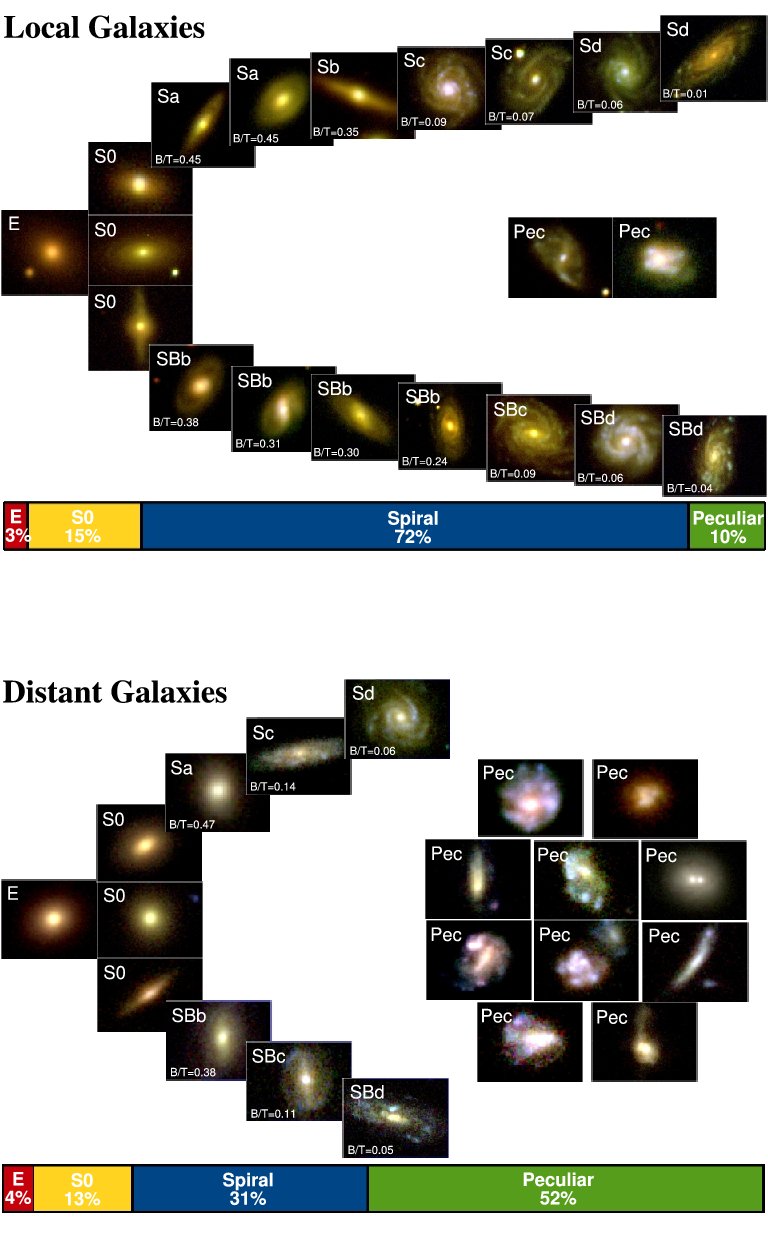}
\caption{{\small Morphologies of galaxies at z=0 (top) and z=0.65 (bottom) from \citet{2010A&A...509A..78D}. Each stamps represents 5\% of the $M_J < -20.3$  galaxy population and are 3-colors composite of u, g and r or of v, i and z bands, respectively. The fraction of regular spirals has significantly increased since 6 Gyr, while for early-type (E/S0) it appears unchanged. At least half of the present-day spirals have a progenitor which is peculiar at z=0.65.}}
\label{Hubble_Delgado}
\end{figure}

At a first order, galaxies are self-graviting systems, which can be described by at least three fundamental parameters, i.e., their (total) mass, velocity (including both bulk and random motions), and size. Scaling relations between these parameters or their proxies are therefore fundamental tools to characterize galaxy evolution. We have thus derived integrated properties of all the IMAGES galaxies, including their stellar mass, gas fraction \citep{2010A&A...510A..68P}, and gas metallicities \citep{2008A&A...492..371R} and studied the evolution of the scaling relations, such as Tully-Fisher relation \citep{2008A&A...484..173P,2010A&A...510A..68P}, the mass-metallicity  \citep{2008A&A...492..371R} and between specific angular momentum and stellar mass \citep{2007A&A...466...83P}.  From the resolved kinematics observations, we have inferred the dynamical state of the galaxies observed with GIRAFFE \citep{2008A&A...477..789Y}. In addition, we have carried out a meticulous classification of their morphologies \citep{2008A&A...484..159N, 2010A&A...509A..78D}. I summarize below the main results found by IMAGES team about the global properties of intermediate mass galaxies at z$\sim$0.6: 
 \begin{enumerate}
\item Among the 63 z=0.4-0.75 galaxies observed by GIRAFFE, most of them show anomalous kinematics \citep{2008A&A...477..789Y}, see Figure \ref{VF_yang}. Accounting for the whole population of $z\sim0.6$ galaxies, this reveals 33\% of rotating disks, while 41\% have anomalous kinematics, including 26\% with complex velocities. Thus galaxy kinematics evolve strongly since the last 6 Gyr. 
\item In \citet{2010A&A...509A..78D}, we have compared a representative sample of local galaxies from the SDSS to the HST/ACS observations of IMAGES galaxies. Inspections of the morphologies evidence that galaxies have strongly evolved. Peculiar galaxies represent half of z=0.65 galaxies, while the fraction of early type galaxies is similar to that at z=0 (see Figure \ref{Hubble_Delgado}). It means that more than half of the progenitors of present-day spirals are peculiar galaxies at z=0.65. 
\item The stellar-mass Tully Fisher relation found shows a much larger scatter (0.63dex) than that from local spirals (0.12dex), similarly to what is found from slit measurements \citep{2006A&A...455..107F,2006A&A...455..131P}. It has been convincingly shown that a tight Tully Fisher relation still hold for rotational disks and then that all the scatter is related to galaxies with anomalous kinematics, principally those with complex kinematics. Adding the gas mass for the sample of 33 z=0.65 galaxies, we have been able to derive, for the first time, the evolution of the baryonic Tully Fisher relation \citep{2010A&A...510A..68P}. Although the result is based on only 7 distant rotating disks, they fall exactly onto the local relation. If confirmed, this result is quite fundamental to link local disks to their progenitors, 6 Giga-years ago. Rotating disks at z=0.65 could have already assembled all their baryonic mass7 (gas and stars), and may have evolved into the present-day spirals simply by transforming two-thirds of their gaseous content into stars during the 6 Giga-years elapsed time. 
\end{enumerate}

The study of the fundamental quantities of the IMAGES samples have put on evidence the strong evolution of intermediate-mass galaxies at z=0.6 in terms of morphologies \citep{2008A&A...484..159N,2010A&A...509A..78D}, kinematics, star formation and gas \& metal content (current work and \citet{2008A&A...492..371R}) and reveals that nearly half of them are systems not yet dynamically relaxed. These observations suggest that merging is an important process that shaped the early evolution of the present-day galaxies. 

\section{Modeling individual galaxies}
\begin{figure}[!h]
\centering
\includegraphics[width=0.55\textwidth]{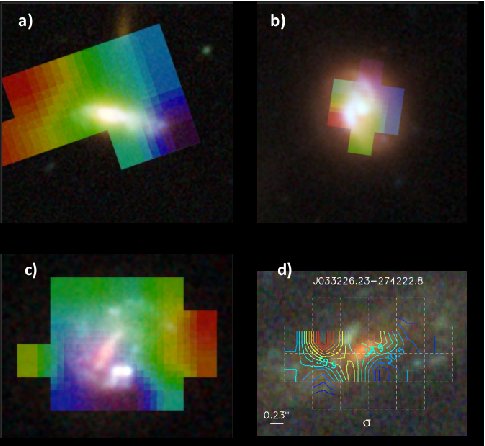}
\caption{{\small Three color images from ACS/HST and superimposed velocity field from GIRAFFE/VLT for four modeled IMAGES galaxies. \textbf{a)} \citet{2009A&A...493..899P} demonstrated that the presence of ionized gas without stars near a highly asymmetric disk can be only reproduced by a remnant of a merger. \textbf{b)} \citet{2009A&A...496..381H} identified a compact LIRG dominated by a dust-enshrouded compact disk that surrounds a blue, centered helix (so-called 'two arms-plus-bar' structure). We interpret this helix structure as regulating the exchanges of the angular momentum and possibly stabilizing the new disk. Indeed, gas inflows along a helix are usual in simulations of mergers, especially in inclined and polar orbits. This gas-rich galaxy appears to be an archetype of disk rebuilding after a 1:1 or a 3:1 merger with an inclined orbit. \textbf{c)} \citet{ 2009A&A...496...51P} identified a giant and star-bursting bar induced by a 3:1 merger, and simulated both morphologies and the off-centre dynamical axis. In this case, the gas pressured in the tidally formed bar has condensed into young and blue stars. \textbf{d)} \citet{2007A&A...476L..21P} have demonstrated that spatially resolved kinematics is sufficiently sensitive to detect the infall of a 1:18 satellite in a z = 0.667 galaxy. The superimposed map is the dispersion map.}}
\label{Individual_studies}
\end{figure}

All this observations have lead us to the following questions: Can we explain the local Hubble sequence by the evolution of merger remanent occurred between 0.6<z<1.4? In particular, can we forme the local spiral galaxies from the perturbed and disrupted population of star-forming galaxies at z=0.6? To answer to these questions we have started a large project of modeling IMAGES galaxies with the same detail than local galaxies to identify the types of merging episodes that end up as a disk galaxy.  The team has already modeled 6 galaxies and more object are in progress \citep{ 2007A&A...476L..21P, 2009A&A...493..899P, 2009A&A...501..437Y, 2009A&A...496..381H, 2009A&A...496...51P, 2010arXiv1001.2564F}, see illustration of four examples in Fig \ref{Individual_studies}. 

We took advantage of the unprecedented amount of observational constrains to model individually distant galaxies using modern simulations with hydrodynamic code. The joint analysis of morphological information with kinematical studies and spectral information is essential to properly characterize distant galaxies to avoid misinterpretations caused by the lack of resolution at higher redshift. We have used modern simulation codes such as GADGET and ZENO \citep{2005MNRAS.364.1105S, 2002MNRAS.333..481B} which account for the hydro-dynamics of the gas as well as for the collision-less star and dark matter components. These models are now mature enough to be able to reproduce nearby mergers accurately (Barnes \& Hibbard 2009).

\section{Is the Hubble sequence compatible with a merger scenario?}
In \citet{2009A&A...507.1313H}, we considered the possibility that the formation of the Hubble sequence relies to a large extent on past merger events. We used a sample of objects around z = 0.65 for which we have both morphology from the HST and high quality kinematics (both velocity field and 2D velocity dispersion maps) and compared them with simulation results varying the viewing angles to obtain the best fits. Although we cannot of course prove that the origin of these objects is a merger, we can safely say that their observed properties are well compatible with them being a merger or their remnants. This result was reached using all available data, i.e. our comparisons included morphology, mean velocities and dispersions.

\begin{figure}[!h]
\centering
\includegraphics[width=1.0\textwidth]{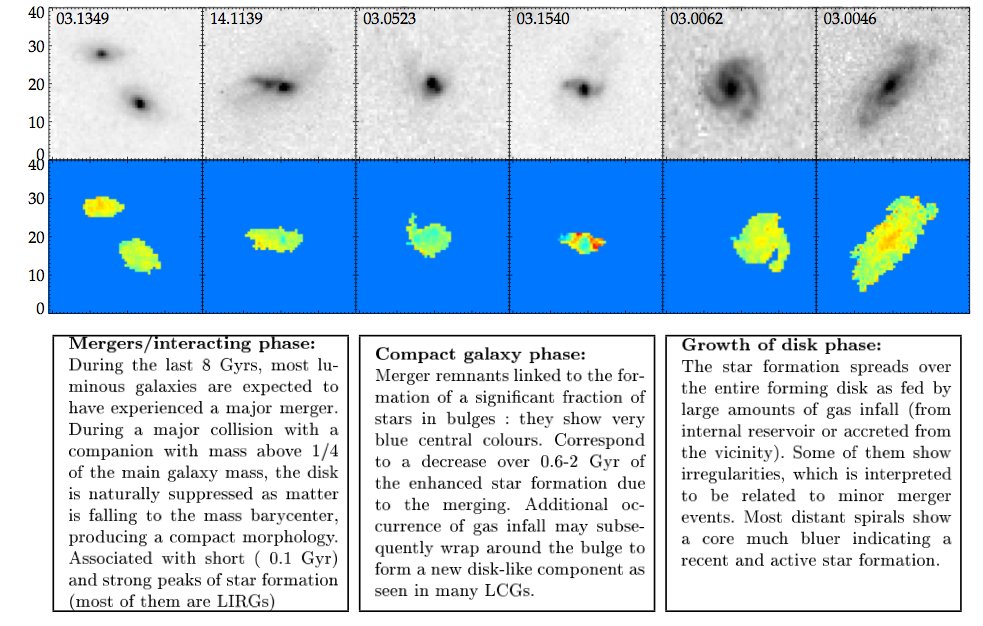}
\caption{{\small The three stages of the formation of local disk according to the \textit{spiral rebuilding scenario}: 
 \textbf{1) Phase of merger}. The progenitor disks are destroyed and the majority of the matter falls toward the center of mass of the interacting system. It produces a short and intense burst of star formation. \textbf{2) Compact phase}.The merger remanent is associated to the formation of a bulge. It may correspond to a Luminous Compact Galaxy (LCG). \textbf{3) Disk rebuild}. The gas expelled in the first stages of the interaction falls into the merger remanent and reformed a disk. It provokes a moderate burst in the rebuilding disk. From \citet{2005A&A...430..115H}.}}
\label{Rebuilding}
\end{figure}

A merger origin of the Hubble sequence is closely linked to the \textit{disk rebuilding scenario}, which has been successfully argued both from the observational point of view and from simulations. The \textit{disk rebuilding scenario} proposes a merger origin for spirals and, by extension, of the whole Hubble sequence, from ellipticals (Toomre \& Toomre 1972; Toomre 1981; Barnes \& Hernquist 1992) to late type spirals, see Figure \ref{Rebuilding}. This scenario has been proposed by \citet{2005A&A...430..115H} to conciliate the coeval evolution of star formation, stellar mass assembly, fraction of peculiar galaxies, interacting pair statistics and gas content. The \textit{disk rebuilding scenario} is supported by the recent advances in numerical simulations of a merger \citep{2002MNRAS.333..481B,2005ApJ...622L...9S, 2006ApJ...645..986R,2008MNRAS.391.1137L,2009MNRAS.397..802H} which show that gas-rich mergers (gas fraction larger than 50\%) may produce a new disk in their remnant phase, and that such a new disk is mainly supported by the orbital angular momentum provided by the collision. From the gas fractions of peculiar galaxies at z$\sim$1, we have quantitatively estimated the gas fractions of their progenitors \citep{2009A&A...507.1313H,2010A&A...510A..68P}. We have concluded that more than a third of the galaxy population about 6 Gyr ago was sufficiently gas rich to rebuild a disk after the merger. Thus our work argues that a merger origin of the Hubble sequence, although it has not yet been proven, is a plausible alternative or channel for the formation of large disks.

\EmptyNewPage

\FrameThisInToc
\EmptyNewPage

\chapter{ FORS2 observations and sample }
During my thesis, I have been in charge of the data reduction, analysis and interpretation of the FORS2 observations of the IMAGES survey.  I describe in this chapter the FORS2 observations and the construction of my main working sample: Sample\,A a sub-sample of the 58 FORS2 targets for which I have measured the metallicity of their gas phase.   
\minitoc
\section{FORS2 Observations}
\begin{figure}[p]
\centering
\includegraphics[width=0.7\textwidth]{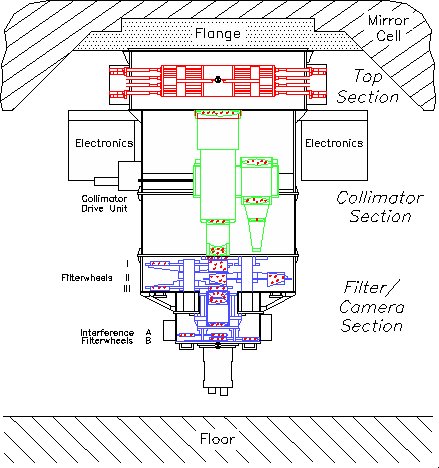}
\caption{{\small Scheme of FORS2. The instrument is composed by three sections : 
 \textbf{Mask section} The mask device for spectroscopic modes is located on the top section of the instrument. It carries a multi-object spectroscope (MOS), a long slit system a unit of tunable mask (MXU) and two calibration units. 
\textbf{The collimator section} The collimator section is composed by two exchangeable collimators with f=1233mm and f=612mm. The spatial resolution of the instrument is defined by the selected collimator. The resulting field of view is 6.8'x6.8' with the standard resolution collimator (SR) and 4.2'x4.2' with the high resolution collimator (HR). 
\textbf{The filter and detector section} This section is composed by a filter wheel, which allows to select among a set of filters during imagery or spectroscopic observations, a grism wheel and a detector. The large number of grism available on the grism wheel allows to have at  disposal a wide number of observational configurations for the spectroscopic mode : Spectral resolution from R=260 to 2600 and wavelength coverage included from 330nm to 110nm. The FORS2 detector is a mosaic of two 2kx4k CCDs with a pixel size of $15\times15\mu\, m^2$ }}
\label{schematic_fors}
\end{figure}

FORS2, FOcal Reducer/low dispersion Spectrograph 2, is a multi mode instrument mounted in the Cassegrain focus of UT1 of the Very Large Telescope (see Figure \ref{schematic_fors} for the mechanic schematic). It offers three modes of utilization : Imagery (IMA, OCC modes), long slit spectroscopy (LSS) and multi-object spectroscopy (MXU and MOS mode). The FORS2 data of the IMAGES survey have been acquired with the MXU mode, one of the two multi-object spectroscopy modes. In this mode, the instrument load in the mask section a pre-prepared mask with slits for a large number of objects. The mask is designed with an appropriate software \textit{FORS Instrumental Mask Simulator - FIMS}. The user can chose the number, position, width and length of the slits. The mask is then manufactured with a laser cutting machine \textit{Mask Manufacturing Unit} of the VIMOS instrument (MMU) and loaded in the  \textit{Mask Exchange Unit - MXU}. 

\begin{figure}[!ht]
\centering
\includegraphics[width=0.9\textwidth]{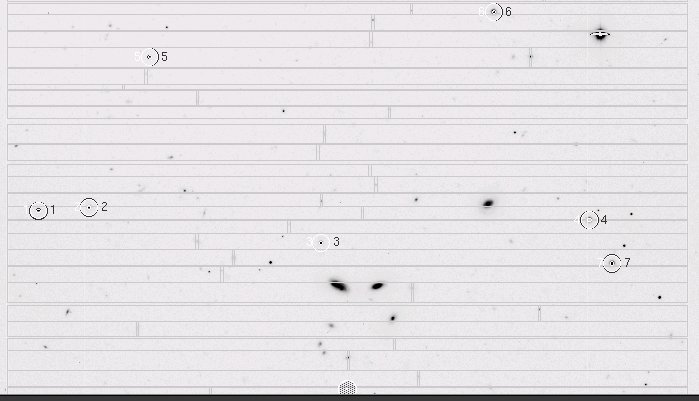}
\caption{{\small FORS2 pre-image with the FIMS mask overplotted.}}
\label{PreFORS}
\end{figure}

A sample of Intermediate mass galaxies ($I_{AB} <23.5\,mag$, see Chapter 1) have been randomly selected in four $6'8\times 6'8$ fields of the CDFS field. The selection of the objects and the preparation of the slit masks have been performed using the GOODS catalogue \citep{2004ApJ...600L..93G} and the FORS2 pre-images (see Fig. \ref{PreFORS}). The arrangement of the slits were optimized to improve the number of objects observed. Among the 270 slits which have been placed, 241 objects have $I_{AB}$ <23.5 mag. In order to sample the rest-frame 3700-5400\AA\, domain from z=0.35 to z=1, all the sample have been observed with two holographic grisms which characteristics are summarized in Table \ref{Grism}. The minimum exposure time of each object in each grism is around 3 hours. The combination of  600z and 600RI grism allows us to detect lines from [OII]$\lambda3727$\AA\, to [OIII]$\lambda 5007$\AA\, which are essential to derive the extinction and metallicity of galaxies gaseous phase. Moreover, the spectral resolution (R$\sim1200$) offered by the two grisms is large enough to analyze the stellar populations and to properly establish the extinction from Balmer lines ratio unaffected by Balmer underlying absorption. The set of 16h of observations have been carried in service mode and with the constrain of seeing ($<0.8''$) and sky transparency (photometric) imposed by the IMAGES survey. Table \ref{Observation_log} register all the FORS2 observations. 

\begin{table} [!ht]
\centering
\begin{tabular}{|c|c|c|c|c|c|c|}
\hline
Grism & $\lambda_{central}$  &$\lambda_{1}$&$\lambda_{2}$ &Dispersion&Resolution&Filter \\
\hline
\hline
GRIS\_600RI+19&6780&5120&8450&0.83&1000&GG435\\
GRIS\_600Z+23&9010&7370&10700&0.81&1390&OG590\\
\hline
\end{tabular}
\caption{{\small Spectroscopic properties of the grism used in the FORS2-IMAGES survey }}
\label{Grism}
\end{table}

\begin{table} [!ht]
\centering
\begin{tabular}{|c|c|c|c|c|}
\hline
Field CDFS &Mask&Filter&Exposure time(min)&N exposures\\
\hline
\hline
Field1&M059P76&600RI&176&4\\
Field1& M059P76&600Z&176&4\\
Field1&M45MASK01&600RI&132&3\\
Field1& M45MASK01&600Z&220&5\\
\hline
Field 2&M060P75&600RI&132&3\\
Field 2&M060P75&600Z&132&3\\
Field 2&M045MASK02&600RI&88&2\\
\hline
Field 3&M062P75&600RI&132&3\\
Field 3& M062P75&600Z&132&3\\
Field 3&M045Masque03&600RI&132&3\\
\hline
Field 4&M043mask04&600RI&132&3\\
Field 4&M056P75M04&600RI&132&3\\
Field 4&M056P75M04&600Z&132&3\\
Field 4&M065P75&600RI&132&3\\
Field 4& M065P75&600Z&132&3\\
\hline
Field 5&M063P75&600RI&132&3\\
Field 5&M063P75&600Z&132&3\\
Field 5&M44mask05&600Z&132&3\\
\hline
\end{tabular}
\caption{{\small Journal of observations }}
\label{Observation_log}
\end{table}

\section{Data reduction}
I have followed the usual procedure to reduce multi-object spectroscopy data. However, as the observed target are faint galaxies, the procedure has been optimized to recover the maximum of photons from the galaxies. I used the \textit{Multired} package for \emph{IRAF} written by  G. Le Fevre and optimized by H.Flores. The reduction \footnote{Each grism is reduced individually} included the following correction and calibration : 
\begin{itemize}
\item Remove cosmic rays from an individual exposure with the \textit{lacos spec} package in IRAF \citep{2001PASP..113.1420V}.
\item Correction of the instrumental distortion
\item Bias and Flat correction. The flat is adjusted by a spline function which order minimize the rms. 
\item Extraction of the 2D spectra from the individual exposures. The slit position has been previously detected with \textit{Sextractor} in the slit image.  Due to an error of fabrication, an offset of the beam of light is induced by the GRISM600RI+19 grism. The position of the spectra in the raw exposure are offset by 140 pixels in the Y direction from the position of the slit on the mask. Around 21'' of the field of view fall outside the CCD in the SR mode.
\item Stack of the individual exposure and mask. The objects can be observed by severals mask configurations of the same field. The position of the object in the slit can vary from a mask to an other. I have implement a IRAF routine in the \textit{Multired} package, \textit{multimask}, which aligns the 2D spectra from various mask and combine the different exposition. I choose the stack the 2D spectra from different mask instead of stacking the 1D spectra because this method increase the final signal to noise. 
\item Sky subtraction. I use the sky subtraction IRAF package for long slit spectroscopy on the individual 2D spectra.
\item Extraction of the 1D spectra in the combined 2D object spectrum and in the 2D calibration spectrum. The spectra have been extracted with a optimized aperture ($3\sigma$ of the light distribution) and the position of the spectrum in the 2D image have been fitted with a polynom. 
\item Wavelength calibration. The calibration spectra are calibrated from the emission line identifications of the Neon-Argon-Helium, using a polynomial relation between pixel and wavelength. 
\end{itemize}

\begin{figure}[!ht]
\centering
\includegraphics[width=1\textwidth]{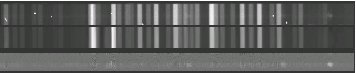}
\caption{{\small \textbf{Up}: Raw 2D spectrum. The object spectrum is sightly visible. \textbf{Middle}: Extracted 2D sky spectrum. \textbf{Down}: Final 2D object spectrum after the sky substraction. The strong emission lines and a weak continum are clearly visible. }}
\end{figure}

\subsection{Flux calibration}
\begin{figure}[!ht]
\centering
\includegraphics[width=1\textwidth]{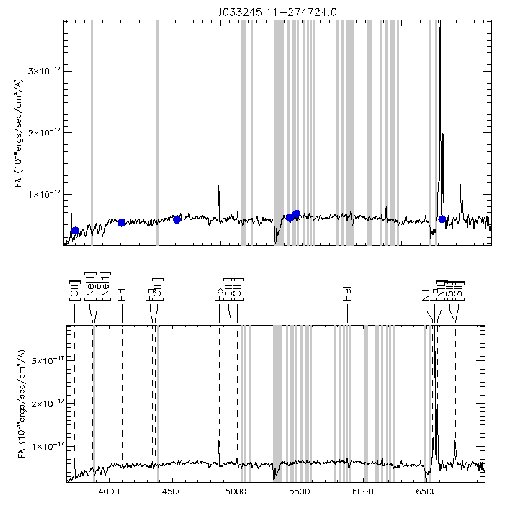}
\caption{{\small  Rest-frame spectrum of J033245.11-274724.0 galaxy after flux calibration and aperture correction. Grey boxes delimit the wavelength region where spectra may be affected by strong sky lines. \textbf{Up}: The blue dots are the photometric point from HST/ACS photometry in V, I, Z bands and V, I, R in the EIS catalogue, see section \ref{Photometry catalogue}. \textbf{Down}: The position of the emission lines are marked with dashed vertical lines }}
\label{photo_calib}
\end{figure}

The two flux calibrated spectra of each grism are combined to set the final spectrum. The spectra were flux calibrated using photometric standard stars. I have compared the spectrophotometry to HST/ACS photometry in V, I, Z bands and V, I, R in the EIS catalogue and found a good agreement, see fig.\ref{photo_calib}. 

\subsection{Aperture correction}
\begin{figure}[!ht]
\centering
\includegraphics[width=0.5\textwidth]{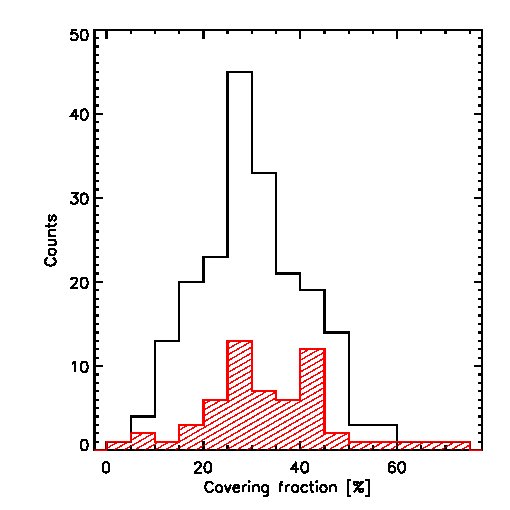}
\caption{{\small  Distribution of the covering fraction (inverse of the aperture correction defined in equation 2.1) for the FORS2 galaxies (black) and the sample used to measure metallicities (red). A minimum covering fractions larger than 20\% may be needed to avoid aperture effects.}}
\label{Aperture}
\end{figure}

In slit spectroscopy, part of the object light can be lost by the aperture. Indeed, the dimension of the object can be larger than those of the slit, see discussion in Chapter 3 of the Introduction part. The SFR estimated from the flux of emission lines do only account for the SFR in the region covered by the slit. To have the total SFR of the object, the line flux has to be corrected by an aperture factor. The correction assumes that the $\Sigma SFR$, the SFR by unit of surface, is constant in all the galaxy. The aperture correction is calculated by comparing the flux measured in the spectrum to those give by isophotal photometry in one or several bands. Figure \ref{Aperture} shows the distribution of the covering fraction in FORS2 data. The mean aperture correction of the FORS2 galaxies is 3.10. 
\begin{equation}
C_{aper}=\langle \frac{f^i_{Photo}}{f^i_{Spectro}} \rangle ^{i\,band} 
\end{equation}

\section{The FORS2/IMAGES sample}

\subsection{Spectroscopic catalogue}
At the end of the data reduction process, the FORS2 sample is composed by 231 targets. Ten targets have not being extracted due to instrumental problems, as target falling outside the slit or dramatic sky extraction. For the remaining spectra, I have classified them into a catalogue according to the redshift of the target and its spectral type. 

\subsubsection{Redshift classification}
\begin{figure}[!h]
\centering
\includegraphics[width=0.60\textwidth]{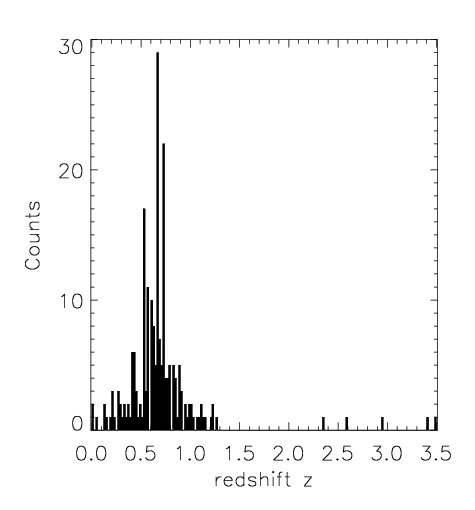}
\caption{{\small Redshift distribution (bin=0.02) of the FORS2--IMAGES sample, including the classes 2(secure), 1(insecure) and 9(single line). The two peaks at  z=0.670 and z=0.735 correspond to structures in the CDFS field.}}
\label{hist_z}
\end{figure}

The redshift has been determined by the position of the severals strong emission lines and other absorption features in the spectrum, using the well known equation : 
\begin{equation}
z=\frac{\lambda_{obs}-\lambda_{rest}}{\lambda_{rest}}
\end{equation}
I have associated to each redshift a quality flag denoting the reliability of the identification, following \citet{2007A&A...465.1099R}: 
\begin{itemize}
\item \textbf{Secure (class = 2)}: contain more than two strong features; 
\item \textbf{Insecure (1)}: contain either a strong feature with not very reliable supporting features, or with multiple features that are not strong enough to confirm the redshift, giving a confidence of only about 50\% to the estimated redshift;
\item \textbf{Single emission-line (9)}: spectra with a single strong emission line without any other features. Redshifts are tentatively assigned for such cases. Typically this is the case where we are unable to differentiate [OII]$\lambda$3727 and $H\alpha$ emission lines.
\end{itemize}
\begin{table} [!ht]
\label{table_obs} 
\centering
\begin{tabular}{c c c c c c c} 
\hline
Class & EL galaxy& AL galaxy& Star & QSO & Total & \%\\  
\hline
2&141&25&19&4&192&83\\
1&14&18&-&-&32&14\\
9&7&-&-&-&7&3\\
\hline 
Total& 162&43&19&4&231&-\\
\hline 
\end{tabular} 
\caption{ {\small The number and type of objects in the redshift quality classes: secure (2) for spectra contain more than two strong features, insecure (1) when only one features was detected but with not very reliable supporting features, and single emission line (9) when the spectra had a single emission line without any features. EL stands for emission lines galaxies and AL for absorption line galaxies.}}
\label{redshift_table}
\end{table}
From the 270 slits, 240 objects have identified redshift, giving a success rate of $\sim$88\%. For objects with $\mathrm{I_{AB}}<$23.5\,mag the success rate reached $\sim$96\%.
In Fig.\ref{hist_z}, I have plotted the redshift histogram for all galaxies and QSO. The redshift distribution has a range of [0.01-- 3.499] and a median value of 0.667. Two main peaks are visible in the redshift distribution at z=0.670 and z=0.735. These peaks correspond to structures in the CDFS field, already identified in \citet{2007A&A...465.1099R}.\\
\subsubsection{Spectral type classification}
\begin{figure}[p]
\centering
\includegraphics[width=0.45\textwidth]{Figures/AGN.jpg}
\includegraphics[width=0.45\textwidth]{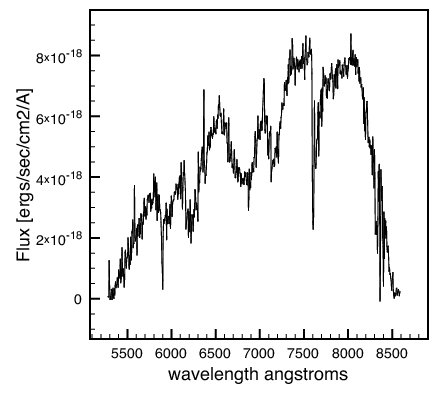}
\includegraphics[width=0.45\textwidth]{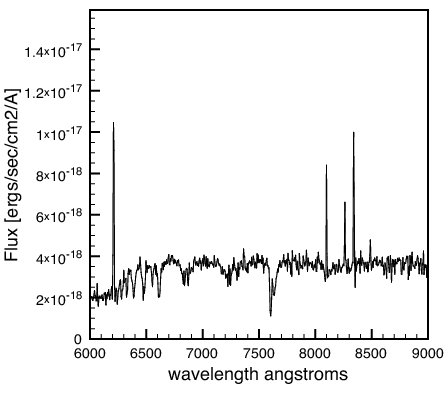}
\includegraphics[width=0.45\textwidth]{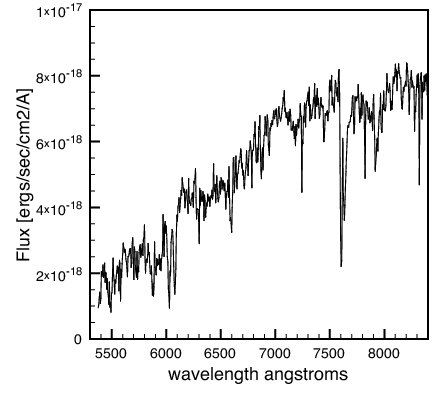}
\caption{{\small Examples from the IMAGES catalogue of the four spectral classes. \textbf{Up-left:} FORS2 spectrum of a typical AGN spectrum J033208.66-274794.4. \textbf{Up-right:} Star spectrum, J033205.58-274638.2. \textbf{Down-left:} Typical spectra of a emission line galaxy (ELG) J033225.26-274524.0. \textbf{Down-right}: Absorption line galaxy, J033230.07-275140.6}  }
\label{Spectral_type}
\end{figure}

I have also classified the target according to their spectral type. The spectral classification is divided into four types (see Fig \ref{Spectral_type}): a galaxy, split into emission line galaxies (ELG) and absorption line galaxies (ALG); quasars and stars. A spectra is classified as a emission line galaxy if the [OII]$\lambda 3727$\AA\, or $H_\beta$ line have a EQW over 15\AA. Quasars are identified by their characteristic continuum, broad MgII emission and Broad Balmer lines. Table \ref{redshift_table} shows the number and type of objects in the different redshift quality classes.
  
\subsection{Photometry catalogue}
\label{Photometry catalogue}
The CDFS field has been widely observed by UV to IR surveys (see Chapter 2). I have gathered photometric measurement for the 208 galaxies of the FORS2-IMAGES sample by cross-correlating the sample with catalogues in public archive or extracting the sources from data release images. The photometry has been extracted in view of spectral energy distribution fitting. Therefore a special care has been taken to homogenize the photometry arising from different instruments and at different wavelengths. The photometry has been extracted within an aperture of 3" of diameter following the methodology of \citet{2009A&A...493..899P} which takes into account PSF correction in the UV and IR band and carefully treats the calculation of the magnitude uncertainties.

\subsubsection{The catalogue and image archives}
For each of the following catalogues, I have extracted the photometry of the FORS2-IMAGES galaxies by cross-correlating their coordinates with those of the catalogues. Only detection within 2" have been selected. All the detections have been verified by hand (and in all bands) and objects with possible flux contamination by neighbor sources have been flagged.  
In UV and mid-IR the direct photometric measurement need to be corrected from the PSF. Indeed, the PSF at these wavelength (GALEX and Spitzer) are larger than the photometric aperture. The PSF correction is derived by the theoretical PSF of the two instruments \citep[See]{2007ApJS..173..682M,2009A&A...493..899P}.
 
\begin{itemize}
\item The near- and far-UV photometry have been extract from the GALEX  catalogue of \citet{2007ApJS..173..682M}.
The PSF of GALEX is larger than the 3" diameter aperture. I have corrected the extracted magnitudes by using the PSF correction given by \citet{2007ApJS..173..682M}: -2.09 in NUV and -1.65 in FUV. 
\item The optical photometry from the ACS/HST in B(F435W), V(F606W), i(F755W) and z(F850LP) have been gathered from \citet{2004ApJ...600L..93G}. The CDFS field has been partially observed by the Ultra Deep Field South (UDF-S). For the objects falling in the UDF-S I have used the photometry of the UDF catalogue \citep{2006AJ....132.1729B}. The isophotal magnitude in z-band have also been extracted. 
\item Additional visible photometry in U', U, B, V, R and I-band from ground-based observation have been extracted from the EIS catalogue \citep{2001A&A...379..740A}. The catalogue gives magnitudes in the Vega photometric system and the photometry is not corrected for extinction. I have converted the photometry into AB system and correct it for extinction following the indications of  \citet{2001A&A...379..740A}. The corrective factors are tabulated in Table \ref{table_EIS}. Note that the photometry in the U' band is affected by large uncertainty due to the UV atmospheric cutoff at $\sim$ 3200\AA.
\item The near-IR photometry in J, H and K bands have been extracted from the GOODS-ISAAC catalogue \citet{2001astro.ph..2300V}. Due to the large uncertainty on the zero-point calibration of the GOODS H-band, we have used the revised calibration proposed by \citet{2008ApJ...682..985W} which add +0.14\,mag in H-band. 
\item The mid-IR photometry can be extracted from the SIMPLE catalogue based on the GOODS-IRAC data release 3 (http://www.astro.yale.edu/dokkum/SIMPLE). However, the error on flux measurements in this catalogue are extremely small and may be underestimated. I have re-extracted a catalogue from the GOODS-IRAC data release 3 images using the software \textit{Sextractor}. Table \ref{table_IRAAC} shows the values of the zero point and the detector gain in each IRAC channel used by \textit{Sextractor}. IRAC/Spitzer has a large PSF, the photometry has been corrected by a PSF correction given in table \ref{table_IRAAC}. The detail procedure to estimate the PSF correction in IRAC is described in \citet{2009A&A...493..899P}. I have compared the magnitudes of my new catalogue with those of SIMPLE and I have found a very good agreement. 
\item The far-IR photometry at $24\mu m$ has been extracted by Hector Flores (private communication) from the GOODS-MIPS data release 3 image. This catalogue is comparable to those of \citet{2005ApJ...632..169L} but reach slightly lower detection limit. Caution, the photometry given in the table is not corrected by the PSF. 
\end{itemize}

\begin{table} [!ht]
\label{table_EIS} 
\centering
\begin{tabular}{c c c} 
\hline 
Band&Vega to AB&Extinction \\ 
\hline
U'&+0.80&-0.05\\
U&+1.04&-0.05\\
B&-0.11&-0.05\\
V&-&-0.03\\
R&+0.19&-0.02\\
I&+0.50&-0.02\\
\hline 
\end{tabular} 
\caption{{ \small Correction to be applied at the EIS photometry to convert the photometry from Vega to AB system and to correct from extinction. From \citet{2001A&A...379..740A} }}
\end{table}

\begin{table} [!h]
\label{table_IRAAC} 
\centering
\begin{tabular}{c c c c c} 
\hline 
Channel&Wavelengh&Zero point&Detector gain& PSF correction \\ 
&\textit{$\mu m$}&\textit{mag (AB)}&\textit{e/ADU}&\textit{mag}\\ 
\hline
1&3.6&22.416&3.3&-0.35\\
2&4.5&22.195&3.7&-0.38\\
3&5.8&20.603&3.8&-0.51\\
4&8.0&21.781&3.8&-0.54\\
\hline 
\end{tabular} 
\caption{{\small Values of the zero point magnitudes and detector gain used for the source extraction with \textit{Sextractor}. From http://irsa.ipac.caltech.edu/data/GOODS/docs/goods\_dr1.html.  The PSF corrections are given by \citet{2009A&A...493..899P} }}
\end{table}

The photometry in all the available bands are tabulated in annex. All the magnitudes have been converted into AB system.  
\subsubsection{Uncertainties}
As our aim is to extract consistent photometry from UV to mid-IR, the uncertainty on the magnitude of each band is not only given by the standard Poisson noise. The uncertainties arising from the different calibration between the instrument such as the zero-point and the uncertainties from the correction of the PSF have to be take into account. The total uncertainty of a photometric point is thus given by the quadratic sum on these three sources of uncertainties \citep{2009A&A...493..899P}:
\begin{equation}
\sigma=\sqrt{\sigma_{Poisson}^2+\sigma_{ZP}^2+\sigma_{PSF}^2}
\end{equation}
\subsubsection{Isophotal photometry}
The isophotal photometry can be derived from the 3" aperture photometry and the isophotal magnitude in z-band using the relation:
\begin{equation}
(m_{band}-m_z)_{ISO}=(m_{band}-m_z)_{APER}
\end{equation}

\subsection{The FORS2-IMAGES database}

I have collected a large number of properties for the galaxies in the FORS2 sample : multi-band photometry and imagery, spectroscopic properties as flux of emission lines fluxes and derived quantities (extinction, metallicity, star formation rate), observation logs, morphology and kinematic properties (Morphological type, bulge to disc ratio, velocity field, etc..). To organize all this information I have designed a database in \emph{MYSQL}, a dedicate language to build databases. Each galaxy of the FORS2 catalogue has an entrance in the database and it is associated to a wide range of observational properties. Insertion, actualization and queries are done with \textit{phpadmin} a \emph{PHP} front-page of MySQL database.

\begin{figure}[!h]
\centering
\includegraphics[width=1\textwidth]{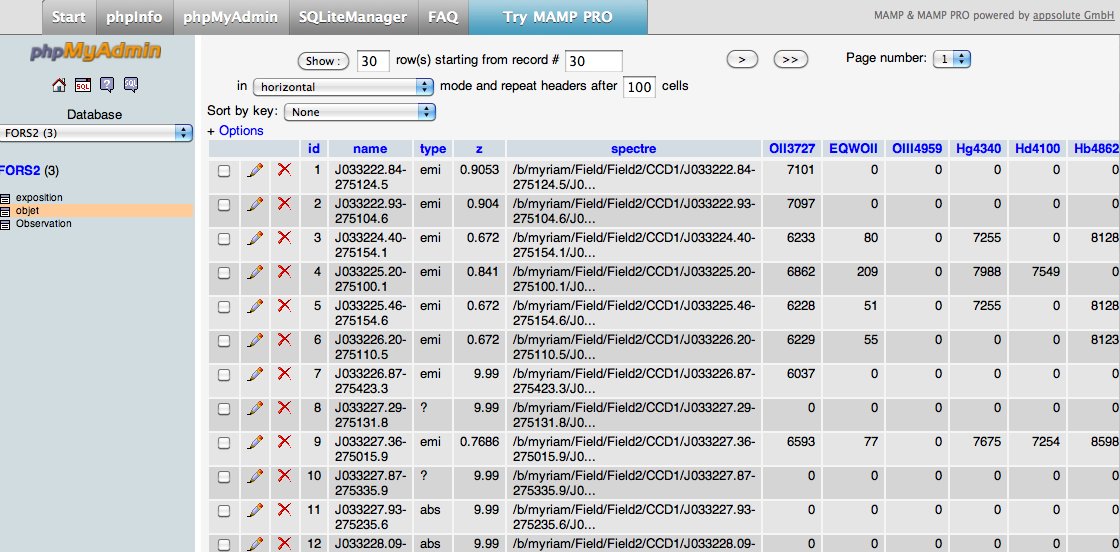}
\label{phpadmin}
\end{figure}

\begin{figure}[p]
\label{individual_db}
\centering
\includegraphics[width=0.9\textwidth]{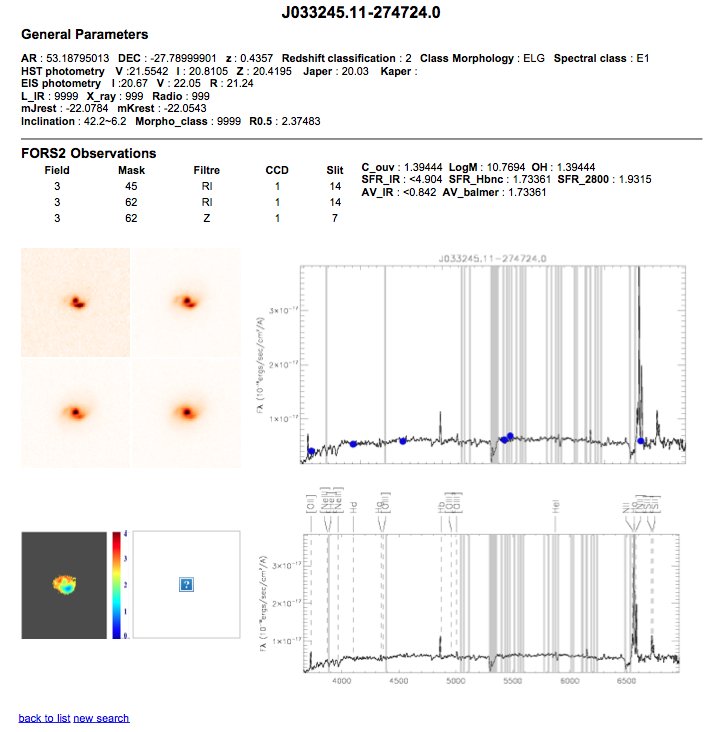}
\caption{{\small First version of the page showing the individual properties of a IMAGES object. In this version, the observational properties are organized by their observational origin: FORS2 observations, GIRAFFE observations, Photometry, etc.. }}
\end{figure}

\begin{figure}[p]
\centering
\includegraphics[width=0.9\textwidth]{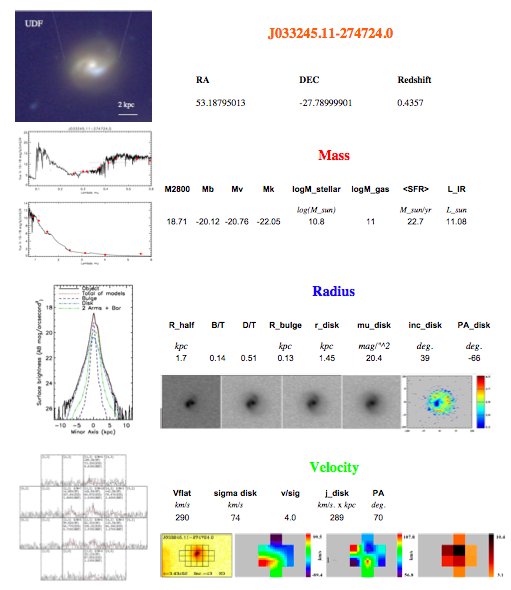}
\caption{{\small Second version of the page showing the individual properties of a IMAGES object. In this version, each observational property is associated to one of the main physical parameters of a galaxy: Mass, Radius, Velocity }}
\label{individual2_db}
\end{figure}

I have also implemented a PHP webpage to allow for a quick presentation of all the main properties of a galaxies. The user can query galaxies by their coordinates, by their name or by their spectral classification. The php script sends the query to the MYSQL database and prints the result of the query in a php page with link to the individual page of each galaxy. I have designed two versions of the page showing the individual galaxy properties, Fig. \ref{individual_db} and \ref{individual2_db}. The final output is still on discussion. The database is not public and it runs in a local apache and MySQL server. 


\section{The Metallicity sample}
\label{The Metallicity sample}
The main purpose of the FORS2 observations is to derive the metallicity of the gas in intermediate mass galaxies at $z\sim0.6$. I have gathered from the FORS2 catalogue a subsample of emission line galaxies for the study of the gas content. 
\subsection{Construction of the sample}
In order to derive metallicities, some specific emission lines are required, such as  [OII], $\mathrm{H\beta}$ and [OIII] emission lines. I have classified the emission line galaxies into 5 subcategories depending on the availability of emission lines. 
\begin{itemize}
\item (E1) : All the strong lines from [OII]$\lambda3727$\AA\, to [SII]$\lambda\lambda6300,6363$\AA\, are detected. Eg. $\mathrm{H\beta}$,$\mathrm{H\alpha}$, [OIII]$\lambda5007$\AA\,
\item (E2); All the strong lines from [OII]$\lambda3727$\AA\, to [OIII]$\lambda5007$\AA\, 
\item (E3); [OII] out of the wavelength range 
\item (E4); $\mathrm{H\beta}$ out of the wavelength range 
\item (E5); one or two emission lines detected 
\end{itemize}
Table 2 illustrates the construction of the working sample by the availability of spectral lines. From the  $\mathrm{I_{AB}}<$23.5\,mag objects I have selected 74 galaxies from the E1, E2 and E3. From the sample of 74 galaxies, 6 objects have been excluded because of bad spectrophotometry or wavelength calibration problems and 10 objects with $\mathrm{H\beta}$ or both [OIII] lines corrupted by strong sky emission or absorption lines. Finally, the sample, hereafter sample A, is composed of 58 galaxies with a median redshift of 0.7. The S/N of the $\mathrm{H\beta}$ line has a mean value of 40. 

 \begin{table} [!ht]
\label{table_sample} 
\centering
\begin{tabular}{| c | c c c c c c | c|} 
\hline 
  & E1 & E2 & E3 & E4 & E5 & Abs &Total \\ 
\hline
\hline
IMAGES&4&65&20&31&49&43&212\\
Sample A&4&65&5&-&-&-&74\\
\hline 
\end{tabular} 
\caption{ {\small Number of galaxies (208 galaxies + 4 QSO) by spectral type in the FORS2 and metallicity sample: [OII], $\mathrm{H\beta}$, [OIII] and $\mathrm{H_\alpha}$ emission lines detected (E1); [OII], $\mathrm{H\beta}$, [OIII]  detected (E2); [OII] out of the wavelength range (E3); $\mathrm{H\beta}$ or/and [OIII] out of range (E4); one or two emission lines detected (E5); galaxies without emission lines (Abs). The E3 galaxies correspond to low-redshift objects. In this work we have only selected the 5 galaxies with z$>$0.4}}
\end{table}

I have completed the sample with 30 galaxies from \citet{2004A&A...423..867L}, referred hereafter as sample B. The authors have observed $I_{AB}<$22.5\,mag galaxies in three cosmological fields : Canada France Redshift Survey (CFRS) 3h, Ultra Deep Survey Rosat (UDSR) and Ultra Deep Survey FIRBACK (UDSF) fields. They have used the FORS2 instrument in the mode MXU with a observational setup similar to IMAGES.  

\subsection{Representativity and completeness}
\begin{figure}[!ht]
\centering
\includegraphics[width=0.50\textwidth]{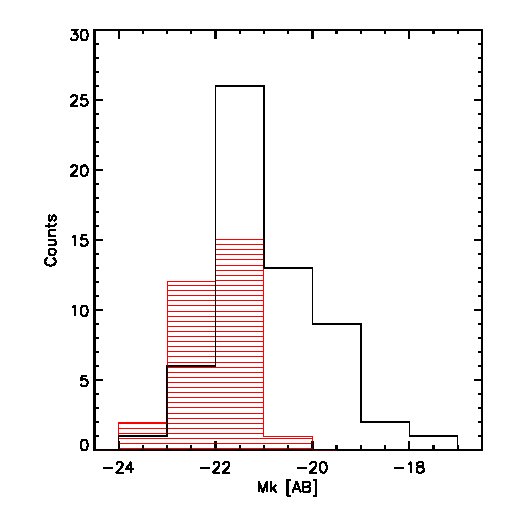}
\includegraphics[width=0.50\textwidth]{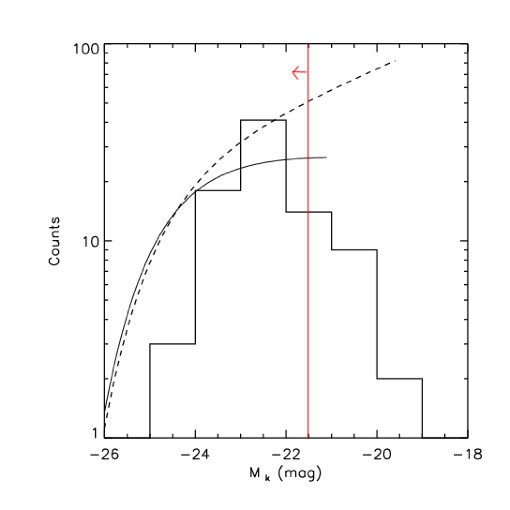}
\caption{{\small \textbf{Top panel}: Distribution of the absolute magnitude in k-band of sample\,A in black and sample\,B in red. \textbf{Bottom panel}: Number counts (on a logarithmic scale) of selected galaxies versus AB absolute magnitude in K-band. The black histogram refers to sample~A+B galaxies. The luminosity function derived from \citet{2003A&A...402..837P} at $\mathrm{z} = 0.50$ is plotted with a dashed line and at $\mathrm{z}=1$ with solid line. The red vertical line represents the limit of 85\% completeness.}  }
\label{mk05}
\end{figure}

The final metallicity sample is then composed by 88 intermediate mass galaxies. The two samples of galaxies span the same redshift range. In terms of absolute magnitude, sample B is composed of galaxies slightly brighter than sample A: sample B has a median absolute magnitude in K-band of -21.89 versus -21.13 for sample A, see histogram of both sample in left panel of fig.\ref{mk05}. However, the combination of these two samples gives a sample representative of the intermediate mass galaxies at z$\sim$0.6. 

The distribution of the K-band absolute magnitudes of sample A+B follows the luminosity function at a redshift of 0.5 and 1 from \citet{2003A&A...402..837P}, see Fig. \ref{mk05}. Kolmogorov-Smirnov tests support that the $M_K$ distribution of sample A+B and those of the luminosity function, in the redshift range z=0.4 and z=0.98 and with $M_K[AB]<-21.5$, have 85\% probability to come from the same distribution. The corresponding stellar mass range of completeness is $\log(M_{stellar}/\mathrm{M_{\odot}})>10$. The aim of the IMAGES large program is to investigate intermediate mass galaxies with $M_{stellar}> 1.5\times10^{10}\,\mathrm{M_\odot}$, which represents 76\% of our sample. Then, the incompleteness of the sample below $\mathrm{M_{K}}=-21.5$ does not affect the conclusions.

\EmptyNewPage
 
\FrameThisInToc
\chapter{The properties of the ISM in z$\sim$0.6 galaxies}
In this chapter I study 88 distant galaxies and derive their main global properties such as extinction, the gas phase metallicity, star-formation rate and gas fraction. The unprecedentedly high S/N of the IMAGES-FORS2 spectra allows us to remove biases coming from stellar absorption lines and extinction, to establish robust estimates of metallicities, instantaneous SFR and gas fraction. This study has led to a refereed  publication: "\textit{IMAGES IV: strong evolution of the oxygen abundance in gaseous phases of intermediate mass galaxies from $z\sim 0.8$"}, by \citet{2008A&A...492..371R}, in Astronomy and Astrophysics, Volume 492, Issue 2, 2008, pp.371-388.

I briefly describe the main steps of the methodology used to derive the physical properties for the sample A. The complete description of the methodology is specified in the Introduction chapter 2, and in  Liang et al. (2004a) and Liang et al. (2006) for the sample B. The methodology used for sample A is very similar to the one used for sample B, thus the two methodologies of analyses are homogeneous.

\subsection{Flux measurement}
In order to derive the metal abundance and $SFR_{H\alpha}$ in galaxies with sufficient accuracy, it is necessary to estimate the extinction and the underlying Balmer absorption. 
I have corrected the Balmer emission lines from the underlying stellar population. For each galaxy, I fitted the observed spectrum, including its continuum and absorption lines, with a linear combination of stellar libraries. I have used a set of 15 stellar spectra from the \citet{1984ApJS...56..257J} stellar library, including B to M stars (e.g B, A, F, G, K and M) with stellar metallicity, see justification in Introduction Section \ref{Extracting the ionized gas}. The best fit was obtained with the STARLIGHT software \citep{2005MNRAS.358..363C}. A \citet{1989ApJ...345..245C} reddening law has been assumed. FORS2 spectra have been pre-processed before proceeding to the best fit with the STARLIGHT software. The pre-process steps are the following :
\begin{enumerate}
\item Emission lines are masked using \emph{DISGAL1D}, a software developed by the team.   
\item The masked spectrum and the templates are smoothed using a smoothing box to a spectral resolution of $\Delta\lambda\sim$8\AA. I have convolved the continuum, except at the location of emission lines, using the software developed by our group \citep{2001ApJ...550..570H}. I have chosen to introduce this step because the smoothing increase the S/N of the absorption features used in the fit. The small decrease of the spectral resolution in the continuum component does not affect the measurement of the emission lines and does not decrease the quality of the fit. 
\item Sampling the spectra at 1\AA\, per pixel. This step is required by the Starlight software and does not introduce any change in the spectrum.
\item Implement a sky line mask. The regions with possible sky contamination are not taken into account during the best fit. 
\end{enumerate}

\begin{figure}[!h]
\centering
 \includegraphics[width=0.8\textwidth]{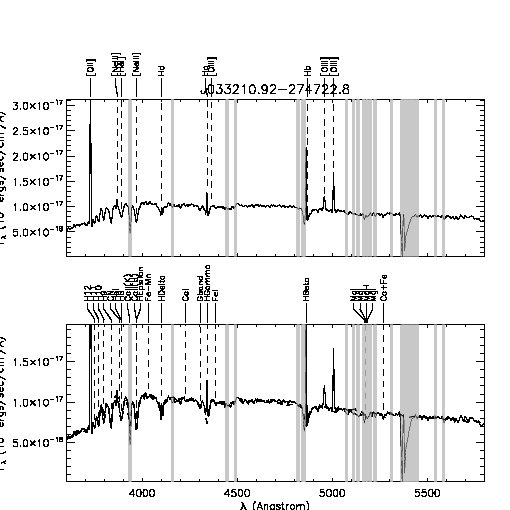} 
 \caption{{\small Rest-frame spectrum of one of the sample~A galaxies, J033210.92-274722.8. Grey boxes delimit the wavelength regions where spectra may be affected by strong sky lines. The continuum has been downgraded to lower spectral resolution than the emission lines in order to enhance the absorption features (see text above). \textbf{Top panel}: Spectrum with the location of strong emission lines. \textbf{Bottom panel}: Spectrum and synthetic spectrum using \citet{1984ApJS...56..257J} templates plotted with dashed line. The positions of the absorption lines are marked with dashed vertical lines.} } 
 \label{fit_cont} 
 \end{figure}

A rest-frame spectrum of one typical galaxy of our sample, J033210.92-274722.8, is given in Fig \ref{fit_cont}. 
After subtracting the stellar component, I have measured the flux of the emission lines using the SPLOT package. When the [OIII] $\lambda 4959$ emission line was not detected, the flux was assumed to be 0.33 times the [OIII]$\lambda 5007$ (the ratio of the transition probability). 

\section{Extinction}

I have measured extinction using two methods: the Balmer decrement (see Chap.\ref{Balmer decrement}) and $\mathrm{IR/H\beta}$ energy balance \citep{2004A&A...423..867L} (see Introduction Section \ref{Extinction from the SFR}). Comparing the two values, I can check the reliability of our methodology and avoid systematic effects on abundances due to a bias in the extinction evaluation. 

\subsection{Extinction estimation}
For 37 objects of the sample, the $\mathrm{H\gamma}$ line was detected and I could derive extinction in the gas phase directly from the spectra, using the observed $\mathrm{H\gamma/H\beta}$ ratio. Assuming a case B recombination with a density of $100\,\mathrm{cm^{-3}}$ and an electronic temperature of 10\,000\,K, the predicted ratio is 0.466 for the $I(\mathrm{H\gamma})/I(\mathrm{H\beta})$ \citep{1989agna.book.....O}. The median extinction of the sample is $A_V=1.53$. 

Because of large uncertainties related to the measurement of the $\mathrm{H\gamma}$ line, I have verified the quality of derived extinction using another method to evaluate $A_V$,  the balance of infrared and emission line SFR.The two SFR have been computed assuming the \citet{1998ARA&A..36..189K} calibrations and both use the same Salpeter  initial mass function (IMF) \citep{1955ApJ...121..161S}. The $L_{\mathrm{IR}}$ is estimated from the \citet{2001ApJ...556..562C} relation and $24\,\mu m$ observations \citep{2005ApJ...632..169L}. The $L_{\mathrm{H\alpha}}$ has been estimated from the flux of $\mathrm{H\beta}$ and assuming $\mathrm{H_\alpha}/\mathrm{H\beta}$=2.87 \citep{1989agna.book.....O}. The $\mathrm{H\beta}$ flux was corrected by an aperture factor derived from photometric magnitudes at $I_{\mathrm{AB}}$ and $V_{\mathrm{AB}}$ and spectral magnitudes. In the case of objects without IR detection, we used the detection limit of 0.08\,mJy of the MIPS catalogue to find an upper limit for $L_{\mathrm{IR}}$ and so an upper extinction limit $A_V(\mathrm{IR_{lim}})$. The median extinction in the sample for $\mathrm{IR/H\beta}$ estimation is $A_V=1.71$. 

\subsection{Comparison between the two derived extinctions}
\begin{figure}[!h]
\centering
\includegraphics[width=0.60\textwidth]{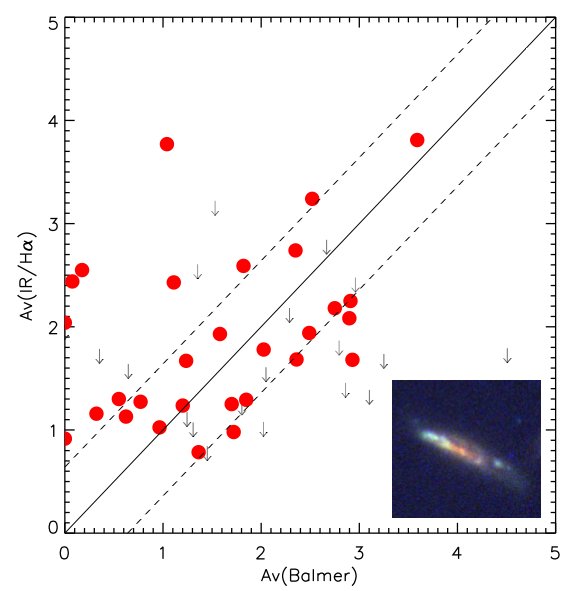}
\caption{{\small Relation between the extinction $A_V$ values derived from the Balmer decrement and by $SFR_{\mathrm{IR}}$ and $SFR_{\mathrm{H_\alpha nc}}$ ratio. We plot the 37 galaxies having reliable $\mathrm{H\gamma}$ detection and the 13 galaxies from Sample~B with $A_V$(Balmer) and $A_V$(IR) measurements. The objects with infrared detection are plotted with filled red circles. The arrows are the 21 objects the Sample~A with only $A_V$(IR) upper limit estimations. The two dashed lines refer to the results with $\pm$0.64\,rms. The lower right image in the graph is the combined ACS/HST image in B,V and I band of the edge-on galaxy J033212.39-274353.6.}}
\label{Extinction_FORS2}
\end{figure}

In Fig \ref{Extinction_FORS2}, I compare the extinction given by the two methods \citep{2004A&A...423..867L}, see Introduction Section 1.2 and 1.7.6. Most galaxies fall within the $\pm1\sigma$ dispersion. This result is consistent with \citet{2004A&A...415..885F} and \citet{2004A&A...423..867L}. 

Some discrepancies between the two measurements of extinction are due to geometrical properties, as in J033212.39-274353.6, J033232.13-275105.5 and CFRS 03.0932 which show $A_V$(IR) greater than $A_V$(Balmer). Indeed, these 3 objects are edge-on galaxies. In such cases, a large fraction of the disk is hidden by the dust in the disk plane. The detected optical Balmer lines trace the star formation of a few optically thin HII regions lying on the periphery of the galaxy. The consequence is an underestimated extinction value. The infrared radiation is not affected by dust so it comes from all regions of the galaxy. Thus, we expect to have higher extinction values when estimated with the IR flux.

\subsection{Uncertainty estimation}

\subsubsection{Balmer decrement extinction}
Due to the logarithmic term in the calculation of extinction (see equation \ref{c_extinxtion_eq} in Introduction), the extinction measured has asymmetrical error bars. The error results from the error on the flux measurement of the emission lines which is dominated by the noise of the sky. Therefore, the relative error of a line flux is the inverse of its S/N. This error propagates in the ratio of lines and results in an upper and lower extinction value. The following equations give the upper and lower value of color excess and extinction when using the $H\beta/H\gamma$ ratio:
\begin{eqnarray}
&E(B-V)_{up}&=\frac{1}{0.15}Log[(H\beta/H\gamma+errH\beta/H\gamma]\times0.466)\\
&E(B-V)_{low}&=\frac{1}{0.15}Log[(H\beta/H\gamma-errH\beta/H\gamma]\times0.466)\\
&\Delta A_{V\,up}&=(E(B-V)_{up}-E(B-V))\times R/1.47\\
&\Delta A_{V\,low}&=(E(B-V)-E(B-V)_{down})\times R/1.47\, ,
\end{eqnarray}

\subsubsection{SFR(Optical-to-IR) extinction}
$A_{\mathrm{V}}$(IR) error has been estimated taking into account the error on the IR luminosity and the Poisson noise of the $\mathrm{H\beta}$ line.

\subsubsection{Tabulated estimation}
The two extinctions do not trace the same quantity. Both methods are very dependent on the star formation history of galaxies. The Balmer decrement is sensitive to the extinction in regions which have experimented a star formation episode in the last 10\,Myr. The SFR(optical-to-IR) extinction is more sensitive to regions with slightly older stellar population (< 100\,Myr). However, when considering the all sample of galaxies, the two methods are statistically equal, see Fig. \ref{Extinction_FORS2}. The dispersion between the two methods is due to the uncertainties on the measurements and on the different star-formation history. As the objective is to correct the spectrum from the mean extension of the galaxy, I have assumed that the mean extinction can be evaluated from the mean value between $A_V$(IR) and $A_V$(Balmer). Adopting the mean values between the two methods allows us to control the systematic errors of each one. For galaxies without IR detection or without $H\gamma$ line, $A_V$ has been calculated according to the conditions summarized in table \ref{table_av}.  

For edge-on galaxies, the difference between the two extinction estimates is geometric as explained above. Since our aim is to correct the optical spectra, I have used $A_V$(Balmer) for these galaxies. 

\begin{table}[h!]
\begin{center}
\begin{tabular}{ l | c| c }
  &$H\gamma$ line detected&$H\gamma$ line not detected\\
 \hline
 IR detection& $A_V$= ($A_{V\,Balmer}$+$A_{V\,IR}$)/2& $A_V$=$A_{V\,IR}$\\
  \hline
 No IR detection& $A_V$= ($A_{V\,Balmer}$+$A_{V\,IR\,lim}$)/2& $A_V$=$A_{V\,IR\,lim}$\\
  \hline
  Edge-on galaxies& $A_{V\,Balmer}$& - \\
 \end{tabular}
\end{center}
\caption{{\small Equation use for the calculation of $A_V$ as function of the detectability of the $H\gamma$ line or IR luminosity.}}
\label{table_av}
\end{table}%
 
As the error on $A_{\mathrm{V}}$(Balmer) is dominated by the error on the $H_\gamma$ line, we can assume that the errors on $A_{\mathrm{V}}$(Balmer) and  $A_{\mathrm{V}}$(IR) are not correlated. Thus, the error of the final extinction depends on the error of the two extinction estimates. 

\section{Contamination by AGN radiation}
Before studying metallicities in galaxies we have to identify the objects whose lines are affected by contamination from an AGN, see Introduction Section 1.3. This identification is essential because the AGN processes affect  especially the [OIII] emission lines and thus also any metallicity estimate based in this line.  
First, I have eliminated from Sample~A two galaxies harboring evident AGN spectral features, like broad MgII and Balmer lines: J033208.66-274734.4 and J033230.22-274504.6. These two galaxies are very compact and are both X-ray detected. Broad Balmer lines suggest that these galaxies are Seyfert~I galaxies. 

\begin{figure}[!h]
\centering
\includegraphics[width=0.55\textwidth]{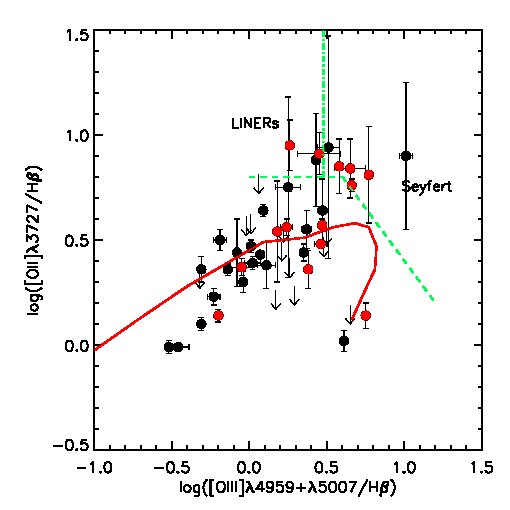}
\caption{{\small Diagnostic diagram for the Sample~A. Galaxies with only $A_V(\mathrm{IR_{lim}})$ are plotted with down arrows. The objects with log([NeII]$\lambda$3868/[OII]$\lambda$3727) $>$ -1.3  are plotted in red dots. The solid line shows the theoretical sequence from \citet{1985ApJS...57....1M} for extra-galactic HII regions. The dashed line shows the photo-ionization limit for a stellar temperature of 60\,000\,K and empirically delimits the Seyfert~2 and LINER from the HII regions. The dot-dashed line shows the demarcation between Seyfert~2 and LINERs from \citet{1989agna.book.....O} .}}
\label{AGN_FORS2}
\end{figure}

Then, the diagnostic diagram log([OII]]$\,\lambda$3727/$\mathrm{H\beta}$) vs log([OIII] $\lambda\lambda$4959, 5007/$\mathrm{H\beta}$) was used to distinguish the HII region-like objects from LINERs and Seyfert galaxies, see Fig. \ref{AGN_FORS2}. We have found 10 objects in the LINER and Seyfert~II regions of the excitation diagram. From these objects we discard 5 galaxies which fall out of the HII region, even if we account for error bars: J033219.32-274514.0, J033222.13-274344.5, J033240.04-274418.6, J033243.96-274503.5, J033245.51-275031.0. The other 5 galaxies are very close to the limit of star-forming galaxies to classify them definitely as AGN. I have chosen to keep them and we have used a different symbol for these objects in the figures. 

I search for evidence of shock processes in our galaxy sample by the presence of the emission line $[NeII]\lambda3868$. Five galaxies present log([NeII]$\lambda$3868/[OII]$\lambda$3727) $>$ -1.3 and fall in the Seyfert~II area. One of them,  J033236.72-274406.4 has IR and X-ray detection. 

\begin{figure}[!h]
\centering
\includegraphics[width=0.55\textwidth]{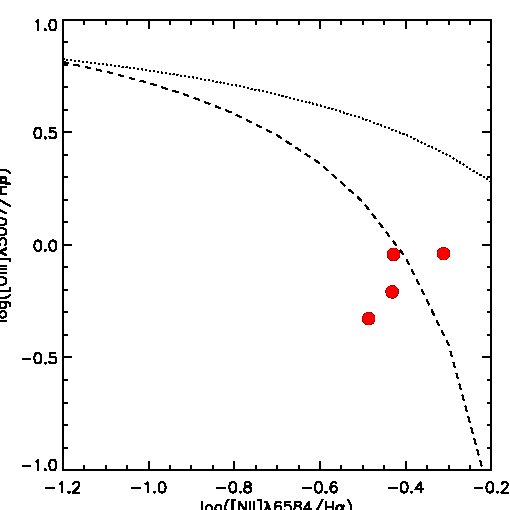}
\caption{{\small Diagnostic diagram for Sample~A. The doted line shows the  limit of the region occupied by star-forming galaxies from \citet{2006MNRAS.372..961K}. The dashed line represents the empirical demarcation separating star-formation galaxies from AGN from \citet{2003MNRAS.346.1055K}}}
\label{AGN2_FORS2}
\end{figure}

For 4 objects, $\mathrm{H_{\alpha}}$ and [NII] measurements are available and we have checked the log([NII]$\lambda$ 6584/$\mathrm{H\alpha})$ vs log([OIII] $\lambda\lambda$4959, 5007/$\mathrm{H\beta}$) diagnostic diagram, see fig. \ref{AGN2_FORS2}: all of them fall in the HII region delimited by \citet{2003MNRAS.346.1055K}.

\section{Outflows}
\label{Outflows sample}
I have determinate the $\Delta v$ in a sub-sample of 20 galaxies of sample\,A from the comparison of the velocity of emission lines with those of absorption lines, see Introduction Section \ref{Shocks and gas fall}. The results have been published in \citet{2010A&A...510A..68P}. Among these galaxies, I found systematic shifts in only three or possibly four of them (at a <100km/s level in J033224.60-274428.1, J033225.26- 274524.0, J033214.97-275005.5, and possibly in J033210.76- 274234.6). We have concluded that large-scale outflows do not play an important role in this representative sample of intermediate mass galaxies. 
\begin{figure}[!h]
\centering
\includegraphics[width=0.60\textwidth]{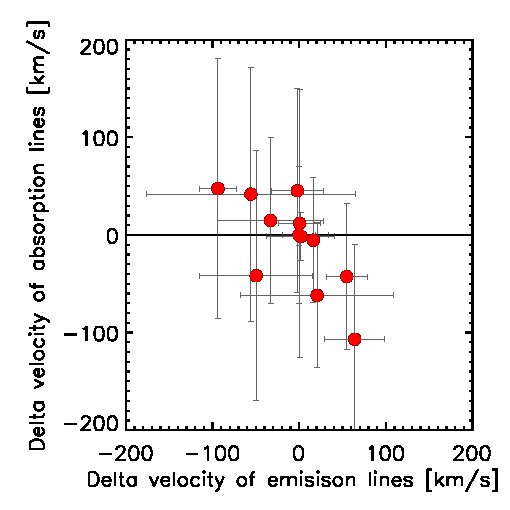}
\caption{{\small Velocities offset of emission and absorption lines compared to the median velocity in 12 galaxies of Sample~A. Delta velocity is defined as being the difference between the mean velocity deviated from all lines and the mean velocity considering only emission or absorption lines. }}
\label{feedback}
\end{figure}

Hints of feedback can be also detected from the emission line profiles, see Introduction Section \ref{Shocks and gas fall}. I have found only three galaxies in Sample\,A which present complex profile. It confirms the previous results: feedback may not be an important process in intermediate redshift galaxies. This is not surprising because all the galaxies considered here have large baryonic masses, in excess of $10^{10}\,M\odot$, and not very large star formation rates: as such the energy developed by massive stars is always negligible when compared to that from gravitational motions. 

\section{Metallicity}

One of the main aims of my thesis was to derived the metal content of distant galaxies and compare it with those of local galaxies in order to constrain the chemical evolution. To avoid bias arising from the metallicity calibration (see Introduction Fig. \ref{Kewley_MZ}), I have used the same strong line ratio and calibration in the FORS2 sample than those used by the reference local sample, \citet{2004ApJ...613..898T}. 

\subsection{Validity of the adopted calibration}
\begin{figure}[!h]
\centering
\includegraphics[width=0.55\textwidth]{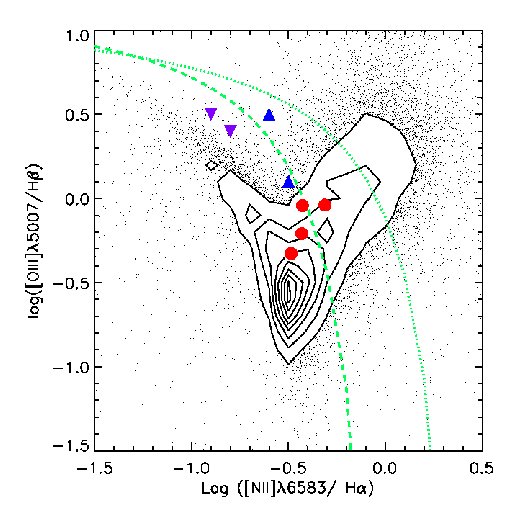}
\caption{{\small The local galaxies from the DR7 of SDSS are represented in the BPT diagnostic diagram in contours and black dots. The four intermediate redshift galaxies with NII detection from the FORS2 sample are plotted in red circle symbols. The up and down triangles represents galaxies at z=1.4 and z=1 respectively from Liu et al. 2008.  The green dotted lines delimit the regions dominated by star-forming galaxies (left) and AGN (right) according to Kauffmann et al. 2003 (long dash) and Kewley et al. 2001 (short dash). The intermediate redshift galaxies lie in the same locus than the SDSS galaxies and thus may have the same physical properties than local galaxies. Whereas, the high-z galaxies are offset to high [OIII]/$H_\beta$ ratio compared to the local sample. Liu et al. 2008 suggest that this offset can be due to higher ionization parameter in the HII region of high-z galaxies. }}
\label{SDSS_diag}
\end{figure}

\begin{figure}[!h]
\centering
\includegraphics[width=0.55\textwidth]{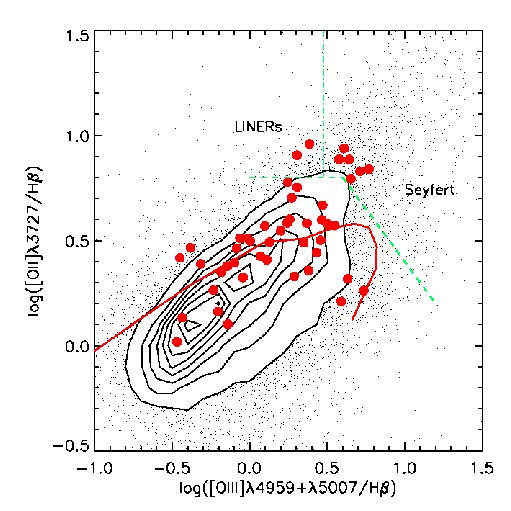}
\caption{{\small  The [OIII]/$H\beta$ vs [OII]/$H\beta$ diagnostic diagram for the local SDSS galaxies (in contours and dots) and the intermediate redshift galaxies from IMAGES (red circles). The solid line shows the theoretical sequence from McCall et al. 1985 for extra-galactic HII regions. The dashed line shows the photo-ionization limit for a stellar temperature of 60\,000\,K and empirically delimits the Seyfert~2 and LINER from the HII regions. The dot-dashed line shows the demarcation between Seyfert~2 and LINERs from \citet{1989agna.book.....O}. The z$\sim$0.6 galaxies lie in the same locus in the diagnostic diagram than the local galaxies and they thus may share the same physical properties. }}
\label{SDSS_diag_OII}
\end{figure}

The \citet{2004ApJ...613..898T} calibration is a strong-line method that uses the $R_{23}$ parameter. It has been derived in SDSS galaxies and may not be valid for galaxies having different properties that the SDSS sample, see discussion in Part I \ref{Remarks_Z}. 
Can we use the strong-line calibration based on SDSS galaxies sample to derive the metallicity of intermediate redshift galaxies? I have searched in the IMAGES sample for evidence of different physical condition in the HII region of z$\sim$0.6 galaxies by comparing their positions with those of local galaxies in diagnostic diagrams. As shown in Fig. \ref{SDSS_diag} and \ref{SDSS_diag_OII}, the intermediate redshifts galaxies lie in the same locus as the SDSS galaxies. There is no evidence that HII region in intermediate redshift have different physical conditions than the local counterparts. Therefore, the metallicity calibration based on SDSS sample can be applied to the FORS2 sample. This result is consistent with the previous work of \citet{1997ApJ...481...49H,1996MNRAS.281..847T} for galaxies at lower redshift 0<z<0.3.

\subsection{Metallicity estimated from the $R23$ parameter }
The \citet{2004ApJ...613..898T} metallicity calibration is defined by:
\begin{equation}
12+log(O/H)= 9.185-0.313x-0.264x^2-0.321x^3 \, , 
\end{equation}
where $x=\log{R_{23}}$. 

\begin{figure}[!h]
\centering
\includegraphics[width=0.5\textwidth]{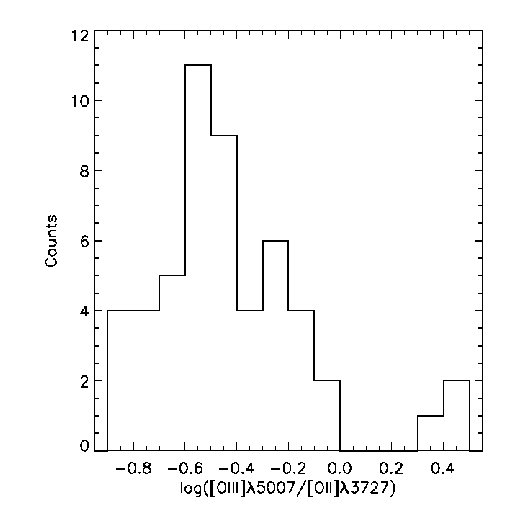}
\caption{{\small Histogram of the [OIII]$\lambda$5007\AA/[OII]$\lambda$3727\AA\, of the Sample A. All the galaxies have $O_3O_2$ ratio inferior to 2 and thus lie in the upper branch of the $R23$ diagnostic according to Maier et al. 2004 and Nagao et al. 2006. The three objects with $O_3O_2$> 2 are AGN candidates.  }}
\label{O2O3}
\end{figure}

\begin{figure}[!h]
\centering
\includegraphics[width=0.5\textwidth]{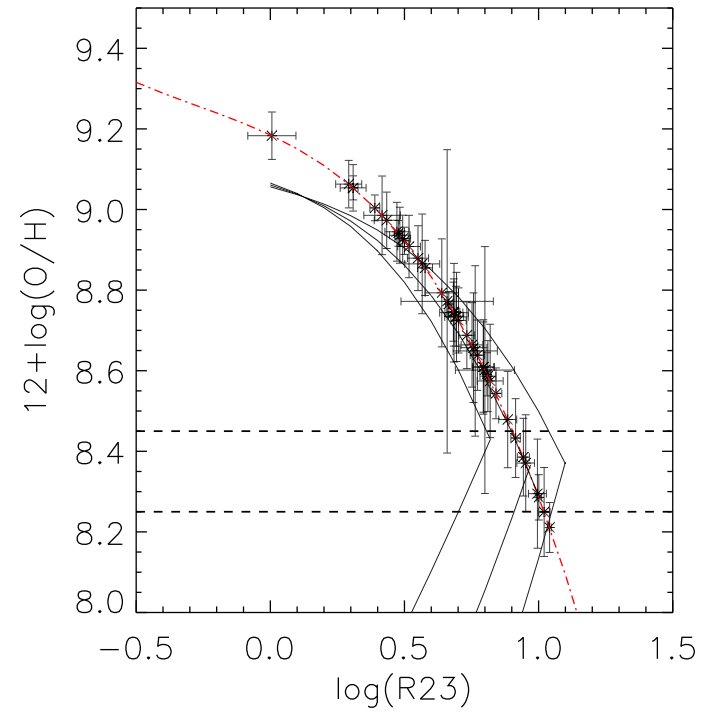}
\caption{{\small R23 vs log(O/H)+12 for the working sample. The red line is the \citet{2004ApJ...613..898T} calibration. The three dark solid lines are the calibration of \citet{1999ApJ...511..118K} for three values of $log(O_{32})$: -1.0, 0 and 1.0 from left to right. The two horizontal dashed lines delimitated from top to down the metal-rich, intermediate and metal-poor regions. For moderate metallicities, near the turning point of the relation, the uncertainties in selecting the appropriate branch are smaller than the uncertainties from sky and extinction.}}
\label{R23_OH_FORS2}
\end{figure}

The calibration from \citet{2004ApJ...613..898T} is valid only for the upper branch of the $R_{23}$ vs $12+log{O/H}$ relation.  I have assumed that our sample of intermediate-mass galaxies lies in the upper branch. In fact, galaxies in the extreme end of the lower branch are extremely rare at the given range of stellar mass and are associated with dwarf galaxies. For moderate metallicities, near the turning point of the relation, the uncertainties in selecting the appropriate branch is smaller than the uncertainties from sky and extinction, see Fig.\ref{R23_OH_FORS2}. I test this hypothesis with the N2O2 indicator. There are only 4 galaxies in our sample with [NII] measurements: all have $\log{f([NII])/f([OII])}> -1$ and therefore belong to the upper-branch \citep{2002ApJS..142...35K}. Maier et al. (2004) and later Nagao et al. (2006) have suggested to use the $O_3O_2$ ratio to select the valid branch of the $R_{23}$ calibration. Confronting observations from the SDSS to grids of photoionization models, Nagao et al. 2006 have shown that the observed $R_{23}$ belongs to the upper-branch of the $R_{23}$ sequence if $O_3O_2< 2$. Indeed, the ratio [OIII]$\lambda$5007\AA)/[OII]$\lambda$3727 is also a good metallicity diagnostic, thanks to the small dispersion of the ionization parameter at a given metallicity. The Figure \ref{O2O3} shows the distribution of the log($O_3O_2$) ratio for the FORS2 sample. Expect the AGN candidates, all galaxies verify  $O_3O_2$<2 and therefore the validity of using the upper branch of the $R_{23}$ sequence is confirmed. I think that, with a large level of confidence, all the objects lie in the upper branch of the $R_{23}$ vs $12+log(O/H)$. Nine objects have $\log{R_{23}}> 1$ where the \citet{2004ApJ...613..898T} calibration is not defined. In such cases I have adopted the limit given by $\log{R_{23}}= 1$.


\subsection{Metallicity estimated from the $R_3$ parameter }
For the 5 lowest redshift galaxies (type E3 in Table 2), the [OII] emission line falls out the wavelength range. I have used in this case the $R_3$ parameter defined by \citep{1984MNRAS.211..507E} :
\begin{equation}
R_3=1.35\times([OIII]\lambda5007/H\beta) 
\end{equation}

The oxygen abundance was estimated with the calibration proposed by \cite{1992ApJ...401..543V}:
\begin{equation}
12+log(O/H)=-0.60\times\log{R_3}-3.24 
\end{equation}
I have not found any evidence of a systematic bias between the $R_{23}$ and the $R_{3}$ calibrations in the Sample\,A. Liang et al. 2004a have arrived to the same conclusion for similar targets (Sample\,B). 

\subsection{Uncertainties}
The uncertainties in the data were assumed to be Poisson distributed. The error budget of emission line flux includes contributions from the sky and the object. The errors of line ratios and metallicity have been calculated using Monte-Carlo simulations taking into account the error on extinction and the sky error on line flux.

\section{Star formation rate}
The three SFR rates have been calculated using the calibration of \citet{1998ARA&A..36..189K} that assume a \citet{1955ApJ...121..161S} IMF.
 \subsubsection{SFR from the $H\alpha$ luminosity}
At high-z, the $H\alpha$ line is redshifted outside the optical window. In such a case, the SFR can be estimated from the higher order Balmer lines such as $H\beta$ line or $H\gamma$. The $H\alpha$ flux is estimated from the $H\beta$ flux by applying the theoretical ratio between $H\alpha$ and $H\beta$, see Part I section\ref{Balmer decrement}. In the FORS2 sample, the SFR($H\alpha$) have been estimated from the $H\beta$ luminosity since only 4 galaxies have a spectral range covering the $H\alpha$ line. The aperture correction has been determined using 2 or 3 bands (mean value), see Fig. \ref{fit_cont} of the introduction part. 

The uncertainties have been calculated by propagating the error on the measurement of the $H\beta$ flux and the error on the extinction.

 \subsubsection{SFR from the ultraviolet luminosity}

The SFR(UV) has been estimated in the sample using the rest frame luminosity at 2800\AA\, and equation 1.12 of the introduction part. The derived SFR(UV) has not been corrected from dust extinction. Thus, it only traces the unobscured star-formation. The absolute luminosity at 2800\AA\, has been evaluated by fitting the galaxy SED  with a linear combination of 6 stellar populations from Charlot \& Bruzual 2007 models and a two components extinction law of \citet{1989ApJ...345..245C}, see Chapter 5 for more details. This method does not give reliable derivation of global properties such as the stellar mass or the extinction, but it gives very robust absolute magnitudes. It should be seen as an interpolation of the SED. The conversion of the absolute magnitude in UV to luminosity is given by: 
\begin{equation}
L_{\nu}(UV)=10^{-0.4\times(M_{2800}-51.63)}
\end{equation}
The uncertainty on the SFR(UV) is derived by propagating the error on the determination of the $M_{2800}$ coming from the uncertainty of the photometry and on the SED fitting. The galaxies of this sample have U and B band observations from ACS/HST and in some cases UV photometry from GALEX. Therefore, the estimation of the magnitude at 2800\AA\, is very robust, random uncertainty < 0.3\,mag, because it results from the interpolation between the blue bands. 

\subsubsection{SFR from the infrared luminosity}
\label{SFR from the infrared luminosity}
To estimate the IR luminosity in the sample I have used the calibration of \cite{2001ApJ...556..562C}. These authors have implemented calibration between the 12$\mu m$, 15$\mu m$ -band luminosity and the bolometric IR luminosity. At z$\sim$0.6, the mean redshift of the FORS2 sample, the observed $24\,\mu m$ band coincide with the $15\mu m$ rest frame band. I thus use the flux at $24\mu m$ from the DR3 MIPS catalogue \citep{2005ApJ...632..169L} to calculate the luminosity at $15\mu m$ rest frame\footnote{The flux is given in $mJy$ in the  \citet{2005ApJ...632..169L}catalogue. The conversion of $mJy$ to $erg\,s^{-1}\,cm^{-2}$ is : \begin{equation}
  f_{24\mu m}(erg\,s^{-1}\,cm^{-2})=\frac{f_{24\mu m}(mJy)(1+z)}{L_\odot }\frac{c}{\nu_{24\mu m}}
\end{equation} }:
\begin{equation}
L_{(1+z)15\mu m}=4\pi(3.086\times10^{24}D_L)^2f(24\mu m)
\end{equation}
and then convert it into $L_{IR}$ using the calibration of \cite{2001ApJ...556..562C} : 
\begin{equation}
L_{IR}=11.1^{+5.5}_{-3.7}\times L^{0.998}_{15\mu m}
\end{equation}
For the 17 objects without MIPS detection, I have computed the upper limit to $SFR_{IR}$ from the $L_{IR}$ limit of detection given in Fig \ref{LeFloch2005}. The associated error has been calculated by propagating the flux error through the SFR calibrations. 

 
\subsubsection{Total SFR}
The total recent star formation can be estimated by the sum of $SFR_{IR}$ and $SFR_{UV}$. Indeed, SFR(UV) traces the UV light which has not been obscured by the dust and SFR(IR) traces the UV light thermally reprocessed by the dust. The sum of the two SFR estimators account for all the UV light emitted by the young stars. For the 17 objects without MIPS detection, I defined the total SFR as being the sum between $SFR_{H\alpha}$ and $SFR_{UV}$. 

\section{The gas radius}

\subsubsection{The gas radius from $R_{half}$}
For a thin exponential disk, the optical radius is 1.9 times the half light radius\footnote{The half light radius, $R_{half}$, is the radius of a circle drawn on the sky that included half of the galaxy light.} (Persic \& Salucci 1999). In a first approximation $R_{gas}$ is proportional to 1.9$\times R_{half}[UV]$. The $R_{half}$ has been estimated by Y. Yang from the rest-frame UV images. We have added the B and V bands HST images and derived for each galaxy an inclination, the PA and the $R_{half}$. The complete description of the method can be found in Delgado's PhD thesis and \citet{2008A&A...484..159N}.
 
 \subsubsection{The gas extension in distant galaxies}
 \label{Rgas}
\begin{figure}[!h]
\centering
\includegraphics[width=0.6\textwidth]{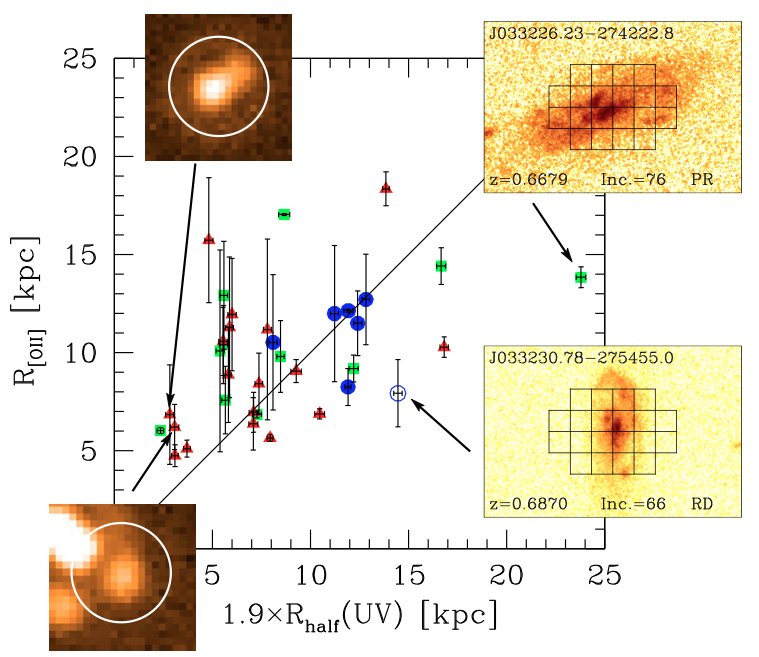}
\caption{{\small Comparison between the [OII] and UV radii for the IMAGES subsample of galaxies. The two insets on the left (30-arcsec wide) show MIPS imaging at 24 $\mu$m of J033228.48-274826.6 (upper one) and J033244.20-274733.5, while the two insets on the right show z-band HST/ACS images of J033230.78-275455.0 and J033226.23-274222.8 superimposed with the GIRAFFE IFU grid. The colors code the kinematic category defined by Yang et al. (2008): rotation disks are shown as blue dots, perturbed rotation as green squares, and complex kinematics as red triangles. The open blue circles correspond to the rotation disks+ galaxy for which the velocity measurement are more uncertain (see Puech et al. (2008)). From Puech et al. (2009)}}
\label{Rhalf_Puech09}
\end{figure}

Comparing the stellar extension from UV images to the extinction of the ionized gas, Puech et al. 2009 have found that the ionized gas is about 1.3 times more extended that the stellar component for z$\sim$0.7 galaxies, see Fig.\ref{Rhalf_Puech09}. Using the optical radius in UV light as a proxy of the $R_{gas}$ would thus lead to an underestimation of the gas fraction. 

 \subsubsection{Calibration $R_{half}$ to $R_{gas}$ for distant galaxies}
\begin{figure}[!h]
\centering
\includegraphics[width=0.6\textwidth]{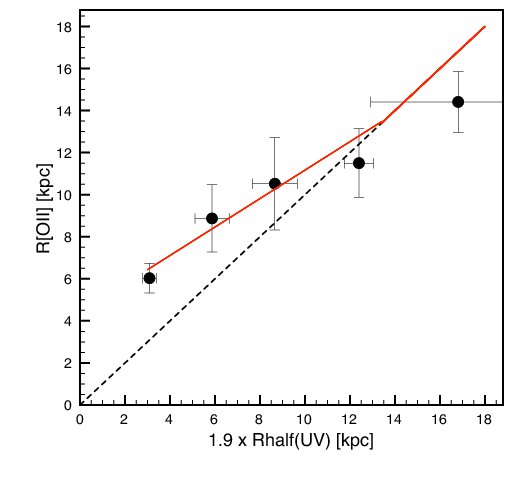}
\caption{{\small Conversion between the $R_{half}$ derived from UV observations and those derived from the extension of the ionized gas using IFU observations of the [OII] line. The data from Puech et al. 2009. have been binned in five interval of $R_{half}$ [UV]. A power law function have been fitted to the data (red line). Compact objects $R_{half} [UV]<5.9Kpc$ have ionized gas that extend at higher distance than their stellar component trace by their UV light. At higher $R_{half}$ [UV] the lower extension of the ionized gas compare to the stellar disk is not real but is due to the dimension of the IFU used to probe the [OII] emission. }}
\label{COnversion_UVtoOII}
\end{figure}

I have derived a corrective relation for using $R_{half}[UV]\times1.9$ as a proxy of $R_{gas}$. I used the sample of Puech et al. 2009 for which we have $R_{half}[UV]$ from (B+V) band images  and $R_{OII}$ from IFU observations.  I have binned the data in five intervals of $R_{half}[UV]$, see Fig.\ref{COnversion_UVtoOII} and adjusted a power law function to the data. The best fit is : 
\begin{equation}
R_{gas}=0.67\times (1.9\,R_{half}[UV]) + 4.40.
\end{equation}
Error bars on each bin have been evaluated by bootstrap resampling. This correction is valid for a compact system below $R_{half}[UV] < 6.8\,Kpc$. At larger $R_{half}[UV]$, the $R_{[OII]}$ is underestimated because galaxies extend father than the GIRAFFE IFU FoV (3$\times$2 arcsec). We then defined $R_{gas}$ as in Puech et al. 2009 : 
\[
\begin{array}{ccc}
R_{gas} =&1.27\,R_{half}[UV] + 4.40 &\, for\, R_{half}[UV] < 6.8\,kpc;    \\
R_{gas}=&1.9\, R_{half}[UV]&\, for\, R_{half}[UV] > 6.8\,kpc.    \\
\end{array}
\]
The error of $R_{gas}$ has been calculated adding the dispersion of the relation and the propagated error of $R_{half}[UV]$ on the calibration.

\section{Gas fraction}
\label{Gas fraction}
I have applied the inverse K-S law to estimate the gas fraction, see Introduction Section \ref{Gas fraction}. I have used the local K-S of Kennicutt (1998) with a power law indix of n=1.4. The $M_{gas}$ is given by:
\begin{equation}
M_{gas}(M_{\odot})=5.188\,10^8\times SFR_{Total}^{\,0.714} \times R_{gas}^{\,0.572}
\label{eq_mgas}
\end{equation}
with $R_{gas}$ in $Kpc$ and $SFR_{Total}$ in $M_{\odot}\, yr^{-1}$. The uncertainties on $M_{gas}$ have been calculated by propagating the error on the $SFR_{total}$ and on $R_{gas}$ in equation \ref{eq_mgas}.

\begin{figure}[!h]
\centering
\includegraphics[width=0.7\textwidth]{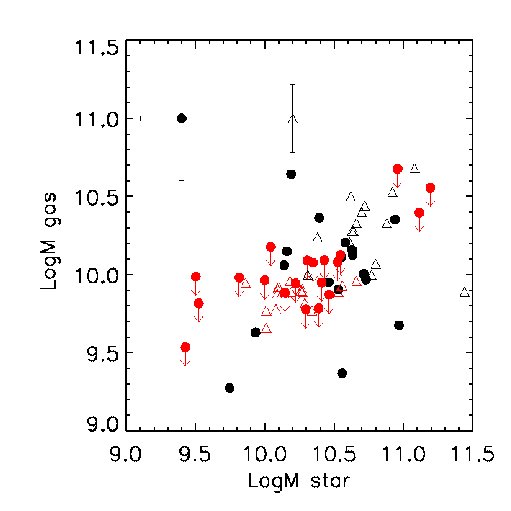}
\caption{{\small The gas mass derived from the inversion of the Kennicutt-Schmitt law as a function of the stellar mass derived from the absolute magnitude in K band. The red dots represent the objects not detected in the 24$\mu m$ observations. The sample of \citet{2010A&A...510A..68P} have been overploted with triangles. The red triangles are galaxies of the \citet{2010A&A...510A..68P} sample without 24$\mu m$ observations and for which $SFR_{IR}$ have been estimated using the MIPS detection \citep{2005ApJ...632..169L}. The typical error bars for the two samples are plotted in the top-left corner.}}
\label{Mstar_Mgas}
\end{figure}

 In a first approximation I have used the absolute K band magnitude derived from the SED fitting with $M_{stellar}/L_K$ expected by observed rest frame B-V color (Bell et al., 2003) and converted to a Salpeter (1955) IMF. The convertion between \citet{2001MNRAS.322..231K} and \citet{1955ApJ...121..161S} is described in \citet{2003ApJS..149..289B}. I will discuss in detail on the stellar mass in the next chapter. See also Hammer et al. (2010) for the detailed discussion on the effect on the gas fraction of the chosen semi-empirical model used to derive the stellar mass. Figure \ref{Mstar_Mgas} shows the gas mass as a function of the stellar mass for the current sample (in circle) and the sample of \citet{2010A&A...510A..68P} in triangles. 

  \EmptyNewPage

\FrameThisInToc
\chapter{The chemical evolution of the ISM}
\minitoc

The metallicity of the gas is a fundamental property of galaxies. It reflects the amount of gas reprocessed by stars (star formation history) and the exchange with environment. Studying the evolution with time of the metallicity then enable to disentangle the contributions of several processes taking place during galaxy evolution such as star-formation, outflows powered by supernovae or stellar winds and infall of gas by mergers or secular accretion. 
In the last five years, a large amount of galaxy surveys had collected gas metallicities in galaxy from low Universe to very high redshift, opening the study of the metal content evolution over a large lookback time interval. 

\section{Selection effect}

The z=0.6 sample has been selected in absolute luminosity in the near-IR, which is a good proxy of the stellar mass \citep{2010A&A...509A..78D}. I have shown in Chapter 2 that the sample is also representative of the luminosity function at the given redshift. However, since the aim of this work is to study the properties of the ionized ISM, only star-forming galaxies with emission lines EW([OII])>15\AA\, have been selected. Quiescent galaxies are not included in the sample since they have EW([OII])<15\AA. \citep{1997ApJ...481...49H} have shown that quiescent galaxies represent 40 \% of galaxies at z=0.6 and more than 80\% in the local Universe. However, quiescent galaxies have small fraction of gas and may thus have small impact in the chemical evolution of galaxies at the given redshift. Quiescent galaxies are thus a controlled source of systematics. 
However, an important bias due to selection effect can arise from the population of Low Surface Brightness galaxies (LSB). The studies in the local Universe have shown that LSB have low density mass, small SFR and low metallicity but have surprisingly high gas fraction (more than 50\% of their baryonic mass). Despite their high gas content, the gas is rather extended in LSB and their gas density are below the star-formation threshold of the Kennicutt-Schmidt law and explain why LSB have low SFR. At z=0, LSB represent a tiny fraction of galaxies, few percents, but their density number may increase with redshift. A strong evolution of LSB could have important consequences on the models of galaxy evolution.  

\section{The stellar mass-metallicity relation}

\subsection{Introduction}
\subsubsection{The origin of the M-Z relation}
The evolution of the metal content in the gaseous phase of galaxies is an important tool to understand the physical processes controlling the evolution of galaxies. On the one hand, the metallicity of the gas reflects the amount of gas reprocessed by stars and the exchange with the environment. On the other hand, the stellar mass reflects the amount of baryons locked into stars. Therefore, the study of the stellar mass-metallicity relation (M-Z) allows us to disentangle the contribution of several processes taking place during galaxy evolution such as star-formation history, outflows powered by supernovae or stellar winds and infall of gas by mergers or secular accretion. 

\begin{figure}[!h]
\centering
\includegraphics[width=0.50\textwidth]{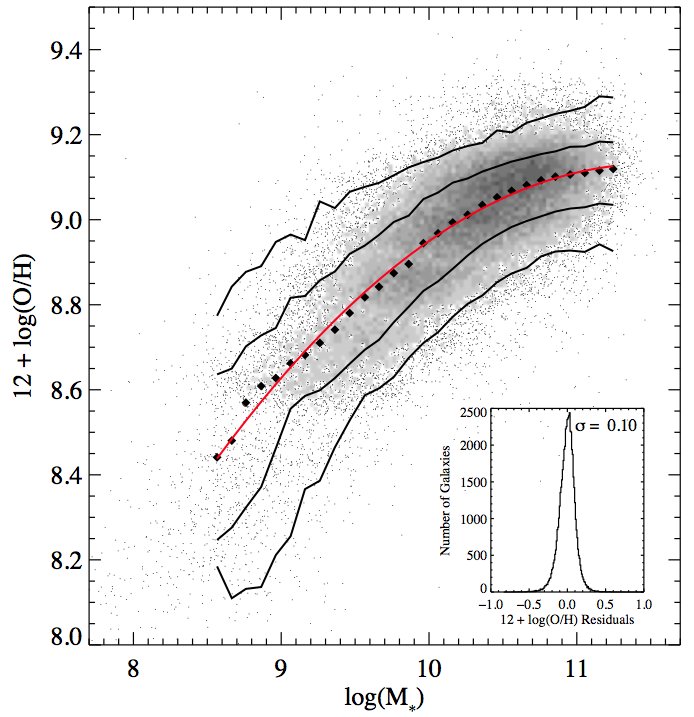}
\caption{{\small The relation between stellar mass, in units of solar masses, and gas-phase oxygen abundance for $\sim$53,400 star-forming galaxies in the SDSS. The large black points represent the median relation in bins of 0.1 dex in mass, which include at least 100 data points. The solid lines are the contours which enclose 68\% and 95\% of the data. The red line shows a polynomial fit to all the data. The inset plot shows the residuals of the fit. From \citet{2004ApJ...613..898T}. }}
\label{M-Z_tremonti04}
\end{figure}

The correlation between stellar mass and gas metallicity has been widely studied since \citet{1979A&A....80..155L}. Figure \ref{M-Z_tremonti04} shows the correlation found by  \citet{2004ApJ...613..898T} using $\sim$53\,000 local star-forming galaxies from the SDSS. This strong correlation reflects the fundamental role that galaxy mass plays on galaxy chemical evolution.  
There are two scenarios explaining the origin of the M-Z relation:
\begin{itemize}
\item The high metallicities in massive galaxies is caused by higher star-formation efficiencies in these galaxies. In other words, galaxies with higher masses have converted more rapidly their gas into stars than low mass galaxies. In this scenario, low mass galaxies are still converting their gas into stars and thus still enriching the ISM with metals. The local M-Z relation is thus a snapshot of a sequence in astration\footnote{The cycle in which interstellar material forms into stars, is enriched with heavy elements as a result of nuclear reactions, and is then returned to interstellar space via stellar winds, planetary nebulae or supernovae.} \citep{2004ApJ...613..898T}. This scenario is supported by gas fraction observations \citep{2001AJ....121..753B, 1997ApJ...481..689M, 2008AIPC.1035..180S}: low mass galaxies have larger gas fraction (40\% for $log\,M_{stellar}\sim9.5$, according to \citet{2008AIPC.1035..180S}) while massive galaxies have already exhausted their gas reservoir ($f_{gas}$=10\% at $log\,M_{stellar}\sim11$).
\item The low metal content of low mass galaxies is due to the loss of metals by gas outflows. Low mass galaxies have small gravitational potentials which cannot retain the ejection of metals from supernova winds \citep{1974MNRAS.169..229L}.  This scenario is supported by the observation of strong stellar outflows in local and distant starburst galaxies \citep{2000ApJS..129..493H} and by the presence of metals in the intergalactic medium \citep{1997MNRAS.290..623G, 2002ASPC..254..292H}. 
\end{itemize}
Nevertheless, these two scenarios are not incompatible since they are both driven by the gravity: massive galaxies evolve more rapidly due to their short gravitational collapse time scale and are not affected by outflows because their potential well prevent the ejection of gas outside the systems. On contrary, low mass galaxies have very long gravitational collapse time scale and they thus enriched their gas in metal more slowly. In addition, their potential well are not enough strong to retain the gas expelled by supernovae. 

There are alternative scenarios to explain the shape of the mass-metallicity relation. For example, \citet{2007MNRAS.375..673K} have suggested a scenario based on a variable integrated stellar initial mass function. They argued that at times of low star formation, the stellar initial mass function does not extend to as high masses as during high star formation epochs. This leads to both higher oxygen abundances and higher effective yields in strong starbursts

\subsubsection{The local M-Z relation}
An homogeneous estimation of oxygen abundance and stellar mass for a large sample of local galaxies by \citet{2004ApJ...613..898T} have enable to robustly establish the M-Z relation in the local Universe. However, the origin of the correlation is still in debate. According to \citet{2004ApJ...613..898T} the origin of the local M-Z is the preferential loss of metal in galaxies with $\log{M_{\mathrm{stellar}}}< 10$ (See Part II \ref{Yield}). On the contrary, \citet{2009MNRAS.396L..71V} have claimed that the M-Z relation is mainly driven by the star formation history and not by inflows or outflows for galaxies with present-day stellar masses down to $\log{M_{\mathrm{stellar}}}< 10$. These authors have used the same SDSS sample that \citet{2004ApJ...613..898T} but have derived stellar metallicities from a full-spectra-fitting method instead of gas metallicities derived from emission lines, which could explain the difference. 

\subsubsection{High-z studies}
In the past years, a large amount of spectroscopic observations of distant galaxies have been achieved. These surveys have enable the study of the M-Z relation up to $z\sim3$, which has shed more light on the following questions: Is the chemical evolution of distant galaxies driven by the same mechanism that local galaxies? Is the evolution of the metallicity in agreement with the theoretical scenarios of galaxy evolution? Several studies have characterized the relation at higher redshift: $0 < z <1$ \citep{2004ApJ...617..240K, 2006A&A...447..113L, 2005ApJ...635..260S, 2007IAUS..235..408L}, 1$\leq$z $\leq$2 \citep{2008ApJ...678..758L, 2006ApJ...644..813E,2004ApJ...612..108S, 2006ApJ...639..858M} and z$\geq$3 \citep{2008arXiv0806.2410M}. Most of these studies agree on a evolution toward lower metallicities at higher z of the M-Z relation. However, the evolution of the shift as a function of redshift and the possible change of the M-Z relation shape is still in debate. The discrepancies between studies may arise from the inhomogeneities on the methods used to measure metallicities, the spectra quality, and sample selection. 

I have shown in chapter 4 part I that reliables estimations of the metal content can only be achieved with spectra having at least moderate spectral resolution (R>1\,000) and a good signal-to-noise. At intermediate redshifts, there are still few studies reaching the spectral quality necessary to estimate with good accuracy the M-Z relation. As instance, the data from the Keck Redshift Survey \citep{2004ApJ...617..240K} are limited by the absence of flux calibration, making it impossible to estimate the extinction and thus metallicities \citep{2006A&A...447..113L}. I will discuss below on the possible systematics in the studies of \citet{2005ApJ...635..260S} and \citet{2009A&A...495...53L}.

\subsection{The stellar mass-metallicity relation at z$\sim$ 0.6}
\label{M-Z relation}
The aim of this work is to obtain a robust M-Z relation for a reasonably large sample of intermediate mass galaxies at z$\sim$0.6 and compare it with the local relation of  \citet{2004ApJ...613..898T}. I have followed the methodology proposed by \citet{2006A&A...447..113L}, which consists in a accurate estimate of the extinction and underlying Balmer absorption in order to obtain reliable $R_{23}$ abundance determinations, as described in the previous chapter. 

\begin{figure}[!h]
\centering
\includegraphics[width=0.70\textwidth]{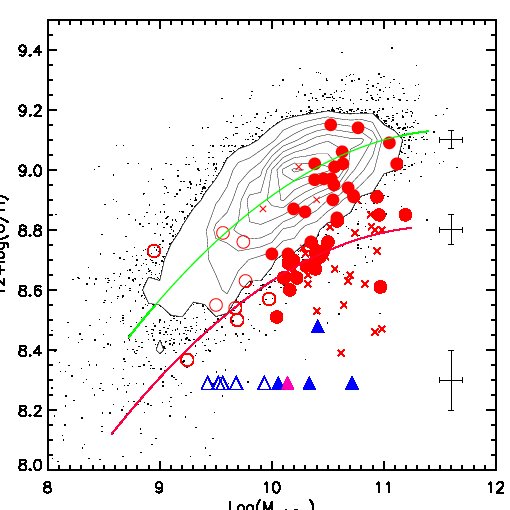}
\caption{{\small Stellar mass-metallicity relation for the SDSS galaxies in contours and dots (z=0, \citet{2004ApJ...613..898T}, \citet{2003ApJS..149..289B} see text ) and the 88 intermediate-z galaxies: sample A is shown as red circles and sample B as red crosses. Galaxies with masses lower than the completeness limit ($\log{M_{\mathrm{stellar}}}< 10$) are plotted with open symbols. The objects in sample~A with $\log{R_{23}}>1$ are plotted as small pink triangles and those detected as possible AGN are plotted as blue triangles. The typical uncertainty bars are plotted on the right. The red line indicates the median of the intermediate-z relation.The green line represents the local relation fit found by \citet{2004ApJ...613..898T}. The shift between the local relation and our data sample is $\Delta$[12+log(O/H)]= -0.31\,dex$\pm$0.03.}}
\label{M-Z_FORS2}
\end{figure}

I have compared the metal abundance of our sample of 88 distant galaxies (derived in previous the Chapter) with those of local starbursts from SDSS \citep{2004ApJ...613..898T}. I have found that distant galaxies are metal deficient compared to local starbursts, as shown in Fig. \ref{M-Z_FORS2}.  At z$\sim$0.7 the local relation found by \citet{2004ApJ...613..898T} is shifted toward lower metallicity by $\Delta$[12+log(O/H)]=0.31\,dex$\pm$0.03. 
The shift has been determinate by calculating for each galaxy the difference between its metallicity and the metallicity given by the \citet{2004ApJ...613..898T} relation for its stellar mass. The offset of the z$\sim$0.7 relation is the median of this difference for all galaxies with stellar masses over the limit completeness. The uncertainty was estimated using a bootstrap re-sampling method. There is a large dispersion of the data around the median relation at z$\sim$0.7, about $\pm$0.45\,dex which is intrinsic to the objects (see discussion in section 4.1 of \citet{2008A&A...492..371R} in annex). I discuss the origin of this dispersion in Sect. \ref{Toward a chemical enrichment scenario}

Given the small range in stellar mass covered by the sample, it was not possible to constrain the evolution of the shape of the M-Z relation at z$\sim$0.7. Thus, I have assumed that the slope of the M-Z relation remained unchanged compared to the local relation. 

The shift is assumed to be only due to a metallicity evolution. Indeed, \citet{2006A&A...447..113L} argued that the evolution in the M-Z relation is due to a diminution of the metal content in galaxies rather than an increase of their stellar mass. In fact, the amount of star formation necessary to increase the stellar mass by a factor of ten would imply that almost all the light in distant galaxies is associated with a very young stellar population. However, the spectra of these galaxies show a mixture of young, intermediate and old stellar populations (see next chapter). 


\subsubsection{Comparison with the other intermediate redshift studies}
\citet{2005ApJ...635..260S} have derived the M-Z relation at 0.4<z<1.0 for a sample of 56 galaxies. Half of the galaxies have been observed by the Gemini Deep Deep Survey and have flux-calibrated spectra. The rest of the sample have been selected from \citet{2003ApJ...597..730L} which used observations of the Canada-France Redshift Survey. They have found a temporal evolution in the M-Z relation with galaxies at z$\sim$0.7 tending toward to lower metallicity than local galaxies of similar masses. However, the shift is smaller than those found in the current study: $\sim-0.1 dex$ at $\log{M_{\mathrm{stellar}}}= 10$. They have also observed that the average metallicity of the most massive galaxies has not evolved from z$\sim$0.7 to the present day and thus have claimed this as evidence for downsizing effect in chemical evolution. The current work is in disagreement with those of \citet{2005ApJ...635..260S}, since I have found a stronger evolution of the relation toward lower metallicities at the same stellar masses: $-0.1\,dex$ in \citet{2005ApJ...635..260S} versus $-0.3\,dex$ in this work, for galaxies at $log\,M_{stellar}\sim10$. The effect of extinction may be one of the reasons for the difference between our observations and those of \citet{2005ApJ...635..260S}. In fact, extinction and stellar absorption under the Balmer lines have not been estimated for each galaxy. These authors have measured an average extinction and absorption correction in a composite spectrum of all galaxies in the GDDS sample and applied this average extinction, $A_V=1$, in all galaxies to correct emission lines. This extinction value is underestimated by about $\sim$1\,mag for at least one third of the \citet{2003ApJ...597..730L} sample, which corresponds to LIRGs. \citet{2006A&A...447..113L} have estimated that this underestimation of the extinction in the \citet{2003ApJ...597..730L} sample leads to an overestimation of $\sim$0.3\,dex on metallicity, which can explain the discrepancy with our sample.  Furthermore, the average extinction used by \citet{2005ApJ...635..260S} may be underestimated because the combining spectra used to measure $\mathrm{H\gamma}$ and $\mathrm{H\beta}$ lines is dominated by low extinction systems. Moreover, the majority of massive galaxies in the \citet{2005ApJ...635..260S} sample come from the measurements of \citet{2003ApJ...597..730L} which are from spectra with low S/N and spectral resolution. The galaxies in the GDDS sample in \citet{2005ApJ...635..260S} have low luminosities and also can be affected by the low S/N of the spectra which does not allow one to retrieve proper extinction from Balmer lines for individual galaxies. These effects are amplified for evolved massive galaxies, especially those experiencing successive bursts and containing a substantial fraction of A and F stars. Unfortunately these stars are predominant in intermediate mass galaxies, and thus contaminate extinction measurements by their large Balmer absorption. 

\citet{2009A&A...495...53L} have derived the mass-metallicity relation of star-forming galaxies up to z$\sim$0.9, using data from the VIMOS VLT Deep Survey.  At z$\sim0.77$, they have found that low mass galaxies, $log\,M_{stellar}=9.4\,M_\odot$, have metallicities  $-0.18\,dex$ lower than galaxies of similar masses in the local Universe. On the contrary to \citet{2005ApJ...635..260S}, they have observed that massive galaxies ($log\,M_{stellar}\sim10.2\,M_\odot$) show the stronger evolution and have much lower Z than their local counterparts ($-0.28\,dex$). They have concluded that the mass-metallicity relation is flatter at higher redshifts. 
The \citet{2009A&A...495...53L} sample and the IMAGES-FORS2 sample are not directly comparable because they do not cover the same range of stellar mass. The \citet{2009A&A...495...53L} sample is composed of galaxies with stellar masses in the range $8.9<log\,M_{stellar}<10.2\,M_\odot$ while stellar masses in the IMAGES sample have $log\,M_{stellar}>9.8\,M_\odot$. The larger evolution found in our sample, $-0.31\,dex$ for a mean $log\,M_{stellar}=10.5\,M_\odot$ at z$\sim$0.6, is compatible with their conclusion of a flatter mass-metallicity relation at higher redshifts. However, I have shown in Introduction Section \ref{New study comparing moderate and low spectral resolution spectroscopy} that with low resolution spectra, such as those of the VVDS sample (R=230), the metallicity can be underestimated. The effect is higher at low metallicities than in high metallicities due to the shape of the $log(O/H)+12$ vs. $R_{23}$ relation. This bias may explain part of the flatness of the M-Z relation found in \citet{2009A&A...495...53L}. 

\subsubsection{Influence of morphological and environment properties}
The effect of the environment and morphology on the M-Z relation has been recently investigated in the local Universe. Kewley et al. (2006) have found that the metallicity in the nuclear regions of galaxies in very close pairs is displaced to lower levels than those measured in more distant pairs and isolated galaxies. The authors have claimed that this effect is caused by gas inflows triggered by interactions. I have searched for systematic effects due to environment properties by isolating the sub-sample of galaxies standing in the CDFS structures, situated at z=0.670 and z=0.735. Metal abundance may be higher in these galaxies because of higher merger rates and then faster galaxy evolution. Nevertheless, I have found that the galaxies in the CDFS structures are located in the same locus as the rest of the sample and thus there is no evidence of any bias. This is probably due to an effect of data selection. The working sample majority includes starburst galaxies with intermediate mass. The evolved galaxies in structures are expected to be massive galaxies with poor star-formation.

During my thesis, I have also investigated the influence of other galaxy properties in their locus in the M-Z relation.  In \citet{2010A&A...514A..24H} (see article in annex), we have investigated the properties of long-duration gamma-ray burst host (GRB) and in particular the location of GRB host in the M-Z relation. We found that GRB host show obvious discrepancies from M-Z relation derived from low redshift galaxies from several samples: they have lower metallicities compared with other galaxies from those samples. This observation and the presence of Wolf-Rayet stars in the spectra of GRB host have enable us to shed more light on the GRB progenitors. In fact these observations may indicate that long-duration GRB are associated to the first stages of star formation in pristine regions of the host galaxy and thus support the "core-collapsar" model for this kind of GRB.

\subsection{Evolution of the stellar mass-metallicity relation}
\subsubsection{Evolution in the IMAGES sample}
\begin{figure}[!h]
\centering
\includegraphics[width=0.90\textwidth]{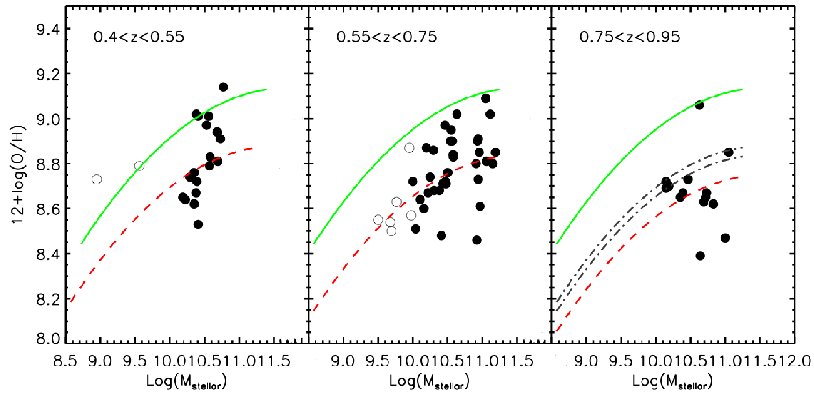}
\caption{{\small Evolution of the M-Z relation inside the Sample~A+B. \textbf{Left Panel}:The 17 galaxies from the $0.4\leq z<0.55$ bin. \textbf{Middle Panel}:The 38 galaxies from the $0.55\leq z<0.75$ bin. \textbf{Right Panel}: The 15 galaxies from the $0.75\leq z\leq 0.95$ bin. The galaxies with masses under the completeness limit are plotted with open circles. Objects with $log(R23)>1$ and AGN are not plotted. The median of each bin is marked in dashed red line, see Table \ref{offset_evo} for the $\Delta$[12+log(O/H)]. The local relation from \citet{2004ApJ...613..898T} is marked with a green line. Only galaxies over the completeness limit $\log{M_{\mathrm{stellar}}}< 10$ have been taken into account to calculate the median shift. The two black lines in the right panel are the median of the two previous bin.}}
\label{M-Z_multi}
\end{figure}

I searched for a redshift evolution of the metal content inside the sample of intermediate redshift galaxies. As the sample covers a large range in redshift, from z=0.4 to z=0.95, I have split it into 3 redshift bins: $0.4<z<0.55$, $0.55<z<0.75$ and $0.75<z<0.95$. In each redshift bin, I have measured the median shift of the local relation toward lower metallicity. The uncertainty on the median offset was calculated by a bootstrap resampling method. Fig. \ref{M-Z_multi} shows the evolution of the M-Z relation along the 3 redshift bins. The median offsets from the local relation in the 3 bins are given in table \ref{offset_evo}. I find an evolution of the M-Z relation from z$\sim$0.45 to z$\sim$0.85 of $\Delta$[12+log(O/H)]=$-$0.13\,dex. 
\begin{table} 
\label{offset_evo} 
\centering
\begin{tabular}{c c c} 
\hline
Redshift bin & $\Delta [12+log(O/H)]$& $ error \Delta [12+log(O/H)]$ \\  
\hline 
$0.40\leq z<0.55$ & -0.26 & 0.08 \\
$0.55\leq z<0.75$ & -0.30 & 0.04 \\
$0.75 \leq z \leq 0.95$ & -0.39 & 0.09 \\
\hline 
\end{tabular} 
\caption{Offset between local relation and the 3 redshift bin.} 
\end{table} 

\subsubsection{Evolution in the literature}
Using samples from the literature at different redshifts, I have evaluated the evolution of the M-Z relation from z$\sim$3 to local. The comparison between different samples needs consistent measurements of oxygen abundance and stellar mass. In fact, particular care has to be taken when comparing metallicities from different metallicity estimators, especially when the aim is to evaluate the offset between samples. The systematic error due to the use of different calibrations be as high as 0.5\,dex and thus can be higher than the metal evolution, cf. \citet{2008ApJ...674..172R,2008arXiv0801.1849K}. I have thus trying to minimize the possible sources of bias induced by the unhomogeneous estimation of stellar mass and metallicity. 

The samples used in the comparison are described below:  
\begin{itemize}
\item \citet{2009A&A...505..529L} have derived the M-Z relation of massive ($log\,M_{stellar}>10.5\,M_\odot$) star forming galaxies from SDSS-DR5. They have found a decrement of $\sim0.1\,dex$ in 12 + log(O/H) from redshift 0 to 0.4. The metallicity has been estimated using the $R_{23}$ method with the \citet{2004ApJ...613..898T} calibration. The stellar masses have been derived from full-spectra fitting method using the STARLIGHT software. According to \citet{2006MNRAS.370..721M}, this method gives stellar masses $0.2\,dex$ higher than those derived from bayesian method \citep{2003MNRAS.341...33K,2004ApJ...613..898T}.
\item \citet{2008ApJ...678..758L} published the mass-metallicity relation for 20 galaxies from the Deep 2 Galaxy Redshift Survey at $1.0<z<1.5$. We only considered the 7 galaxy spectra having individual oxygen measurements. As only [OIII], $\mathrm{H\beta}$, [NII] and $\mathrm{H_\alpha}$ emission line measurements were available from the NIRSPEC/Keck observations, the $N2$ parameter was used with the \citet{2004MNRAS.348L..59P} calibrator to estimate their metallicity. The stellar masses were calculated from K-band photometry and the procedure proposed by \citet{2005ApJ...625..621B}.
\item At z$\sim$2, \citet{2004ApJ...612..108S} observed 7 star-forming galaxies. Their oxygen abundance was estimated using the $N2$ parameter and the \citet{2004MNRAS.348L..59P} calibration. Their stellar masses were calculated by fitting their spectral energy distribution. 
\item At the same redshift, $\langle z\rangle$=2.26, \citet{2006ApJ...644..813E} have measured the metal abundance of 6 composite spectra from 87 star-forming galaxies selected by ultraviolet rest-frame luminosity and binned by stellar mass. They used the same calibration for estimating $12+log(O/H)$ and the same stellar mass estimates as \citet{2004ApJ...612..108S}.
\item \citet{2008arXiv0806.2410M} presented preliminary results from the AMAZE large program to constrain the M-Z relation at z=3, see also \citet{2009MNRAS.398.1915M}.The oxygen abundance was measured by the $R_{23}$ parameter and following the procedure described by \citet{2006A&A...459...85N}. We only selected the 5 objects above our stellar mass completeness limit and used the $R_{23}$ measurement to estimate the metal abundance using the \citet{2004ApJ...613..898T} calibration. The stellar masses have been derived by fitting the spectral energy distribution to multi-band photometry using \textit{Hyperzmass} \citep{2007A&A...474..443P}.
\end{itemize}

\begin{figure}[!h]
\centering
\includegraphics[width=0.60\textwidth]{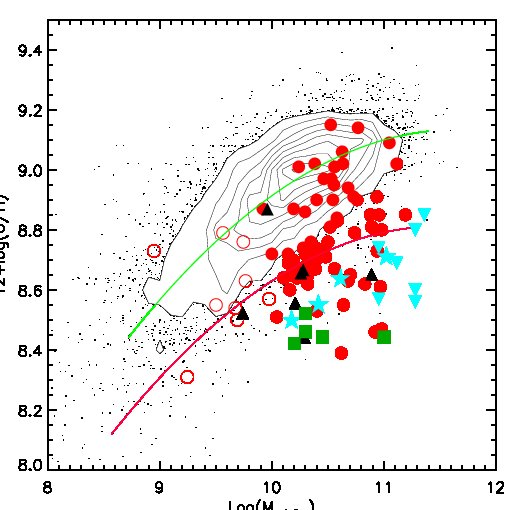}
\caption{{\small Stellar mass-metallicity relation for the SDSS galaxies in contours and dots (z=0, \citep{2004ApJ...613..898T, 2003ApJS..149..289B} ) and the 58 intermediate-z galaxies above the completeness mass limit in red filled circle. Galaxies below the completeness mass limit are plotted with red open circles. The local relation is plotted in green \citep{2004ApJ...613..898T} and the median of the intermediate-z relation is in red. The 7 individual galaxies at z$>$1 of \citet{2008ApJ...678..758L} are plotted as black triangles. The two z$>$2 samples are plotted in blue: inverted blue triangles indicate the 7 star-forming galaxies of \citet{2004ApJ...612..108S}; blue stars indicate the 5 stellar mass bin of \citet{2006ApJ...644..813E}. The $\mathrm{z}>3$ sample of Maiolino et al. (2008) is plotted in green squares.}}
\label{Z_evol}
\end{figure}

Three samples use the $N2$ parameter to evaluate the oxygen abundance. Unfortunately, metallicity estimation based on $N2$ can give strongly discrepant results compared to those derived from $R_{23}$ and thus cannot be directly compared. To compare the galaxy metallicities using the $N2$ parameter I decided to re-evaluate the metallicities using the \citet{2002MNRAS.330...69D} calibration. This calibration is similar to the \citet{2004ApJ...613..898T} one and other $R_{23}$ based calibrations in the working range $8.4\leq12+log(O/H)\geq8.7$ \citep{2008arXiv0801.1849K,2004ApJ...613..898T}. For the galaxies in the sample having [NII] measurements, I compared the metallicity to those of several $N2$ calibrations. I confirmed that the \citet{2002MNRAS.330...69D} calibration gives similar results to the abundances estimated in this work. The M-Z diagram for the local, intermediate and high redshift samples are plotted in Fig. \ref{Z_evol}. See \citet{2008A&A...492..371R} in annex for discussion of systematics. 

\subsubsection{The chemical evolution over the past 8 Gyr}
\begin{figure}[!h]
\centering
\includegraphics[width=0.50\textwidth]{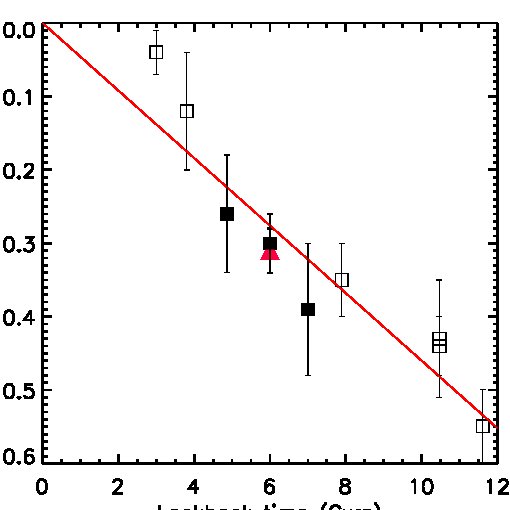}
\caption{{\small The metallicity shift from the local relation of the 4 high-z samples as a function of lookback time is plotted as black open squares. The metallicity shift for the 3 redshift bins of the IMAGES sample are in black squares and the median of the 3 bins is a red triangle.}}
\label{lookbacktime}
\end{figure}

I have measured the median offset of the local relation and the associated uncertainty in all the high-z samples. The offset $\Delta$(O/H) for the 3 redshift bins of the intermediate galaxies and for the four high-z samples are plotted in Fig. \ref{lookbacktime} as a function of lookback time. I found that the evolution of the mean metallicities from the local Universe to a lookback time of $\sim$12\,Gyr is linear with a slope given by:
\begin{equation}
\log{Z_{\mathrm{tb}}/Z_0} = -0.046\times t_{\mathrm{Gyrs}} 
\end{equation}
I have assumed that the shape of the M-Z relation remains constant over the past 8\,Gyr. Notwithstanding, several authors have claimed a change of the slope in the relation at higher redshift \citep{2005ApJ...635..260S, 2009A&A...495...53L, 2006ApJ...644..813E}, but the effect seems not to be very strong within this range of stellar mass and the change of slope is a source of intense debate in the community.

The main source of systematic uncertainty is the one induced by the metallicity calibrations, mainly the difference between the calibrations based on $N2$ and $R_{23}$. However, even using the \citet{2004MNRAS.348L..59P} calibration instead of the \citet{2002MNRAS.330...69D} calibration, we find a linear evolution of the metal content in galaxies with a similar slope. As the evolution of the M-Z relation is strongly constrained at z$\sim$0.6, any bias in the high-z sample will not change the shape of the $\Delta(O/H)-z$ relation, but only its slope. 
 
\section{The star-formation efficiency}
\label{The star-formation efficiency}
I have estimated the doubling time, defined by $t_{\mathrm{D}}=M_{\mathrm{stellar}}/SFR_{\mathrm{Total}}$, for emission line galaxies in the IMAGES sample. The SFR for galaxies without IR detection was estimated using the $L_{\mathrm{IR}}$ limit of detection of MIPS/Spitzer and the $SFR_{\mathrm{2800\AA}}$ (see previous the Chapter). 

I confirm that massive galaxies convert less gas into stars than lower mass galaxies at $0.5<z<1$: e.g the $t_{\mathrm{D}}$ for a massive galaxy ($\log{M_{\mathrm{stellar}}}$=11) is above 3\,Gyr whereas a less massive galaxy (log$M_\mathrm{\mathrm{stellar}}$=10.3) doubles its stellar content in less than 1\,Gyr. This effect is usually called "downsizing": low mass galaxies form stars at later epochs than massive galaxies. 

\begin{figure} [!h]
\centering
\includegraphics[width=0.60\textwidth]{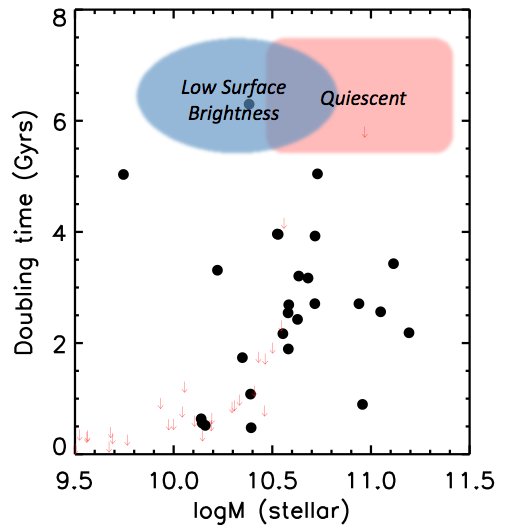}
\caption{{\small Doubling time, defined by $t_{\mathrm{D}}=M_{\mathrm{stellar}}/SFR_{\mathrm{IR}}$, for emission line galaxies in the IMAGES sample. The downward arrows are galaxies without $24\,\mu m$ detection for which the SFR has been estimated using the the detection limit $L_{\mathrm{IR}}$ limit of detection of MIPS/Spitzer. The approximated location of quiescent and LSB galaxies, which are not included in the sample, are represented by red and blue region.}}
 \label{double_hist} 
 \end{figure} 

The small values of the doubling time in intermediate mass galaxies at z$\sim$0.6 (see Figure \ref{double_hist}) confirm the strong evolution of these sources at this epoch. I have  found that the doubling time diminishes with redshift which is mainly due to the star-formation activity increase with redshift.
 
\section{Evolution of the gas fraction}
In this section I derive the gas fraction of a representative sample of $z\sim0.6$ intermediate mass galaxies. I have gathered high-z samples with gas fraction estimation and compared them with the $z\sim0.6$ galaxy sample. In order to compare homogenous gas fractions and stellar masses at different redshifts, I have converted all the gas fraction and stellar masses to the same IMF. The stellar masses are all given in Salpeter 'diet' IMF \citep{2003ApJS..149..289B} and the SFR used to derive the gas mass are in 'regular' Salpeter IMF. \citet{2009A&A...507.1313H} have argued that the stellar masses derived with 'diet' Salpeter IMF with the \citet{2003ApJS..149..289B} recipe are roughly equal to those derived from \citet{2003MNRAS.344.1000B} models and a 'regular' Salpeter IMF. I have thus not corrected the stellar masses and SFR for this two different IMFs.   

\subsection{The gas fraction at z$\sim$0.6}
\label{The gas fraction}
\subsubsection{Description of the data sample}
In Section \ref{Gas fraction}, I have derived gas fractions for a sample of 45 intermediate mass galaxies, at $z\sim0.6$, hereafter Sample\,A. I have completed this sample with 36 galaxies from \citet{2010A&A...510A..68P}. The galaxies in \citet{2010A&A...510A..68P} have been observed with 3D spectroscopy with VLT/GIRAFFE in the framework of IMAGES survey, see Section \ref{Philosophy_IMAGES}. The targets have been selected with the same selection criteria that those of Sample\,A: $M_J>-20.3$ and $EW(OII)>15$\AA. The two samples have homogeneous measurements of SFRs and $M_{stellar}$. Moreover, the gas radius of Sample\,A galaxies have been derived using a calibration established from the extension of the ionized gas measured in the 3D observations of \citet{2010A&A...510A..68P}, see section \ref{Rgas}. The final sample (i.e., FORS2+GIRAFFE) is composed of 81 galaxies with a mean redshift of 0.666.   

\begin{figure}[!h]
\centering
\includegraphics[width=0.5\textwidth]{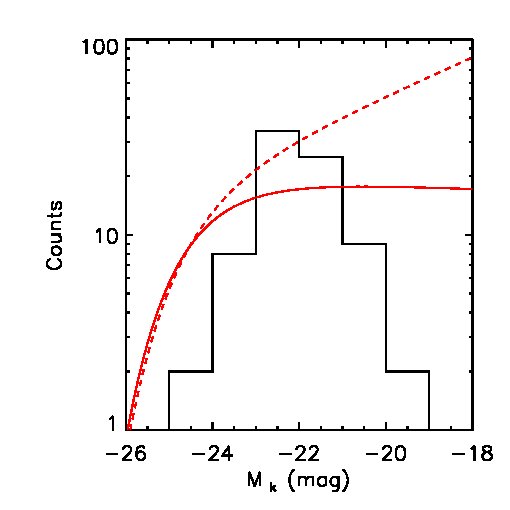}
\caption{{\small Distribution of the absolute magnitude in K-band for the gas sample (in black). The red lines are the luminosity function from \citet{2003A&A...402..837P} at z$\sim$0.5 in dashed-line and at z$\sim$1 in solid line.}}
\label{hist_Mk_yield}
\end{figure}

I have verified the representativity of the IMAGES sample by comparing the distribution of their AB absolute magnitudes in K-band with the K-band luminosity function at z=0.5 and z=1 from \citet{2003A&A...402..837P}, see fig. \ref{hist_Mk_yield}. Kolmogorov-Smirnov test supports that the $M_K$ distribution of sample A+IMAGES and those of the luminosity function, in the redshift range z=0.4 and z=0.98, have 88\% probability to arise from the same distribution. The sample is representative in the mass interval $log\,M_{stellar}$>9.8.

\subsubsection{Gas fraction vs. stellar mass}
The properties of the final sample and the two sub-samples are summarized in table \ref{gas samples}. The uncertainty on the median gas fraction has been estimated by bootstrap resampling. 
\begin{table} [!h]
\centering
\begin{tabular}{l | c c c c c} 

Sample &N&$<z>$&$<R_{gas}>$&$<log\,M_{stellar}>$&$<\mu>$ \\ 
&&&$kpc$&$dex$&\% \\ 
\hline 
Sample\,A&45&0.666&8.58&10.30& 34.3$\pm$5.5\\
$log\,M_{stellar}>9.8$&35&0.669&9.17&10.43&27.4\\
\hline
Puech et al. 2010&36&0.668&10.09&10.38& 30.8$\pm$9\\
\hline 
Sample IMAGES&81&0.666&9.17&10.38&32.9$\pm$3 \\
$log\,M_{stellar}>9.8$&70&0.668&9.23&10.41&30.3$\pm$13\\
\end{tabular} 
\caption{ {\small Main properties of the gas sample.}} 
\label{gas samples}
\end{table} 

\begin{figure}[!h]
\centering
\includegraphics[width=0.7\textwidth]{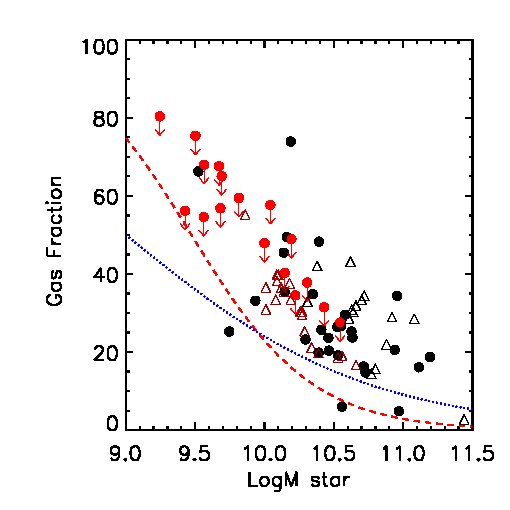}
\caption{{\small The gas fraction derived from the inversion of the Kennicutt-Schmitt law as a function of the stellar mass derived from the absolute magnitude in K band. The two sub-samples composing the IMAGES sample have been plotted with different symbols: Sample\,A with circle and \citet{2010A&A...510A..68P} with triangles. The color red codes the availability of $24\mu m$ observations: red symbols are galaxies without MIPS detections. For these galaxies the total SFR has been estimated using the $SFR_{H\beta}$ (extinction and aperture corrected) for the Sample\,A data and $SFR_{lim}$ from the MIPS detection limit at z$\sim$0.6 (Le Floc'h 2005) for \citet{2010A&A...510A..68P} galaxies. The red dashed-line represents the effect of the limit detection of the SFR on the gas fraction. }}
\label{GasFrac}
\end{figure}

Fig.\ref{GasFrac} puts into evidence the strong correlation between the gas fraction and the stellar mass: lower stellar mass galaxies have higher fraction of gas. This relation have already been observed in the local Universe \citep{1997ApJ...481..689M,2000MNRAS.312..497B, 2004ApJ...611L..89K} and at higher redshift, e.g. \citet{2006ApJ...646..107E} at z$\sim$2 and \citet{2005A&A...433..807M} at z$\sim$3. 
 
\subsubsection{The $\mu$ vs. $logM_{stellar}$ relation}
\begin{figure}[!h]
\centering
\includegraphics[width=0.6\textwidth]{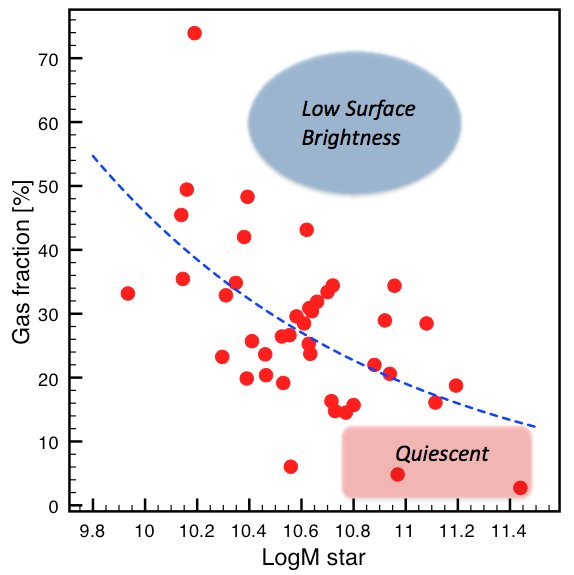}
\caption{{\small  The gas fraction $\mu$ vs $logM_{stellar}$ for the IMAGES sample having IR detection. The data has been fitted with an exponential law, see text above. The 1$\sigma$ dispersion of the data is about $\pm$ 14\%. The approximated location of quiescent and LSB galaxies, which are not included in the sample, are represented by red and blue region.}}
\label{GasFrac_fit}
\end{figure}

I have fitted the $\mu$ vs $logM_{stellar}$ relation with an exponential function. If only galaxies having secure total SFR (with IR detection) and stellar mass over the limit of completeness are taken into account, I get: 
\begin{equation}
\mu = 3.10^5\,e^{-0.88\times logM_{stellar}}.
\label{equation_gas}
\end{equation}
The solution of the best-fit is shown in fig.\ref{GasFrac_fit}. The 1$\sigma$ spread of the data is about $\pm$ 14\% in $\mu$.

 \subsubsection{Comparison with local sample}
The gas fractions in z$\sim$0.6 intermediate mass galaxies are larger than in the local Universe.  For reference, the  Milky Way has a gas fraction of 12\% (Flynn et al. 2006) and M 31 has $\mu$=5\% (Carignan et al. 2006). In the local Universe, the gas fractions derived from the K-S law are twice lower than those of z$\sim$0.6 galaxies (ALFALFA survey by \citet{2008AIPC.1035..180S}). Compared to the local relation of \citet{2008AIPC.1035..180S}, the $z\sim0.6$ galaxies have $\sim$10\% (20\%) higher gas fraction than their local counterparts at $log\,M_{stellar}=10.43\,M_\odot$ (9.8$\,M_\odot$). 

\subsubsection{Is the sample SFR-limited?}
By construction (see \ref{The Metallicity sample}), the IMAGES sample is made up of emission lines galaxies with EW($H\beta$)>15\AA\, and hence it is SFR-limited. This limit in SFR induces a lower limit on the $M_{gas}$ which can be detected. Such a limit affects the low mass region in which only the very gas-rich objects can be detected. 
The $\mu$ detection limit has been evaluated using $SFR_{lim}$=2$M_\odot$/yr. As the dependence of $R_{gas}$ on the stellar mass is small, a possible evolution of the mass-size relation does not dramatically affect the calculation of the $\mu$ detection limit. In fig.\ref{GasFrac}, I have overploted the $\mu$ detection limit as a function of the stellar mass using a red dashed line. The $z\sim$0.6 galaxies are well over the SFR limit threshold. I have concluded that the shape of the $\mu$ vs. $logM_{star}$ relation and the higher gas fraction found are real and are not due to any bias from the SFR-limit. 

\subsection{Gas fractions at z>1.5 from the literature}

\subsubsection{\citet{2010ApJ...713..686D}}
The authors have estimated the gas fractions of normal, near-IR selected galaxies at z$\sim$1.5 using CO observations. They have used dynamical models of clumpy disk galaxies to derive dynamical masses and then estimate the gas mass. To be consistent with the gas fractions derived in the $z\sim0.6$ sample, I have re-estimated the gas fractions in the \citet{2010ApJ...713..686D} sample from the tabulated $R_{CO}$, SFR and $logM_{stellar}$ and the K-S law. This new derivation leads to a median gas fraction of 61\%$\pm 6$ for a median $log\,M_{stellar}$=10.72. This value is in very good agreement with the gas fraction derived by \citet{2010ApJ...713..686D} from the simulation. 

\subsubsection{ \citet{2006ApJ...647..128E}}
The galaxies discussed in this paper are drawn from the rest-frame UV-selected z$\sim$2 spectroscopic population described by \citet{2004ApJ...604..534S}. The sample is not necessarily representative of the UV-selected sample as a whole, because objects were chosen for near-IR spectroscopy for a wide variety of reasons, such as: galaxies near the line of sight of a QSO; elongated morphologies in most cases, with a few more compact objects; galaxies with red or bright near-IR colors or magnitudes; galaxies with excellent deep rest-frame UV spectra \citep{2006ApJ...646..107E}. Even more, due to the criteria of selection, they may have selected red and bright objects and therefore mainly massive galaxies at these redshift (see figure 5 of \citet{2006ApJ...646..107E}). Given the distribution of stellar mass, the sample of \citet{2006ApJ...646..107E} is representative of massive galaxies with $log\,M_{stellar}>10.8\,M_\odot$). The representativity of the full sample of UV-selected galaxies at z$\sim$2 is however difficult to evaluate.

In order to have homogeneous measurements and methodologies between the $z\sim2$ and $z\sim0.6$ samples, I have re-derived the gas fraction of \citet{2006ApJ...644..813E} from the tabulated $SFR_{UV}$, $SFR_{H\alpha\,c}$, $R_{gas}$ and $log\,M_{stellar}$ in \citet{2006ApJ...644..813E,2006ApJ...647..128E,2006ApJ...646..107E}. 
The SFRs and stellar masses have been derived in Erb's papers assuming a \citet{2003PASP..115..763C} IMF. I have converted these values to a \citet{1955ApJ...121..161S} IMF by multipling them by 1.8 \citep{2006ApJ...644..813E}. The total SFR is defined as the sum of the dust extincted $SFR_{UV}$ and the $SFR_{H\alpha}$ corrected for extinction and aperture\footnote{The $SFR_{H\alpha}$ does not trace the same population than the $SFR_{IR}$, however \citet{2004A&A...415..885F} have shown that when corrected from aperture and extinction the two SFR estimates give similar results.}. The extinction has been estimated from SED fitting and then applied to the stellar UV continuum and to the nebular emission line $H\alpha$. \citet{2006ApJ...646..107E} derived gas masses considering a gas surface of $FWHM^2_{gas}$ where $FWHM_{gas}$ is the extension in the slit of the $H\alpha$ emission. The gas surface of $z\sim$0.6 galaxies has been estimated assuming a circular distribution of the light equal to $\pi\,R^2_{gas}$. Following \citet{2010MNRAS.tmp..689P}, I have assumed the $R_{gas}$ in \citet{2006ApJ...646..107E} to be $2\sigma$ of the $H\alpha$ emission distribution. Assuming that the gas surface is Gaussian, $R_{gas}$ is then equal to $(2/2.35)\,FWHM_{H\alpha}$. This new derivation leads to a median gas fraction of 53.1\%$\pm 5$ for a median $log\,M_{stellar}$=10.41. 

There are not tabulated errors for $SFR_{H\alpha}$ and $SFR_{UV}$ in Erb's papers. I have thus taken the typical error bars for $\epsilon_\mu$=20\% and $log\,M_{stellar}=0.3\,dex$. These error bars take into account the uncertainty on the estimation of $R_{H\alpha}$ ($\sim$30\%), on the $H\alpha$ flux measurement and the aperture and extinction corrections (factor 2), and the uncertainty due to the scatter in the Schmidt law itself ($\sim0.3\,dex$). 

\begin{figure}[!h]
\centering
\includegraphics[width=0.7\textwidth]{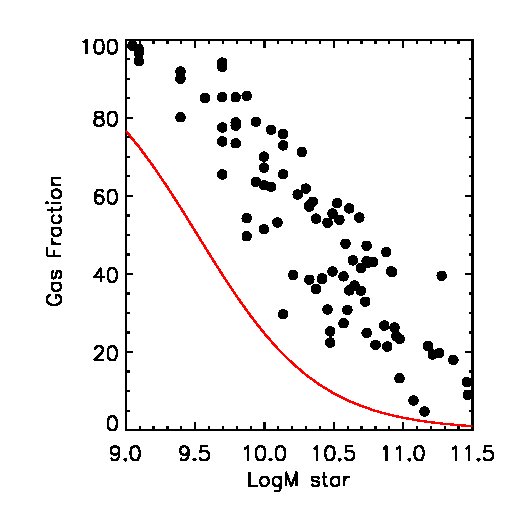}
\caption{{\small The gas fraction vs stellar mass of the $z\sim2$ sample of \citet{2006ApJ...647..128E}. The stellar mass has been converted into a 'diet' Salpter IMF and the gas fraction have been re-derived as explained in the text. The solid red line is the limit of detection of $\mu$ derived from the limit of detection of the $H\alpha$ emission line. The sample is only completed for $log\,M_{stellar}$ above 10.7$M_\odot$.}}
\label{Erb_fgas}
\end{figure}

In fig.\ref{Erb_fgas}, I have plotted $\mu$ as a function of $log\,M_{stellar}$ for the $z\sim2$ galaxies, with the new estimation of $\mu$. The $\mu$ limit detection has been plotted with a red solid line. The sample is not biased by the SFR-limit and thus the large gas fraction found for low mass galaxies is not due to the effect of the limit of SFR detection. However, it is difficult to infer any conclusion about the z=2 galaxies because the sample is not complete for massive galaxies with $log\,M_{stellar}<10.7\,M_\odot$.

\subsubsection{z$\sim$3 sample \citet{2009MNRAS.398.1915M}}
In \citet{2009MNRAS.398.1915M}, the authors present preliminary results for the sizes, SFRs, morphologies, gas-phase metallicities, gas fractions and effective yields of a sample of z$\sim$3 Lyman-break galaxies. However, they have not considered companions and secondary peaks of emission when measuring the gas extension. Therefore the gas mass is only consistently derived for the central region and the gas fraction estimation may suffer of strong underestimation. I have thus decided not to keep this sample in the comparison.

\subsection{Evolution of the gas fraction during the past 8 Gyr}
\begin{figure}[!h]
\centering
\includegraphics[width=0.60\textwidth]{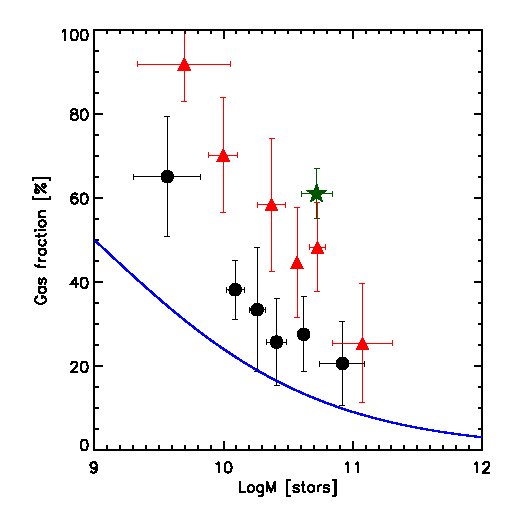}
\caption{{\small The gas fraction $\mu$ vs. stellar mass in local, intermediate and high-z samples. The local galaxies from \citet{2008AIPC.1035..180S} are represented as a blue line. For clarity the $z\sim0.6$ (black dots) and $z\sim2$ (red triangles) samples have been represented in bins of $\sim$12 and $\sim$15 galaxies respectively. The gas fraction at $z\sim1.5$ from the CO observation of \citet{2010ApJ...713..686D} is plotted with green stars.}}
\label{Fgas_evo}
\end{figure}

Figure\,\ref{Fgas_evo} shows the evolution of the $\mu$ vs. $log\,M_{stellar}$ of intermediate mass galaxies at different lookback times. The local galaxies from \citet{2008AIPC.1035..180S} are represented as a blue line. For clarity the $z\sim0.6$ (black dots) and $z\sim2$ (red triangles) samples have been represented in bins with equal number of galaxies,  $\sim$12 and $\sim$15 galaxies respectively. Each point represents the median gas fraction and stellar mass in the bin. The errors bars are the quadratic sum of the dispersion in each bin and the error on the evaluation of the median value calculated by bootstrap resampling.  

The \citet{2006ApJ...644..813E} sample is only complete for massive galaxies with $log\,M>10.7\,M_\odot$. Thus the comparison between the local, $z\sim0.6$ and $z\sim2$ sample is only valid for galaxies above this limit. I found a significant evolution of the gas fraction of $9-14$\% in the high mass galaxies. In the local Universe the mean gas fraction in the 10.6<log\,M<11.2 interval is 10\%, and evolves to 19-24\% at $z\sim0.6$ and reaches 34-45\% at $z\sim2$. 

\begin{figure}[!h]
\centering
\includegraphics[width=0.60\textwidth]{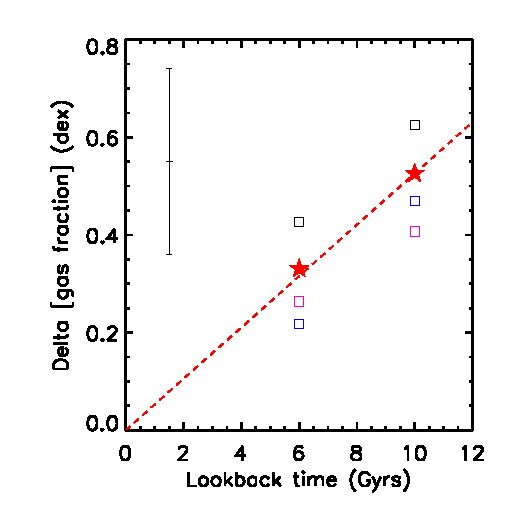}
\caption{{\small The evolution of the gas fraction with lookback time: $log(\mu(t)/\mu(t_0))$ as a function of lookback time for logM>10.6 (black symbol), 10<logM<10.6 (pink symbol) and logM<10 (blue symbol). Star symbols include galaxies with stellar masses in the completeness range of IMAGES sample (logM>10).}}
\label{Fgas_lookback}
\end{figure}

In figure\,\ref{Fgas_lookback}, I have plotted the evolution of the gas fraction normalized to the local gas fraction as a function of lookback time for three ranges of stellar mass and the range of stellar mass in which IMAGES sample is complete. Only the most massive range, with $log\,M_{stellar}$ above 10.6, corresponds to the evolution of a complete sample at the three lookback times. I found that the evolution of the mean gas fraction from the local Universe to a lookback time of $\sim$10\,Gyr is linear with a slope given by
\begin{equation}
\log{\mu_{\mathrm{tb}}/\mu_0} = 0.052\times t_{\mathrm{Gyrs}}.
\end{equation}
This result has to be put in parallel with the evolution of metallicity in the gaseous phase for star-forming galaxies, see fig\,\ref{lookbacktime}. The two quantities have complementary evolution: gas decreases by a ratio of 0.052\,dex per giga year while metallicity increases by 0.046\,dex per giga year. It suggests that all the evolution of the gas content in those galaxies is due to the conversion of gas into stars. 


\subsubsection{Systematics}
The comparison between the gas fraction of different samples can be affected by several systematics, such as the estimation of the gas radius and the SFR, and the representativity of the samples. The uncertainties related to the exact slope of the KS relation are certainly marginal because the distributions of SFR surface densities in the samples are concentrated within a quite small range, $\Sigma\,SFR=0.01-0.1\,M_\odot/yr/kpc^2$. In fact, \citet{2010A&A...510A..68P} have discussed that the variation on the power law of the KS does not affect this range of $\Sigma\,SFR$. Therefore, a possible evolution of the power-law index of the K-S law, between z=0 and z$\sim$0.6, will weakly affect the results. 
Uncertainties in estimating the gas mass are probably dominated by the determination of the gas radii. Indeed, the gas radius is difficult to measure with accuracy for severals reasons. To overcome the lack of data giving the distribution of the ionized gas, the UV-band has been used as a proxy of the ionized gas. However, \citet{2010A&A...510A..68P} have shown that the extension of the ionized gas can be much more extended than the UV light in intermediate redshift galaxies. The gas radius  derived from UV half radius can thus be underestimated in distant galaxies. Moreover, the gas surface is calculated by assuming a circular overture, while galaxies at high redshift have clumpy morphologies. The complex distribution of the gas in high-z galaxies may introduce large uncertainties in the derivation of the gas mass. 


\section{The Yield }
\label{Yield}
\subsection{Introduction}
\subsubsection{Definition}
In a Closed-box model (Searle \& Sargent 1972, Pagel \& Patchett 1975, Pagel 1981) in which galaxies evolve passively without any exchange of gas with the environment, the enrichment of the ISM is strictly due to the feedback from star formation. The total mass of a galaxy remains constant with time.  Searle and Sargent, 1972 had introduced the equations describing the Closed-Box model.  In a galaxy with an initial gas mass $M_{gas}^{i}$, the gas is converted into stars at the star formation rate. One part of the gas transformed into stars is injected in the ISM when stars finish their life. The evolution of the gas in this model can thus be written :
\begin{equation}
\frac{dM_{gas}}{dt}=-\phi(t) + E(t)
\end{equation}
where $\phi$ is the star formation rate and  E(t) is the rate at which dying stars restore material into the ISM at time t. 
In the same way, the evolution of metals is :
\begin{equation}
\frac{d(Z\,M_{gas})}{dt}=-Z\phi(t) + E_Z(t)
\label{closebox_equa}
\end{equation}
where $E_Z$ represent the rate at which metals are ejected into the  ISM.  
Searle and Sargent have demonstrated that the solution of equation \ref{closebox_equa} is 
\begin{equation}
Z=y_Zln(1/\mu)
\end{equation}
where $y_Z$ is the true nucleosynthetic yield, defined as the mass of newly produced metals that is locked up in long-lived stars and stellar remnants and $\mu$ the gas fraction. 
The model assumed the following assumptions : 
\begin{enumerate}
\item the system is closed, no inflow or outflow;
\item the initial gas is pristine;
\item the metals produced are instantaneously mixed in the ISM.
\end{enumerate}

To evaluate the deviation of the evolution of a galaxy to a closed-box model, the effective yield defined by $y_{eff}=Z_{gas}/ln(1/\mu)$  is compared to the true yield. In fact, any infall of pristine gas or expulsion of metals through an outflow will decrease the value of the effective yield, see analytic models from \citet{1990MNRAS.246..678E}. As instance, \citet{2007ApJ...658..941D} have investigated the influence of gas fall in the effective yield using analytical models. They have demonstrated that: (1) metal-enriched outflows are the only mechanism that can significantly reduce the effective yield, but only in gas-rich systems; (2) that it is nearly impossible to reduce the effective yield of a gas-poor system, no matter how much gas is lost or accreted; (3) any subsequent star formation rapidly drives the effective yield back to the closed-box value.

\subsubsection{Yield in the local Universe}
\begin{figure}[!h]
\centering
\includegraphics[width=0.50\textwidth]{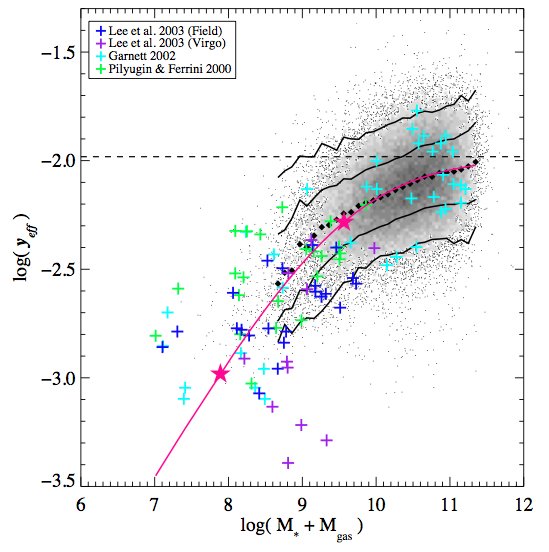}
\caption{{\small Effective yield as a function of total baryonic mass (stellar +gas mass) for 53,400 star-forming galaxies in the SDSS. The large black points represent the median of the SDSS data in bins of 0.1 dex in mass which include at least 100 data points. The solid lines are the contours which enclose 68\% and 95\% of the data. The colored crosses are data from Lee, McCall, \& Richer 2003, Garnett 2002, and Pilyugin \& Ferrini 2000. Both the metallicities and the gas masses used to derive the effective yield have been computed differently in the SDSS data and the samples from the literature. The agreement nevertheless appears quite good. From \citet{2004ApJ...613..898T} }}
\label{Yield_Tremonti}
\end{figure}

\citet{2004ApJ...613..898T} have derived the yield of 53\,000 local star-forming galaxies from the SDSS and found that low mass galaxies have lower effective yield than massive galaxies. This trend can be explained by the presence of strong outflows in low mass galaxies. Indeed, the intensity of an outflow is expected to be more important in low mass galaxies in which the gravitational potential well is small, allowing metals to escape from the system. The autors argue that outflow is the main driver of the mass-metallicity relation shape in the local Universe(Tremonti et al. 2004), see Fig.\ref{Yield_Tremonti}. 

\subsubsection{Yield in high-z galazies}
In the last five years, numerous surveys have collected gas metallicities and gas fraction for a large number of local to very high redshift galaxies. These observations have allowed to investigate the contribution of outflows and inflows at different epochs. Contrary to what is found in the local Universe, \citet{2008ApJ...674..151E} and \citet{2009MNRAS.398.1915M} have shown that the effective yield decreases with the stellar mass at redshifts above 2: Massive galaxies are more distant to a closed-box evolution than less massive ones. These surprising results can be explained by considering the infall of pristine gas. \citet{2008ApJ...674..151E} have recently proposed an analytic model to describe the effect of infall and outflow on the evolution of metals, assuming a star formation given by the Kennicutt-Schmit law. She found that the strong star formation in $z=2$ galaxies is sustained by the accretion of gas at approximately the gas processing rate, the outflow rate is roughly equal to the SFR, and metal enrichment is modulated by both outflows and inflows. Using a similar approach, \citet{2009MNRAS.398.1915M} have reached the same conclusion for $z=3$ Lyman break galaxies. 

\subsection{The yield at z$\sim$0.6}
\label{yield at z=0.6}
\begin{figure}[!h]
\centering
\includegraphics[width=0.60\textwidth]{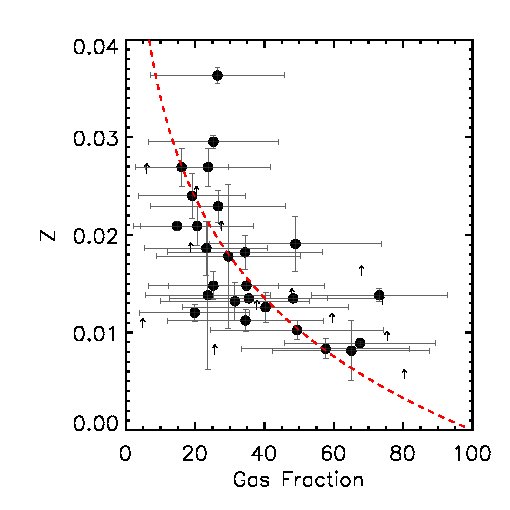}
\caption{{\small Metallicity vs gas fraction of Sample\,A (dot symbols). The AGN candidates have not been plotted. The upward arrows are Sample\,A galaxies with  lower limit of metallicity. The theoretical relation in the case of a closed-box model with yield=0.0148 is represented by a red solid line. }}
\label{Yield_SampleA}
\end{figure}

Using the metallicities and the gas fractions derived in the previous chapter, I have estimated the yields of sample\,A, a representative sample of $z\sim0.6$ intermediate mass galaxies. Figure \ref{Yield_SampleA} shows the relation between gas fraction and metallicity. I have overploted with a red solid line the theoretical model for a closed-box model with yield=0.0148, see equation above. The $z\sim0.6$ galaxies are well fitted with a closed-box model. I fit the data with the models with outflow and infall of \citet{2008ApJ...674..151E}. The data is compatible with models close to the closed-box model having small contribution of outflow and infall. 

\section{Toward a chemical enrichment scenario}
\label{Toward a chemical enrichment scenario}

The quality of the FORS2 data have allowed us to investigate the main ingredients of the chemical evolution: metals, gas mass, stellar mass, and the SFR. The main results are: 
\begin{itemize}
\item The M-Z relation at $z\sim0.6$ puts on evidence the strong evolution ($\sim0.3\,dex$) of metal content in the gaseous phase of galaxies (section \ref{M-Z relation}). 
\item The M-Z relation at $z\sim0.6$ has a larger dispersion than the local relation (section \ref{M-Z relation}). 
\item The gas fraction doubles from z=0 to $z=0.6$ (section \ref{The gas fraction}). 
\item The effective yield of $z\sim0.6$ galaxies is concordant with a closed-box model. However, small contributions of outflows and infalls are not excluded (section \ref{yield at z=0.6}). 
\item There is no evidence of outflow from $z\sim0.6$ galaxy spectra and therefore, that large-scale outflows do not play an important role in intermediate mass galaxies at this redshift (section \ref{Outflows sample}). 
\end{itemize}
In this section, I propose to assemble all these observations together into a consistent picture of the chemical evolution in intermediate mass galaxies.

\subsection{A closed-box evolution of galaxies during the last 6 Gyr}
The observations of the effective yield and the no detection of outflows in the integrated spectra support that $z\sim0.6$ galaxies are well represented by a closed-box model. The validity of the closed-box model can also be tested from the evolution of the metal content and the gas fraction. Indeed, the amount of gas converted into stars assuming a closed-box model can be estimated from the M-Z relation, following the same methodology as \citet{2004ApJ...617..240K} and \citet{2006A&A...447..113L}. The metal abundance Z is related to the gas mass fraction by
\begin{equation}
Z(t)=Y\,ln\frac{1}{\mu(t)} \, ,
\end{equation}
where $\mu=M_{\mathrm{gas}}/(M_{\mathrm{gas}}+M_{\mathrm{stellar}})$ is the fraction of gas. As the yield is constant, the variation of metallicity depends only on the variation of the gas fraction :
\begin{equation}
\mathrm{d}(\log{Z})/\mathrm{d}\mu=0.434/\mu\mathrm{\ln}{\mu} 
\end{equation}
The metals produced by a given fraction of gas in a closed-box model is thus given by:
\begin{equation}
\Delta[O/H]=[O/H]_{ini}+ \frac{0.434\, \Delta \mu}{\mu\, ln(\mu) - 1/\mu}
\end{equation}
Given the observed evolution of the gas fraction, I have estimated the evolution of the local M-Z relation of \citet{2004ApJ...613..898T} at $z\sim0.6$ in the frame of a closed-box model. At a fix stellar mass, [O/H] is the metallicity at z=0 given by \citet{2004ApJ...613..898T} , $\mu$ is the fraction of gas at $z\sim0.6$ and $\Delta \mu$ is the evolution of the gas fraction between z=0.6 to z=0. I have assumed that galaxies follow the $logM_{stellar}$ vs $\mu$ relation of \citet{2008AIPC.1035..180S} in the local Universe and the equation \ref{equation_gas} for intermediate redshift galaxies. 
\begin{figure}[!h]
\centering
\includegraphics[width=0.65\textwidth]{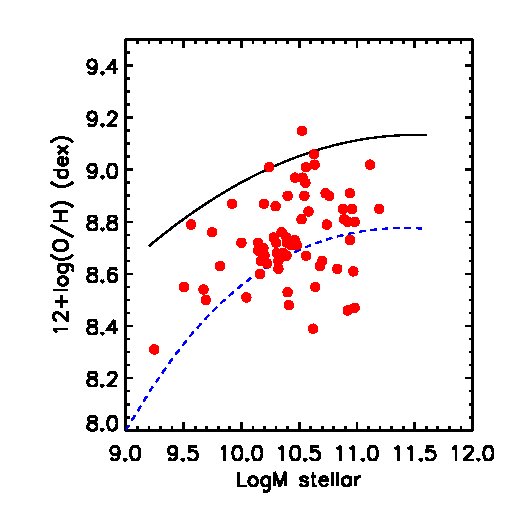}
\caption{{\small  The observed M-Z relation at $z\sim0.6$ (red symbol) and the predicted M-Z relation assuming a closed-box model and the observed evolution of the gas fraction (blue dashed line). }}
\label{M-Z_closebox}
\end{figure}
Figure \ref{M-Z_closebox} shows in blue dashed line the expected M-Z relation at $z\sim0.6$ given the observed evolution of the gas fraction in the case of a closed-box model. The local \citet{2004ApJ...613..898T} relation (solid black line) and the $z\sim0.6$ galaxies (red dots) are also shown. The observed M-Z relation at $z\sim0.6$ is compatible with a closed-box model. However, the observed relation is slightly over the theoretical model. The shift can be due to a systematic effect in the measurement of the metallicity  or in the gas fraction. An overestimation of the gas fraction in the z$\sim$0.6 sample or an underestimation in the local universe can lead to the observed shift. However, the effect can be physical and may be due to a small amount of gas infall (not pristine). I am currently investigating the effect of such an infall using the analytic models of \citet{2008ApJ...674..151E}. 

A closed box model is a very crude approximation for intermediate mass galaxies, which can be affected by gas outflows, inflows and minor or major mergers. Indeed, a large number of results of IMAGES reveals the importance of merger events at intermediate redshift. A closed-box scenario, where galaxies evolve secularly, seem to be in contradiction with the violent exchanges of gas between galaxies during interaction and merger events. However, within a statistical argument, the whole intermediate mass galaxy population can be assumed to characterize most of the gas to mass ratio of the Universe \citep{2006A&A...447..113L}. There are no infall or outflow, merger event with object outside the population of intermediate mass galaxies. For instance, in \citet{2010A&A...509A..78D}, we have shown that the probability of a merger between two intermediate galaxies which will result in a remnant outside the range of mass is very small. The validity of closed-box model for the population of intermediate mass galaxies favors the conclusion of \citet{2009arXiv0911.3247H}.

\subsection{The effects of mergers in the dispersion of M-Z at $z\sim0.6$}

\begin{figure}[!h]
\centering
\includegraphics[width=0.55\textwidth]{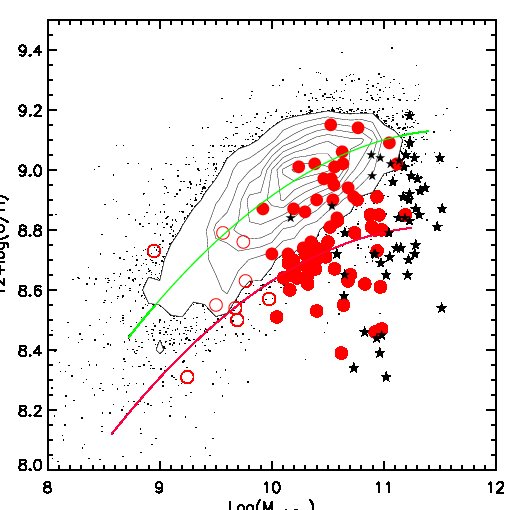}
\caption{{\small M-Z relation for starburst galaxies at z$\sim$0.6 and local LIRGs from \citet{2008ApJ...674..172R}. The symbols used in this plot are the same as in Fig. 9. The local LIRGs are plotted as black stars.} } 
\label{M-Z_LIRGS} 
\end{figure}

\begin{figure}[!h]
\centering
\includegraphics[width=0.31\textwidth]{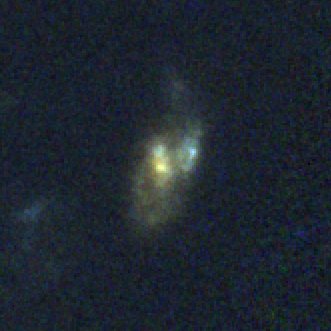}
\includegraphics[width=0.39\textwidth]{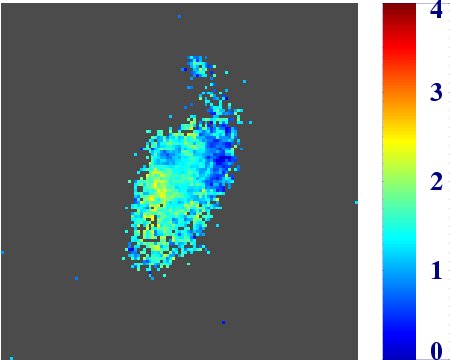}
\includegraphics[width=0.31\textwidth]{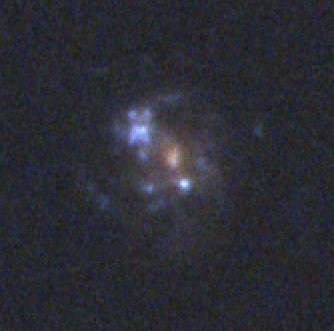}
\includegraphics[width=0.39\textwidth]{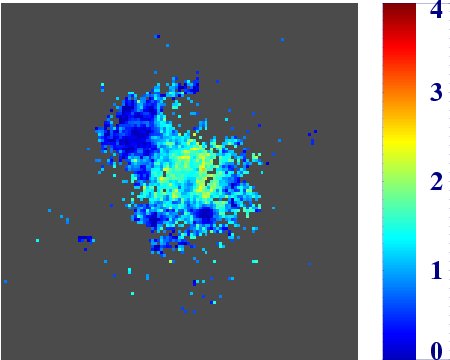}
\caption{{\small Composite image (B+V, i and z bands) of two IMAGES galaxies taken with the Advanced Camera for Surveys (ACS) of the HST (left column) and their B-z color map. The size of field-of-view is 5.0 arcsec$\times$5.0 arcsec ($\sim$30 kpc$\times$30 kpc at z$\sim$0.6). Both galaxies have metallicities below the average of $z\sim0.6$ galaxies and present evidence for merger. \textbf{Up}: J03324.60-274428.1 has two cores and can be probably classified as merger during the first passage. \textbf{Down} J03337.96-274652.0 have a very perturbed morphology. Its morphology is very similar to J033214.97-275005.5 which have been classified as a merger near fusion by \citet{2009A&A...507.1313H}.  } } 
\label{merger_lowZ} 
\end{figure}

Merger events induce redistributions of the gas \& stars, and can power strong star-formation rate. During the past years, several authors have investigated the effects of interactions and mergers on the chemical content of galaxies. \citet{2006AJ....131.2004K} have found that the metallicity in the nuclear regions of galaxies in very close pairs is displaced to lower levels than those measured in more distant pairs and isolated galaxies. The authors have claimed that this effect is caused by gas inflows triggering by the interactions. \citet{2008MNRAS.386L..82M} have found that for pairs showing signs of strong interactions, the mass-metallicity relation differs significantly from that of galaxies in isolation. In such pairs, the mean gas-phase oxygen abundances of galaxies with low stellar masses ($log\,M_{stellar}<9$) exhibit an excess of $0.2\,dex$. Conversely, at larger masses ($log\,M_{stellar}>10$) galaxies have a systematically lower metallicity, although with a smaller difference ($-0.05\,dex$). Using N-body/SPH numerical simulations of equal-mass mergers, \citet{2010ApJ...710L.156R} have shown that the low metal content found in the nucleus of interacting galaxies is due to radial inflow of low-metallicity gas from the outskirts of the two merging galaxies. Between the first and the second passage of the merger, low-metallicity gas from outer regions of the galaxy is tidal torqued into the high-metallicity galaxy center, resulting in gas with a lower average abundance. This results has been confirmed by \citet{2010arXiv1003.1374M} who have estimated a timescale around $10^8$ years or less for these strong inflow episodes. As the merger proceeds, the star formation depletes the gas and increases the metal content. The metallicity of the central region increases to a level higher than the initial metallicity of the merging galaxies. This scenario was first suggested by \citet{2008ApJ...674..172R} to explain the low metallicities of local LIRGs, see figure \ref{M-Z_LIRGS}. The large scatter in metallicity observed at intermediate redshifts, similar to those of local LIRGs, can be reproduce by the diversity of galaxy building histories within the framework of a merger scenario. 
I have found a similar trend in the IMAGES sample: LIRGs have lower metallicities than the star-forming galaxies at the same stellar mass. Unfortunately, the majority of LIRGs and  the most metal poor galaxies in the sample do not have deep imagery because these galaxies arise from \citet{2006A&A...447..113L} sample which does not have ACS imagery. However, two galaxies from CDFS sample with lower metallicities with regards to median M-Z relation at z$\sim$0.6 seem to be mergers or interacting galaxies. For instance, J03324.60-274428.1 has two cores and can be probably classified as a merger during a first passage, see figure \ref{merger_lowZ} left. J03337.96-274652.0 (right) has a very perturbed morphology. The galaxy has a large amount of faint objects in its outskirts and an intense star-formation activity in the outer region of its disk (from the color map). Many of these faint objects have similar colors to the galaxy outskirt, and examination of its image strongly suggests that they are associated to it. The numerous surrounding satellites make us to suspect that the galaxy is apparently experiencing large gas infall/outflows in its halo. Its morphology is very similar to J033214.97-275005.5 which has been classified as a merger near fusion by \citet{2009A&A...507.1313H}. However, only kinematical observations will enable us to class with a good confidence level these two galaxies. I have also investigated the properties of over-metal rich galaxies in the M-Z relation. According to \citet{2010arXiv1003.1374M} these galaxies should be late stage of merger. The luminous J033245.11-274724.0 galaxy seem to be a good example of such kind of object. It has very high metal content, $Log(O/H+12)=9.14$ and the detail simulation of its morpho-kinematics has proved that it corresponds to a final stage of major merger, see section \ref{dust-enshrouded disk}.  

\subsection{'Downsizing' and hierarchical scenario}
The specific SFRs and the gas fractions of $z\sim0.6$ galaxies show evidence of 'downsizing': low mass galaxies form stars and evolve at later epochs compared to massive galaxies. Several authors have pointed out that downsizing is in contradiction to the hierarchical scenario. Massive galaxies form later than low mass galaxies by accretion of small dark halos, in the bottom-up scenario. However \citet{2006A&A...459..371M} and \citet{2006MNRAS.372..933N} have demonstrated that the downsizing of star-forming galaxies is inherent to gravitational processes in the hierarchical scenario. Low mass galaxies have very long gravitational collapse time-scales due to their small potential well. Thus they start to produce stars later than massive galaxies.

\subsection{Chemical evolution of galaxies at higher redshift}
According to \citet{2008ApJ...674..151E} and \citet{2009MNRAS.398.1915M}, high-z galaxies are very far from a closed-box model. 
Both authors claim that the accretion of gas at approximately the gas processing rate is the main driver of distant galaxies evolution. The metallicity vs. gas fraction relation in distant galaxies also suggests the importance of outflows in these galaxies. However, these observations can also result from systematic effects due to the incompleteness of the high-z samples. The high-z samples selected by emission line only gathered the most active galaxies where outflows and infalls from merger can have strong contributions. This may systematically select galaxies with higher metal contents than the mean distant galaxy population. The emission line selected samples at high redshifts may not be presentative of a complete population of galaxies at a given lookback time. For example, the observation of Damped Lyman Absorber, absorption-selected, gives a quite different picture of the chemical enrichment at high-z. DLAs have lower metal content compared to emission selected galaxies and have a wider scatter in metallicity. \citet{2006fdg..conf..319P} discusses the metallicities measured in different systems at high redshift, as a function of their typical linear scales. He claims that it is the depth of the potential well shaped by baryons that drives the pace at which gas is processed though stellar nucleosynthesis at early epochs of galaxy evolution. 

   \FrameThisInToc
\chapter{Stellar mass and stellar populations}

\minitoc

\section{The stellar mass}
\subsection{A poorly constrained quantity}
Many works have put on evidence the fundamental role of stellar mass in galaxy evolution. Stellar mass is found to correlate with many galaxy properties, such as luminosity, gas metallicity, color and age of stellar populations, star-formation rate, morphology, and gas fraction, to enumerate a few of them \citep{2000ApJ...536L..77B,2003ApJS..149..289B,2003MNRAS.341...33K,2004ApJ...613..898T,2007ApJ...663..834B}.  Ironically, the stellar mass is one of the most poorly constrained quantities in distant galaxies. This difficulty arises from the fact that one of the main contributors to stellar mass, the low-mass stars on the main sequence (type G to K stars) are extremely faint and almost do not contribute to the integrated light of a whole stellar population. The strong dominance of young massive stars in any stellar population is responsible from the strong sensibility of the SED shape to the star-formation history. 

\subsection{The stellar mass-to-light ratio}
\begin{figure}[!h]
\centering
\includegraphics[width=0.45\textwidth]{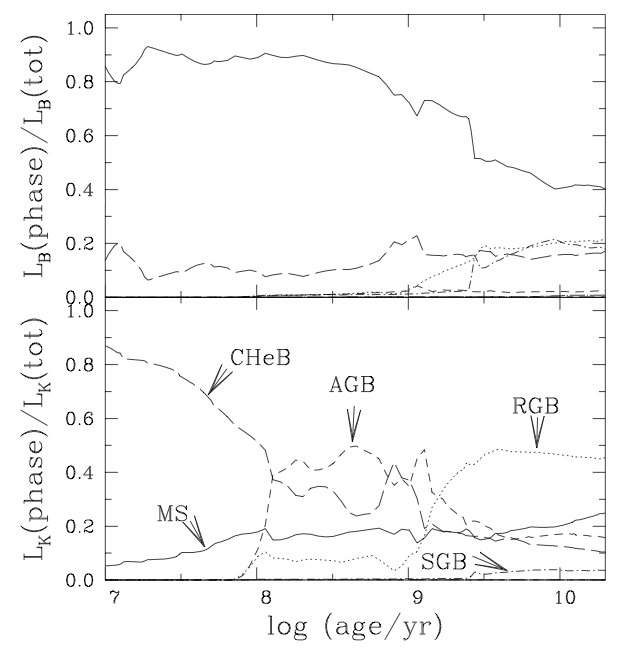}
\caption{{\small  Fractional contributions to the B and K light of an instantaneous burst stellar population with solar metallicity and a Salpeter IMF, by stars on the main sequence and various post-main sequence evolutionary phases. From Charlot 1997}}
\label{Charlot1}
\end{figure}

\begin{figure}[!h]
\centering
\includegraphics[width=0.7\textwidth]{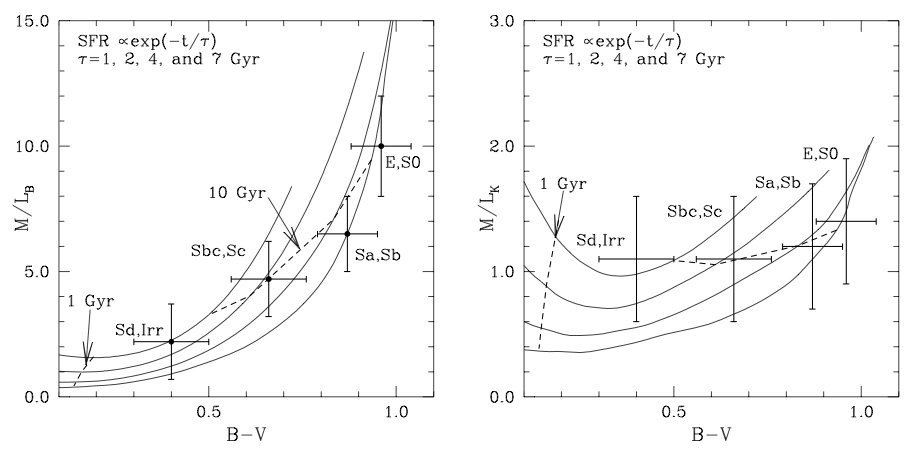}
\caption{{\small Mass-to-blue light ratio (left) and mass-to-infrared light (right) ratio versus B-V color for models with various exponentially declining star formation rates ($\tau$=1,2,4,7\,Gyr), compared to the values observed in nearby galaxies of varios morphological types. From Charlot 1997.}}
\label{Charlot2}
\end{figure}

The stellar mass of a galaxy can be estimated from its luminosity and its mass-to-light ratio (M/L). The value of the M/L ratio depends on the observed band and the star formation history of each galaxy. In the B-band, main sequence stars represent more than 80\% of the light, while the K-band light is dominated by post main sequences star, such as supergiant, AGB and RGB stars, see Fig.\ref{Charlot1}. According to this, the B-band seem to be the most appropriate color band to measure the stellar mass. However, the luminosity in the B-band is dominated by massive young stars from the main sequence (type O to F) which turn the $M/L_B$ very sensitive to the star-formation history and thus the estimation of the stellar mass in this band very uncertain.  Fortunately, the mass-to-light ratio in near IR has a small dependence on the star formation history and it thus well adapted to estimate the stellar mass of galaxies, see Fig \ref{Charlot2}. This is a intriguing result since most of the near IR light is not coming from the main sequence stars which make most of the galaxy mass, see Fig. \ref{Charlot1}. It leads many authors to suspect that large uncertainties are related to these estimates, and wonder if these can apply equally to starbursts and early type evolved galaxies. 
Several methods have been introduced to take into account the influence of the SFH on the M/L ratio. They rely on the evaluation of the SFH of each galaxies by comparing absorption lines features or broad band-photometry to synthetic evolutionary models. Distant galaxies have usually low quality spectra which unable the estimation of their stellar populations and thus the estimation of their stellar mass from full-spectra fitting or lick-indices methods. The study of their stellar population rely thus on broad-band photometry which gives considerably less observational constrains than the study of their spectra. There is a extend literature about the estimation of stellar masses and SFH in distant galaxies. Mainly, all methods rely on the SED fitting from a grid of models with different parametrization of the star formation history, see Part I Chapter 3:
\begin{itemize}
\item Simple CSP. The SFH is parametrized by a exponential decaying SFR with an e-folding time $\tau$. A simple dust model is assumed. The best-fit of the data constrain the $\tau$, age, metallicity and extinction \citep{2001ApJ...559..620P, 2001ApJ...562...95S, 2004ApJ...616...40F,2006ApJ...646..107E}. 
\item Two component CSP. The SED is best-fitted by the combination of two CSP: a old stellar population modeled by CSP having an exponential decay SFH with a long e-folding time and a young stellar population also modeled by a CSP \citep{2001ApJ...559..620P, 2001MNRAS.326..255C,2007MNRAS.382.1415S}.  
\item Complex CSP with random burst. A grid of models with various SFH is generated by Monte-Carlo realization. The parametrization of the SFH is similar to the previous one, i.e. it is defined by an old stellar population with exponential decay SFR and a superimposed burst. The difference with the previous parametrization arise from the Bayesian approach to the problem. Indeed, the stellar mass, dust extinction and other quantities are estimated by the likelihood distribution \citet{2003MNRAS.341...33K,2004MNRAS.351.1151B,2005ApJ...619L..39S,2005MNRAS.362...41G}. 
\end{itemize}
However, even with the wealth of UV to mid-IR data from IMAGES, it is difficult to disentangle effects of metallicity, age, extinction, and past star formation history. In addition, all the derived quantities depend strongly on the models, in particular by the IMF, star formation history, and on the recipe for the extinction law assumed to generate the models, see discussion in \citet{2008A&A...484..173P} and \citet{2009A&A...507.1313H}.

\section{Testing the algorithms from the literature}
In this section, I have tested the reliability of several methods available in the literature to estimate the stellar mass and the stellar populations from broad-band photometry. The objective of this study is to estimate the possible systematic of SED methods when they are applied to distant galaxies having high-star formation rate and young stellar populations. This is a preliminary study and thus does not aim to be exhaustive of all available methods based on SED fitting. All the methods use a grid of stellar population templates from evolutionary synthetic populations models. I have used in all of them the models of Charlot \& Bruzual 2007. 

\subsection{Mock galaxy library}
\begin{figure}[!h]
\centering
\includegraphics[width=1.0\textwidth]{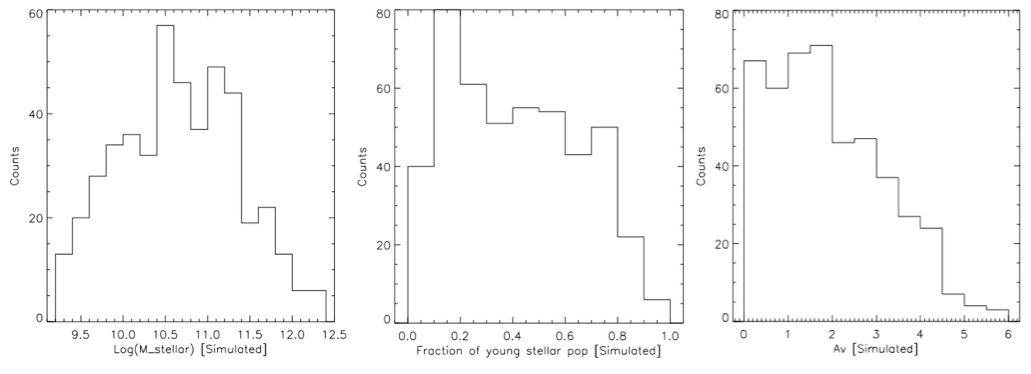}
\caption{{\small  Histogram of the stellar mass, mass fraction of young stellar population, and extinction of the mock galaxies.}}
\label{Mock_galaxy}
\end{figure}

To test these methods, I have generated a SED library of fake galaxies having similar properties to the IMAGES galaxies. Each galaxy has been generated by mixing six CSP from CB07 models having an exponential decaying star formation history with an e-folding time $\tau=100\,Myr$. The oldest population has an age of 7\,Gyrs, which corresponds to the age of the Universe at $z\sim0.6$, and the other population have logarithmic spaced ages: 5\,Myr, 200\,Myr, 500\,Myr, 1\,Gyr and 4\,Gyr. I have not assumed prior on the star formation history. The fraction of light of each CSP follows a uniform random distribution, see Fig. \ref{Mock_galaxy} central-panel. The library is thus compounded by galaxies having very different SFH: from galaxies dominated by old star as well as pure starburst galaxies having almost of their stellar mass created recently. The range of absolute magnitude, controlled by a normalization parameter, has been chosen to reproduce the stellar mass range of IMAGES galaxies, see Fig. \ref{Mock_galaxy} left-panel. I have also added a recipe for dust extinction. I have assumed that the dust extinction affects principally the new born stars and thus applied a \citet{1989ApJ...345..245C} extinction law only to the two youngest stellar populations. The \citet{1989ApJ...345..245C} law is parametrized by two variables: $E(B-V)$ and $R_V$. The ranges of value assumed by these variables are set to follow the observed values of $E(B-V)$ and $R_V$ in local group galaxies \citep{2007ApJ...663..320F}: $E(B-V)$ ranges 0 to 1 and $R_V$ from 2 to 6. The standard value of $R_V$ is 3.1, however we have found in the IMAGES sample that some galaxies can have extreme $R_V$ values, such as  J033245.11-274724.0 which have an Orion extinction law $R_V\sim4$. As the metal content of the gas decrease with redshift, see previous chapter, I have chosen to use CSP with sub-solar metallicity $0.40\,Z_\odot$. 

\subsection{The \citet{2003ApJS..149..289B} recipe}

The \citet{2003ApJS..149..289B} recipe is the method that has been used to derive all the stellar masses in the IMAGES survey. This method takes advantage of the tight correlation found between rest-frame optical colors and $M_{stellar}/L_K$ ratios, assuming a universal IMF. These correlations are found to be relatively insensitive to the details of galaxy SFH, dust content, and metallicity  \citep{2000MNRAS.312..497B,2003ApJS..149..289B}, which implies that they are invaluable for deriving stellar mass without being too sensitive to the details of the stellar population synthesis models. In this method, $M_{stellar}/L_K$ ratios are corrected for the amount of light due to red-giant stars using $g-r$ colors. According to \citet{2003ApJS..149..289B}, the total random uncertainty on $log(M_{stellar}/M_\odot)$ at z$\sim$0 should be lower than $0.1\,dex$, and the systematic uncertainties due to galaxy ages, dust, or bursts of star-formation can reach 0.1\,dex. 

\subsection{Single complex stellar population fitting}
This method assumes a very simple star formation history: an exponential declining SFH with an e-folding time $\tau$. It consists on fitting the galaxy SED with a grid of CSP templates covering a wide range of ages (from 10\,Myr to the age of the Universe), metallicities (Z=0.005-5.0$Z_\odot$) and e-folding time $\tau$ (from SSP to 500\,Myr).  The best extinction parameters and the normalization parameter are fitted to each template using a $\chi^2$ minimization algorithm. The best solution is the template with the smallest $\chi^2$. 

The results of the test realized with the fake galaxy library are in agreement with those of Fontana et al. 2004. These authors have found that the stellar masses estimated by this method are underestimated by $\sim$25\% when the object is observed during a burst. In fact, the SED distribution is dominated by the light produced by the young stars which hide the old stellar population that represent the bulk of the stellar mass. The simple star formation history assumed by this method is not valid for distant star-forming galaxies that are experimenting a strong star-forming burst. Nevertheless, 1-2\,Gyr after a burst, the stellar mass can be properly recovered by this method. 

\subsection{Maximum mass method}
\subsubsection{Description}

\begin{figure}[!h]
\centering
\includegraphics[width=1.0\textwidth]{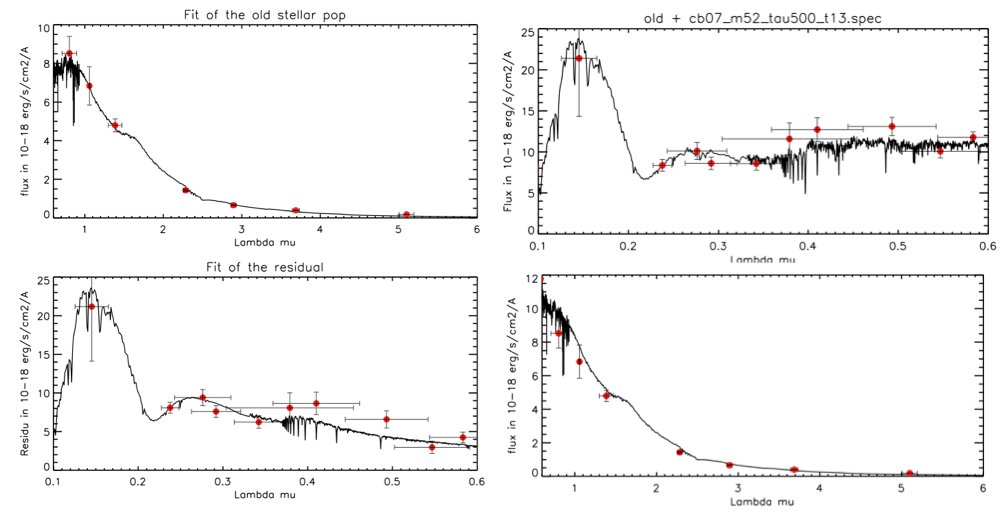}
\caption{{\small The maximum mass method applied to an IMAGES galaxy. \textbf{Top row, left panel}: Fit of an old stellar population model to the red-part of the SED. \textbf{Bottom row, left panel}: Best-fit of the residual with a young stellar population. \textbf{Right panels}: Best-fit of the SED by the two components.}}
\label{J033219.68_maxmass}
\end{figure}

This method is based on a two-component parametrization of the star-formation history, i.e. an old component which models the formation of the bulk stellar population in the early Universe and a young component that represents a recent star-formation episode. The old stellar component is modeled by a CSP with an age close to the age of the Universe at the observed redshift, an exponential decaying star formation history with an e-folding time $\tau=100\,Myr$, and a sub-solar metallicity ($Z=0.4Z_\odot$). In a first step, the red part of the SED (J-band to near IR) is fitted by the old component with the normalization and the two parameters of extinction as free variables. The best fit is subtracted to the observed SED. The second step consists on evaluate the contribution of a recent burst from the residual of the SED. The residual SED is compared to a grid of CSP models spanning a wide range of metallicities, ages and e-folding time. Each model of the library is compared to the residual SED by computing the $\chi^2$ statistic where extinction and the normalization flux are the free parameters of the fit. The optimal solution is the model of the grid with the lower $\chi^2$, see figure \ref{J033219.68_maxmass}. The confidence level of the best solution can be derived from the distribution of the $\chi^2$ as a function of the age, $\tau$ and Z of the library templates. This method aims to estimate maximum limit of the stellar mass since it minimizes the fraction of young and intermediate stellar populations \citep{2001ApJ...559..620P}. 

\subsubsection{Test of the method}
\begin{figure}[!h]
\centering
\includegraphics[width=0.55\textwidth]{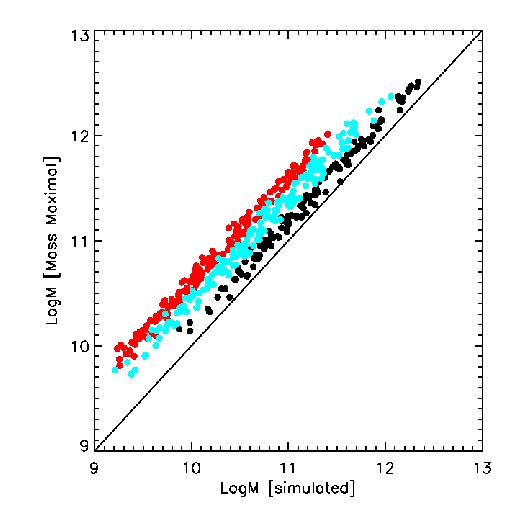}
\caption{{\small  Stellar mass derived by maximal mass method as a function of the real stellar mass in the fake galaxy sample. The color code the mass fraction of old stellar population (age>3\,Gyr) in the galaxy: black (blue) symbols are galaxies with more than 80\% (50-80\%) of their stellar mass locked into old stellar population. The red symbols corresponds to very young system with 50\% of stars younger than 3\,Gyr.}}
\label{MaxMass}
\end{figure}

I have tested the accuracy of this method to derive the stellar mass using the library of fake galaxies described in the previous section. Results are shown in figure \ref{MaxMass}. The method overestimates the stellar mass for all kind of SFH with an increasing bias towards galaxies with recent star formation: $+0.25\,dex$ for galaxies having 30\% of their mass in stars younger than 1\,Gyr, and $+0.4\,dex$ when half of the stellar mass is located in young stars. This systematic effect is expected since the method maximizes the stellar mass. 
\subsubsection{The effect of TP-AGB}
\begin{figure}[!h]
\centering
\includegraphics[width=0.45\textwidth]{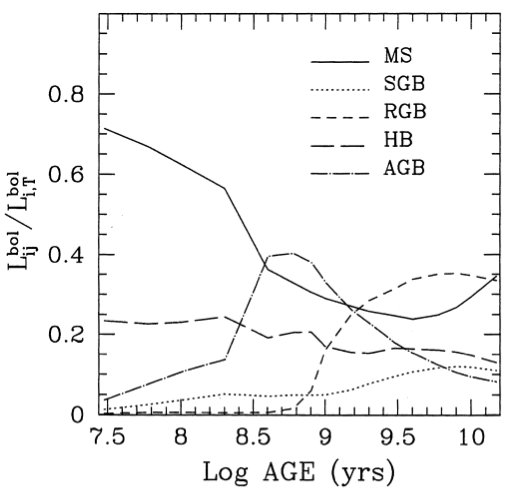}
\caption{{\small The time evolution of the relative contributions to the bolometric luminosity of stars in the various evolutionary stages. From \citep{1998MNRAS.300..872M}}}
\label{TP-AGB}
\end{figure}

The overestimation of the stellar mass in blue galaxies is due to the main assumption of the method that all the near IR is produced by old stars. However, such an assumption is not valid for objets which have undergone a star formation episode during the past 1\,Gyr. Indeed, from 0.2 to 2\,Gyr after the beginning of a burst, the contribution of TP-AGB star at these wavelengths can be very large, see fig \ref{TP-AGB}. They can be even dominant in the near-IR: up to 80\% in K-band \citep{1998MNRAS.300..872M}. 

\subsection{The impact of the assumed SFH on the stellar mass estimation}

In the local Universe, the method described above is robust enough to recover accurate stellar masses. Indeed, the majority of local galaxies has low SFR and the amount of young and blue stars produced in past bursts has almost disappeared. Their SED is therefore no more dominated by the light of young stars. A simple stellar population (e.g a CSP) or a SFH parametrized by a young and old component are enough to retrieve reliable estimation of the stellar mass. 
At higher redshifts, galaxies are in a phase of intense star formation. In starburst galaxies the light emitted by the young and massive stars dominates the photon budget along most of the spectral energy distribution and hides the old and intermediate stellar population. The fraction of old stars is systematically underestimated by current methods \citep{2009ApJ...696..348W}. I have thus investigated a new method to retrieve stellar masses and stellar populations in distant galaxies using more complex assumptions on the star formation history. The idea is to decompose the galaxy into its different components: young, intermediate and old stars and the dust. 

\section{Adding a more complex star formation history}
In a first attempt, I have tried to introduce a more complex star formation history parametrization when estimating the stellar mass of star-forming galaxies. The method is similar to those used in stellar population synthesis, such as full-spectra fitting algorithm. Galaxy is modeled by a combination of several single stellar population templates or complex stellar populations with exponential SFH decay. I am sorry to disappoint here the expectation and curiosity of the reader by revealing in advance the failure of this method on retrieving stellar masses. The failure of such a method is expected because of the degeneracy between age-metallicity and extinction, see details hereafter.  However, this method has turned out to be a very efficient way to measure absolute magnitude and it is now used by the group to derive the absolute magnitudes of IMAGES galaxies.  

\subsection{Description of the algorithm}
\begin{figure}[!h]
\centering
\includegraphics[width=0.70\textwidth]{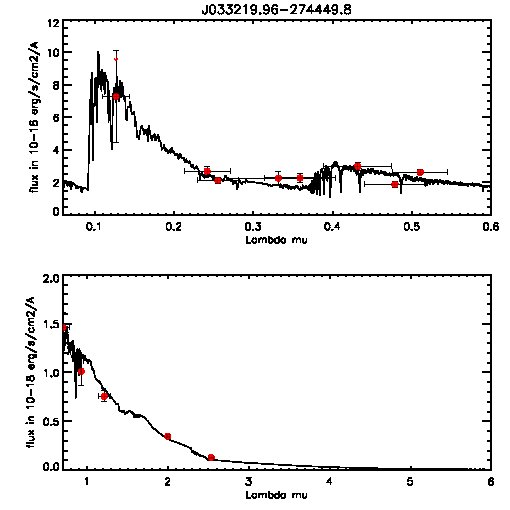}
\caption{{\small Rest-frame SED of J033219.96-274449.8 for a 3-arcsec fixed aperture photometry (red symbols). A SED compiled using a combination of 6 CSP with exponential decline $\tau$=100Myrs and a two component extinction law has been fitted to the object (solid line).}}
\label{J033219.96-274449.8.SED7}
\end{figure}

I have fitted the SED by a set of $N_\star$ single metallicity population synthesis templates $T^i(\lambda)$ (i=1,..., N). The best-fitting model SED $F(\lambda)$ is given by:
\begin{equation}
F(\lambda)=\sum^{N_\star}_{i=0}x_i\,T^i(\lambda)\otimes Dust(E(B-V),R_V)
\label{equation_6csp}
\end{equation}
where $x_i$ are non-negative coefficients representing the fraction of light of the template $T^i$, and Dust is the two parameter extinction law of  \citet{1989ApJ...345..245C}. I have assumed that only the young stellar population (age < 500\,Myr) is affected by the dust and the same E(B-V) and $R_V$ have been applied to these templates. 
To find $x_i$, E(B-V), $R_V$, I have minimized the $\chi^2$ using MPFIT (Levenberg-Marquardt least-square minimization) where
\begin{equation}
\chi^2(x_i, E(B-V),R_V)=\sum^{N_{filter}}_{j=1} \frac{(f_j^{obs} - f_j^{model})^2}{\sigma ^2 {f_j^{obs}}},
\label{equation_chi6csp}
 \end{equation}
and $f^{obs}_j$ is the rest-frame flux of the object in filter $j$ and its associated uncertainty $\sigma fobs_j$, $f^{model}_j$ is the flux of the synthetic spectrum in the j-band. The number of templates used in the fit depends on the number of photometric points. For the IMAGES galaxies observed by 17 filters, I have used $N_\star$=6. The number of free variables are thus eight (six from stellar population and two from the extinction) and the number of constrains is 17. An example is given for a IMAGES galaxy in figure \ref{J033219.96-274449.8.SED7}. 

\begin{figure}[!h]
\centering
\includegraphics[width=0.60\textwidth]{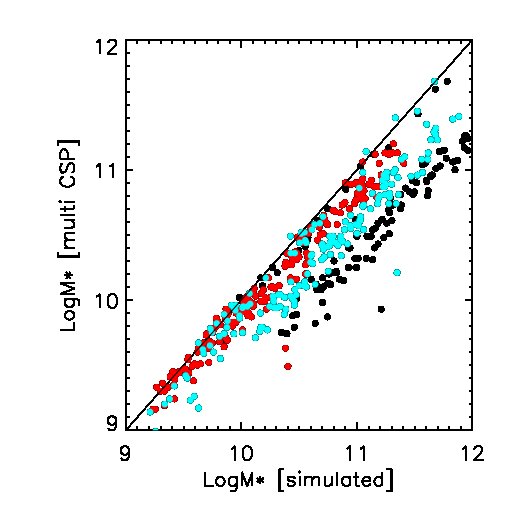}
\caption{{\small Fit of broad band photometry from UV to near IR with the combination of 6 CSP with exponential decline $\tau$=100Myrs and a two component extinction law. The degeneracy is important in galaxies with old populations because the old stellar population is hidden by the young. The color codes the mass fraction of old stellar population (age>3 Gyr) in the galaxy, see fig. \ref{MaxMass}}}
\label{multi_CSP_deg}
\end{figure}

The fit gives low values of $\chi^2$ but this does not guaranty that the result is physical. Are the derived stellar mass and $x^i$ reliable? In figure \ref{multi_CSP_deg}, I show the stellar mass derived by the algorithm as a function of the real stellar mass for the sample of fake galaxies. The figure puts on evidence the degeneracy of the problem. In galaxies composed in majority of old stellar populations, the stellar mass is underestimated because of an overestimation of the extinction. However, the best-fit provides us with an accurate interpolation of the SED and thus enables us to derive absolute magnitudes. 

\subsection{Testing the accuracy on derived absolute magnitudes}
\begin{figure}[!h]
\centering
\includegraphics[width=0.60\textwidth]{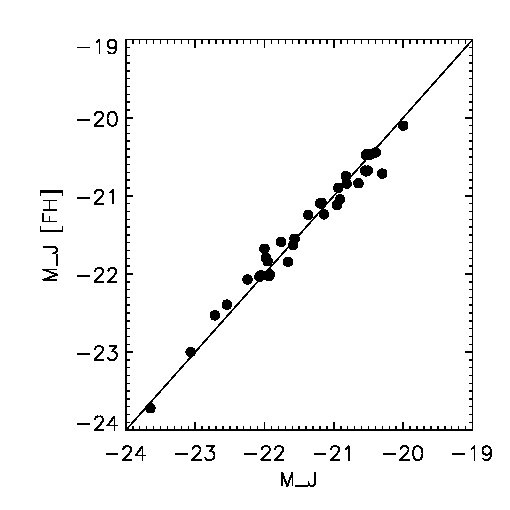}
\caption{{\small  Comparison between the absolute magnitude in J-band derived by \citet{2001ApJ...550..570H} and those derived using a combination of 6 CSP and a two-parameter extinction law.}}
\label{M_j_test}
\end{figure}

I have compared this method to those used previously in the team \citep{2001ApJ...550..570H} to estimate absolute magnitudes. The absolute magnitude of IMAGES galaxies were derived using Bruzual \& Charlot 2001 stellar-population models, assuming a CSP template with solar metallicity and an exponential star-formation history with $\tau$ = 0.5\,Gyr, which describes the properties of most galaxies between z = 0.4 and 1 (Hammer et al. 2001). For each galaxy, an optimal SED was found for each pairs of observed colors. Figure \ref{M_j_test} shows the comparison between the absolute magnitude in J-band derived by the previous method and those derived by the method described in the current section. I have found a good agreement between the two estimations: the two methods give equal results within $\sim$0.1\,mag. The small differences come from the models used by the two methods: In \citet{2001ApJ...550..570H}, Bruzual \& Charlot 2001 have been used while the new method runs Bruzual \& Charlot 2007 models. The main difference between the two models in the near IR is the treatment of the TP-AGB stars in the Bruzual \& Charlot 2007 models. The influence of TP-AGB stars is expected to be important at these wavelengths in galaxies having intermediate age stars. However, I found larger discrepancy when comparing blue-bands. The discrepancy can be explained by the more complex shape of the SED in blue wavelengths, which is not take into account in \citep{2001ApJ...550..570H}. This new method improves in particular the estimation of the absolute magnitudes in the blue bands. I am currently testing the accuracy on the derived absolute magnitude using the library of fakes galaxies.

\section{New constrain: the SFR }
The method available in the literature suffers from systematic effects due to the simplicity of the assumed star formation history. As described above, the decomposition of the SED by a linear combination of stellar population templates with different age is not possible because of the degeneracy of the problem. To break the degeneracy, I have added a new constrain in the optimization problem: the total star-formation rate ($SFR_{UV}$ and $SFR_{IR}$). 

In starburst galaxies the principal issue to recover properly the stellar populations and the stellar mass is the fact that the young stars dominate the light budget in all SED. Adding the total SFR constrain allows to resolve the problem by limiting the amount of young stars with ages younger than 100 Myr. In particular, it induces strong constrains in the blue part of the spectra, in particular the amount of light due to young stars and permits to better evaluate the dust obscuration from the blue slope of the SED. Moreover, adding the SFR constrain solves the problem encountered in spectrum synthesis \citep{2005MNRAS.358..363C, 2006MNRAS.365..385M}, i.e. the difficulty to recover the mass fraction of intermediate age population. This difficulty arises from a fundamental limitation of the spectral interpretation of galaxies for which the optical signature of intermediate age stars (0.5 - 4\,Gyr) is masked by those of younger and older stars. Indeed, an intermediate stellar population can be reconstructed by the combination of a young and old stellar population. By constraining the SFR, the amount of young stars is limited and the contribution of intermediate age stars in the blue part of the spectrum can thus be retrieved, as well as its contribution to the red part. 

\subsection{An optimization problem}
The problem is defined as in the previous section, by equations \ref{equation_6csp} and \ref{equation_chi6csp}, but I add a penalty function to impose the constrain on the SFR: 
\begin{equation}
P(SFR)=\left(\frac{SFR_{obs}-SFR_{model}}{\sigma SFR_{obs}}\right)^2.
\end{equation}
The optimization function to minimize is thus:
\begin{equation}
\chi^2(x^i,E(B-V),R_V,SFR_{obs})=\chi_{photo}^2(x^i,E(B-V),R_V)+P(SFR).
\label{equation_swarm}
\end{equation}
The grid of templates is composed by six CSP with $\tau=100\,Myr$ and same metallicity. I have associated to each template the mean SFR of the CSP in the last 100\,Myr. The observed SFR is the total SFR defined by the sum of $SFR_{IR}$ or $SFR_{UV}$. Note that the $SFR_{obs}$ depends on the assumed IMF and thus it is mandatory to use the same IMF for the stellar population models.  

\subsubsection{Multiple local solutions}
I have first tried to minimize the $\chi^2$ function using a non-linear least squares curve fitting algorithm based on gradient: the MPFIT IDL routine by Markward 2008. Unfortunately, the method was enable to find a global minimum, falling rapidly in a local minimum near the initial conditions. Adding the SFR penalty function in the fitting process has introduced a plethora of local minimum in the parameter space. The global minimum cannot be reached by the usual optimization algorithm based on the gradient or Hessian function. Though gradient-based methods are computably efficient, they provide us with a locally optimal solution and fail in general to provide the global optimal solution to a problem. 

\subsubsection{Grid of possible solution}
In a first attempt, I have tried to search for a global minimum in a grid of parameters which verifies the condition $SFR_m=SFR_{obs}\pm\epsilon_{SFR}$, i.e. the combination of $x_i$ which verify the SFR constrain. The preliminary results of this approach applied to the fake galaxy library were very promising. The stellar mass was retrieved within 0.1\,dex. However, the use of the grid introduces error due to the limited resolution of the grid and it is very time-consuming (severals hours by galaxy). I have thus searched for an optimization algorithm able to find the global solution of a non-linear problem. 

\subsection{The Swarm intelligence algorithm}
One way to solve problems with multiple local minima is to use an algorithm which employs some randomization as part of its logic. The randomization allows the algorithm to jump in other regions of the space parameter instead of setting locked directly in a local minimum as gradient based algorithms. There are several algorithms which include randomized search steps, or are 'heuristic' in nature, such as: simulated annealing, genetic algorithms, ant colony optimization, particle swarm optimization, to name a few. The idea is to leave the best solution untouched while allowing other states to explore the search space. I have also tested a genetic and a swarm intelligence code and they gave very similar performance in term of convergence and execution time.  I have chosen the 'particle swarm optimization' to solve the current optimization problem because this algorithm is easier to implement since it has less fitting parameters to optimized than the genetic code. 

\subsubsection{Description of the algorithm}
\begin{figure}[!h]
\centering
\includegraphics[width=0.90\textwidth]{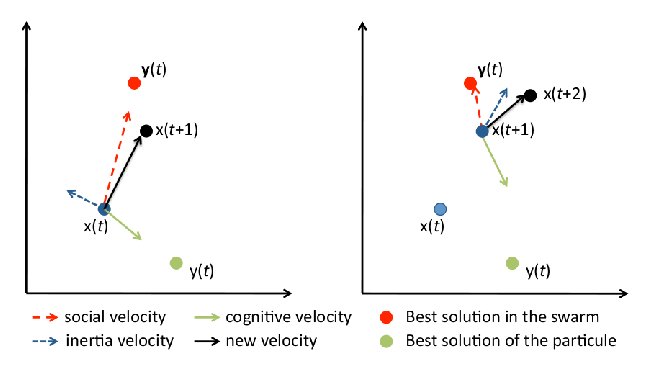}
\caption{{\small  Geometrical illustration of velocity and position update for a single two-dimensional particle at time step $t$ (left) and $t+1$ (right).}}
\label{Swarm velocity}
\end{figure}

Particle swarm optimization (PSO) is a population based stochastic optimization technique developed by Eberhart and Kennedy in 1995, inspired by the social behavior of bird flocking or fish schooling. A PSO algorithm maintains a swarm of particles, where each particle represents a potential solution. The particles are flown through a multidimensional search space, where the position of each particle is adjusted according to its own experience and that of its neighbors. Let $\textbf{x}_i(t)$ denote the position of particle $i$ in the searched space at time step $t$. The position of the particle is changed by adding a velocity, $\textbf{v}_i(t)$, to the current position, i.e.:
\begin{equation}
\textbf{x}_i(t+1)=\textbf{x}_i(t)+\textbf{v}_i(t+1)
\end{equation}
The velocity vector $\textbf{v}_i$ drives the optimization process. It depends on the experiential knowledge of a particle -\textit{the cognitive component}- and a socially exchanged information - \textit{the social component}. The cognitive component is proportional to the distance of the particle from its own best position found since the first time step. The social component reflects the information from all the particles in the swarm and is proportional to the distance of the particle to the best position found by the swarm, defined as $\textbf{y}(t)$.
The velocity of a particle $i$ is calculated as:
\begin{equation}
\textbf{v}_{ij}(t+1)=w\textbf{v}_{ij}(t)+C_1r_{1j}(t)[\textbf{y}_{ij}(t)-\textbf{x}_{ij}(t)]+C_2r_{2j}(t)[\textbf{y}_j(t)-\textbf{x}_{ij}(t)],
\end{equation}
where $\textbf{v}_{ij}(t)$ and $\textbf{x}_{ij}(t)$ are respectively the velocity and position of the particle $i$ in dimension $j=1,\,\ldots\,,n_x$ at time step $t$, $C_1$ and $C_2$ are the cognitive and social components, $w$ is the inertial component, and $r_{1j}(t)$ \& $r_{2j}(t)$ are random values in the range [0,1], sampled from a uniform distribution. The random values introduce a stochastic element to the algorithm. The inertia weights decreases from 1 at each time step so that after k updates, the value is $w^k$. Figure \ref{Swarm velocity} shows a geometric illustration of the effect of the velocity equation in a two-dimensional vector space. 
\subsubsection{Implementation of the algorithm}
\begin{figure}[!h]
\centering
\includegraphics[width=0.80\textwidth]{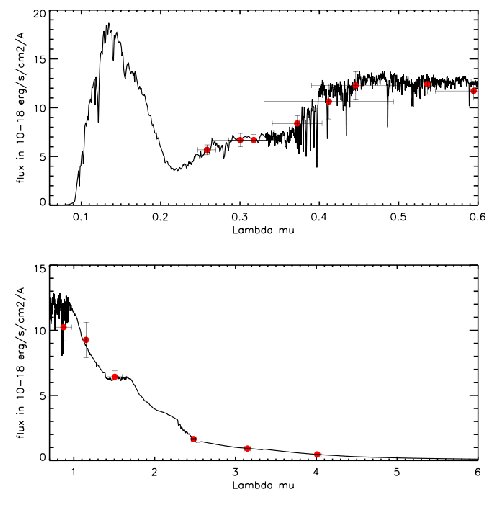}
\caption{{\small  Synthesis of a real galaxy SED from photometry and observed SFR using a combination of 6 CSP with exponential decline $\tau$=100\,Myrs and a two component extinction law. The minimization of the $\chi^2$ statistic has been realized with a particle swarm optimization algorithm. }}
\label{Swarm_fake}
\end{figure}

I have used a PSO algorithm developed in IDL, \textit{rmd\_pso}, by the NIST center for neutron research (http://www.swarmintelligence.org/). The optimization function to be minimized is defined in equation \ref{equation_swarm} and the free parameters have the following constrains: $x^i$ $\in$[0,1], E(B-V) $\in$ [0,1] and $R_V$ $\in$ [2,6]. The model templates have been normalized to the magnitude in z-band of each object. For each galaxy, I have ran 10 experiences\footnote{Re-run the search from different initial condition of the swarm.} with 50 swarm particles and a maximal number of 80 iterations. The initial inertia parameter is set to 0.95 and the cognitive and social parameters are both 2.0. The best fit result is the experience with the lower $\chi^2$. Confidence level interval can be estimated from the $\chi ^2$ distribution generated by the several restart experience (similar to a Monte-Carlo iterations). An example of fit of the SED is shown in figure \ref{Swarm_fake}. 

The optimization of the convergence and time execution of the algorithm is adjusted by the number of particles in the swarm, the number of time step (iteration), the number of experiences, the inertia, cognitive and social parameters. The optimization of these variables is still in progress. 

\subsection{Validation of the method}
\begin{figure}[!h]
\centering
\includegraphics[width=0.60\textwidth]{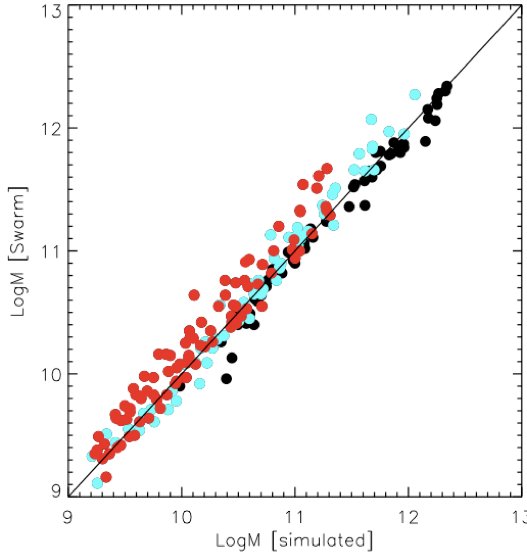}
\caption{{\small Fit of the rest-frame SED and the constrain on SFR by a combination of 6 CSP with exponential decline $\tau$=100Myrs and a two component extinction law.The color codes the  the mass fraction of old stellar population (age>3 Gyr) in the galaxy, see fig. \ref{MaxMass} }}
\label{Swarm_results}
\end{figure}

I have tested the accuracy of the algorithm to retrieve the stellar mass using the fake galaxy library. The results are shown in figure \ref{Swarm_results}. The algorithm recovers the stellar mass within a random uncertainty of 0.2\,dex. There is a small systematic of $+0.1\,dex$ in the derived stellar mass and it is more pronounced in galaxies with high fraction of young stars. However, the objects with the worst stellar mass estimation coincide with those having the higher $\chi^2$. I have concluded that in some cases the algorithm has not finished to converge. To test this hypothesis, I have modeled the SED of some galaxies with a different set of algorithm parameter: larger number of swarm particles and experiences. The execution time increases, around one hour per galaxy, but the quality of the fit improves and the stellar masses are recovered within a random uncertainty of $0.1\,dex$ for all SFH. 

\subsection{Systematics and limitations}
The preliminary results show that the constrain on the SFR improves the accuracy on the derived stellar mass and stellar population in star-forming galaxies. The degeneracy on age and extinction can be broken by using the SFR as a new constrain. However, the test realized does not account for the uncertainties due to template mismatches. Indeed, I have used the same models for simulating fake galaxies and for extracting the stellar population. I am planning to realize more tests using fake galaxies with more complex SFHs, such as model galaxies taken from hydrodynamical merger simulations \citep{2009ApJ...696..348W} and to evaluate the uncertainties due to the models \citep{2009ApJ...699..486C,2010ApJ...708...58C,2010ApJ...712..833C}.

\section{Application to individual galaxies}
I have applied the swarm algorithm and full-spectra fitting methods to individual galaxies of the IMAGES sample. I present hereafter the study realized in two CDFS galaxies in the context of a modeling campaign of distant galaxies started by the group.

The photometric measurements come from the catalogue described in \ref{Photometry catalogue}. I have shifted the observed quantities to the rest frame, $\lambda= \lambda^{obs}/(1+z)$, $f_\lambda=f^{obs}_\lambda \times(1+z)$ and $\sigma_f =\sigma_f^{obs}/(1+z)$.

\subsection{A forming, dust-enshrouded disk at z = 0.43}
\label{dust-enshrouded disk}
 In "\textit{A forming, dust-enshrouded disk at z = 0.43: the first example of a massive, late-type spiral rebuilt after a major merger?}" \citet{2009A&A...496..381H}, we have modeled a compact LIRG galaxy J033245.11-274724.0 from the IMAGES survey.

\subsubsection{Morpho-kinematic observations and physical model}
\begin{figure}[!h]
\centering
\includegraphics[width=0.70\textwidth]{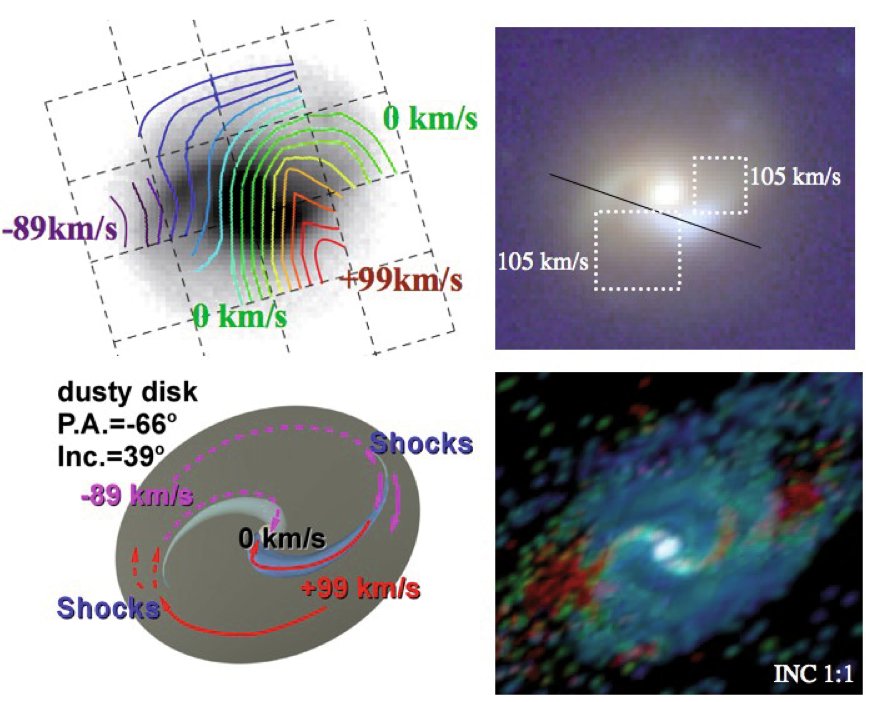}
\caption{{\small  \textbf{Upper left}: v-band image of J033245.11-274724.0 on which the GIRAFFE iso-velocities have been superimposed. Colour coding is the same as inYang et al. (2008). \textbf{Upper right}:b?v+i?zcolour- combined image of J033245.11-274724.0 on which the dynamical axis (full lack line) and the velocity dispersion peaks (dotted squared boxes) at $105\,km\,s^{-1}$ have been superimposed. The entire galaxy shows large velocity dispersion \citet{2008A&A...477..789Y}. The figure in the electronic version is of higher contrast. \textbf{Bottom left}: a sketchy model of the galaxy showing the dusty disk, and the gas motions (purple: approaching; red: recessing) from the background (dotted lines) to the foreground of the disk (full lines). The bar close to the centre is inclined with respect to the disk plane, explaining its disappearance at the top-right of the bulge as well as the asymmetry between the two arms. The resulting dynamical axis is thus caused by the gas motion both through the bar and around the disk. \textbf{Bottom right}: result of a galaxy collision (1:1 mass ratio) with an inclined orbit (close encounters) from \citet{2002MNRAS.333..481B}, 0.5\,Gyr after the merger. The gas wraps up inside the disk, creating a bar and a symmetric-arm system whose motion does not follow the gas rotation, creating shocks and high velocity dispersions (seen as red particles in the simulation), which are consistent with observations. From \citet{2009A&A...496..381H}}}
\label{Hammer_ind1}
\end{figure}

J033245.11-274724.0 is a luminous infrared galaxy with a stellar mass similar to the Milky Way. It has a very compact morphology: $R_{half}\sim1.7\,kpc$. The color images from the Ultra Deep Field have revealed a central bulge with a small bar prolonged by very luminous arms, which are surrounded by a red and dusty disk, see Fig.\ref{Hammer_ind1} upper-right panel. The disk account for the bulk of the stellar mass and star formation in the galaxy. 
The arms have different colors - one red and the other blue - due to the dust distribution. It suggests that the arms are not aligned with the disk: the blue arm emerges above the disk while the red arm is hidden below the disk.  
The kinematical axis is strongly offset ($\sim45^o$) from the photometric major axis of the disk and is slightly offset from the galaxy centre (Fig. 4, upper panels), arguing for a non-axisymmetric, heavily perturbed and out of equilibrium disk.  Morphology and kinematics show that both gas and stars spiral inwards rapidly to feed the disk and the central regions.
Figure \ref{Hammer_ind1} (Bottom-left panel) shows a sketchy representation of the galaxy, which reproduces both its morphology and kinematics.

\begin{figure}[!h]
\centering
\includegraphics[width=0.75\textwidth]{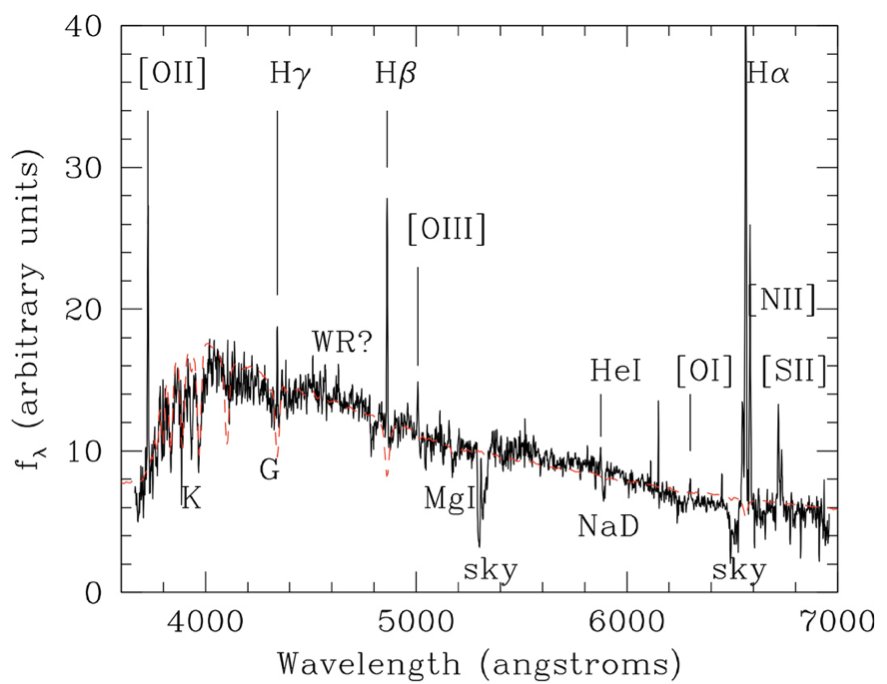}
\caption{{\small  FORS2 spectrum of J033245.11-274724.0 after de-reddening the continuum by $c_{extinction}$ = 1.0 (Orion extinction curve). The spectral energy distribution can be reproduced (red line) by a set of models with stellar ages t from 0.25 to 1.2 Gyr assuming $\tau$ (SFR$\sim$exp(-t/$\tau$)) from 0 to 5\,Gyr. From \citet{2009A&A...496..381H}}}
\label{Hammer_ind2}
\end{figure}

According to the merger simulations of Barnes 2002, the particular structures bar+arms observed is typical of galaxy mergers well after the collision. This observation has led us to model the galaxy as a merger of two equal-mass gaseous-rich galaxies on an inclined orbit. The morphology and kinematics are consistent with the properties of the remanent galaxy observed 0.5\,Gyr after the merger, see Figure \ref{Hammer_ind2}. 

\subsubsection{ISM properties and stellar population modeling}
FORS2 spectroscopy from the IMAGES survey is available for this object (\citet{2008A&A...492..371R} and previous chapters). The spectrum has a high quality and it is at least comparable to the spectrum of a typical galaxy from the SDSS. A first look to the spectrum allows us to argue that it is dominated by the light from young and intermediate age (A/F) stars. Indeed, the strong emission lines and strong Balmer absorption lines indicate the presence of such intermediate age stars. On the other hand, the detection of the He5875\AA\, line\footnote{Surprisingly, this line is produced in low-metallicity medium while the global metallicity of the galaxy is over-solar. The line may come from the outer regions of the galaxy which can have lower metal content as suggested by the scenario proposed by \citet{2008ApJ...674..172R}.}, and possibly Wolf-Rayet features  such as the blue bump, and possible detections of emission lines produced by high ionization species, such as [SiIII]4565\AA\, and [HeII]4686\AA\, indicate the presence of extremely young and hot stars. In addition, the presence of metal absorption lines (CaIIK, G, MgI \& NaD) indicates the presence of relatively old stars.

In order to constrain the contributions of young, intermediate and old stellar populations to the total stellar mass, we have used the STARLIGHT code (see Part I chapter 3). The best-fit provides a distribution of stellar ages including a large mass-component of young to intermediate age stars, the rest being associated with older ages (>2 Gyr), see table \ref{Stellar pop disk rebuild}. It suggests that a significant part of the stellar mass in J033245.11-274724.0 has been formed recently. This result has been confirmed by a independent method, see table \ref{Stellar pop disk rebuild}. We have modeled the spectral energy distribution of the galaxy with a mixture of six stellar populations and a two-parameter extinction law. The optimization of the stellar population decomposition has been performed with the Swarm intelligence algorithm described in previous chapter. According to the simulation, the first impact would occur at 0.5 Gyr, the merging of the nuclei at 1.1 Gyr, the appearance of the non rotating two-armed system at 1.6 Gyr, and its disappearance at 1.9 Gyr. These numbers are quite consistent with the ages of the stellar populations which dominate the J033245.11-274724.0 continuum. 

\begin{table} 
\centering
\begin{tabular}{l c c c} 
\hline
Method &Young&Intermediate&Old \\  
&age<300\,Myr&300\,Myr<age<1\,Gyr&age>1Gyrs \\  
\hline 
Full-spectra fitting&20\%&33\%&47\%\\
SED fitting& 5\%&15\%&80\%\\
\hline 
\end{tabular} 
\caption{The contribution fraction of the young, intermediate and old stellar population on the total stellar mass of J033245.11-274724.0. }
\label{Stellar pop disk rebuild}  
\end{table} 

I have discarded stellar outflow as a possible process to explain the large misalignment of the kinematics. I have compared the velocity of the emission-line system with that of the absorption-line system, and by examining the line profile of the NaD absorption lines, i.e. the two methods outlined by Heckman \& Lehnert (2000). By measuring the position of ten robust emission lines ($H\gamma$, $H\beta$, [OIII]4959 and 5007, HeI, [OI], [NII]6548 and 6583 and $H\alpha$), I find a remarkable agreement between those lines with a scatter of only $15\,kms^{-1}$. We excluded the two [SII] lines, since they exhibit a much larger shift, presumably due to the fact that they lie at the edge of the spectrum, where wavelength calibration may be more uncertain. Examining the absorption lines with a reasonable S/N (Balmer H12 and H11, CaII K, CaI, NaI-D1 and D2), I find a larger scatter of $52 \,kms^{-1}$ as expected by the larger uncertainty on measuring absorption lines. The absorption and emission line systems appear to have the same velocity, within an uncertainty of $33\,kms^{-1}$, a value far below the expected uncertainties on absorption measurements. The galaxies do thus present no evidence of outflows. Examination of the emission-line profiles also show no evidence of an outflow. An absence of a significant outflow contribution is unsurprising given the high mass of the galaxy, whose gravitational forces prevent stellar super-winds dominating the velocity field.

\subsection{A face-on merger at z=0.7: J033227.07-274404.7}
In \textit{Determining the morpho-kinematic properties of a face-on merger at z=0.7}, \citet{2010A&A...513A..43F}, we have presented the morphological and kinematical analysis of a system at z=0.74. J033227.07-274404.7 displays an unusual elongated arc-like morphology and a very peculiar kinematics. 
\subsubsection{Morpho-kinematic observations and physical model}
\begin{figure}[!h]
\centering
\includegraphics[width=0.80\textwidth]{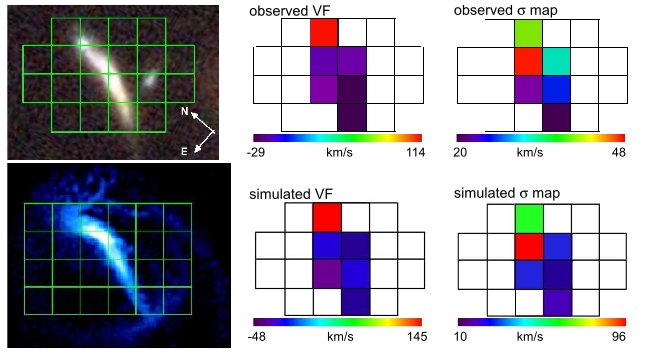}
\caption{{\small  This figure, extracted from \citet{2010A&A...513A..43F}, illustrates the principle of individual galaxy modeling. The observables (in the top row) such as the kinematics and the morphology impose strong constrain on models and allow to break the degeneracy between a large number of model scenarios.  \textbf{Top row, left panel}: Composite image (B+V, i and z bands) of J033227.07-274404.7 taken with the Advanced Camera for Surveys (ACS) of the HST. Size of FoV is 2.0 arcsec $\times$ 2.6 arcsec (14.4 kpc $\times$ 18.8 kpc). Superposed grid shows the position of the GIRAFFE IFU.  \textbf{Top row, middle panel}: Line-of-sight velocity map of the object derived from GIRAFFE observations of the [OII]3726,3729 \AA\, doublet.  \textbf{Top row, right panel}: velocity map and $\sigma$-map of the object derived also from GIRAFFE observations.  \textbf{Bottom row, left panel}: Projected distribution of gas of the simulated encounter using the ZENO code (Barnes 2002). Snapshot shows the encounter just after the second peri-passage (12 Myr after) and 1.93 Gyr after the first peri-passage. The FLAMES IFU grid is superposed to simulate the GIRAFFE observations.  \textbf{Bottom row, middle panel}: Line-of-sight velocity field derived from the ZENO simulation considering an IFU similar to that of GIRAFFE. Bottom row, right panel: $\sigma$ map of the same simulation. From }}
\label{Isaura_galaxy}
\end{figure}

The object consists of an elongated structure about 1.52 arcsec (11kpc) long and a smaller structure (2.45kpc) located to the north-northwest of the main structure. A small faint structure is also seen to the southwest of the elongated structure. Bluer colors are seen all the way from the center of the main galaxy to the companion. The smaller region to the southwest has very blue colors. The colors are consistent with those of pure starburst and Sbc galaxy \citep{2008A&A...484..159N} and thus suggest that the whole galaxy is actively forming stars. The elongated structure show a very uniform VF, see Figure \ref{Isaura_galaxy} upper-row. For these pixels, the global velocity gradient is shallower than $14\,km\,s^{-1}$ . The extremely small velocity gradient is a surprising observation for this galaxy. In fact, small velocity gradients are expected in face-on galaxy, whereas the observed morphology suggests that the galaxy is on contrary seen edge-on. On the other hand, the northwest structure (upper northernmost pixel) has a velocity that differs by more than $130\,km\,s^{-1}$ from that of the other regions.
We have identified this system as resulting from a 3:1 merger activating a giant bar in a previously face-on disc.

\subsubsection{ISM properties and stellar population modeling}
The galaxy has a FORS2 spectrum from the GOODS spectroscopic survey, see Fig. \ref{Isaura_ind1}. From the study of the emission lines, we have found that the galaxy is almost devoided of extinction (from Balmer decrement)  and that the ISM may be affected by shocks according to the strong emission of the [NeIII]$\lambda$3869\AA\, line. To constrain the stellar population content of the galaxy, I have performed a fit to the observed spectrum with a linear combination of stellar libraries. I have used a base of 39 template single-stellar-populations (SSPs) from Bruzual \& Charlot (2007), of 13 ages (from 10 Myr to 5 Gyr) and 3 metallicities (subsolar, solar, and supersolar) using the STARLIGHT software \citep{2005MNRAS.358..363C}. A \citet{1989ApJ...345..245C} extinction law was assumed. The best-fit relation indicates that a high fraction of the galaxy spectrum is dominated by young stellar features typical of 13 Myr, 100 Myr, 200 Myr, and intermediate age stellar population of around 1 Gyr. However, the quality of the spectra, and the limited wavelength range does not allow us to provide strong constrains on the fractional contribution of light from stellar populations of different ages. To constrain the stellar population properties more reliably, we have investigated the spectral energy distribution (SED) based on photometric measurements.
 \begin{figure}[!h]
\centering
\includegraphics[width=0.70\textwidth]{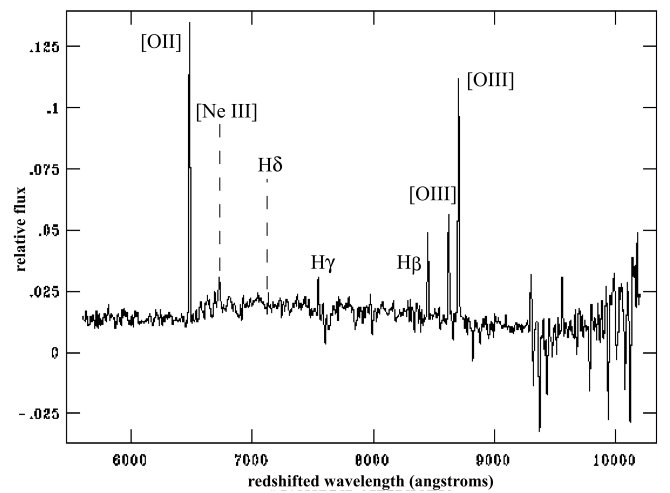}
\caption{{\small  The GOODS spectrum of J033227.07-274404.7. The main emission lines are indicated. The relative flux is in energy per unit area per unit time per wavelength. From \citet{2010A&A...513A..43F}}}
\label{Isaura_ind1}
\end{figure}

 I have adjusted a synthetic SED consisting of a linear combination of composite stellar populations (CSP) and a two parameter extinction law from Cardelli (1989). The base consists of 6 CSP from Charlot \& Bruzual 2007 models with a Salpeter IMF (Salpeter 1955) and a $\tau$-exponentially declining star formation history with $\tau$ = 100 Myr. According to the low metallicity detected in the gaseous phase, I have selected CSP with sub-solar metallicity Z = 0.5 $Z_\odot$. The ages of the 6 stellar populations are 13 Myr, 200 Myr, 500 Myr, 1 Gyr, 4 Gyr, and 7 Gyr, respectively. The results for the bestfit relation ( $\chi$2 = 2.09) are shown in Table \ref{SED_isaura_table}. 
 \begin{figure}[!h]
\centering
\includegraphics[width=0.70\textwidth]{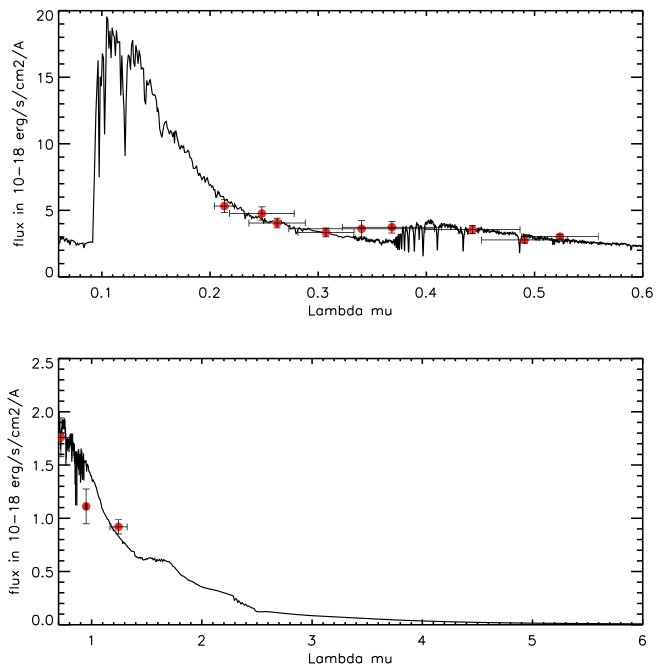}
\caption{{\small  Synthesis of J033227.07-274404.7 SED from photometry and observed SFR using a combination of 6 CSP with exponential decline $\tau$=100\,Myrs and a two component extinction law. The minimization of the $\chi^2$ statistic has been realized with a particle swarm optimization algorithm. From \citet{2010A&A...513A..43F}}}
\label{Isaura_ind1}
\end{figure}

\begin{table} 
\label{Stellar pop disk rebuild} 
\centering
\begin{tabular}{l c c c} 
\hline
Method &Young&Intermediate&Old \\  
&age<200\,Myr&300\,Myr<age<1\,Gyr&age>1Gyrs \\  
\hline 
SED fitting&17\%&13\%&70\%\\
median 68\%&13\%&32\%&55\%\\
\hline 
\end{tabular} 
\caption{The contribution fraction of the young, intermediate and old stellar population on the total stellar mass of J033227.07-274404.7 from SED fitting with SFR constrain. The second row is the median stellar-population mass fraction for solution in the 68\% confidence interval.}
\label{SED_isaura_table} 
\end{table}

    \NumberThisInToc
\FrameThisInToc
\part{Future instrumentation on E-ELT}
 \begin{flushright}
\textit{"Keep your eyes on the stars,\\
 and your feet on the ground."\\
 by Theodore Roosevelt}
 \end{flushright}
 \EmptyNewPage
\FrameThisInToc
\chapter{Preparing the E-ELT era}

\begin{figure}[!h]
\centering
\includegraphics[width=0.70\textwidth]{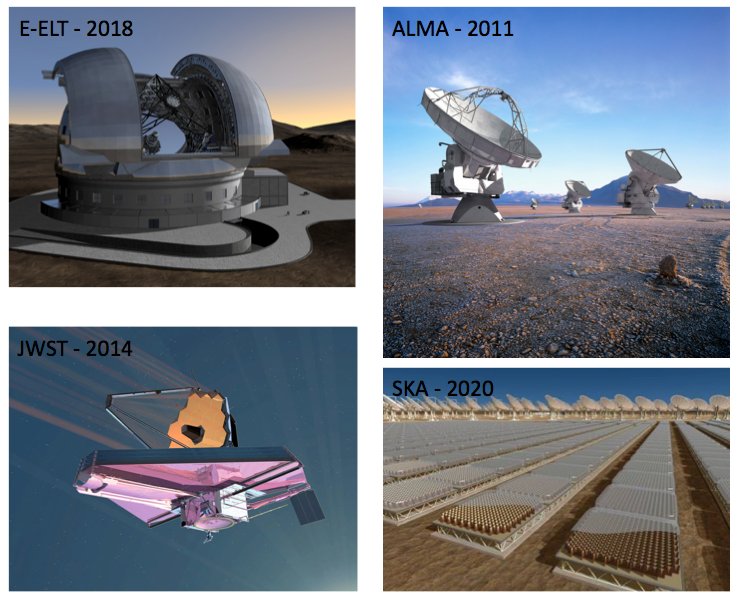}
\caption{{\small The artistic view for the future facilities planned for the next decade, with the respective expected year of operation.}}
\label{E-ELT_era}
\end{figure}

Despite the great progress in our understanding of galaxy evolution and formation during the past decade, the key question of the stellar mass assembly - how and when galaxies assembled across cosmic time - remains a main issue. Pushing the actual 8m class telescopes and the Hubble Space Telescope to their limits, it is only possible to study galaxies up to z=1 with the same detail than local studies. Although galaxies at higher redshift have been detected, the limitation on the spatial resolution and the low signal-to-noise of the observations and the strong bias on the representativeness of these galaxies prevents from having a consistent snapshot of galaxy properties at lookback time superior to eight Gyrs. A consistent picture of the mass assembly and star-formation across the entire history of the Universe requires deeper observations at higher spatial resolution. In the next decade, such observation will be possible thanks to the advent of a new generation of facilities (see Fig \ref{E-ELT_era}): 
\begin{itemize}
\item \textbf{James Webb Space Telescope}. This spatial telescope is planned to replace the Hubble Space Telescope in 2014. It is an optical and near infrared telescope and it has a 6.5-meter primary mirror. 
\item \textbf{Extremely Large Telescope}. Three optical-to-near IR ground-based telescopes with a primary mirror greater than 20m of diameter are planned to be constructed in the next ten years. The most ambitious project is the European-Extremely Large Telescope (E-ELT) promoted by the European Southern Observatory (ESO), which will have a  primary mirror of  42 meters. 
\item \textbf{Atacama Large Millimeter/submillimeter Array}. ALMA is a millimetre-/submillimetre-wavelength interferometer array. It will mainly observe the molecular gas of distant galaxies.
\item \textbf{Square Kilometer Array}. SKA will be a radio array with a total collecting area of approximately one square kilometer. It will map out the cosmic distribution of neutral hydrogen at high cosmological distances. 
\end{itemize}

The Galaxy Etoile Physique Instrumentation lab (GEPI) at the Observatoire de Paris, my hosting lab, is involved in several E-ELT projects, such as the telescope design (the M4 deformable mirror), adaptive optic module (the laser tomography AO module ATLAS) and instruments (OPTIMOS-EVE and EAGLE). In this nourishing environment, I had the opportunity to participate in a phase-A E-ELT instrument: OPTIMOS-EVE, a fiber-fed visible-to-near infrared multi-object spectrograph. My work in the consortium is to define a strategy to sample the sky around each scientific target and develop specific software to subtract the sky. In this chapter, I will start to quickly describe the E-ELT project and then present the driving science cases and the technical design of OPTIMOS-EVE. In the last sub-section I will discuss on the pros and cons of fiber-fed and multi-slit spectrograph. I will particularly emphasize the issue of sky extraction which, as I will show, is one of the technical challenges for faint spectroscopy with fiber-fed instrument. 

\section{European Extremely Large Telescope}
The European Extremely Large Telescope (E-ELT) is an ambitious project for a 42m primary mirror optical-to-near infrared telescope, operated by the European Southern Observatory. The E-ELT is expected to reveal a revolutionary view of the Universe enabling the study of extra-solar planets (in the Galaxy and nearby dwarfs), of stellar populations in external galaxies, and of faint distant galaxies tracing the early history of the Universe. 

\begin{figure}[!h]
\centering
\includegraphics[width=0.90\textwidth]{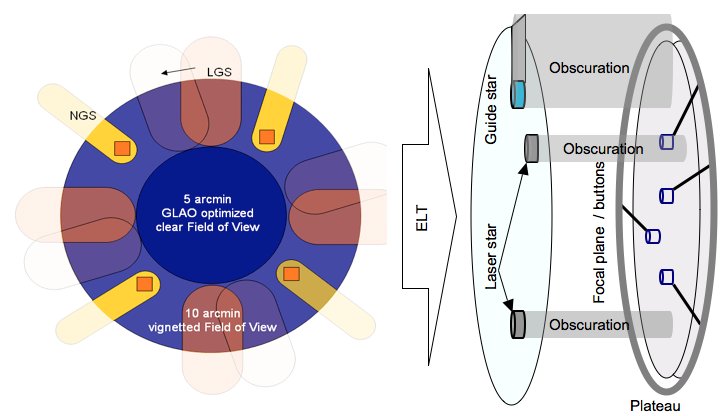}
\caption{{\small \textbf{Left}: accessible FoV as seen from OPTIMOS-EVE. Laser Guide stars are locked with the pupil in this example and therefore rotate with respect to the image plane. \textbf{Right}: projected view of the Focal plane showing the plane of the plate and the obscuration of the LGS and Guide star. From the OPTIMOS-EVE Operational Concept Definition document.}}
\label{fov_elt}
\end{figure}

Since January 2007, the project is in phase B and will lead to the proposal for construction. The phase B includes contract with industry to design and manufacture prototypes of key elements (primary mirror, M4 and mechanical structure). The first light is expected in 2018. The current opto-mechanical concept of the E-ELT (B. Delabre, 2008) is composed by five mirrors which feed two Nasmyth foci and a coude type focus. The correction of the atmospheric turbulence is directly included into the optical design by the M4 and M5 mirror, respectively an adaptive mirror of 2.6m and a tip/tilt mirror. In addition, four to nine laser-stars will allow to analyze the wave front in the E-ELT field of view ($\sim$10 arcmin). Fig. \ref{fov_elt} shows the configuration of the laser guide stars and the effective field of view of ELT seen by the instruments. The instruments will be distributed in permanent way among the different focal stations to permit rapid switching between instruments at any time. 

In parallel to the E-ELT study, the "E-ELT Design Study", partially funded by the European commission, have defined a complete list of instruments able to fulfill the science cases of the E-ELT (DRM E-ELT/ESO \emph{http://www.eso.org/sci/facilities/eelt/science/drm/}). As in the case of the VLT, most of the instruments design and construction will be carried out by the community. In September 2008, eight instrument concepts and two post-focal AO modules (MCAO and LTAO) have been defined and committed to consortia (see Table \ref{Instrument_ELT}, for the characteristic of the future instruments). The Phase\,A of the instruments are now complete and the selection of Phase B instruments are planned to be announced this year. Two instruments are planned to be operational for first light, from which  OPTIMOS is a possible candidate.

\begin{table}[!h]
\begin{center}
\begin{tabular}{|c|c|c|}
\hline
Name&Instrument type &Science case \\
\hline
CODEX&High Resolution&Exoplanet\\
Pasquini et al. (2008)&High Stability &cosmological constant\\
&Visual Spectrograph&cosmology\\
\hline
EAGLE&Wide Field, Multi IFU NIR &high-z galaxy\\
Cuby et al. (2009)&Spectrograph with MOAO&resolved stars\\
\hline
EPICS&Planet Imager and &Exoplanet \\
Kasper et al. (2008)& Spectrograph with&proto-stars\\
& Extreme Adaptive Optics&\\
\hline
HARMONI&Single Field&high-z galaxies\\
Tecza et al. (2009)&Wide Band Spectrograph&Quasars\\
&&Stellar population\\
\hline
METIS&Mid-infrared Imager &Exoplanets \\
Brandl et al. (2008)&and Spectrograph with AO&proto-stars\\
\hline
MICADO&Diffraction-limited &Galactic center \\
Davies et al. (2010)&NIR Camera&stellar population\\
\hline
OPTIMOS&Wide Field Visual MOS &high-z galaxy survey\\
&&stellar population\\
&&exoplanet\\
\hline
SIMPLE& High Spectral Resolution&Chemical abundances\\
Oliva et al. (2009)&NIR Spectrograph&\\
\hline
\end{tabular}
\caption{{\small The 8 instruments planned for E-ELT, from D'Odorico et al. 2008.}}
\label{Instrument_ELT}
\end{center}
\end{table}

\section{OPTIMOS-EVE}
OPTIMOS-EVE is a fibre-fed, optical-to-infrared multi-object spectrograph designed to explore the large field of view provided by the E-ELT at seeing-limited conditions. Therefore, OPTIMOS-EVE is well-suited to be used in the early operational phase of the E-ELT and also beyond, when the atmospheric conditions will not be optimal to provide full adaptive optics corrections. OPTIMOS-EVE is one of the few E-ELT instruments under study that will explore the visible to near-infrared wavelength region.
\subsection{Consortium}
The OPTIMOS phase A study combines two parallel studies, one investigating a fibre-fed MOS concept (i.e. this study regarding the Extreme Visual Explorer, PI F. Hammer, GEPI and co-PI Lex Kaper, UvA), the other one exploring a slit-mask MOS concept (DIORAMAS, PI O. Le Fevre, LAM). Both studies prepared their own science cases and development plan. The shared work packages Detectors (RAL), Instrument Control Software (OAT-INAF) and Instrument Control Electronics (RAL) support both studies. The OPTIMOS phase A study was coordinated by ESO responsible S. Ramsay and study manager F. Zerbi.
The OPTIMOS-EVE consortium builds on the expertise of the FLAMES/GIRAFFE (optical fibre-fed multi-object spectrograph on the ESO Very Large Telescope), VLT/X-shooter (wide-band optical-to- near-infrared spectrograph) and Subaru/FMOS (fibre-fed near-infrared spectrograph) consortia. The following partners constitute the OPTIMOS-EVE consortium:
\begin{itemize}
\item GEPI: Galaxies, Etoiles, Physique et Instrumentation, Observatoire de Paris, France
\item NOVA: Nederlandse Onderzoekschool Voor Astronomie, University of Amsterdam, Radboud University Nijmegen, NOVA-ASTRON, Dwingeloo, The Netherlands
\item RAL: Rutherford Appleton Laboratory, Oxford, United Kingdom
\item NBI: Niels Bohr Institute, Copenhagen University, Denmark
\item INAF:Instituto Nazionale di AstroFisica (Osservatorio astronomico di Trieste, Osservatorio Astronomico di Brera), Italy
\end{itemize}
\subsection{Science cases}
The OPTIMOS-EVE phase A study Science Team studied a number of key science cases covering the three main science themes motivating the development of the E-ELT (Science Case for the E-ELT, Ed. I. Hook): (i) Planets and Stars; (ii) Stars and Galaxies; (iii) Galaxies and Cosmology. The Science Team explored 5 key science cases from which the scientific and technical requirements for OPTIMOS-EVE have been derived:
\begin{itemize}
\item \textbf{Planets in the Galactic bulge and stellar clusters, and in external dwarf galaxies}. 
In spite of the fact that over 400 extra-solar planets are known, these are mostly hosted around stars in the solar vicinity. The few distant planets known do not have orbits, masses etc. We expect, on theoretical grounds, that the environment plays a significant role in the process of planet formation. Therefore, it is important to detect and characterize planets in environments different from the solar vicinity, like the Galactic Bulge and Local Group dwarf spheroidals. With its capability of obtaining a radial velocity precision of 10 m/s for giant stars down to the 20th magnitude OPTIMOS-EVE will make such a study possible. Its multiplex capabilities will allow us to monitor a significant number of stars in each observed field.
\item \textbf{Resolved stellar populations in nearby galaxies}. 
With VLT we are beginning to study in detail the stellar populations of the Local Group galaxies. However, many galaxy types are not present in the Local Group. In order to make real progress in our understanding of galaxy formation and evolution we need to study in detail all the different types of galaxies which can be found in the groups of Sculptor and Cen A. The high efficiency low-resolution mode of OPTIMOS-EVE and its high multiplex grant this possibility.
\item \textbf{Tracking the first galaxies and cosmic re-ionization from redshift 5 to 13}. 
We know, from the fluctuations of the microwave background, that at z=10 the Universe was already largely re-ionized. Nevertheless, we know little about the objects that produced this re-ionization. The search for the "first lights", the sources of the photons responsible for the re-ionization can be carried out with OPTIMOS-EVE, which, thanks to the IR arm, holds the promise of tracing these up to z=13.
\item \textbf{Mapping the ionized gas motions at large scales in distant galactic haloes}.
The observations of the local Universe have shown that galaxies are surrounded by extended haloes of ionized gas. These haloes are the interfaces of the galaxies to the intergalactic medium and the way to enrich it of metals. The study of these haloes also allows understanding the history of galaxy-galaxy interactions, since these always leave recognizable signatures on the halo kinematics. Such studies could be conducted with an instrument like GIRAFFE at the VLT up to a redshift of z$\sim$0.5; OPTIMOS-EVE will make this possible up to a redshift of 3.5.
\item \textbf{3D reconstruction of the Intergalactic Medium}.
We know that the space between galaxies and galaxy clusters is not empty, but is filled with a very low density warm medium. Such a medium shows up as Ly alpha absorption in the spectra of distant quasars. Although this affords a 'cut through' of the structure of the IGM along the line of sight, nothing is known about its transverse structure. Cosmological simulations show that the IGM has a filamentary structure and filament crossings correspond to the locations of galaxy clusters. Some information on the transverse structure can be obtained in the case of pairs of gravitationally lensed quasars. OPTIMOS-EVE will provide sufficient resolution and sensitivity to use Lyman- break galaxies of 25th magnitude as background sources. These galaxies have a sufficient space density to allow a 3D reconstruction (tomography) of the IGM, a real 3D picture which may be compared to cosmological simulations.
\end{itemize}

\subsection{System overview}
\begin{figure}[!h]
\centering
\includegraphics[width=0.90\textwidth]{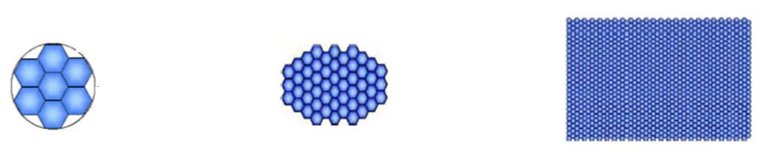}
\caption{{\small The three observational modes of OPTIMOS-EVE. From left to right: the multi-object mode MO, the medium Integral Field Unit MI, and the large integral field unit mode. Each hexagon represents a single fibre, assumed to be 0.3".}}
\label{EVE_mode}
\end{figure}

OPTIMOS-EVE has been designed for the E-ELT Nasmyth focus and provides low-, medium- and high-resolution spectroscopy (R$\sim$5\,000 -30\,000) from the ultraviolet to the near-infrared (0.37 to 1.7 $\lambda$) for multi-object studies of sources nearby and at cosmological distances. The instrument consists of three main sub-systems: 
\begin{itemize}
\item A positioner that consists in a carousel containing four focal plates. One focal plate (so-called 'active') contains the fibers buttons previously positioned and is aligned with telescope axis while the opposite focal plate is being configure.
\item Micro-lenses and fibers that transmit light from the active focal plate to the slit of the spectrographs. Micro-lenses and fibers are gathered in different kind of bundles, with different apertures and different fibre sizes in order to address the five science cases, see Fig. \ref{EVE_mode}. One kind of bundle is active at a time.
\item Two dual beam spectrographs that can cover three spectral resolutions (low (LR), medium (MR) and high (HR)) thanks to exchangeable gratings and a slit width adaptation (see Table \ref{EVE_spectro}).
\end{itemize}
\begin{table}[!h]
\begin{center}
\begin{tabular}{c|c|c}
&\textbf{VIS arm}&\textbf{NIR arm}\\
\hline
\# of spectrographs&2&2\\
Wavelength&370-930nm&940-1700nm\\
Spectral coverage&$\lambda/3$&$\lambda/10$\\
Resolution MO& 5000 15000 30000&5000 15000 30000\\
Resolution MI LI& 5000&5000\\
\hline
\end{tabular}
\caption{{\small Main characteristics of VIS and NIR arms spectrograph}}
\label{EVE_spectro}
\end{center}
\end{table}

\begin{figure}[p]
\centering
\includegraphics[width=1.0\textwidth]{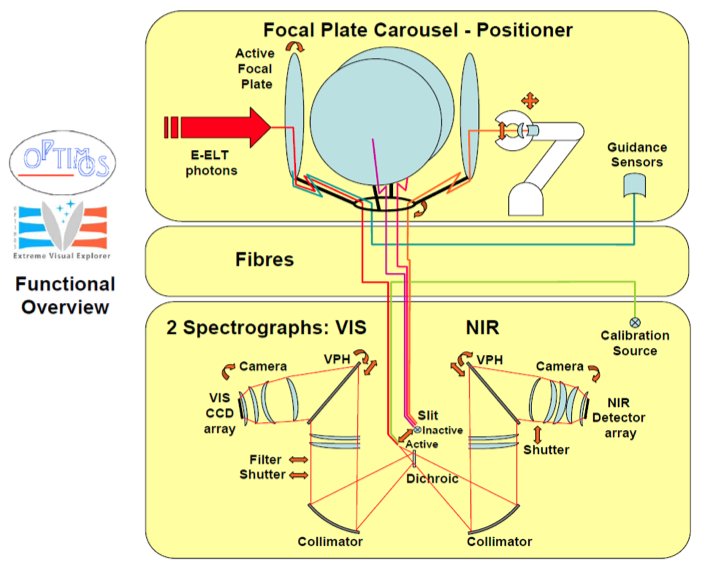}
\caption{{\small Schematic overview of OPTIMOS-EVE. The 3 main subsystems of the instrument are clearly indicated: The Focal Plate Carousel - Positioner, containing 4 Focal plates with various Mono Objects Fibre inputs and IFUs as well as a robot positioner; the fibres, to transport the collected light to the spectrograph, guidance sensors and from the calibration source; the spectrographs, consisting of a visible and an infrared arm, separated by a dichroic.}}
\label{EVE_schema}
\end{figure}

A schematic overview of this system is shown in Fig. \ref{EVE_schema}. It is designed to fit within the volume and mass limits of the focal plane station. OPTIMOS-EVE will allow to observe simultaneously multiple scientific targets in a 7" field of view. Five fibre setups are provided, see Fig. \ref{EVE_mode}:
\begin{itemize}
\item Mono object (MO), that comes in three different sub-modes providing different spectral resolutions. The MO LR mode will permit to observed at low spectral resolution (R=5\,000) 240 objects with a spatial resolution of 0.3". The medium and high resolution modes will allow to observe respectively 70 and 40 objects within an aperture of  0.9" and 0,72" and a spectral resolution of 15\,000 and 30\,000.   
\item Medium integral field unit (MI): 30 IFU composed by 56 fibers covering a surface of 2"x3". The spatial resolution is 0.3" and the spectral resolution is fixed to R=5\,000. 
\item Large integral field unit (LI): one large IFU of 15"x9" composed by 1650 fibres. The spectral and spatial resolution are the same as for the MI mode. 
\end{itemize}

\subsection{OPTIMOS: Multi-slit vs fibers }
As explained previously, the OPTIMOS project is composed by two parallel studies, which investigates the fiber-fed technology (EVE) and the slit spectroscopy (DIORAMAS) in the frame of multi-object spectroscopy. In this sub-section, I describe the key parameters that define the performance and efficiency of a multi-object spectrograph and compare them in the case of fiber and slit technology instruments. The weight given to these factors depend on the scientific targets and goals. I will focus on the comparison of two OPTIMOS science cases: high-z galaxies and resolved stellar population of nearby galaxies. 
\subsubsection{Global efficiency}
Historically, fiber-fed instruments have been considered to be unsuitable for faint object spectroscopy. In fact, until very recently the low efficiency of througput in fibers had limited the total efficiency of fiber-fed spectrographs. To give an example, the multi-slit spectrograph FORS2/VLT has an effective efficiency of $\sim$ 25\% while GIRAFFE/VLT, a fiber-fed spectrograph, has an efficiency around 8-10\%. However, thanks to important progress in fiber technology, it is now possible to design multi-fibers spectrographs with efficiencies close to multi-slit technology. OPTIMOS-EVE will have an average throughput of 26\% in the R=5000 mode, for the mono-object fibers, including an estimated 67\% throughput of the fibre system.
\subsubsection{Aperture loss}
Both technologies suffer from aperture loss. However, the issue was considered more critical for fiber-fed instruments. Fibers have smaller collecting area than slit and any uncertainty on the positioning induces dramatically aperture losses. Early fiber-fed instruments suffered from large uncertainties on the positioning of the fibers, mainly due to the mismatch on astrometry of input catalog, the low precision of the fiber positioning system, curvature of the focal surface, bad telecentring angle of fibers and the effect of the atmospherical refraction (Newman et al. 2002). However, after two decades during which more than 40 fiber spectrographs have been built or proposed, these issues have been in majority successfully solved. As an example, the GIRAFFE positioning robot have a precision of $10\mu m$. Actual astrometric catalogues have precision of 0.15" but the forthcoming generation of catalogue will have precision reaching 0.05" (e.g VISTA). I have investigated the aperture loss in both systems arising from the only geometric factor. In other words: Does the use of a rectangular or a circular aperture significantly affect the aperture loss? 

\begin{figure}[p]
\centering
\includegraphics[width=0.50\textwidth]{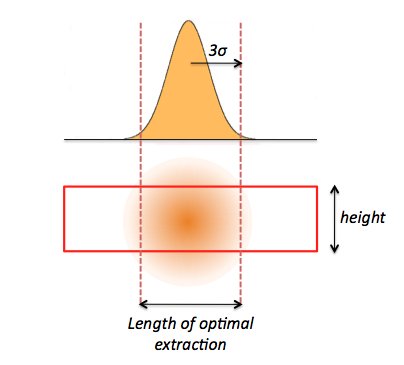}
\includegraphics[width=0.45\textwidth]{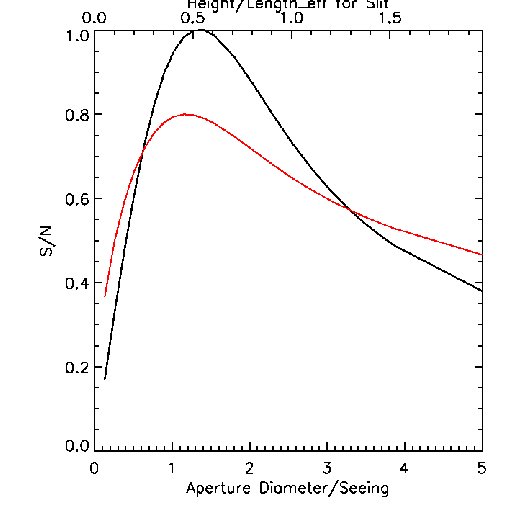}
\caption{{\small \textbf{Left}: Optimal extraction of a target in a slit. The area of aperture considered in the S/N calculus is the height of the slit $\times$ the length of optimal extraction. \textbf{Right}: S/N as a function of the aperture diameter/Seeing for fiber (in black line) and height over the length of extraction for slit (in red line). The aperture can be optimized by fixing the typical seeing expected. For reference, a typical seeing in Paranal, at blue wavelength, is around 0.78" and 0.47" with GLAO. }}
\label{Aperture_psf}
\end{figure}

In the case of point source observations, as in resolved stellar population science cases, the dimension of the aperture is optimized to minimize the losses of light at a given wavelength. It results in a necessary trade-off between S/N and aperture size. The optimum fiber aperture or slit height is determined from the computed S/N changes versus aperture and seeing. I have simulated images of point source objects, modeled by a 2D gaussian\footnote{The PSF follows a Moffat function. However, since the Moffat function is well approximated by a gaussian function in the central region, I have modeled the PSF by a gaussian function.}, at different seeing condition- from seeing-limited to GLAO. I then modeled slit and fiber observations with different aperture dimensions, assuming a perfect centering of the apertures, and calculate the S/N in each case. The signal to noise is defined as the fraction of flux falling into the aperture ($\propto$ to encircled energy, EE) divided by the square root of the aperture area (flux from the sky). It corresponds to a S/N ratio: $EE\sqrt{Sky}$. In the case of slit, the area of aperture considered corresponds to the height of the slit vs the $6\times\sigma$ of the light distribution, which corresponds to an optimized extraction of the object, see Fig. \ref{Aperture_psf}. From the simulations, I have concluded that the optimum dimension for the slit height is $\sim1.2\times\, seeing$ and $\sim1.3\times\, seeing$ for the diameter of fibers. In the optimal regime, slit reaches lower S/N than fiber, around 20\% less, because more sky is sampled in a rectangular aperture. 

\begin{figure}[p]
\centering
\includegraphics[width=0.50\textwidth]{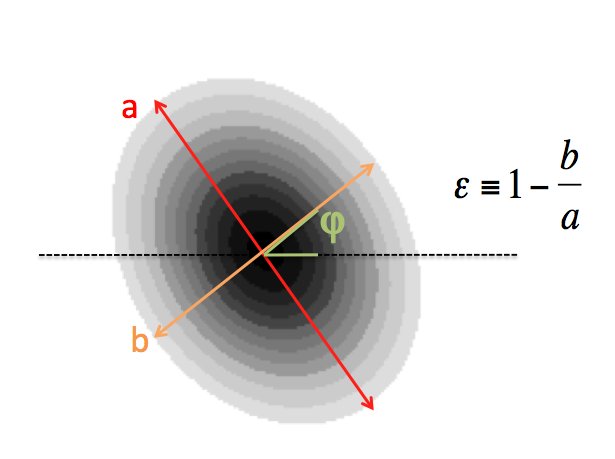}
\includegraphics[width=0.45\textwidth]{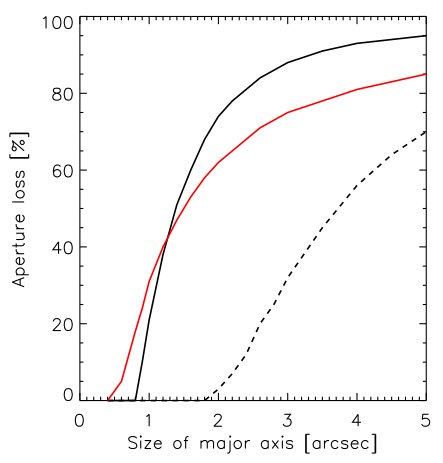}
\caption{{\small Left: Example of simulated galaxy, a/b=1.3 and $\phi$=$32^{\circ}$ Right: The red line shows the aperture loss in a 0.5" height slit in function of galaxy size define as the major axis of the disk. The black solid line shows the aperture loss in an IFU with the characteristic of OPTIMOS-EVE MI IFU (2"x3") and the black dashed line represents the case of a MO fiber of 0.9".}}
\label{Galaxy_aper}
\end{figure}

For extended sources, I have simulated disk galaxies of several sizes observed with a slit of 0.5" height and evaluated the fraction of light losses by the aperture. Galaxies are modeled by ellipses of ellipticity $\epsilon$, size of major axis $a$ and principal angle $\phi$, see Fig.\ref{Galaxy_aper}. The light distribution follows a bivariante gaussian distribution with $\sigma_1$ and $\sigma_2$ equal to $a/6$ and $b/6$ respectively. For a given galaxy size $a$, I have simulated 1000 galaxies with ellipticity following a lognormal distribution (distribution of the $\epsilon$ of SDSS galaxies by Ryden et al. 2003) and a uniform distribution of phase angle from 0 to $\pi$. Fig. \ref{Galaxy_aper} shows the fraction of light loss by the aperture as function of galaxy size for a 0.5" slit. High-z galaxies have typically half-light radii around 0.5 arcsec, as indicated from HST imaging of objects at z$\sim$5-8 (Ellis et al. 2001, Richard et al. 2008, Oesch et al. 2010b). Taking into account the effect of seeing \footnote{0.5" in seeing-limited observations and 0.3" GLAO at $\lambda$=1.2$\mu m$, based on Paranal statistics.} the characteristic size of high-z galaxy is between $\sim$1.6". A 0.5" slit will loss around ~50\% of the target light due to the aperture loss. In fiber instruments, the aperture loss can be considerably decrease by using integrated field unit. OPTIMOS-EVE will have two IFU modes which will match the size of distant galaxies (Medium IFU (MI) = 1.8"x2.9") and extended sources (Large IFU (LI) = 7.8"x13.5"). The sampling of the IFU is settled to 0.3" for a seeing limited instrument covering the UV to the NIR.

\subsubsection{Multiplex capability}
An important point for multi-object instruments is to maximize the number of objects observed in a scientific exposure. It depends on the size of the field of view, the number of spectra which can be sampled on the detector (multiplex) and the space density of target in the field. Technically the multiplex of slit and fiber instruments are nearly identical, but the cost of multiplex is higher in fiber-fed instruments due to the complexity of their opto-mechanic design. This budget reason have led to construct multi-fiber instruments with less effective multiplex than the slit counterparts. However, fiber spectrograph have the advantage to be able to observe large number of objects in large fields of view. Indeed, fiber-fed spectrographs are particularly stable and are not affected by mechanic problems from the self-deformation of large (and thus heavy) mask used in multi-slit spectroscopy. Moreover, in crowded field, fiber-fed instrument have higher effective multiplex. In fact, the selection of the targets is very flexible compared to multi-slit spectrograph. It is only restricted by the minimal separation possible between two fiber buttons. In multi-slit observations, the selection of targets have to take into account the avoidance of spectra overlapping, which can be very challenging in a field with high density of targets.

\subsubsection{Spectral resolution and coverage}
\begin{figure}[!h]
\centering
\includegraphics[width=0.60\textwidth]{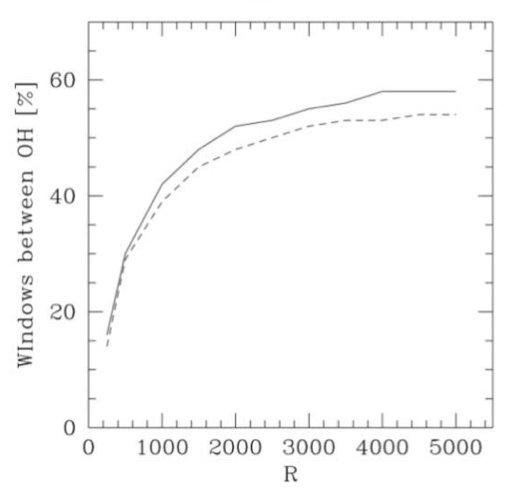}
\caption{{\small Percentage of the spectral band available between OH sky lines in the J band (solid line), and in the first half of the H band (dashed line), as a function of the spectral resolution. From OPTIMOS-EVE executive summary document.}}
\label{Sky_R}
\end{figure}

A main difference between the two technologies is the use of different dispersion systems. Slit-instruments use grisms which allow an easy switch between imagery and spectroscopy mode, but have the drawback to be limited to a maximal spectral resolution around 3000 \citep{1998SPIE.3355..409E}. Fiber-fed instruments use in general grating which allow to reach very high spectral resolution $\sim$100\,000. The optimal range of spectral resolution for a spectrograph depends obviously on the scientific cases. However, in the near-IR, the abundance of strong OH sky lines turns is a deciding point. In a large number of science case the observations of emission or absorption lines will need to be done between the skylines, implying a minimal spectral resolution. A spectral resolution larger than 4\,000 is mandatory in the NIR to provide enough spectral regions that are only limited by sky background and not affected by strong OH skylines, see Fig. \ref{Sky_R}. 
In addition, contrarily to multi-slit spectrograph in which the spectral coverage of each spectrum depends on the position of the object in the field of view, all spectra obtained by a fiber-fed instrument have the same spectral coverage.  

\subsubsection{Sky subtraction}
Multi-fibers spectrograph are accused to suffer from a major drawback: the low precision of sky subtraction. This is a critical issue in the case of faint object spectroscopy. In slit observations, the sky is directly sampled in the adjacent side of the object and can be extracted with an accuracy beneath 1\%. The sky subtraction is more problematic in fiber-fed observations because the sky cannot be sample in the immediate vicinity of the target. This comparison has led to a commonly held view that precise sky subtraction cannot be achieved with a multi-object fibre spectrograph. In spite of all the advantage of fiber-fed instrument cited previously, the quality of the sky-subtraction is the primary justification for the construction of slit system.  
Several cons have been cited in literature against accurate sky subtraction in fiber-fed instrument, such as: spatial variations of the sky background, spatial variations of the quantum efficiency of the detector, variation of the response fiber-to-fiber (transmission, instrumental flexure), scattered light, cross-talk between fibers \citep{1992MNRAS.257....1W,1998ASPC..152...50W}. To overcome these difficulties, two standard observational procedures have been implemented to treat the sky subtraction in fiber-fed instruments: 
\begin{itemize}
\item \textbf{Simultaneous background}. Several fibers are dedicated to sampling the sky in the region of observations. The number of sky fibers depends on the wavelength domain observed, the dimension of the field of view and the required on quality of sky-subtraction (see Chapter 3). Observations have to be previously corrected from the individual response of the fiber and scattered light. 
\item \textbf{Beam-switching}. Each fiber is switched between an object and a sky position (nodding). This strategy has the advantage to observe the sky background with the same fibers used for the targets (Watson 1987). The sky and the instrumental signal can be subtracted simultaneously. However, this procedure implies to dedicate half of the observation time to sky sampling and does not account for temporal variation of the sky \citet{1992MNRAS.259..751R,2005NewA...11...81L}.
\end{itemize}
Both procedures have advantages and drawbacks. Sky-subtraction algorithm and procedures for fiber-fed spectrographs, available in literature, hardly reach quality of 1\%. Solution based on observational strategies have been found for fiber transmission, scattered light and cross-talk and are described in the OPTIMOS-EVE scientific analysis report.  

\subsubsection{Summary}
In the last paragraph, I have investigated the key difference between fiber-fed and slit spectrograph. It focuses on some points and therefore it does not aim to be an exhaustive and complete study on the subject. The following table summarizes the principal pros (\smiley)\, and cons (\frownie)\,  of fiber and slit multi-object spectrograph. This preliminary study shows that fiber-fed instruments are more competitive than slit instruments in terms of the flexibility of observations in multi-object mode (e.g. less limitation on the distribution of target in the field and the support of integrated-field mode) and of spectral resolution. I have shown that the characteristic of OPTIMOS-EVE fit the technical specifications of the observation of faint and distant objects, in term of aperture loss, multiplex, global efficiency and spectral resolution. Its main drawback is the accuracy of the sky-subtraction. In the frame of the phase A, we have developed a new technique that overcomes the apparent limitation of fiber-fed instrument to recover with high accuracy the sky. The next chapter contains a complete definition of the problem and the description of the implemented sky subtraction algorithm.

\begin{table}[!h]
\begin{center}
\begin{tabular}{l|l|l}
&\textbf{Fiber}&\textbf{Slit}\\
\hline
Efficiency&Low but increasing \frownie &High \smiley\\
Aperture loss&Flexibility on aperture &Critical for extended objects \frownie \\
&size with IFUs \smiley&\\
Multiplex&Optimum in crowded field \smiley&Optimum in low density field \smiley \\
Spectral Resolution&From low to high \smiley&Until 3000. Critical in NIR \frownie\\
Sky subtraction& 1\% in litterature \frownie & Less than 1\% \smiley\\
\hline
\end{tabular}
\caption{{\small The pros and cons of slit and fiber multi-object spectrographs.}}
\label{EVE_spectro}
\end{center}
\end{table}

\EmptyNewPage
  \EmptyNewPage

\FrameThisInToc
\chapter{OPTIMOS-EVE sky subtraction: A new technique }
\minitoc
As referred in the previous chapter, the question of the sky subtraction is a critical point for OPTIMOS-EVE. The currently existing algorithm and observation strategy of sky subtraction in a fiber-fed instrument can hardly reach 1\% of accuracy for the sky background (REF). The EVE science case requires though an accuracy in the sky background extraction $<1\%$. In the frame of the phase A, we have developed a new technique that overcomes the apparent limitation of fibre-fed instrument to recover with high accuracy the sky. This new algorithm allows us to subtract the sky background contribution in an FoV of 7x7 arcmin2 with an accuracy of 1\% in the MO mode, and 0.3-0.4\% for integral-field-unit observations. The method is based on the reconstruction of the spatial fluctuations of the sky background (both background and emission).  Knowing how the sky background fluctuates in space and time is crucial for the implementation of an optimal sky subtraction strategy. What are the spectral characteristics of the sky? How does the sky fluctuate spatially and temporally? Indeed, the extent to which sky is changing during integration time will constrain the spatial and temporal frequency with which the sky needs to be sampled. I have thus dedicated a section of this chapter to a full description of the sky properties. In the next subsection I describe the new algorithm and discuss its performance in the different observing modes of the instrument.
 
\section{What is the \textit{sky}?}
\begin{figure}[p]
\centering
\includegraphics[width=0.8\textwidth]{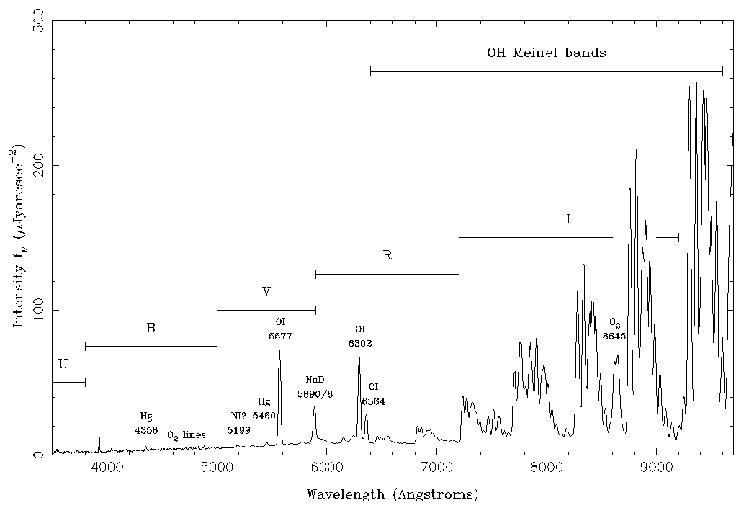}
\caption{{\small Night spectrum at La Palma observatory. The principal emission lines and bands of the airglow are indicated. From \citet{1998NewAR..42..503B}}}
\label{NightSkySpectrum}
\end{figure}

The sky is commonly defined as all the light arriving to the detector which doesn't proceed from neither the observed astronomical source nor the instrument. The nightsky light is composed by a mixture of radiation produced by several sources, see Fig. \ref{NightSkySpectrum}, from which the main contributors are:
\begin{itemize}
\item Moonlight. The contribution obviously depends on the moon phase. It can increase the night brightness by several order of magnitude, in particular in the blue bands;
\item Zodiacal light. It is sunlight scattered by planetary dust. As such, it has a solar spectra multiplied by the scattered efficiency as a function of the wavelength.  The brightness of the zodiacal light decreases as a function of ecliptic latitude \citep{1960ApJ...131...15B,1996ApJ...464..412C,2008srvz.book.....M}\footnote{Yes, It is him! The most famous astronomer on earth: Brian May the guitarist of Queen};
\item Airglow. It is the non-thermal emission from the upper atmosphere. It is the source of the well-known OH lines in the near-IR, see complete description above; 
\item Aurora. The auroral emission is produced by the interaction of solar wind particles with the components of the upper atmosphere.  It is only relevant when observation are conduced at observatories located at polar latitudes; 
\item Thermal emission. There are three components of thermal emission: the telescope, the atmosphere and the instrument. The radiation follows a grey body distribution \citep{1996ApJ...464..412C}. The contribution of the thermal emission from the telescope becomes significant in the K band and the atmosphere contributes significantly at wavelength longer than 2.35$\mu m$; 
\item Unresolved background astrophysical sources \citep{1973QB817.R54,1992MNRAS.257....1W};
\item Scattered light from the human activity in sites with light contamination.  
\end{itemize} 
The contribution of each of these sources varies from site to site, with time and as a function of the wavelength. I will focus the study on the conditions characteristic of faint object spectroscopy observations: dark sky, without moonlight contamination and observation at high latitudes to avoid contamination from zodiacal light. I have also assumed that the site of observation is situated far from poles (that is the case of the possible E-ELT sites) and thus the contributions of aurora to the sky is negligible. In such observational condition, the main contribution to the nightsky brightness is the airglow in UV to near-IR wavelength and the thermal radiation from the telescope and atmosphere at wavelength long-ward to 2$\mu$m.

\subsection{The sky spectrum}

\subsubsection{Emission lines}
\begin{figure}[p]
\centering
\includegraphics[width=0.7\textwidth]{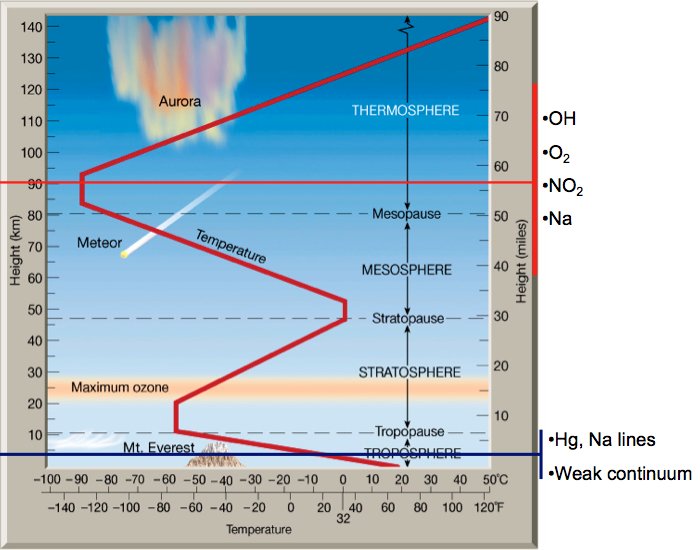}
\caption{{\small Schematic view of the atmosphere with the different layers, the profile of temperature and the emitting layer of several sky features. From Patat presentation (http://www.eso.org/$\sim$fpatat/science/skybright/index.html).}}
\label{Atmosphere_layers}
\end{figure}

In dark skies, the sky spectrum, in particular the emission line component, is dominated by the contribution of the airglow light. The existence of what is now termed airglow was established for first time photometricaly by \citet{1909PGro...22....1Y}. It corresponds to the light radiated by electronically or/and vibration-rotationally excited atoms and molecules located in the upper part of the atmosphere, the mesosphere\footnote{A layer of the atmosphere situated around $\sim$100km of altitude and corresponding to a minimum in the vertical temperature profile.}, see Fig. \ref{Atmosphere_layers}. During the day energetic UV photons dissociate molecules and remove electrons from atoms and molecules. Later at night, the dissociation or ionization products interact with other atmosphere components and radiate light mainly by chemiluminescence process \citep{1961paa..book.....C,1973QB817.R54}. Table \ref{Airglow_components} shows the various species, which are involved in the airglow emission, and their associated emission lines. The intensity of the lines are expressed in Rayleighs, the flux unit used in atmospheric physics: 1R=3.72$\times10^{-14}\,\lambda^{-1}($\AA$)\,erg\,s^{-1}\,cm^{-2}\,arcsec^{-2}$ 
\begin{table}[!h]
\begin{center}
\begin{tabular}{lclclc|c|c|}
Emitter&Wavelength&Emission&Absolute&Remarks\\
& &height(km)&intensity(R)&\\
\hline
Hydroxyl OH&red and near-IR &86&4\,500\,000&Produced by the reaction\\
&&&&of hydrogen with ozone\\
\hline
$O_2$&1.27,1.58,1.9$\mu m$&90&80\,000&IR atmospheric bands\\
&7619,8645\AA&80&6\,000&Atmospheric bands\\
&2600-3800\AA&90&600&Herzberg bands. Contribute\\
&&&&to the blue pseudo-continum\\
\hline
O&5577\AA&90&250&Very intense line. Accounts\\
&&&&for 20\% of the sky brightness\\
&&&&in the V-band\\
&6300,6364\AA&250-300&150&\\
\hline
N&5200\AA&260&1&Blend of several lines\\
\hline
Na&5890,5896\AA&92&50&Strong seasonal variation\\
\hline
\end{tabular}
\caption{{\small Nightglow radiation due to excitation by chemical associations ordered by emission intensity, based on \citet{1973QB817.R54} and \citet{2008A&A...481..575P}.}}
\label{Airglow_components}
\end{center}
\end{table}
\subsubsection{Continuum}
The main sources of the sky continuum are the moon, the zodiacal light, the galactic dust, unresolved galactic sources and the thermal emission from the earth, atmosphere and the instrument. For this reason, the background sky at a given wavelength and at a given position in the sky depends on a number of parameters such as: the position of the observed field with respect to the galaxy, to the ecliptic and the moon, the moon phase, the phase of the solar cycle, the season and the airmass. Compared to very dark night, these systematic effects increase the background by up to several tenths of a magnitude. In addition to these well-known sources of continuum, a significant contribution to the sky background proceeds from the airglow. The airglow continuum was discovered by \citet{1953AnAp...16...96B} and the question of its origin has led to a multitude of theories. Its strong variations with wavelength and time suggests than more than one source is involved. In blue band, the airglow continuum is a pseudo-continuum composed by unresolved and blended $O_2$ Herzberg bands. According to \citet{1992P&SS...40..211B}, the green continuum is due to metastable oxygen molecules colliding with ambient gas molecules and forming complexes that dissociate by allowed radiative transitions. This theory is supported by the strong correlation observed between the intensity of the green continuum and those of the 5577\AA\, OI line and the Herzberg system. However, other theory based on the emission of $NO_2$ molecules are not totally discarded \citep{1962P&SS....9..883K}. In the red and near-IR, recent works have shown that the continuum airglow is due to the contribution of unresolved OH doublet in the interlines regions \citep{2008MNRAS.386...47E}. The scattered instrumental light of the strong OH lines are also considered a significant contribution of the near-IR airglow continuum. Contribution of a pseudo-continuum emission from $O_2$ radiative association is not excluded \citep{1977P&SS...25..787W,1996ApJ...464..412C}. 

\subsection{Temporal and spatial fluctuation of the sky}
If we observe the sky brightness in a given field of view during all the night, in the condition of observation given above -dark sky, without moonlight and the field observed is at high ecliptic latitude- we will see that it varies spatially and temporally. These variations are due to the dynamical nature of the atmosphere. Indeed, the atmosphere is a complex system in which the main properties at a given layer -composition, density, temperature - do vary with time and space. 

\subsubsection{Long-scale variations}
The main contributor of this variation is the evolution of the chemical composition and density variation in the mesosphere, involved by harmonic periods of diurnal tides. The amplitude of these variations depends on the evolution of the concentration of species and on the coefficient rate of the reaction involved in the emission of the radiation. For illustration, the variation of the OH lines due to the diurnal oscillation is around 50\% \citep{2008ExA....22...87M}. The OH radicals are produced by the reaction of hydrogen with ozone. As soon as the upper atmosphere is no more illuminated by the sun, the concentration of ozone increase, while the abundance of atomic hydrogen, which is produced during the day, decreases. It results on the increase of the OH radical at the beginning of the night and then a decrease in the remaining 2/3 of the night \citep{1992MNRAS.259..751R,1996ApJ...464..412C}. The variation of intensity of OH lines due to diurnal variation is well approximated by a function $1/(1+t)^{0.57}$, where $t$ is in hours \citep{2008MNRAS.386...47E}.  
The variations induced by the diurnal oscillation are not very problematic in the framework of our study of sky subtraction. Indeed, the fluctuations are very smooth and have long temporal and spacial scale. Its effect in observation in a 7'x7' field of view and with a typical exposure time of one hour is negligible.  
\subsubsection{Small-scale variations}
\begin{figure}[!h]
\centering
\includegraphics[width=0.45\textwidth]{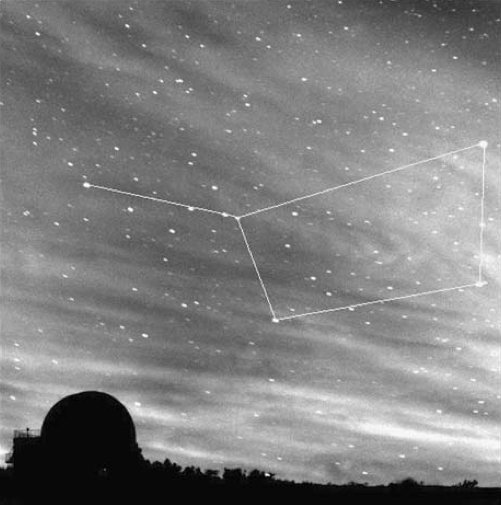}
\includegraphics[width=0.50\textwidth]{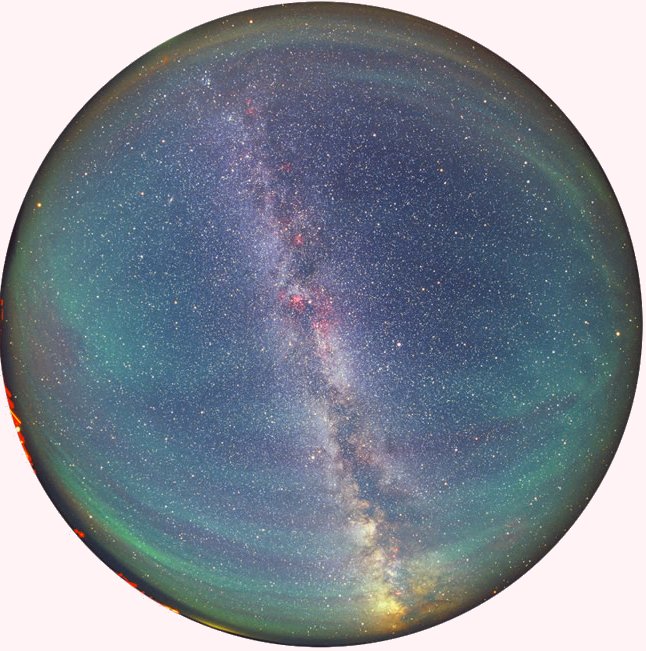}
\caption{{\small \textbf{Right}: Image of the airglow distribution in near-IR (mainly OH Meinel bands) during a cloudless night. A series of stripes extend over the $19^o\,\times \,19^o$ field of view. From \citet{2008ExA....22...87M}. \textbf{Left}: The optical airglow, mainly from the contribution of the green OI line, in an all-sky photography. The optical airglow is also structured in bands stretch. From Doug Zubenel.}}
\label{Airglow_distribution}
\end{figure}

Superposed to this smooth diurnal variation, the airglow emission is also affected by short-scale variations that are more problematic for our observations. In the near-IR, even during a cloudless night, the sky is structured by wide strips filling the all sky, see Fig. \ref{Airglow_distribution}. If the human eye were sensitive to near-IR radiation, we would see a glowing and changing sky, as impressive as aurora. In the optical, such bands are also visible, but their are less prominent because the intensity of the sky at these wavelengths is less intense than in near-IR. 
Such short-period variation, of the order of few minutes to an hour, is originated by the passage of a density and temperature perturbation through the upper atmosphere. The existence of such gravity waves has been well documented and modeled \citep{1980JGR....85.1297Y}. They are caused by a variety of sources as the passage of wind across terrestrial landforms, the interaction of the polar jet stream or the radiation incident from space. The passage of gravity waves induces a local perturbation of the density which produces changes in the reaction rates and thus variation on the emission of the airglow. In the 90's several observational campaigns have been undergone to constrain the properties of gravity waves in the mesosphere \citep{1991GeoRL..18.1337T,1992MNRAS.259..751R,1995GeoRL..22.2881T,1997GeoRL..24.1699T}. Gravity waves are sinusoidal waves with characteristic horizontal wavelength $\lambda_c$, period T and horizontal propagation velocity $v_{c}$. They induce in the sky two main features: bands originated by waves of $10\,min$ period, $\lambda_c\sim25\,km$ and $v_{c}\sim50\,m\,s^{-1}$; rippled arising from waves with smaller period ($\sim5\,min$) and shorter wavelengths  $\lambda_c\sim10\,km$.  Converting these properties into spatial angle unit, we found that gravity waves induce spatial fluctuation on the sky of the order of 15$^o$ for bands and 6$^o$ for rippled which propagated at the velocity of 2'/s respectively (at a mean altitude of $\sim\,90km$).

\begin{figure}[p]
\centering
\includegraphics[width=0.45\textwidth]{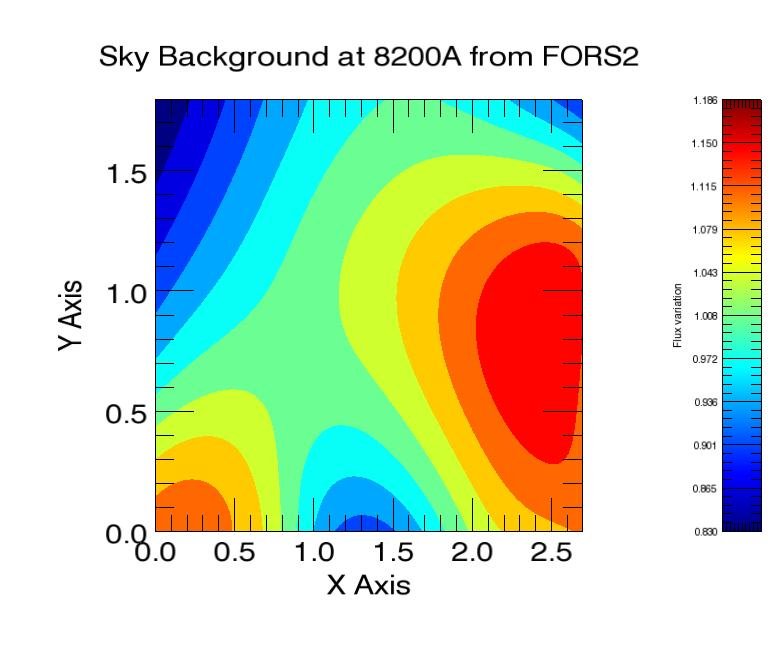}
\caption{{\small Intensity variation of the sky background at 8200\AA\,(left) in a 2"x2" field of view from FORS2 observation.}}
\label{FORS2_var}
\end{figure}

The intensity of the fluctuation of the airglow emission induced by a gravity wave varies as a function of wavelength. Indeed, the intensity variation of a sky spectra features - line, band or continuum- depends principally on the altitude of the emitting layer and on the reaction rate and density of the species that originate the emission. In order to evaluate the intensity of the fluctuation in the optical domain, we have undergone a quick characterization of the spatial variation of the sky features in a field of view of 2"x2". We have used the MXU/FORS2 observations (600RI and 600z) described in Part II Chap. 2, to sample the intensity of the sky continuum at 8200\AA\, and the flux of OI and OI sky lines in flux calibrated sky spectra arising from several slits of a FORS2 mask (see Fig.\ref{FORS2_var}). We found that the sky background varies slightly across the field: 5-10\% between two positions on the sky. The intensity of skylines does also vary spatially from 10\% to 20\%. The spatial variations are smooth and have a spatial scale of 1.0' which is coherent with the passage of a gravity wave. In the near-IR (1-1.8$\mu m$), \citet{1992MNRAS.259..751R} and \citet{2008ExA....22...87M} have shown that the fluctuation of the OH emission bands can have variations of 5-10\% and sky background fluctuates with an amplitude of 15\%, see also \citet{2010PASP..122..722H}. All the OH bands vary in phase, but without the same intensity \citep{1992MNRAS.259..751R}.  

 \begin{figure}[p]
\centering
\includegraphics[width=0.7\textwidth]{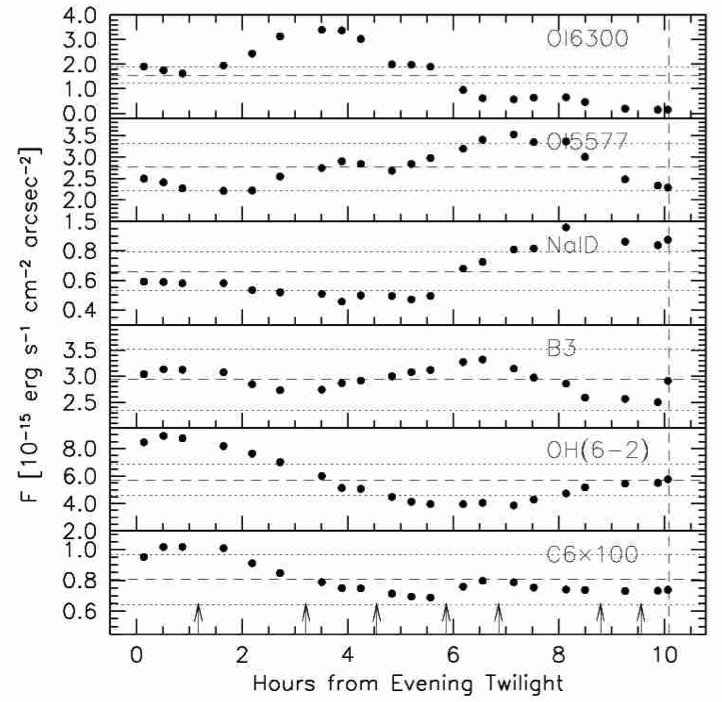}
\caption{{\small  Sequence of spectral measurements for some sky features measure in FORS1 spectra by \citet{2008A&A...481..575P}. Exposure times ranged from 10 to 20 minutes. The vertical dashed line marks the start of morning astronomical twilight. The horizontal dashed lines mark the average value, while the dotted lines indicate $\pm$20\% levels.}}
\label{Fluc_correlation_sky}
\end{figure}

Severals observations have put on evidence the correlation of the fluctuations between sky spectral features \citep{1961paa..book.....C,2008A&A...481..575P}, see Fig. \ref{Fluc_correlation_sky}. The spectral features can be regroup into three co-variance group, inside which all features vary in phase:
\begin{itemize} 
\item The green-line group : [OI]$\lambda$\, 5577\AA, $O_2$ Herzberg bands, blue pseudo-continum, green continuum and $O_2$ atmospheric bands;
\item Na\,D$\lambda$\,6300\AA\, and OH Meinel bands;
\item Red OI line and N $\lambda$\,5200\AA.
\end{itemize}
Such correlations rise from common emitters or same emitting layer altitude shared by the members of a co-variance group. For instance, all the features of the green-line group are produced by $O_2$ molecules emitting in a $\sim90\,km$ layer. Likewise, the red OI line and N $\lambda$\,5200\AA\, are both produced in high altitude mesospheric layer ($\sim250\,km$).

\section{A new algorithm for sky extraction}
\begin{figure}[p]
\centering
\includegraphics[width=1.0\textwidth]{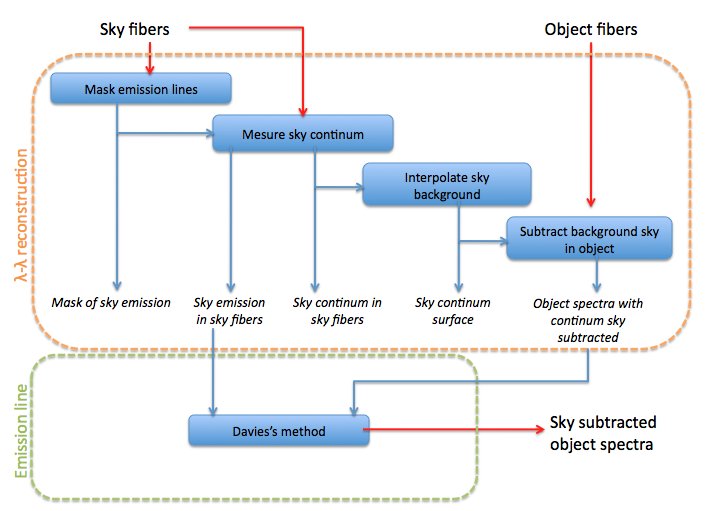}
\includegraphics[width=1.0\textwidth]{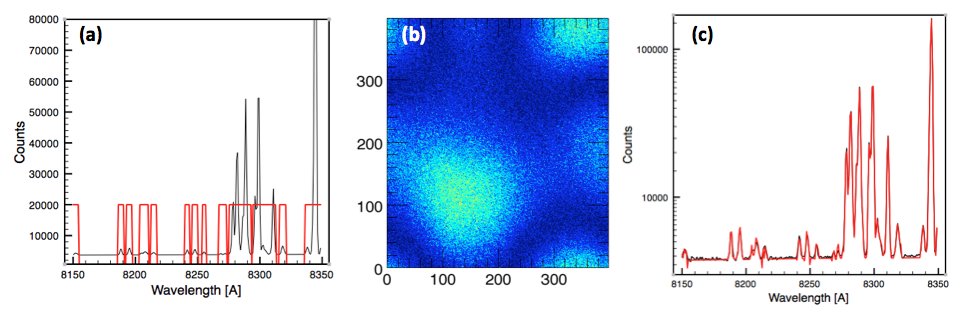}
\caption{{\small Flowchart of the sky subtraction algorithm. Examples are shown in the lower panels: (a) Sky emission mask (red boxes) produced by kappa-sigma-clipping. (b) Residual from the subtraction of the continuum background in a $7'\times7'$ field using 120 sky fibres in 1h exposure time. (c) Modeled sky (red line), continuum background and emission lines derived by the Davies's method, over-plotted to the input sky spectrum (black line), in logarithmic scale.}}
\label{Sky_flowchart}
\end{figure}

As the two components of the sky-emission and continuum- do not vary in phase, we have chosen to approach the sky subtraction issue by treating  the two terms independently. The algorithm is divided into two steps: the determination of the sky background using a $\lambda-\lambda$ surface reconstruction and the extraction of the sky emission lines with the Davies's method \citep{2007MNRAS.375.1099D}. The flowchart of the sky extraction algorithm is presented in Fig \ref{Sky_flowchart}. The bulk of the algorithm is programmed in IDL. 

\subsection{The sky background : $\lambda-\lambda$ reconstruction}
To reach an extraction of the sky background within 1\%, the sky background has been reconstructed at different wavelengths. The principle of the method is to sample the level of the background sky in different points of the field of view, the sky fibers, and reconstruct the whole surface using an interpolation method on an irregular grid. The reconstruction of the sky background surface is a new approach to the sky extraction issue. To our knowledge, there is no such approach in the literature. The algorithm of sky background subtraction is divided into three steps which are described below.
 
\subsubsection{Decouple the continuum from the emission lines in sky spectrum}
The goal of this first step is to mask all the emission lines in a stacked sky spectrum in order to properly evaluate the sky continuum. 
All the sky lines are stacked into a single sky spectra in which a sigma-clipping algorithm detects the emission lines (see Numerical Recipes for the description of the sigma-clipping algorithm and Erwin ). The user can choose the sigma threshold and the wavelength range where the sigma of the sky continuum is measured. This step gives a sky emission mask.

In some cases, the sky continuum spectrum can be contaminated by instrumental scattered light. The continuum has a complex profile as function of wavelength and the emission lines are difficult to extract with the sigma-clipping algorithm. In prevision of such cases I have added an optional step: a first guess of the continuum is evaluated by applying a low pass filter or a wavelet decomposition on the stacked sky spectrum \citep{1994A&A...288..342S}. The order of the wavelet decomposition and the cutoff frequency of the low pass filter can be defined by the users. I am also planning to implement an interactive IDL routine which will allow the users to create his own mask by masking the emission lines by hand. The position of usual sky emission lines \citep{2003A&A...407.1157H} will be indicated.  

\subsubsection{Sky background}
On each sky fibre, the emission lines are masked by the sky emission mask. The regions without emission lines are smoothed using wavelet decomposition. The smoothed continuum is then interpolated into the entire wavelength range in order to have a background pure sky spectrum in each sky fibre. Here again the order of the wavelength decomposition can be selected by the user depending on the used instrument, nevertheless the default value of such parameters (wavelet order, sigma-clipping, etc..) have been optimized for the OPTIMOS-EVE case. 

\subsubsection{$\lambda-\lambda$ surface reconstruction}
The surface of the sky background level is reconstructed in several layers of $\lambda$. The sky continuum is binned in wavelength spectral resolution element of width $\Delta \lambda$. For each $\lambda-\lambda$ layer, the surface is reconstructed using a triangulation interpolation method, the IDL routine GRIDDATA, from the position of the sky fibre and the value of the continuum in the given $\lambda$ bin. I have used the GRIDDATA routine with the method of natural neighbor interpolation, in which each interpolant is a linear combination of the three vertices of its enclosing Delaunay triangle and their adjacent vertices. 

At the end of this step, the value of the sky background is constrained in all the field of view and at different wavelength bins. The continuum at the position of the sky and object spectra is extracted from the $\lambda-\lambda$ sky background cube and subtracted to the sky and object spectra. The sky background subtracted spectra are then sent to the algorithm of emission lines subtraction. 

\subsection{Sky emission lines: the Davies's method}

The method proposed by \citet{2007MNRAS.375.1099D}, briefly summarized hereafter, removes the OH lines in the near infrared (1 to 2.5 $\mu m$). The reader is refered to \citet{2007MNRAS.375.1099D} for a detailed description of the algorithm. 

\begin{figure}[!h]
\centering
\includegraphics[width=1.0\textwidth]{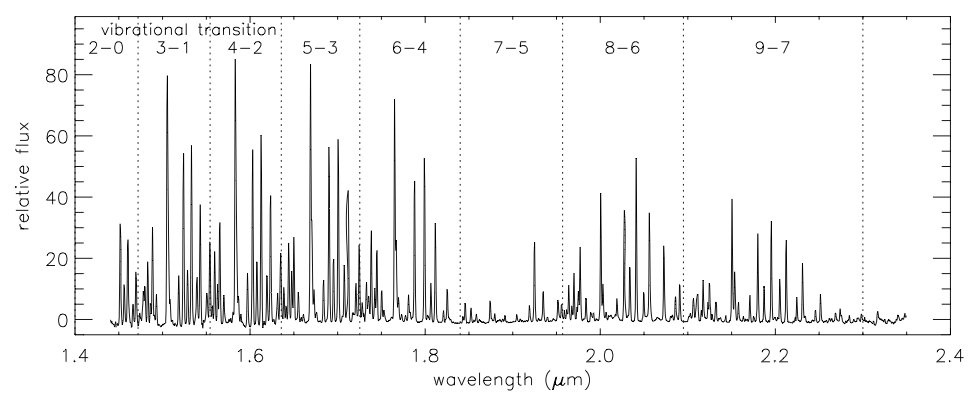}
\caption{{\small Spectrum of the OH emission across the H and K bands (the thermal background has been subtracted). The various spectral ranges corresponding to the different vibrational transitions (labelled) are indicated, from \citet{2007MNRAS.375.1099D}.}}
\label{Davies}
\end{figure}

As described in section 2.1, the main difficulties in subtracting sky emission lines, in particular the Meinel OH lines, are the fluctuations in their absolute and relative flux both temporally and spatially, see Fig. \ref{Davies}. The Davies's algorithm compensates these changes and also correct the effect of instrumental flexure which can lead to 'P-Cygni' type residual due to small mismatch in the wavelength scaling. The algorithm is based on the physical reason of the absolute and relative flux change in OH lines : the fluctuation in the vibrational and rotational temperatures. The effect of the vibrational temperature fluctuations is the main contributor of the OH flux variation. As the emission lines produced in a given vibrational transition lie in well defined bands, it is possible to divide the sky spectrum into section corresponding to a single vibrational transition. In each section, the spectrum to be sky subtracted is fitted to a reference sky spectrum in order to find a scaling factor which corrects the difference of vibrational temperature between the reference sky and the sky at the object position. After iterating in all the section, a scaling factor function is derived and the combination of this scaling factor function with the reference sky spectrum is subtracted to the object spectrum. The changes on the rotational temperature are treated in a similarly way. We have extended this method to optical wavelengths. 

We have also tested to subtract the sky emission with principal component analysis, a statical method based on the statistical description of random variables. The PCA is applied on the data sampled by the skylight fibers to obtain the modules of sky emission lines of this observation. The components are then linearly combined to remove the sky lines from the continuum subtracted object spectra. This method has been proposed by \citet{2005MNRAS.358.1083W} to remove residual of sky lines and implemented to the SDSS data by \citet{2008ChA&A..32..109B}. Despite the very good results of PCA in SDSS data, the method was not very convincing in the case of faint object spectra. The residuals on the extraction of the line is still too high for very small S/N spectra and the main limitation is the property of this residual. As the PCA is trained always with the same array, the residual do not vary from one exposure to an other. It always have the same profile and same intensity. Contrary to the random noise, the residual from PCA will not diminish when the different exposures are combined.  Moreover, the PCA method doesn't manage to compensate the issue of the instrumental flexure which introduce more 'P-Cygni' type residual in the final spectrum. 

\subsection{Applying the algorithm to data}
\subsubsection{Input file format}
\begin{figure}[!h]
\centering
\includegraphics[width=.8\textwidth]{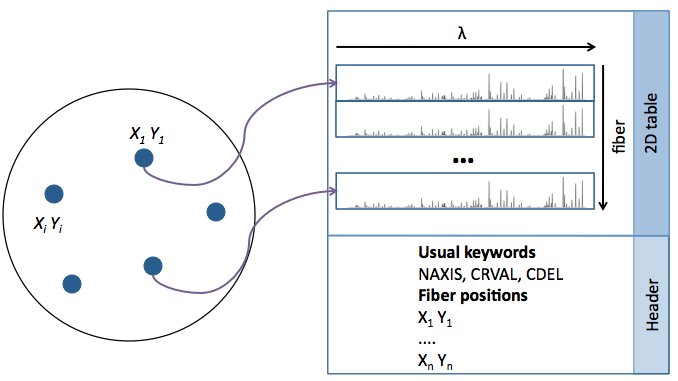}
\caption{{\small Format of the input file data used by the sky extraction software. The spectra from the fibers and their positions are stored in a FITS file.}}
\label{Data_input_sky}
\end{figure}

At the present time, the software works with two input data files in FITS\footnote{FITS stands for 'Flexible Image Transport System' and is the standard astronomical data format endorsed by both NASA and the IAU.} format: a 2D sky file and a 2D target file. These files store the \textit{n} sky or target spectra coming from the fibers during an observation. Their header contains the usual information- as the spatial and spectral dimension scaling and reference pixel- and the X-Y positions of the fibers in the field-of-view, see Fig. \ref{Data_input_sky}. In the future, specific formats of existing instrument will be made available. 

\subsubsection{Defining user's parameters}
The algorithm can be applied to any fiber-fed spectrograph by changing the user's parameters. Table \ref{Users_param_sky} summarizes the default parameters of the sky background subtraction algorithm used for OPTIMOS-EVE.
\begin{description}
\item[$\sigma_{thres}$] - It is the sigma threshold used by the sigma-clipping algorithm for detecting the emission lines in the sky spectrum. All spectral pixels with sigma superior to $\sigma_{thres}\times\sigma_{cont}$ will be flagged as emission lines in the mask. 
\item[$\lambda_{1}-\lambda_{2}$] - It is the region where the sigma of the continuum is measured. The user must select a region free of emission lines in the sky spectrum. 
\item[$N_{wavelet}$] The continuum is evaluated by smoothing with wavelets the sky spectra after masking the emission lines. The masked sky spectrum is decomposed into seven different wavelet scales using the undecimated (keeping an identical sampling in each wavelet scale) wavelet transform. The user can choose the number of scale which will be used in the recombination. In the wavelet space, the first scale is the highest frequency (index on scale 0) and the lowest frequency is in our case the wavelet scale number 6. The algorithm will recombine the $N_{wavelet}$ lowest frequency wavelet scale to produce the smoothed continuum. 
\item[$\Delta \lambda$] - It is the spectral resolution at which the sky background surface is reconstructed. The $\Delta \lambda$ have to be optimized in order to well sample changes of slope in the sky spectrum. The ratio spectral-range/$\Delta \lambda$ needs to be kept to reasonable values $<100$. Indeed, an increase of the spectra resolution (decrease of $\Delta \lambda$) will increase drastically the execution time. The software will need to interpolate more surface layers. 

\end{description}

\begin{table}[!h]
\begin{center}
\begin{tabular}{| l |c| l |}
\hline
Parameter&Value&Remarks\\
\hline
$\sigma_{thres}$&3&$\sigma$ threshold for the sigma-clipping\\
$[\lambda_{1}-\lambda_{2}]$&-&Interval in which the sigma\\
&&is measured for sigma-clipping\\
$N_{wavelet}$&1/7&Order of the wavelet decomposition \\
&&used in continuum fitting\\
$\Delta \lambda$&$\sim$10-50&Spectral resolution of the $\lambda-\lambda$ sky cube\\
\hline
\end{tabular}
\caption{{\small Default parameters of the sky background subtraction algorithm used for OPTIMOS-EVE}}
\label{Users_param_sky}
\end{center}
\end{table}

\section{Sky subtraction on EVE-OPTIMOS}
In order to test and optimize the algorithm to the EVE instrument, I have constructed a simulator of fiber- and IFU mode EVE observations.  Armed with this simulator, it is possible to explore the full potential of the new algorithm in the frame of EVE observation and investigate different aspects of the sky subtraction issue in fiber-fed instrument, such as:
\begin{itemize}
\item the accuracy of the sky subtraction in the two modes, mono-object and medium integral field unit;
\item the number of sky fibers which are necessary to reach 1\% of accuracy on the extraction of the sky-background;
\item evaluate the gain on the sky-background extraction accuracy from using MI or slit;
\item define the better observational strategy for specific science case;
\item verify the feasibility of the challenging EVE science cases, e.g.: high-z galaxies;
\item investigate the optimization of the position of sky fibers with respect to the object fibers. 
\end{itemize} 
In this section I will first describe how the simulated EVE observations have been constructed. In a second part, I will present the tests which have been performed and I will provide some answers to the above questions.  

\subsection{Tests on simulated EVE data}
\begin{figure}[!h]
\centering
\includegraphics[width=0.80\textwidth]{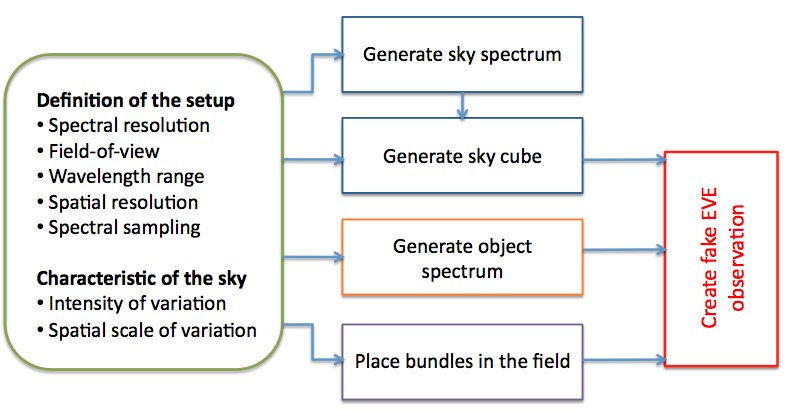}
\caption{{\small Flowchart of the simulator for the fiber- and IFU mode EVE observations.}}
\label{Flowchart_skysimulator}
\end{figure}

The simulator is compounded of a sky generator, which creates sky datacubes, an object generator and an interactive routine that allows placing IFU or fibers in an EVE field of view. The parameters of the simulation have been chosen to be the closer to the observational setup: same spectral resolution, spatial scale, $\lambda$ sampling and field-of-view as EVE setups. The architecture of the simulator is described in the flowchart in Fig. \ref{Flowchart_skysimulator}.

\subsubsection{Simulating sky}
The input sky cube is the critical point of the simulator. Our sky subtraction algorithm is based on the spatial and temporal properties of the sky. It is thus crucial to test it with simulated sky as realistic as possible. Armed with the characterization of the sky properties presented in section 2.1, I have constructed fake sky cube with the following methodology: 
\begin{description}
\item[- Define the parameters of the simulation:] the spectral resolution R, the wavelength range, central wavelength, exposure time, the spatial resolution. 
\item[- Create a theoretical sky spectrum] using the parameters defined above. The spectra is generated from the combination of a sky continuum, given by the Standard ESO ETC sky brightness and \citet{2000Msngr.101....2C}, with a library of sky lines from \citet{2003A&A...407.1157H}. The sky background is interpolated over the defined wavelength range considering the magnitude given per band. 
\item[- Generate a random sky background fluctuation surface] according to the properties of the sky at the given wavelength. A surface with smooth gaussian variation having the same dimension as the EVE field-of-view is generated, see Fig. \ref{Sky_surface_simu}. The spatial scale of the variation is set to 1' and the intensity peak-to-peak is 10\% in optical and 20\% in near-IR, see justification in Section 2.1. 
 \begin{figure}[!h]
\centering
\includegraphics[width=0.40\textwidth]{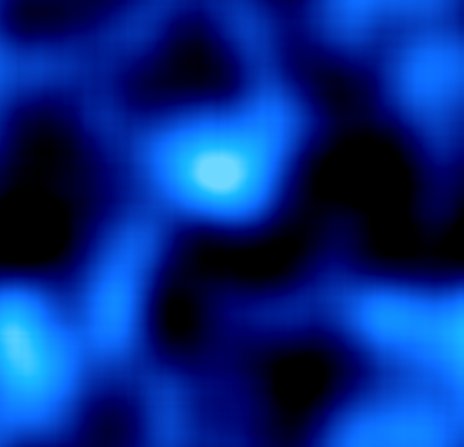}
\caption{{\small Simulated sky background fluctuation surface. The dimension of the field-of-view is the same as those of EVE. The fluctuation has a spatial scale of one arcmin and an intensity peak-to-peak of 20\% (near-IR observation).}}
\label{Sky_surface_simu}
\end{figure}
\item[- Build the sky cube.] In each spaxel the value of the sky background fluctuation is multiplied to the theoretical sky spectrum and a Poisson noise is added. The final product is a three dimensional file with the dimension of the field-of-view for the spatial axis and the same dimension as the theoretical sky spectrum for the wavelength axis. 
\end{description}

\begin{figure}[!h]
\centering
\includegraphics[width=0.90\textwidth]{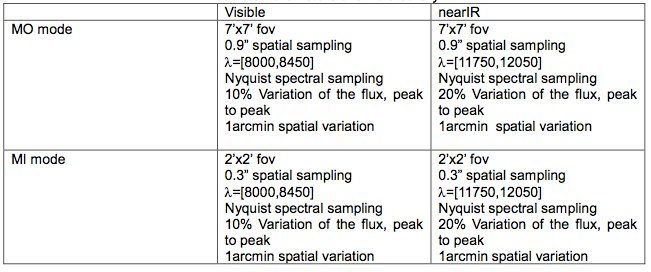}
\caption{{\small Input parameters for the EVE observation simulator in function of the mode - MO or MI- and the wavelength domain.}}
\label{Sky_simu}
\end{figure}

Sky cubes have been generated for the MI and MO modes in optical and near-IR. The input parameters of the sky datacube model are summarize in table \ref{Sky_simu} as a function of the observation mode and the spectral domain.

\subsubsection{Simulating observations}
The simulated observations are composed by a series of sky spectra coming from sky fibers or spaxel positioned at a given position in the field and sky+object spectra from object fiber or spaxel. To generate a simulated observation I proceed the following steps:
\begin{description}
\item[- Define the position of the fibers or IFU in the field-of-view.] I have created two IDL interactive routines, one for the MO mode and another for the MI mode to make this task easier. In the MO mode, the user can select the position of sky and object fibers in a $7'\times7'$ field-of-view with a spatial scale of 0.9", see Fig \ref{Widget}. In the MI mode, the interactive widget have two setup: a low spatial resolution with a spatial resolution of 0.9" and a normal spatial resolution with a spatial resolution of 0.3" which is the resolution of the MI IFU. The user draws in the field-of-view the shape of the IFU and can select which spaxel is associated to an object spaxel or to a sky spaxel. I have chosen this solution, instead to directly draw the IFU as a block because it gives more flexibility to test questions as the geometry of the IFU and the optimization of sky background subtraction using no illuminated spaxels in the algorithm, see Fig \ref{Widget}. The widget saves a mask file with the position of the sky and object fibers or spaxels. 
\begin{figure}[!h]
\centering
\includegraphics[width=0.40\textwidth]{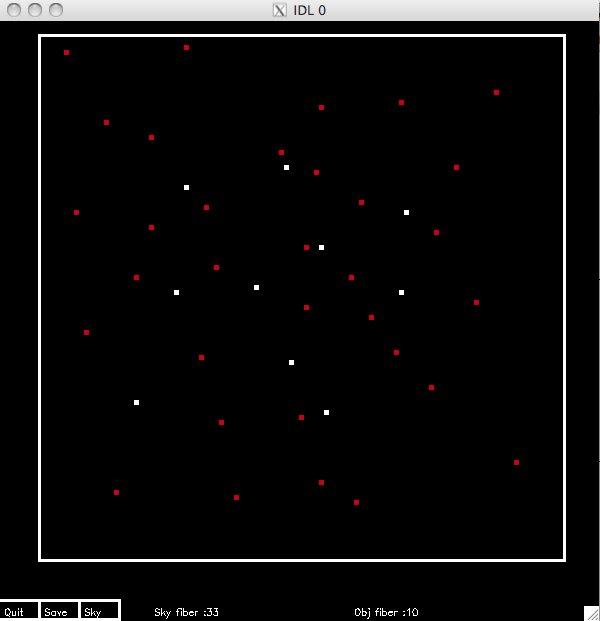}
\includegraphics[width=0.48\textwidth]{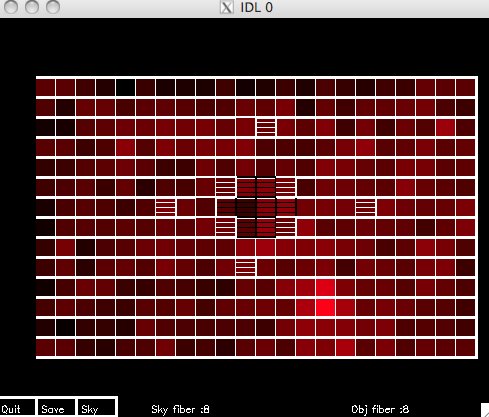}
\caption{{\small Interactive widget for the construction of fake MO and MI observations. Left: Widget for the MO mode. The red dots are sky fibers and the white fibers dedicated to object observations. The white squares represent the limit of the EVE field-of-view. Right: Widget for the mode MI. The white dashed spaxel are associated to sky and the dark are associate to an object.}}
\label{Widget}
\end{figure}

\item[- Simulate target spectra.] I have simulated two kinds of astrophysical objects based on two EVE science case: a stellar template and an emission line galaxy. 

The stellar science case is planned to be observed with the MO mode. The spectrum is constructed from a standard stellar template from Pickles library, which is resampled to the right spectral resolution and wavelength range and converted into photon counts. The user gives the magnitude in a band and the time of exposure. The conversion from flux to counts is the same that those of the sky simulation described previously. However, due to the atmospheric turbulence, only some fraction of the object's flux will fall within the fiber aperture. This fraction depends on the PSF and hence on the chosen AO mode: Seeing limited (i.e. no AO) or Ground Layer AO (hereafter GLAO). The fractional energy encircled by the MO fiber (0.9" of aperture) is given in table \ref{EE_ETC} for different bands and in the case of seeing limited and GLAO observation. The final object spectrum (in counts) is multiplied by the fraction of energy enclosed at the given band and AO mode. 

The high-z galaxy science case is modeled by a constant continuum of magnitude $mag_{con}$ and an emission line defined by a gaussian function centered at the position $\lambda_{emi}$ with a total flux of $flux_{emi}$ and a velocity dispersion $\sigma_{emi}$. This science case can be observed with the MO and MI mode. For the MO mode, the galaxy has been assumed to be a point source and the same conversions as the stellar template have been applied. In the MI mode, the galaxies have been assumed to have a spatial dimension. The users can define the number of illuminated pixel in the IFU, $N_{spaxel}$. The spectrum in each spaxel corresponds to the total flux of the galaxy divided by the $N_{spaxel}$.
\end{description}
\begin{table}[!h]
\begin{center}
\begin{tabular}{ccccccccc}
&U&B&V&R&I&J&H&K\\
\hline
Seeing limited&57&59&62&65&69&73&76&80\\
GLAO&61&64&67&70&73&77&81&83\\
\hline
\end{tabular}
\caption{{\small Percentage of energy enclosed in the S/N reference area of aperture 0.9". Measured from PSF simulations for a D = 42 m telescope. From \textit{E-ELT Spectroscopic ETC: Detailed Description},Liske.}}
\label{EE_ETC}
\end{center}
\end{table}

\subsection{Results of the benchmark study}
\subsubsection{Accuracy of the sky background extraction}
As refereed previously, the main motivation to the implementation of a new sky subtraction algorithm is to achieve the extraction of the sky continuum with an accuracy $<1\%$ required by the EVE science cases. I have thus tested if the design algorithm can reach this mandatory requirement in both observational modes, MO and MI. The quality of the extraction is defined as the residual between the input and the recovered sky surface. 

\begin{figure}[p]
\centering
\includegraphics[width=1.0\textwidth]{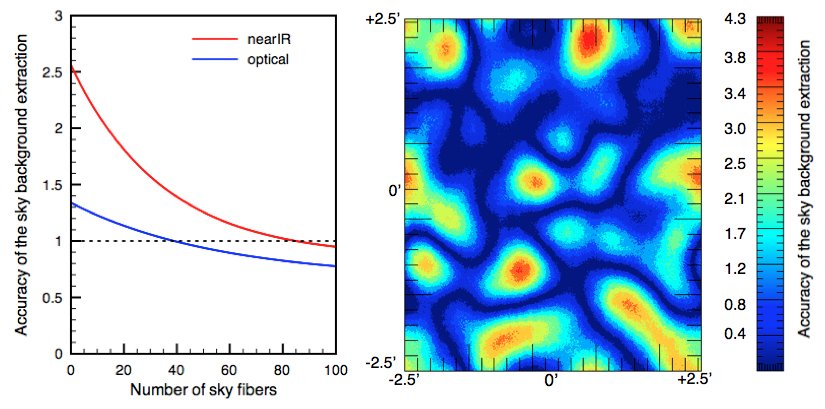}
\caption{{\small Right panel: Accuracy of the sky background subtraction as a function of the number of dedicated sky fibres for the optical (blue) and near IR (red) wavelength domain. Left Panel: Quality of the extracted sky background in the MO mode at visible wavelengths (\%). Simulations were done on a field of 7x7arcmin with 91 fibres randomly distributed on the sky. 110 objects fibres have been also distributed in the field of view. The accuracy of the sky background extraction in the object fibre is less than 0.9\%.}}
\label{Sky_MO}
\end{figure}

In the MO mode, the quality of the sky-continuum surface reconstruction depends on the number of dedicated sky fibers and how they are distributed in the field-of-view. I have tested the quality of the sky background extraction as a function of the number of dedicated sky fibers. For a given number of sky fibers, I have simulated 50 observations of one hour exposure with sky fibers randomly distributed in the field of view and measure the accuracy of the sky-background extraction in the central $5'\times5'$ region of the field-of-view. Fig. \ref{Sky_MO} illustrates the results of the Monte-Carlo simulations. The median value of the accuracy over the 50 iterations has been plot at each given number of sky fibers. At visible wavelengths, the sky background can be easily retrieved with accuracy of 1\% using 30-40 sky fibers. In the near IR, due to the large variation of the sky background (20\% peak to peak), a least 80 fibers are needed to properly recover the sky background with the same accuracy.

\begin{figure}[p]
\centering
\includegraphics[width=0.55\textwidth]{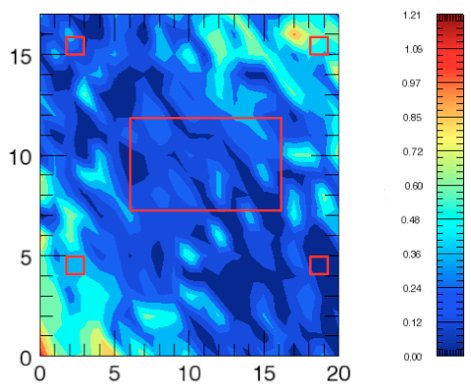}
\caption{{\small Simulated MI button including the 4 sky fibres. The color bar on the right side indicates the quality of the interpolation.}}
\label{Sky_MI}
\end{figure}

In the MI mode, we have simulated the case of an object illuminating nine central spaxels. The number of illuminating spaxels has been estimated from the apparent size of a z=8 Lyman alpha emitters, around 0.2", observed in GLAO mode. The remaining free fibers in the IFU were used as constraints for interpolating the sky background, in addition to four dedicated fibers sampling the sky contribution at larger spatial scales (see \ref{Sky_MI}). The MI mode allows us to reach an accuracy of sky extraction of 0.3\%. It is thus recommended for observations for which the sky background has to be recovered with very high accuracy.

\subsubsection{Application to the High-z galaxy science case}
The goal of this science case is to study the sources of the reionization. OPTIMOS-EVE is the ideal instrument to address this problem. Given its stability and spectral resolution, it is well suited to observe a large number of targets during very long exposure times. However, this is also one of the most challenging cases. Indeed, it requires being able to recover the very faint emission, down to $1\times10^{-19}\,ergs/cm^2/s$, from objects with a continuum reaching m(AB)= 28 or more. I have verified the feasibility of this science case in term of sky subtraction, in the MO and MI mode. The modeled z=8.8 galaxy has an apparent size of 0.2", which is typical of galaxies at this redshift. At z$\sim$9, the Lyman alpha line falls in a spectral window devoid of strong sky emission lines. The test has been performed using a Monte-Carlo simulation of 40 independent observations of 1h exposure (40 different sky cubes) for each mode. 

For the mode MO, 82 targets and 120 sky fibers have been distributed in the EVE field-of-view. Forty independent observations of 1h exposure have been generated by the EVE observation simulator. Each individual exposures have been sky subtracted using the sky subtraction algorithm and then combined together with a median minmax rejection algorithm (IRAF \textit{scombine} task). Fig. \ref{High_z_MO} summarizes the performance of the MO mode in recovering a distant galaxy with an AB magnitude of 28, an emission line flux of $1\times10^{-19}\,ergs/cm^2/s$, and a $FWHM_{obs}$=150km/s, in 40 hr of integration time. The emission line is recovered with a mean S/N ratio of 8.
 \begin{figure}[!h]
\centering
\includegraphics[width=1.0\textwidth]{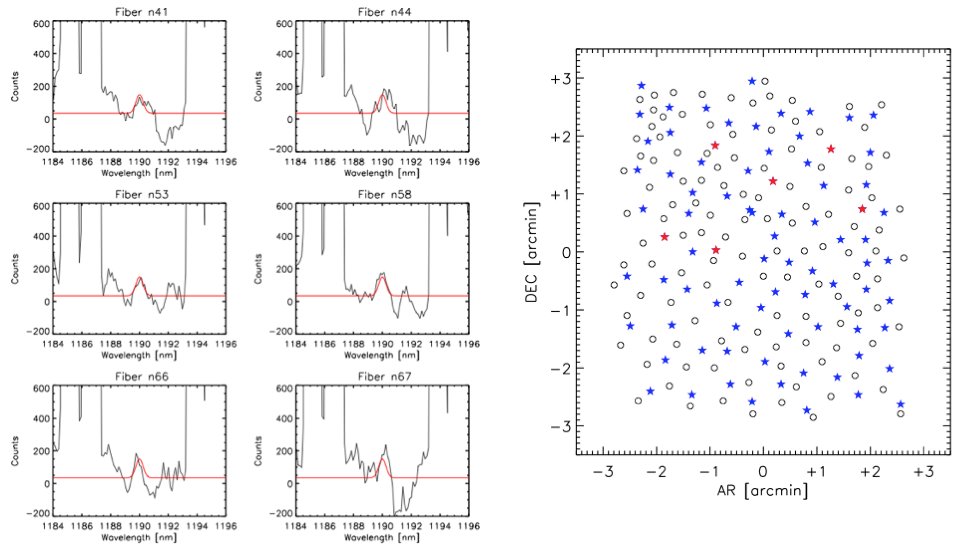}
\caption{{\small Simulation of 82 galaxies at z=8.8 in the J-band observed during 40 hours in the MO mode with GLAO. All galaxies have a continuum in the J-band of m(AB)=28 and the Lyman alpha line has an integrated flux of $1\times10^{-19}\,ergs/cm^2/s$ and a fwhm=150km/s. Left panels: The Lyman alpha line detected after sky background subtraction in different fibers (black line). The input spectrum is shown as reference (red line). The quality of the detection depends on the quality of the sky background subtraction at the location of the fibre. Sky emission lines have not been subtracted. Two sets of sky lines are visible at 1184-1187nm and 1193-1195nm. Right panel: Show the EVE field-of-view and the observational setup. The sky has been sampled with 120 fibers (half of the total number of available MO bundles; see open circles). The 82 simulated galaxies are shown as blue stars, while the red stars correspond to the fibers shown in the left panels.}}
\label{High_z_MO}
\end{figure}

\begin{figure}[!h]
\centering
\includegraphics[width=0.8\textwidth]{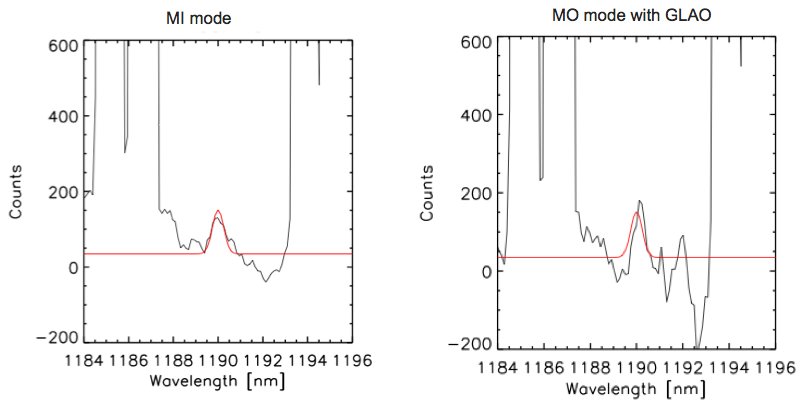}
\caption{{\small Simulation of a z=8.8 galaxy observed in the J band during 40 hours in the MI (left panels) and MO+GLAO (right panel) modes. The galaxy has a J-band continuum of m(AB)=30, 28, and 28, respectively, and the Lyman alpha line has an integrated flux of $1\times10^{-19}\,ergs/cm^2/s$ with FWHM=150km/s. Left and middle panels: the Lyman alpha line is detected in the binned central spaxels of the IFU after sky background subtraction (black line). The target illuminates nine central spaxels that have been summed up. The remaining spaxels are used for sky sampling. Right panel: simulation of observations in the MO mode of the Lyman alpha line. The observational setup is the same as the one described in Fig. \label{High_z_MO}.}}
\label{High_z_MI}
\end{figure}

In MI mode, the galaxy illuminates nine spaxels of the IFU. The other spaxels of the IFU are used for the determination of the sky, in complement of the four surrounding sky fibers. As the MO mode, 40 independent observations of one hour exposure have been simulated. The sky has been extracted in each individual exposure, using an optimal extraction to subtract the sky contribution. The observations have been binned in order to sum up all the flux of the object and then sky subtracted. The MI mode allows us to reach an S/N of 5 and 16 for faint emission lines ($1\times10^{-19}\,ergs/cm^2/s$) and an underlying continuum of 30 mag and 28 mag, respectively. The MO mode, for the same faint emission line and an underlying continuum of 28mag in J-band the galaxy is detected with an S/N=8. Fig. \ref{High_z_MI} shows the detected Lyman alpha line after sky background subtraction (exposure time of 40 hours) in the MI and MO modes. In the MI mode, the sky background is better subtracted and the line is detected with a better S/N than with the MO mode. 

\subsubsection{Observational strategy}

The algorithm described above will be applied in staring mode, when only the visible arm is used. In the case of NIR data, the sky is observed using offset exposures. These frames (on and off) are subtracted one from the other in 2D: \textit{A} - \textit{B} (this removes dark current contribution and sky contribution). In a following step, the residual of the sky (due to temporal variations between \textit{on} and \textit{off}) is removed by applying the aforementioned algorithm, using the fibers dedicated to the sky in the 'A' configuration.

\section{ Conclusion and prospective}
The sky-subtraction algorithm do answers the mandatory requirement of the EVE science case: the accuracy $<1\%$ in sky-background subtraction. The sky background contribution can be extracted in an FoV of 7x7 arcmin2 with an accuracy of 1\% in the MO mode, and 0.3-0.4\% for integral-field-unit observations.  I verified the performance of the algorithm in faint object spectroscopy observations and found very promising results: Emission lines with total flux $1\times10^{-19}\, erg/s/cm^2$ over a 30mag continuum can be extract from the sky background noise with the $\lambda-\lambda$ surface reconstruction algorithm. This study demonstrate the ability of fiber-fed instruments to proceed accurate sky-subtraction and thus show that they are serious competitors to slit instruments for faint object spectroscopy. In the future, I aim to develop severals points of this study :
\begin{itemize}
\item Create a user-friendly interface for the sky-subtraction algorithm and adapt it to reduce data from several fiber-fed spectrograph. The software will be make public. 
\item Test the algorithm in existing fiber-fed instrument, such as GIRAFFE/VLT and PKAS/Caha. A preliminary test have been done with GIRAFFE data. I found that the algorithm is able to extract the sky background with an accuracy of 1.8\% for 18 sky fibers. This results coincides with the expected accuracy on sky extraction with 18 using the $\lambda-\lambda$ reconstruction algorithm. 
\end{itemize}

      \NumberThisInToc
\FrameThisInToc
\part{The study of Host galaxies as a proxy SN Ia progenitors}
 \begin{flushright}
\textit{ "I' am a shooting star leaping through the sky\\
Like a tiger defying the laws of gravity\\
I' am  a racing car passing by like lady Godiva\\
I' am gonna go go go\\
There's no stopping me\\
I' am  burning through the sky yeah!\\
Two hundred degrees\\
That's why they call me mister Fahrenheit\\
I' am  traveling at the speed of light\\
I wanna make a supersonic man out of you"\\}
Queen, Don't stop me know
 \end{flushright}
 \EmptyNewPage

\FrameThisInToc

\chapter{ Introduction}
\section{ Supernovae Ia}
\begin{figure}[h!]
\centering
\includegraphics[width=8cm]{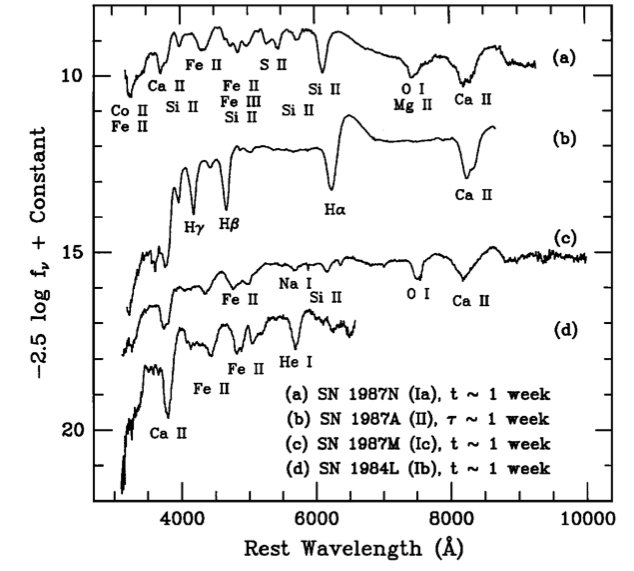}
\caption{{\small Spectra of SNe showing the main characteristics of  the four Supernovae types and subtypes. $t$ is the time  after $B$ max, $\tau$ is the time after core collapse and the units of $f_\nu$ are in $ergs\,s^{-1}\,cm^{-2}\,Hz^{-1}$. From \citet{1997ARA&A..35..309F}, fig. 1.}}
\label{SN_type_spectra}
\end{figure}

Supernovae are extremely luminous events originated by a stellar explosion. According to their spectral features supernovae are classified into two main categories:  Type II, which show hydrogen features in the spectra and type I with no hydrogen lines. Type I have a sub-classification according to the presence of specific spectral features. For example, type Ia supernovae (SNe Ia) are characterized by the presence of the Si II 615.0\,nm absorption line, see figure \ref{SN_type_spectra}. Contrary to the other supernova types, which are originated by core-collapsing stars, SNe Ia are generally considered to be the result of the thermo-nuclear explosion of a white dwarf(s) \citep{1960ApJ...132..565H,1961ApJ...134.1028}. SNe Ia produce consistent peak luminosities around $M_{V}=-19\,mag$ and have characteristic light curves (luminosity as a function of time). Due to the uniformity of their observed properties, Type Ia Supernovae have been used as standard candles and have played a central role in modern cosmology in the last decade. I will discuss in more detail on the progenitor nature of SNe Ia in section 1.3 of this chapter. 

\section{Supernovae Ia : a tool for Cosmology}
The main goal of observational cosmology is to constrain the cosmological parameters that describe the Universe. Due to its homogeneity and isotropy, the Universe can be described by a Friedman-Roberston-Walker model defined by the following cosmological parameters : 
$H_0$ the Hubble constant, $\Omega_m$ the energy density in matter, $\Omega_r$ the energy density in radiation and $\Omega_\Lambda$ the vacuum energy density and the curvature.  $H_0$ and  $\Omega_r$ have already been determined from the measurement of redshift and luminosity distance in galaxies and from the energy density in Cosmic Background respectively \citep{2009ApJ...699..539R,2010arXiv1001.4538K}. To determine $\Omega_m$  and $\Omega_\Lambda$, the space geometry of the Universe must be measured on large scales. One way to do it, is based on the use of standard candles to probe, to high redshifts, their redshift-magnitude relation (also called Hubble diagram). A standard candle is an astrophysical object which intrinsic luminosity can be determined through its physical properties, such as the luminosity-period relation of Cepheids stars.  Considering,
\begin{equation}
f=\frac{L}{4\pi d_{L}},
\end{equation}
where $L$ is the intrinsic rest-frame luminosity, $f$ the apparent flux and $d_{L}$ is the luminosity distance expressed in terms of the redshift and the cosmological parameters (see for instance \citet{1992ARA&A..30..499C}), we can obtain the redshift-magnitude relation for standard candles and derive from it the cosmological parameters.

\par In early 90's, astronomers have discovered that SN Ia can be considered as standard candles, due to the relationship linking their intrinsic peak luminosity and their light curve decay time \citep{1984SvA....28..658P,1993ApJ...413L.105P,1995ApJ...438L..17R,1996ApJ...473...88R,1995AJ....109....1H}. However, the scatter of apparent B magnitude at the peak can reach $\sim 1\, mag$. To overcome this limitation, empirical calibration based on the light curve width-luminosity, e.g. $m_{\Delta15}$, \citep{1993ApJ...413L.105P} and intrinsic colors \citep{1995ApJ...453L..55V} have been implemented. Currently used methods to calibrate SNe Ia allow to a reduced scatter of 0.13 mag \citep{1996AJ....112.2438H,2006A&A...447...31A}, making SNe Ia valuable cosmological standard candles (see review, e.g. \citet{1998ARA&A..36...17B}). 

\begin{figure}[h!]
\centering
\includegraphics[width=10cm]{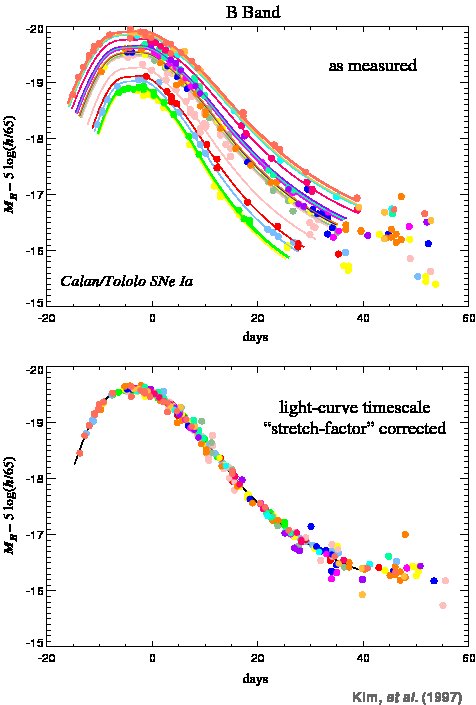}
\caption{{\small \textbf{Upper panel}: light-curves in B band of the Calan/Tololo SNe Ia sample. SNe Ia have a very homogenous light-curve shape. The observed scatter of peak luminosity can be reduced by the stretch correction. \textbf{Lower panel}: light curve corrected by the stretch-factor. From \citet{2003PhT....56d..53P}}}
\end{figure}

\par During the past decade, new telescopes, new search techniques, and improvement of calibration from local SN Ia survey and the systematic detection and photometric observation of high-z SNe, allow to construct the magnitude-redshift relation at high redshift where the effect of cosmological constant is significant. Two independent groups the \textit{Supernova Cosmology project} (SCP) \citep{1999ApJ...517..565P} and the \textit{High redshift SN Search Team} (HZT) \citep{1998AJ....116.1009R} performed the Hubble relation for SN Ia.
Their results led to a striking conclusion. Contrary to the predictions based on the Einstein-de Sitter Universe, both groups found that the Universe is accelerating. The data put on evidence a nonzero dark energy driving the expansion of the Universe.  Independent observations from the Cosmic Microwave Background CMB \citep{2003ApJS..148....1B,2010arXiv1001.4538K} and large scale structures \citep{2001Natur.410..169P} have confirmed the discovery.  At the present date, the assumed set of cosmological parameters correspond to a flat $\Lambda$-CDM cosmological model with $\mathrm{H_0}$=70$\mathrm{km~s^{-1}Mpc^{-1}}$, $\Omega_\mathrm{m}=0.263\pm0.042(stat)\pm0.032(sys)$,  for a $\Lambda$ CDM Universe (\citet{2006A&A...447...31A}). These results are in agreement with the 7 year WMAP results \citep{2010arXiv1001.4538K}.

\begin{figure}[htb]
\centering
\includegraphics[width=12cm]{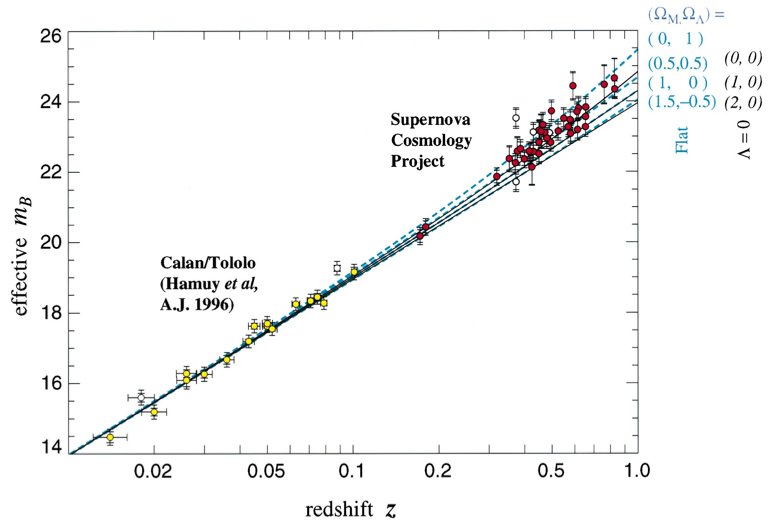}
\caption{ {\small The redshift-magnitude relation for type Ia Supernovae measured by the Supernovae Cosmology Project \citep{1999ApJ...517..565P}. Three theoretical curves are shown for various values of $\Omega_M$ and $\Omega_\Lambda$. Assuming a flat Universe, the best fit to the data does favor a non zero cosmological constant.}}
\end{figure}

\par Current and future SNe Ia surveys are focussed on the determination of the nature and properties of dark energy \citep{2006A&A...447...31A,2007ApJ...664..675E}. To differentiate between the number of dark energy models, the expansion history of the universe has to be mapped with high precision and small systematic uncertainties. Accurate photometry of distant high-z SNe are very challenging because of their small angular sizes and their faint apparent luminosities. Such kind of observations requires very stable detectors with high spatial resolution, high quantum efficiency and arrays almost devoid of bad pixels. The new generation of earth and space-based surveys, such as the \textit{Pan-STARRS}, \textit{Dark Energy Survey} and \textit{Joint Dark energy mission} (JDEM), will provide a better control on photometry and possible systematics. Indeed, currently the major obstacles to fully exploit the potential of SNe Ia in cosmology are the subtle systematic uncertainties from the insufficient knowledge of the SN Ia physics: nature of their progenitor stars, physics of the explosion, possible evolutionary effects, effect of the dust dimming in the host galaxy (see for instance, \citet{2008ARA&A..46..385F,2010Natur.463...35H}).

\section{Supernovae Ia progenitor scenarios}
Years of observational and theoretical studies have led to the general agreement that SNe Ia are the result of thermonuclear disruption of carbon/oxygen white dwarfs (C+O WD), in a binary system, that reach the Chandrasekhar mass of $1.39 M_\odot$ \citep{1960ApJ...132..565H,1961ApJ...134.1028}. The evolution of their luminosity with time (the light-curve) is driven by the radioactive decay of the $^{56}Ni$ produced during the explosion into $^{56}Co$ and $^{56}Fe$ \citep{1994ApJS...92..527N,2000ApJ...530..757P}.  However, up to the date there is no direct observations of a Sn Ia progenitor and thus on what is the exact evolutionary scenario that leads to the explosion. Theory proposes two main scenarios for the origin of SNe Ia: 
\begin{itemize}
\item \textbf{Single degenerate scenario - SD}\\ In this scenario, the C+O WD accretes mass from its companion, a main sequence star \citep{1973ApJ...186.1007W}. Different models predict different delay times between the episode of star formation producing the progenitor system and the time of explosion. According to \citet{2006MNRAS.370..773M}, the delay time is around $670\,Myr$. However, prompt delay shorter than 100\,Myr are possible if the companion is a massive main sequence star (6-7$M_\odot$) \citep{2008ApJ...683L.127H,2000ApJ...530..757P} or a helium star \citet{2008ApJ...686..337W}. Very long delay are also possible if the companion is a low-mass red giant  \citep{2008ApJ...683L.127H}. 
\item \textbf{Double degenerate scenario - DD}\\  According to this scenario, a binary system, composed by two white dwarfs, merges to a WD surrounded by a heavy disk or by a rotating envelope around the more massive WD. When the mass of the remaining WD reaches $\sim1.4\,M_\odot$, due to the mass accretion from the envelope or disk a SN Ia occurs \citep{1984ApJS...54..335I}. The delay time depends on the timescale of the formation of the WD binary system and ton the decrease of the angular momentum of the system due to the emission of gravitational waves \citep{1983bhwd.book.....S}. \\
\end{itemize}

\subsection{Influence of the nature of the progenitor on the properties of SN Ia}
\begin{figure}[htb]
\centering
\includegraphics[width=16cm]{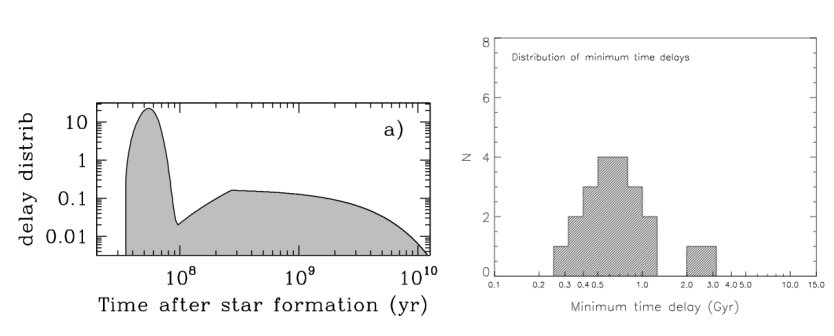}
\caption{ \textbf{Left}: The Delay Time Distribution (DTD) for a two-channels scenario, in which the contribution of a prompt and a long channels are equal, from \citet{2006MNRAS.370..773M}. \textbf{Right}: Observed DTD for a sample of local late-type SN Ia host, see \citet{1997AAS...191.8504P} }
\label{DTD}
\end{figure}

The progenitor age or the delay time distribution (DTD) is a powerful tool to disentangle among the progenitor scenarios. Different explosion models and progenitors predict different delay time distributions, and can be confronted with the observations. From the observation of the evolution of the SNe Ia rate with redshift, \citet{2005A&A...433..807M,2006MNRAS.370..773M} found evidence of a bimodality of the DTD. As shown in fig.\ref{DTD} part of SNe Ia explode shortly after the star formation, < 0.6 Gyr, and should be related to young, massive short-lived stars ($\sim$ 6-8 $M_\odot$). The other part of SNe Ia explode $\sim$ 2-4 Gyr after the star formation \citep{2004ApJ...613..200S} and are probably related to old low mass stars. The authors show that the SN Ia rate can be expressed in terms of two components, one proportional to the star formation rate of the host galaxy, and the other to the stellar mass of the host. The origin of this bimodality is still unknown. Indeed, both progenitor scenarios can originate such distribution when assuming particular parameter of the explosion. But it can also indicate the existence of  two populations of SNe Ia coming from different evolutionary channels. The existence of two populations SNe Ia, linked to two different age progenitors, can have a big impact on cosmology with SNe Ia \citep{2001ApJ...557..279D,2008ARA&A..46..385F}. For instance, presently, it is not yet known if the two populations will follow the same empirical relations used to calibrate SNe Ia into standard candles. It should be noted that current existing data, including high redshift supernovae does not allow to clarify the existence of bimodality. According to the recent work by \citet{2010ApJ...713...32S}, HST-SN data favors a single mechanism with DTD of 3-4 Gyr. This conclusion seams to contradict the results from the SNLS-Supernova Legacy Survey \citep{2006AJ....132.1126N} and the observed SNe Ia heterogeneity at low redshift (see for instance, \citet{2006MNRAS.370..773M,2006ApJ...648..868S,2009MNRAS.397..717S,2010arXiv1002.0848B}).\\

Another property of the progenitor that might affect the SNe Ia properties is the metallicity of the progenitor. Indeed, the peak luminosity is proportional to the amount of $^{56}Ni$ produced during the explosion \citep{1982ApJ...253..785A,1994ApJS...92..527N,2000ApJ...530..757P}. According to \citet{1999ApJ...522L..43U}, progenitors with low fraction of carbon lead to fainter SNe Ia. In a theoretical work, \citet{2003ApJ...590L..83T} concluded on the existence of an inverse linear relation between the intensity of the peak brightness and the metallicity of the progenitor. According to this work, metallicity variation from a progenitor to another can explain a scatter of 25\% when assuming the metallicity range see in the sun neighborhood. However other simulation do not agree with these conclusions and the dependence of the SNe Ia properties on the metallicity of the progenitor is still an open question \citep{2010ApJ...711L..66B,2001ApJ...557..279D}.\\

\subsection{Impact of progenitor properties on cosmology}
As already described, theoretical explosion models suggest that WD properties, like age and metallicity, may significantly influence the SN Ia peak luminosity, a crucial parameter in cosmology with SNe Ia. Observations of high-z redshift galaxies show that their global properties, as the star formation rate, age of stellar population and metal content \citep[among many other]{1997ApJ...489..559G, 2008A&A...483..107B, 2008A&A...492..371R}, do evolve strongly with lookback time. Any explosion or progenitor parameters that correlate with redshift (age, metallicity, production channel) will bias the determination of the cosmological parameters derived from SNe Ia. \citet{2000ApJ...530..593D} have explored the influence of an evolving luminosity peak with redshift on the determination of the cosmological parameters. They show that the determination of the SNe Ia evolution and the cosmological parameters is degenerated. To break the degeneracy, the dependence of the supernova properties on the progenitor properties, explosion mechanisms and possible evolution with redshift have to be completely understood. For instance, \citet{1998ApJ...495..617H} point out that a change if metallicity (C/O ratio) by a factor of three leads to an exchange of the peak magnitude by $\sim0.3\,mag$. This value is comparable to the effect of changes in $\Omega_M$, $\Omega_\Lambda$ parameters for SNe Ia at z>0.5. However, \citet{2001ApJ...557..279D} found an upper limit of 0.07\,mag for the dependence of SNe Ia magnitude on the progenitors metallicity for z<0.02. The measurement of the slope of the residuals in the Hubble diagram as a function of metallicity for a large SNe Ia data set is of crucial importance to reveal the physics of the explosion and to control possible systematics \citep{2010ApJ...711L..66B}.

\section{Sne Ia Host galaxies}
Type Ia Supernovae are rare events and their progenitors are difficult to identify. One approach to indirectly study the properties of SN Ia progenitors and their effect on the properties of SN Ia is through the analysis of the host galaxies. In the last years, several groups have been studying the correlations of the host properties with the SN Ia Hubble diagram residual. 

\subsection{The impact of the host metallicity}
The metallicity of the host might be correlated to the metallicity of the progenitor and thud affect the luminosity of the SN Ia. Different authors have made different prediction based on model \citep{1998ApJ...495..617H,2010ApJ...711L..66B} or data \citep{2005ApJ...634..210G,2008ApJ...685..752G2009ApJ...691..661H,2009ApJ...707.1449N}. The effect of the age and metallicity are difficult to disentangle because both properties may affect the explosion. 

\subsection{The impact of the age of stellar population}
There are observational evidences that the morphology of the host correlates with the peak luminosity of the SNe Ia \citep{1996AJ....112.2438H,2003MNRAS.340.1057S}. Fainter events are found in early-type galaxies while late-type and irregular galaxies host more bright SNe Ia. The correlation with morphology is strongly connected to the stellar content of host. Early type galaxies are composed by old stellar population and  host almost no star formation activity (Howell 2001). Such observations suggest a long delay channel as the most probable scenario for SNe Ia in early type galaxies. One the other hand, late-type galaxies host star formation activity and a large amount of young stars, which suggest a prompt delay channel. The SN Ia rate per unit of stellar mass is strongly correlated with the host  specific SFR. SNe Ia are ten times more frequent in star forming galaxies than in more quiescent hosts \citep{2005A&A...433..807M,2006ApJ...648..868S}. In passive galaxies, the SN Ia rate is linked to the stellar mass of the host.

\subsection{Evidence for a two channel production of SNe Ia}
The observations suggest the following: a SN Ia might result from one of the two different explosion channels, see table \ref{table_SN1a_channel}. If so, the prompt time delay channel is associated to star forming environment, giving origin to the brightest events. In this case, the explosion rate is dependent on the star formation rate of their hosts and thus may be correlated with the current number of WD. On the other hand, the long delay channel is associated with old stellar population in early type galaxies. Their explosion rate is correlated to the stellar mass of the host and thus to the cumulative number of produced WD. 
\begin{table}[h!]
\centering
\begin{tabular}{l | c c } 
Property & Prompt channel & Delayed channel\\ 
\hline
Time delay&$\sim$600Myrs&2-4Gyrs\\
Host galaxy&Late-type&Early-type\\
Stellar population&Young&Old\\
Rate explosion &SFR&M$_{stellar}$\\
&just formed WD&cumulative WD
\end{tabular} 
\caption{ Observed properties of prompt and delayed SN Ia channel. }
\label{table_SN1a_channel} 
\end{table}

\begin{figure}[h!]
\centering
\includegraphics[width=10cm]{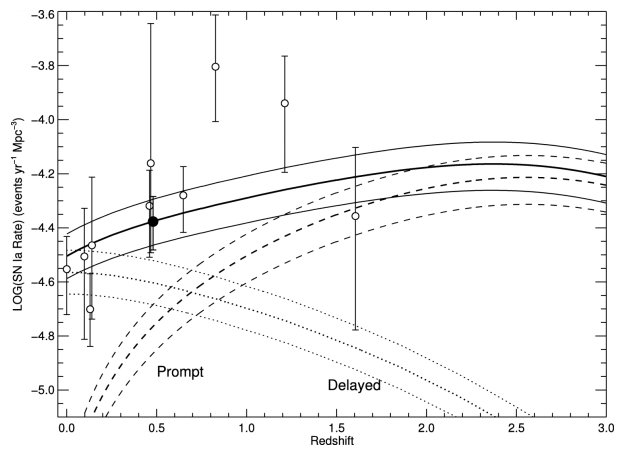}
\caption{ From \citet{2006ApJ...648..868S}. Contribution of the prompt and delayed channel as a function of the redshift. The crossover point between the two channel is expected to be around $z\sim$0.5-0.9}
\label{fig_evol_sn1a}
\end{figure}

As the SFR strongly evolves with redshift,  the contribution from each evolutionary channels can probably change with redshift. A large number of prompt delay SNe Ia is expected to be observed at high-z redshifts, while long delay SNe Ia should decrease with increasing $z$. According to \citet{2006ApJ...651..142H}, the crossover point between the two channels is expected to be around z$\sim$0.5-0.9, see fig \ref{fig_evol_sn1a}. However, no evidence for brighter SNe Ia has been observed in the distant Universe. \\

To date no progenitor of a SN Ia has been identified and/or observed and information about the SNe Ia progenitors can only be obtained indirectly. One promising route to study the SN progenitors is through the analysis of their host galaxies using 3D spectroscopy, e.g. similar to the \citet{2008A&A...490...45C} studies of the host GRB 980425/SN 1998bw. In fact, previous studies on metallicity and stellar population of host galaxies were based on global properties \citep{2005ApJ...634..210G,2008ApJ...685..752G,2009ApJ...691..661H}. The results are inconclusive and contradictory. And, as it is well known, the properties of the gas and stellar population can vary considerably radially \citep{2009MNRAS.395...28M}. It is thus very important to compare the global properties of the galaxy to the local properties. Eventual differences between global and local can be a source of systematic errors if the luminosities of supernovae used in the Hubble diagram are corrected using the global characteristic of the host. To clarify this aspects I am participating in a large project to study hosts of SN Ia as will be explain in the next chapter.  \\
\FrameThisInToc
\chapter{ IFS observations of SN Ia hosts}
\minitoc

This study is the result of the collaboration between my two hosting institution CENTRA-Lisbon and GEPI-Paris. The team is compounded by A. M. Mour\~{a}o and V. Stanishev from CENTRA and H. Flores from GEPI and myself (CENTRA \& GEPI). We have combined the two main competence of the two labs, namely observational cosmology with SN Ia in CENTRA and the observations of galaxies in GEPI, into a synergy project for the study of SNe Ia hosts and the properties of the SN environment. 
We aim to perform a large survey study of local SNe Ia hosts. In this chapter, I present the results of the pilot observational run carried out with PPAK/PMAS integral field spectrograph at Calar Alto observatory in November 2009. My role in the team is to produce and interpret the maps of the physical properties of galaxies (extinction, metallicity, ionization parameter, stellar population, etc..) from the reduced data cube or/and emission line flux maps.   

\section{Designing a large survey for hosts of SNe Ia}
\subsection{Science goals}
The aim of the large survey for SNe Ia hosts is to use 3D spectroscopy at intermediate spectral resolution to study the properties of the gas and the stellar populations of a sample of about 50 galaxies hosting well-observed SNe Ia. Only galaxies in the Hubble flow with redshift z > 0.02 have been selected in order to minimize the effect of peculiar motions. We have selected galaxies hosting well-studied SNe Ia, with good photometry coverage before and after the maximum. 
The two main objectives are: 
\begin{itemize}
\item to study the effect of the metallicity and age of the local stellar population on the SN luminosity. The deviation of SNe from the best fit Hubble diagram will be correlated to the properties of the local environment. Such a correlation, if exists, will help to further improve the use of SNe Ia as distance indicators; 
\item to use the properties of the local SN environment (gas and stellar population properties) to put constrains on the nature of SNe Ia progenitors. 
\end{itemize}

Previous studies on metallicity and stellar population of host galaxies were based on global properties, e.g. SED fitting to board- band photometry \citep{2006ApJ...648..868S,2009ApJ...691..661H,2009ApJ...707.1449N} or integrated galaxy spectra \citep{2005ApJ...634..210G, 2008ApJ...685..752G}. It should be emphasized that the properties of the gas and stellar population can vary radially  \citep{2009MNRAS.395...28M}.  This survey will allow to explore directly the influence of the local properties on SNe Ia for first time. 

The following properties of the host will be derived from the integral field spectroscopy observations:
\begin{itemize}
\item Maps of the gas properties: From the emission lines we will produce maps of the extinction, electronic density, star formation rate, ionization parameter, electronic temperature and abundances of several elements (O, N). We will also search for traces of star formation activity at the position of the SN. \citet{2005A&A...433..807M} suggested that half of the SNe Ia come from young progenitors (< 0.1 Gyr). Therefore, it is very likely that recent active star formation took place in the immediate vicinity of the SN. This may be detected by the presence of $H\alpha$ emission at the SN position.
\item Maps of stellar population: Combining semi-empirical stellar population synthesis code \citep{2003MNRAS.344.1000B, 2006ApJ...652...85M, 2005MNRAS.358..363C} with the measurement of absorption lines we will produce maps of the age, stellar extinction and metallicity.
\end{itemize}

\subsection{Sample selection}
\begin{figure}[htb]
\centering
\includegraphics[width=16cm]{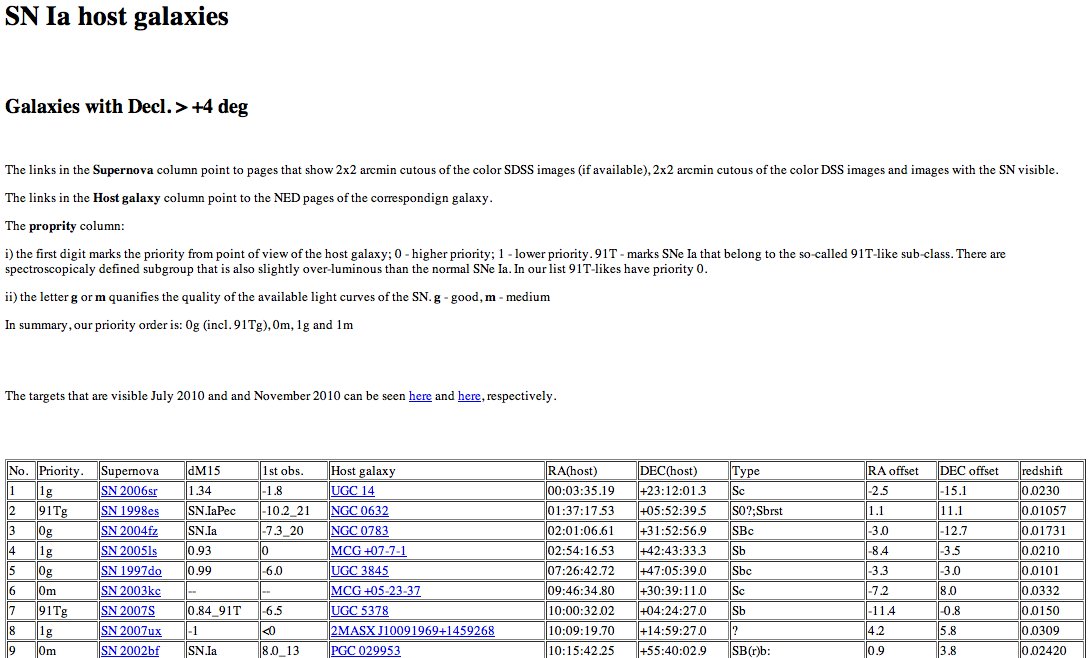}
\caption{ {\small Screenshot of the SN Ia host webpage.}}
\label{Sn1a_web}
\end{figure}

We have carefully selected our targets from a sample of galaxies that hosted well-observed SNe Ia for which the important parameters like luminosity, intrinsic color indices, luminosity decline rates and deviation from the best fit Hubble diagram can be accurately measured. We also require the targets to be in the unperturbed Hubble flow (z > 0.02) in order to minimize the effect of peculiar motions. From this large pool of well-observed SNe we have selected a primary list of about 100 galaxies with Declination> $4^o$ (visible in northen hemisphere observatories). The list is available at the following URL:
http://centra.ist.utl.pt/ $\sim$vall/sn\_hosts/sneia\_hosts 2010.php and includes link to NED and information about the SN, see  Fig.\ref{Sn1a_web}. These galaxies were carefully examined and selected to be nearly face-on in order to minimize the projection effects when correlating the SN and its host galaxy properties. 
We plain to observe around 50 galaxies in the next years using integral field spectroscopy instruments, such as PMAS at CAHA, ARGUS at VLT, Sparse at Wisconsin Observatory. Moreover, the team is part of the proposed CALIFA Legacy Survey (PI: Dr. Sebastian Sanchez). CALIFA is an ambitious project to observe about 500 galaxies at redshift z $\sim$ 0.02 with PMAS/PPAK, a wide-field IFS, over 3 years and our SN science case is included as sub-project. The CALIFA sample includes some galaxies that hosted supernovae (of any kind) which will complete our sample.

\section{PPAK observations}
\subsection{Observational setup}
\begin{figure}[htb]
\centering
\includegraphics[width=12cm]{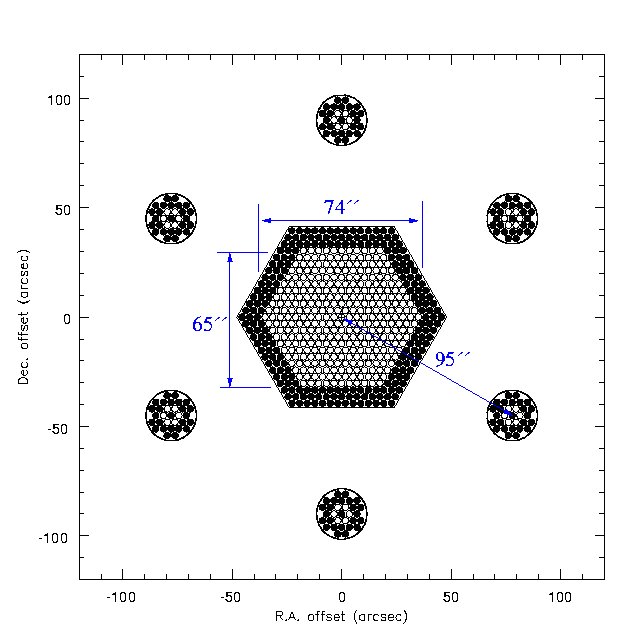}
\caption{ {\small Geometrical layout and size of the central hexagonal IFU (331 fibers) and six surrounding sky fiber bundles (each consisting of 6 fibers). Only open plot symbols are optically active fibers, filled circles are indicating auxiliary fibers which were employed in the manufacturing process for mechanical reasons. Note that contrary to the lens array, which provides a contiguous spatial sampling with negligible edge effects between adjacent lenses, the PPAK bundle has gaps between each fiber and its next neighbors. It is, however, possible to fill these gaps by repeated observations with small offsets (dithering). [From PMAS/PPAK webpage]}}
\label{PPAK_fov}
\end{figure}

We have undergone a pilot observational run of 2 nights using the \textit{Postdam MultiAperture Spectrophotometer} (PMAS) in PPKA mode mounted at the 3.5m telescope of the \textit{Observatorio Astron\'{o}mico Hispano Aleman Calar Alto} (CAHA). PMAS is an integral Field Spectroscope that can operate in two modes with a with a fiber-coupled lens array (LArrn mode) and a fiber bundle IFU(PPAK mode).  In the PPAK mode, the IFU is a hexagonal packed fiber-bundle composed by 331 science fibers, 15 calibration fibers and 36 sky fibers located at about ~80" from the center, see the configuration of fiber in figure \ref{PPAK_fov}. The IFU covers a hexagonal field of view of 74 x 65 arcsec$^2$ with a spatial sampling of 2.7" per fiber. We have used PMAS/PPAK with the V600 grating to obtain medium-resolution spectroscopy ($\sim$5.7\AA\, spectral resolution) of the SNe Ia host galaxies. This grating allows us to cover the spectral range 3700-7000\AA\, required to fulfill the scientific goals. As it can be seen in fig.\ref{PPAK_fov}, the PPAK IFS does not cover continuously the field of view: there are spaces between the fibers. In order to have a full and uniform coverage, we have observed each galaxy using a dithering pattern of 3 pointings. The full description of the dithering procedure can be found in the PMAS cookbook (\emph{http://www.caha.es/sanchez/pmas/index.html}). 

\subsection{Sample}
From the host sample described above, we have selected randomly 6 galaxies visible at CAHA in November. The report of observations is shown in table \ref{table_sn1a_obslog}. In order to detect the low surface brightness regions of the galaxies ($\sim$21.6 mag\,arcsec$^2$) with a S/N>5 per spaxel at 5000\AA\,, we have estimated from the PMAS exposure time calculator that exposure times of $\sim$\,30\,min were needed. Therefore, we have observed each galaxy by three pointings of 30\,min each. 
\begin{table}[htb]
\centering
\begin{tabular}{l c c c c c } 
Object & RA & Dec & Exposure& Date & Seeing\\ 
\hline
NGC 0976&02:34:00.02&+20:58:36.4&$3\times30\,min$&15 Nov 2009& 2"\\
NGC 0495&01:22:56.00&+33:28:18.0&$3\times30\,min$&14-15 Nov 2009& 2"\\
UGC 5129&09:37:57.95&+25:29:39.1&$3\times30\,min$&15 Nov 2009& 2"\\
CGCG 207-042&08:11:43.48&+41:33:18.5&$3\times30\,min$&15 Nov 2009& 2"\\
NGC 105&00:25:16.91&+12:53:01.4&$3\times30\,min$&14 Nov 2009& 2"\\
UGC 4008&07:46:38.20&+44:47:32.0&$3\times30\,min$&14 Nov 2009& 2"\\
\hline 
\end{tabular} 
\caption{{\small SN Ia host galaxies observed with PPAK/PMAS at CAHA observatory during the pilot observation of 14-15 November 2009.}}
\label{table_sn1a_obslog} 
\end{table}

\section{ Data reduction and analysis}

\subsection{Data reduction}
The data has been reduced using a private IDL software designed by V. Stanishev. The reduction procedure included the standard corrections and calibrations for 3D observations, such as: bias subtraction, cosmic ray correction, tracing and extraction, wavelength and distortion calibration, sky subtraction and flux calibration, reconstruction of the datacube. The fluxes have been calibrated using standard spectrophotometric stars. Each individual science pointing has been reduced independently and then combined into a final 3D datacube and finally corrected for atmospheric diffraction. According to \citet{2005PASP..117..620R,2006PASP..118..129K}  the final spatial sampling of the datacube when using a 3 dithered-pattern is 1"$\times$1". However, as the mean seeing during observation was 2", we have set the final spatial sampling of the cube to a square pixel of $2"\times2"$. This operation does not downgrade the spatial resolution since we are limited by the seeing and it has the advantage to increase the signal-to-noise in each spaxel. A complete description of the data reduction procedures will be available in Stanishev et al. (in prep).

\subsection{Analyzing the datacube}

The spectrum of each spaxel is analyzed as the integrated spectra following the method described in the Introduction part of this manuscript\footnote{The reader is referred to this section for a detail description of the methods used}. In good S/N spaxels, the spectrum is composed by a continuum component from the stellar populations and an emission line spectrum from the ionized gas. Before measuring emission lines, in particular Balmer recombination lines, the continuum component has to be evaluated and subtracted. Simultaneously, the fitting of the stellar continuum with semi-empirical models provides the composition of the stellar populations -mean stellar age and metallicity, star-formation history.   

The datacubes have been set to rest-frame and corrected from the galactic extinction. In table \ref{table_sn1a_extin} the redshift and E(B-V) \citep{1998ApJ...500..525S} for the six SNe Ia host galaxies are given. 
\begin{table}[htb]
\centering
\begin{tabular}{l l c l c c}
\hline
Host&z&E(B-V)$_{Gal}$&SN& RA& DEC\\
&&&&offset& offset\\
\hline
NGC\,0976& 0.014327& 0.110&1999dq &-5.1&-6.0\\
NGC\,0495& 0.013723& 0.072&1999ej&+17.7&-20.1\\
UGC\,5129& 0.013539& 0.022&2001fe& -12.8& +2.2\\
CGCG\,207-042& 0.031592& 0.046&2006te&-7.3& -1.7\\
NGC\,105& 0.017646& 0.073&2007A&-3.7&+11.1\\ 
UGC\,4008& 0.031584& 0.047&2007R&-10.1&-9.9\\ 
\hline 
\end{tabular} 
\caption{{\small SN Ia host galaxies }}
\label{table_sn1a_extin} 
\end{table}

\subsubsection{Continuum subtraction}
\begin{figure}[htb]
\centering
\includegraphics[width=14cm]{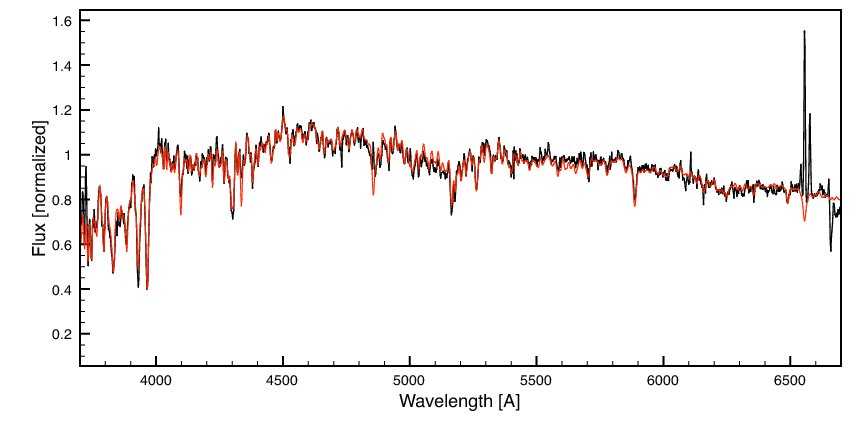}
\caption{ {\small In black, the spectrum from a spaxel of the SN 2007R host. In red, the synthetic continuum spectra estimated from the semi-empirical models of Charlot \& Bruzual using the MILES stellar library (CB07). }}
\label{SN2007R_fitcon}
\end{figure}

As the fitting of the continuum is time-consuming, we have only fitted the spaxels with a S/N in [4730-4780]\AA\,  continuum higher than to 5. Below this S/N, the physical properties derived from the fit are no more trustable \citep{2005MNRAS.358..363C} and the underlying Balmer absorption is negligible. 
Each individual rest-frame spectra has been fitted with the Starlight software \citep{2005MNRAS.358..363C} using a base of 3 metallicities $\times$ 15 ages templates from Charlot \& Bruzual (in prep, hereafter CB07), see fig. \ref{SN2007R_fitcon}. The new models of CB07 use the stellar library of MILES \citep{2007yCat..83710703S} which have the advantage to not be affected by the same flux problem around the $H\beta$ lines as the previous \citet{2003MNRAS.344.1000B} models, see Introduction part section 3.2.1. The regions with emission lines, strong sky lines and the red-end of spectra\footnote{The red-end of the PPAK spectra is affected by vigneting and the low efficiency of the detector at these wavelengths} have been masked during the fit. The continuum has been subtracted in each spaxel spectrum which verify the S/N condition. 

\subsubsection{Measuring emission lines}
Due to the large amount of spectra per galaxy ($\sim$ \,600), we have measured the emission lines with an automatic routine. A complete description will be provided in Stanishev et al. (in prep). Briefly, the IDL-based software fits each line by a single gaussian distribution + a pseudo-continuum, using the $\chi^2$ minimization algorithm MPFIT \citep{2009ASPC..411..251M}, and provides the central wavelenght $\lambda_{center}$, sigma $\sigma_{line}$ and total flux of the line.  The flux uncertainty of the emission lines is estimated by propagating the uncertainties on the flux and on the fitted parameter ($\lambda_{center}$, $\sigma_{line}$, level of the pseudo-continuum). Each line has been visually checked. We verify in several spaxels that our automatic measurements were consistent with manual measurements using the SPLOT/IRAF routine.

Only lines with spectral S/N > 6 have been considered as detected. The spectral S/N is defined as the ratio between the total flux in the line $F_{line}$ and the noise in the pseudo-continuum $\sigma$ (Flores et al. 2006). The S/N is normalized by the number of spectral resolution elements $N$ in which the emission line is spread (3$\sigma_{line}$ from the $\lambda_{center}$).
\begin{equation}
<S/N>=\frac{F_{line}}{\sqrt(N)\times\sigma}
\end{equation}


\subsection{Deriving physical properties from 3D datacube}

\subsubsection{Spatially resolved and integrated informations}

With the 3D datacubes, we are able to measure the properties of the stellar population and of the ionized gas in each spaxel and to produce 2D maps of these quantities for each galaxy. The 2D maps allow us to study the spatial variation of the properties of the galaxies and to probe directly such properties in the region of the SN Ia. 

One of the main goals of this study is to compare the properties of the host, as derived from the integrated spectroscopy, to those derived from spatially resolved spectroscopy. For this purpose, we have produced for each galaxy, 2D maps and for comparison the following integrated spectra: 
\begin{itemize}
\item integrated spectrum in a central aperture of 3\,arcsec. We have summed the spectra of the spaxel localized within 3\,arcsec of the center of the galaxies. This condition corresponds to the SDSS spectroscopy:  3" aperture spectroscopy of the center of the galaxy light distribution.
\item integrated spectrum of the all galaxy from the sum of all spaxel spectra. This simulates an observation of the same galaxy with long-slit spectroscopy as happens with high redshift galaxies.   
\end{itemize}
 
\subsubsection{Properties of the ionized gas}
The study of the emission lines allows us to derive extinction, star-formation rate, presence of shocks, electronic densities, metallicity of the ionized gas. The full description of the method used in the derivation of the physical properties of the ionized gas can be found in Part I Chap. 2. Briefly, I remind the methods and calibrations used to measure the following properties:  
\begin{itemize}
\item \textbf{Extinction}. The extinction in V-band, $A_V$, is derived from the flux ratio $H\alpha / H\beta$ using the Balmer decrement method \citep{1989agna.book.....O} and assuming a \citet{1989ApJ...345..245C} extinction law with $R_V$=3.1. 
\item \textbf{Source of ionization}. Some diagnostics, as the strong lines method to derive metallicity, cannot be applied if the source of ionization is not exclusively arising from the stellar radiation. For this reason, we use the diagnostic diagram [OIII]$\lambda\lambda4959, 5007/H_{\beta}$ vs [NII]$\lambda 6584 /H_{\beta}$ \citep{1987ApJS...63..295V, 2003MNRAS.346.1055K} to search for possible AGN contamination and spatial variations of the ionization parameter. Spaxels which fall in the AGN area of the diagnostic diagram have been excluded in the derivation of the metallicity maps. We use the delimitation of \citet{2003MNRAS.346.1055K} and \citet{2001ApJS..132...37K}.
\item \textbf{Shocks}. We search for the presence of shocks by the detection of the OI 6300\AA\, line and the diagnostic diagram [OII]$\lambda$3727/$H\beta$ vs OI\,$\lambda$6300\AA/$H\beta$ \citep{1985ApJ...292...29F}. 
\item \textbf{Star formation rate}. The SFR has been derived from the flux of the $H\alpha$ line corrected for extinction. We use the calibration of \citet{1998ARA&A..36..189K} between the luminosity of the $H\alpha$ line and the SFR. 
\item \textbf{Presence of Wolf-Rayet stars}. The presence of Wolf-Rayet stars can be detected by the detection of two emission features characteristic of these very young and massive stars: a blue bump at 4600-4680\AA\, (Allen 1976) and a red bump at 5650-5800\AA\,\citep{1986A&A...169...71K}. 
\end{itemize}
In order to compare with previous studies, I have derived the oxygen abundance, $log(O/H)+12$ using several calibrations: four strong line diagnostics and the $T_e$-direct method, namely: 
\begin{itemize}
\item \textbf{Metallicity from $R_{23}$}. Strong line method based on the $R_{23}$ parameter defined as the ratio ([OII]$\lambda$3727 + [OIII]$\lambda\lambda$4959,5007)/$H\beta$ \citep{1979MNRAS.189...95P}. I have used the Kobulnicky calibration which account for the ionization parameter and the popular \citet{2004ApJ...613..898T} calibration. However, these diagnostics are quite sensible to extinction due to the large distance in wavelength between the [OII]$\lambda$3727\AA\, and the [OIII]$\lambda$5007\AA\, lines. 
\item \textbf{Metallicity from $N2$}. Strong line method based on the $N2$ parameter defined as  the ratio  [NII]$\lambda$6584/$H\alpha$ \citep{1994ApJ...429..572S}. We applied the \citet{2004MNRAS.348L..59P} calibration between $N2$ and $log(O/H)+12$. The diagnostic is almost not affected by extinction due to the close vicinity of the two lines, but it has the disadvantage to saturate rapidly at over-solar metallicities, see Introduction part chapter 1. 
\item \textbf{Metallicity from $O3N2$}. We use the O3N2 parameter defined as 
\begin{equation}
O3N2=\log{\frac{[OIII]\lambda5007/H\beta}{[NII]\lambda6584/[OII]\lambda3727}}
\end{equation}
by \citet{2000ApJ...542..224D} and the calibration of \citet{2004MNRAS.348L..59P}. This diagnostic has the advantage to involve ratios of close lines, [OIII]$\lambda$5007/$H\beta$ and [NII]$\lambda$6584/$H\alpha$, and thus is almost not affected by extinction.
\item \textbf{$T_e$-direct method}. In the cases where [OIII]$\lambda$4363\AA\, is detected, we have derived the metallicity directly from the electronic temperature, using TEMDEM \textit{IRAF} routine.
\end{itemize}
The uncertainties on these quantities have been derived by propagating the uncertainties on the measurements of the line fluxes.  
\subsubsection{Properties of the stellar component}
The properties of the stellar components have been derived from the results of the continuum fitting given by the Starlight software\footnote{See Chapter 6 for a complete description and discussion of the limitation of the method.}. Briefly, the mean stellar age and metallicity, extinction, stellar mass and a first guess of the velocity field have been derived following \citet{2005MNRAS.358..363C}, as described bellow:
\begin{itemize} 
\item \textbf{The mean light-weighted stellar age $<Log\,t_\star>_{L}$}. This quantity is very sensitive to the presence of young and massive stars in the galaxy since these stars are the main contributors of the light at optical wavelengths. Low values of  $<Log\,t_\star>_{L}$ indicate for recent star-formation episodes, while high values reflect the presence of very old stellar population without any recent star-formation burst. The mean light-weighted stellar age is derived from the light-fraction population, $x_j$, and the respective age  $t_{\star,j}$ of the stellar population of the model templates, by:
\begin{equation}
<Log\,t_\star>_{L}=\sum^{N\star}_{j=1} x_j log\,t_{\star,j}.
\end{equation}

\item \textbf{The mean light-weighted stellar metallicity $<Log\,Z_\star>_{L}$}. This quantity traces the metallicity of the more recent and luminous stars. As the mean light-weighted stellar age, the mean stellar metallicity can be measured from the metallicity of the stellar population $Z_{\star,j}$:
\begin{equation}
<Log\,t_\star>_{L}=\sum^{N\star}_{j=1} x_j log\,Z_{\star,j}.
\end{equation}

\item \textbf{Stellar extinction, $A_{V\star}$}.
The stellar extinction derived by the Starlight software corresponds to the global extinction inferred from the stellar continuum. In some cases, the stellar extinction is a none physical negative value. Negative $A_{V\star}$ occur in old and dustless galaxies. It is due to the $\alpha$-enhancement ($\alpha$/Fe) in early-type galaxies which is not taken into account in the model used in the fit \citep{2006MNRAS.370..721M}. 

\item \textbf{Stellar mass}.
The stellar mass has been derived from the mass-fraction population, $\mu_j$, and the stellar mass-to-light ratio $M_\star/L_{\lambda0}$ associated to each stellar populations of the template base. 
\item \textbf{Velocity field and dispersion map of the stellar component}. We have also mapped the velocity and velocity dispersion of the spectra in each spaxel. The derived velocity and $\sigma$ fields are a first approximation and can be compared to the velocity field derived from the ionized gas. However, for a more detail study it is preferable to use dedicated software such as PPXF from the Sauron team \citep{2004PASP..116..138C}. 
\end{itemize}
In Table. \ref{table_err_con}, we present the uncertainties tabulated in \citet{2005MNRAS.358..363C} according to the S/N. In this paper, the authors used Monte-Carlo simulations to estimate the uncertainties of these quantities depending on the signal-to-noise. 

\begin{table}[htb]
\centering
\begin{tabular}{l c c c c c}
\hline
\hline
Quantity& \multicolumn{5}{c}{S/N in $\lambda=4020$\AA} \\
&5&10&15&20&30\\
\hline
log $M_\star$&0.11&0.08&0.06&0.05&0.04\\
$<log\,t_\star>_L$&0.14&0.08&0.06&0.05&0.04\\
$<log\,Z_\star>_L$&0.15&0.09&0.08&0.06&0.05\\
$A_{V\star}$&0.09&0.05&0.03&0.03&0.02\\
$v_\star$&17.73&8.55&5.92&4.23&2.81\\
$\sigma_\star$&24.32&12.36&7.71&5.73&3.78\\
\hline 
\end{tabular} 
\caption{{\small Uncertainties on the properties derived from the stellar continuum fitting, according to \citet{2005MNRAS.358..363C}. }}
\label{table_err_con} 
\end{table}

\section{Example of an individual study of a SN Ia host: NGC 976}
In this section, I give the example of the preliminary study of the spiral galaxy NGC\,0976, host of SN1999dq. The complete study of this galaxy and of the other galaxies of the sample is currently in progress. 

The supernova SN1999dq has been discovered on the 2th of September by W. Li (see IAU circular $n^o$7247). This SN Ia is located at AR=-5.1 DEC+-6.0 from the galaxy center (02:34:00.02 +20:58:36.4). 
\subsubsection{Morphology}
\begin{figure}[htb]
\centering
\includegraphics[width=8cm]{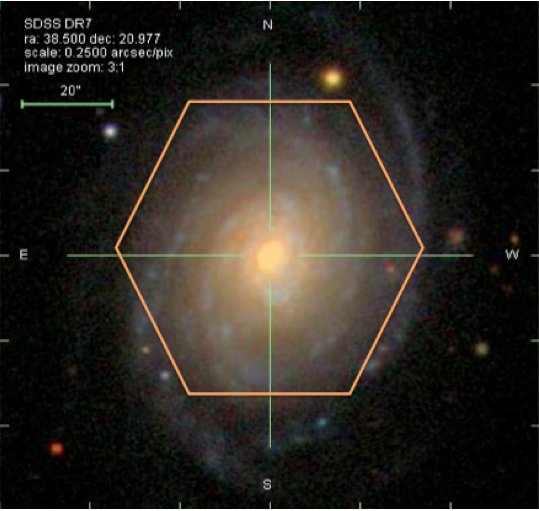}
\caption{ {\small Color image of NGC\,0976 from SDSS. The field-of-view of PPAK is represented by a orange hexagon.}}
\label{sn1999dq_SDSS}
\end{figure}

The NGC\,0976 is a spiral galaxy with a small inner ring. It is classified as a Sc galaxy in the Hubble classification \citep{2006AJ....131..527J} and a Sa(rs)c in De Vaucouleurs classification. The color image of NGC\,976 from the SDSS is shown in fig.\ref{sn1999dq_SDSS}. The field-of-view of the PPKA observation is delimited by a orange line. The inner ring is visible at less than 5" from the red bulge. 
  
\subsubsection{Kinematics}
\begin{figure}[htb]
\centering
\includegraphics[width=6.5cm]{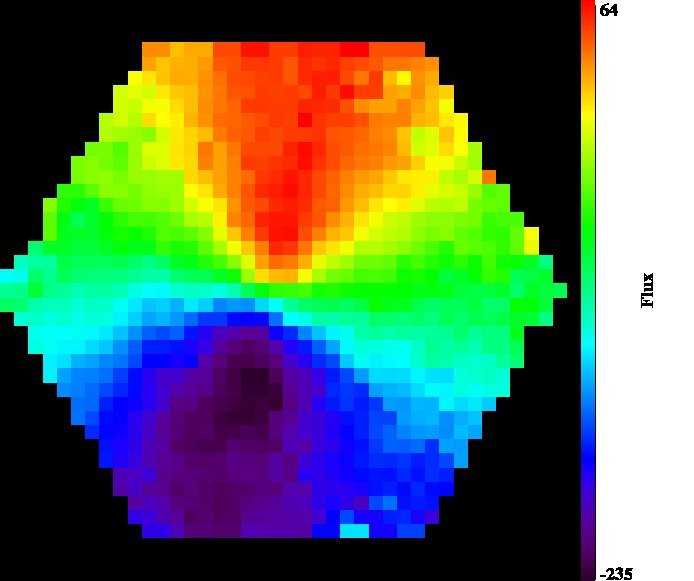}
\includegraphics[width=6.5cm]{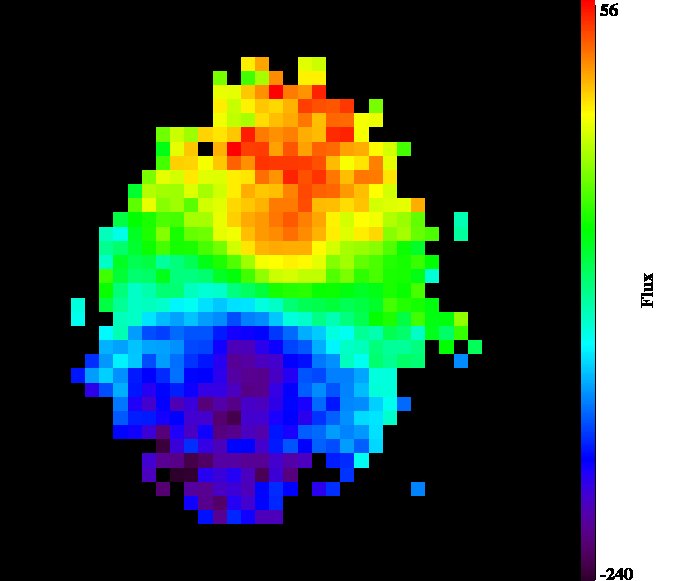}
\caption{ {\small Stellar and ionized gas velocity field, only spaxels with S/N >10 are presented. Left: $H\alpha$ velocity field. Right: stellar velocity field.  }}
\label{sn1999dq_vf}
\end{figure}

\begin{figure}[htb]
\centering
\includegraphics[width=6cm]{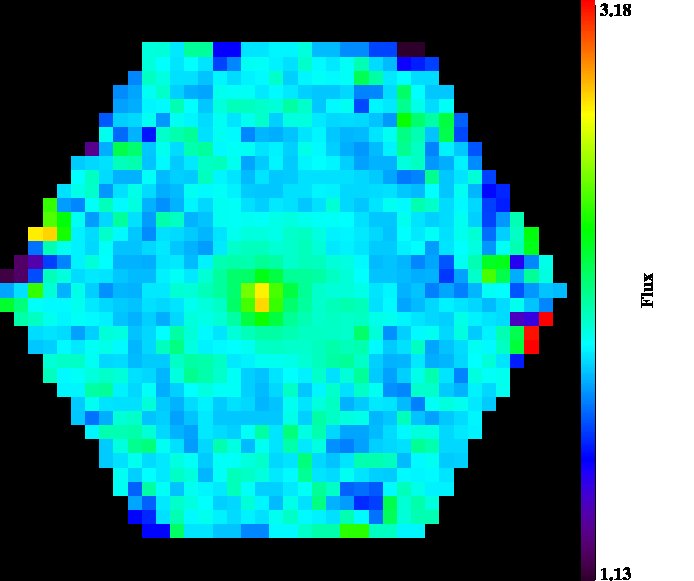}
\caption{ {\small  Velocity dispersion from ionized gas derived from the $H\alpha$ line.  }}
\label{sn1999dq_vds}
\end{figure}

In fig.\ref{sn1999dq_vf}, I show the velocity fields of the ionized gas derived from the $H\alpha$ line and the velocity field derived from the stellar continuum. I have used a S/N cutoff of 10 for the detection of $H\alpha$ line and a S/N> 10 for the stellar continuum in the region [4730-4780]\AA. The two velocity fields are very similar and suggest that the disc of NGC 976 is relaxed. In fig. \ref{sn1999dq_vds} is shown the velocity dispersion from the ionized gas derived from the $H\alpha$ line. The stellar dispersion map, not shown here, presents a very similar distribution. The central peak is due to the presence of the bulge. The external peaks coincide with the inter-arm regions and can be an artifact effect due to the lower signal-to-noise of the spectra in those regions. A detailed study based on bulge dispersion models is needed in order to detect a possible increase on the gas dispersion induced by the SN Ia explosion.  

\subsubsection{Extinction}
\begin{figure}[htb]
\centering
\includegraphics[width=6cm]{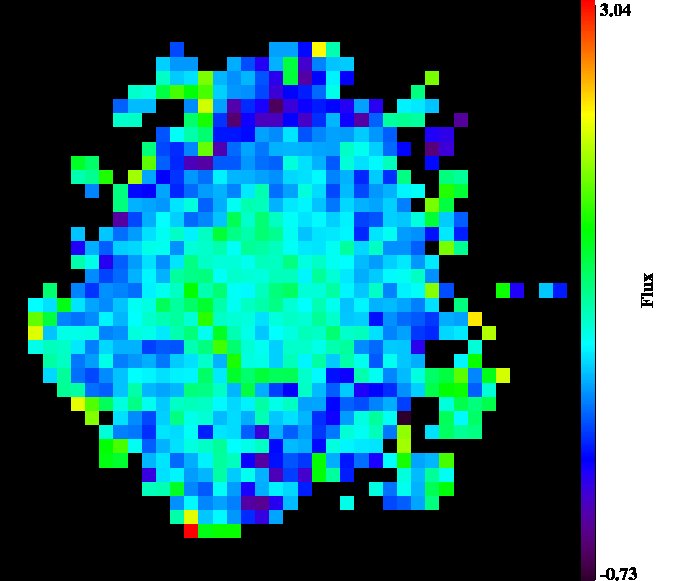}
\includegraphics[width=6cm]{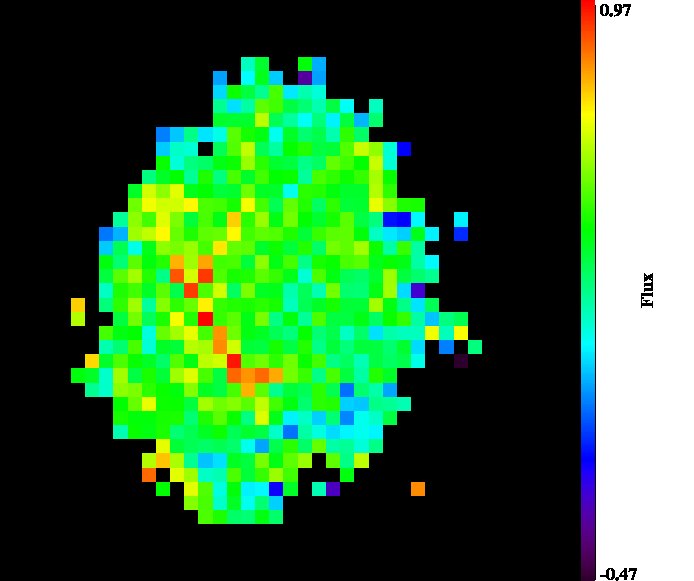}
\caption{ {\small  Left: Extinction from the Balmer decrement $H\beta/H\alpha$. Right: Extinction from the fit of the stellar continuum.}}
\label{sn1999dq_av}
\end{figure}

Fig. \ref{sn1999dq_av} shows the interstellar extinction map from the Balmer decrement (left) and the stellar extinction derived from the fit of the stellar continuum (right). Both maps show a gradient of the dust content, with higher extinction in the central region. The blue ring have the higher dust content of the all galaxy in both maps. The ring has a larger extinction in the left size but it can be due to a projection effect from the galaxy inclination. 
 
 \subsubsection{Electronic density, ionization parameter and shocks }
 
\begin{figure}[htb]
\centering
\includegraphics[width=8cm]{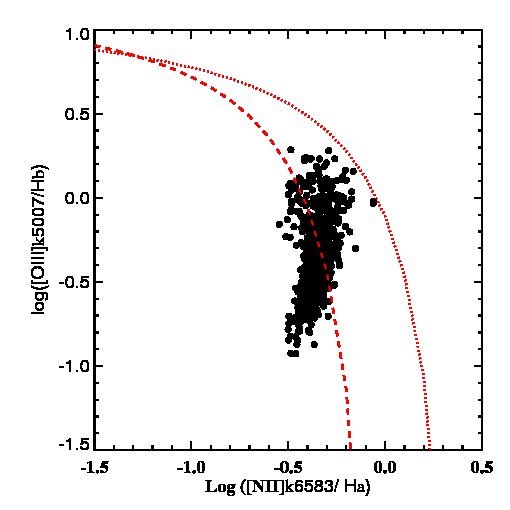}
\caption{ {\small  Diagnostic diagram showing the delimitations of \citet{2001ApJS..132...37K} in dot-line and  \citet{2003MNRAS.346.1055K} in dashed-line. There is no evidence for the presence of a AGN. }}
\label{sn1999dq_BTT}
\end{figure}

Figure \ref{sn1999dq_BTT} shows the location of each spaxel in the diagnostic diagram [OIII]$\lambda\lambda4959, 5007/H_{\beta}$ vs [NII]$\lambda 6584 /H_{\beta}$ \citep{1987ApJS...63..295V, 2003MNRAS.346.1055K}. According to the delimitation of \citet{2001ApJS..132...37K} (dot-line in fig.\ref{sn1999dq_BTT}), there is no evidence for the presence of an AGN in the galaxy. When considering the \citet{2003MNRAS.346.1055K} delimitation (dashed line) some regions fall in the composite AGN-SF area. However, these spaxels correspond to the outskirt regions of the host which have lower S/N spectra (S/N<15). We are now analyzing the data by applying a Voronoi binning \citep{2003MNRAS.342..345C} which has the property to preserve the maximum spatial resolution of general two-dimensional data, given a constraint on the minimum signal-to-noise ratio. We hope with this new method to disentangle between bias and artifacts due to low S/N or real difference of physical properties in these outskirt regions. 

\begin{figure}[htb]
\centering
\includegraphics[width=6cm]{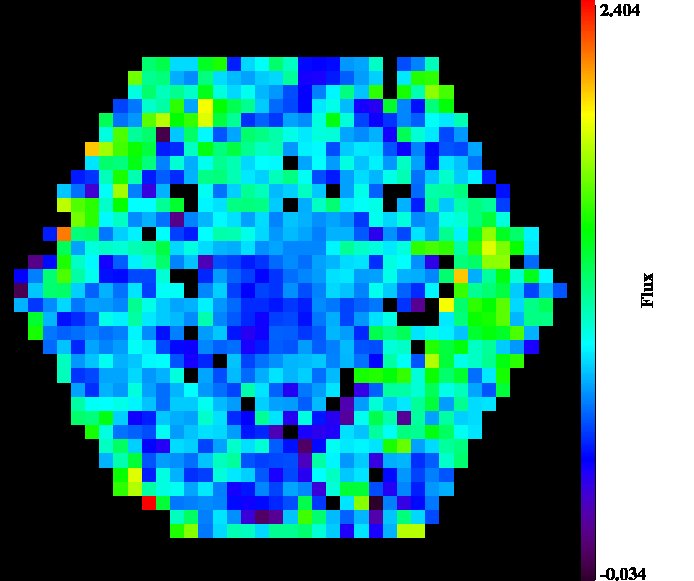}
\includegraphics[width=6cm]{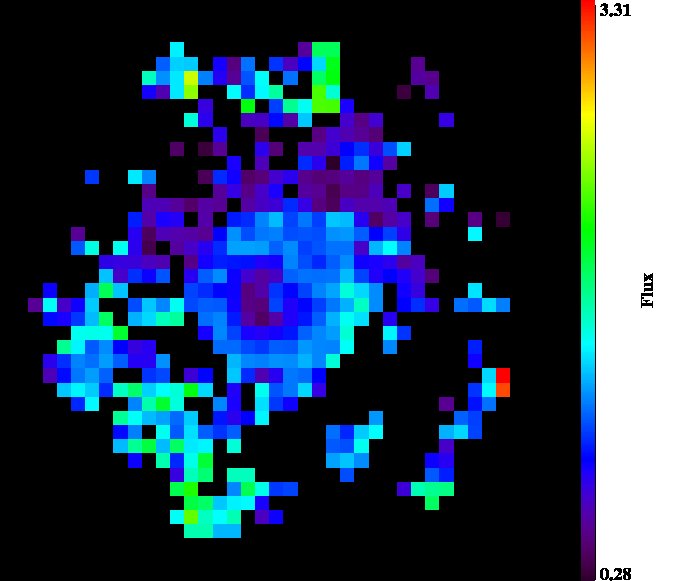}
\caption{ {\small  Right: OII/OIII ratio which is proportional to the ionization parameter. Left: The [SII]$\lambda\lambda$6716,6731\AA\, doublet ratio. This ratio is proportional to the electronic density, higher values correspond to lower electronic densities. }}
\label{sn1999dq_io}
\end{figure}

Figure \ref{sn1999dq_io} shows the [OII]$\lambda$3727\AA/[OIII]$\lambda$5007\AA\, ratio map. This ratio is proportional to the ionization parameter and it has the advantage to be less affected by the extinction than the [OII]$\lambda$3727\AA/NII$\lambda$6784\AA\, ratio, which is also used in the literature for probing the ionization parameter. In figure \ref{sn1999dq_io} right panel, I present the map of the [SII]$\lambda\lambda$6716,6731\AA\, doublet ratio. This ratio is proportional to the electronic density: higher values correspond to lower electronic densities. The ring and the spiral arms have the lower electronic densities, $N_e\sim10^2\,-\,10^3\,cm^{-3}$, while the bulge and the north inter-arm (between the ring and the first arm) have the higher electronic densities $N_e>10^4\,cm^{-3}$ \citep{1989agna.book.....O}. 

\begin{figure}[htb]
\centering
\includegraphics[width=6cm]{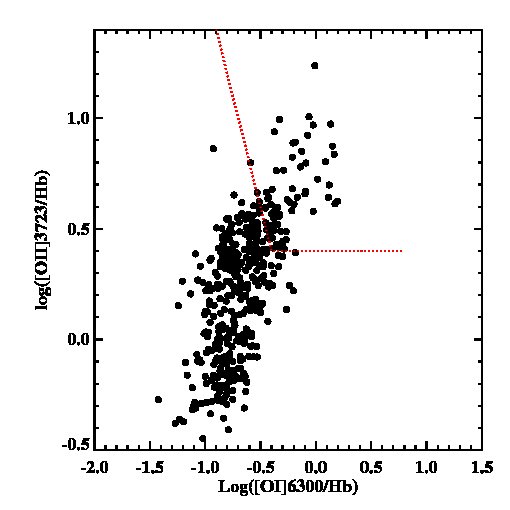}
\includegraphics[width=6cm]{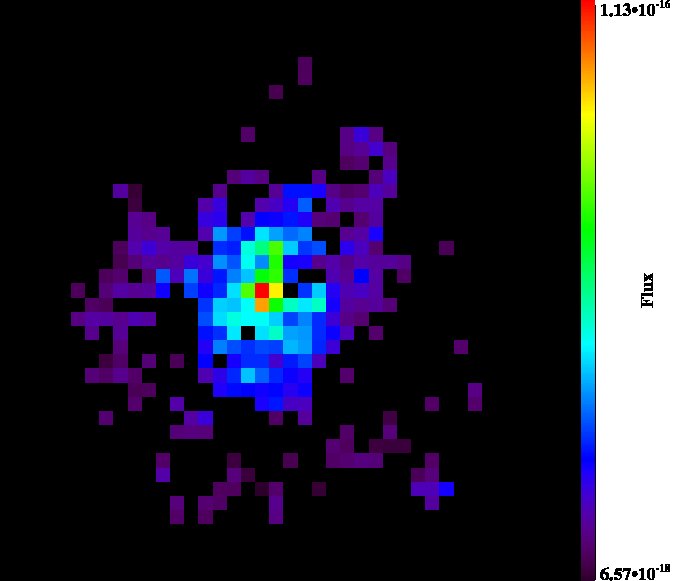}
\caption{ {\small  \textbf{Left}: the [OI]$\lambda$6300\AA/$H\beta\lambda$ vs [OII]$\lambda$3727\AA/$H\beta$  diagnostic diagram proposed by \citet{1985ApJ...292...29F} to detect shocks in the ISM. Points falling in the area at the right of the de-limitation (red dashed line) have shock. All the spaxels with a S/N>10 have been plotted (black circle). \textbf{Right}: The flux map of the [OI]\,6300\,\AA\, emission line. The presence of this line is a good estimation of the presence of shocks in the ISM. The [OI] is detected in the center and star-forming region (arms and ring).  }}
\label{sn1999dq_shocks}
\end{figure}

In figure \ref{sn1999dq_shocks} left panel, I give the flux map of the [OI]\,6300\AA\, emission line. The presence of this line is a good estimation of the presence of shocks in the ISM. The OI is detected in the center and star-forming region (arms and ring). I have also plotted the diagram diagnostic [OI]$\lambda$6300\AA/$H\beta\lambda$ vs [OII]$\lambda$3727\AA/$H\beta\lambda$ ratio map which allow to disentangle between OI lines produced in shocks or by the stellar radiation (see left panel of Fig.\ref{sn1999dq_shocks}). Some spaxel fall in the upper right region of the diagram attesting for the presence of shocks. However, these spaxels belong to the outskirt regions and may be affected by the lower S/N. 

 \subsubsection{Star formation rate and history}
 The SN 1999dq is located in the peak of a star-forming region.
   \begin{figure}[htb]
\centering
\includegraphics[width=6cm]{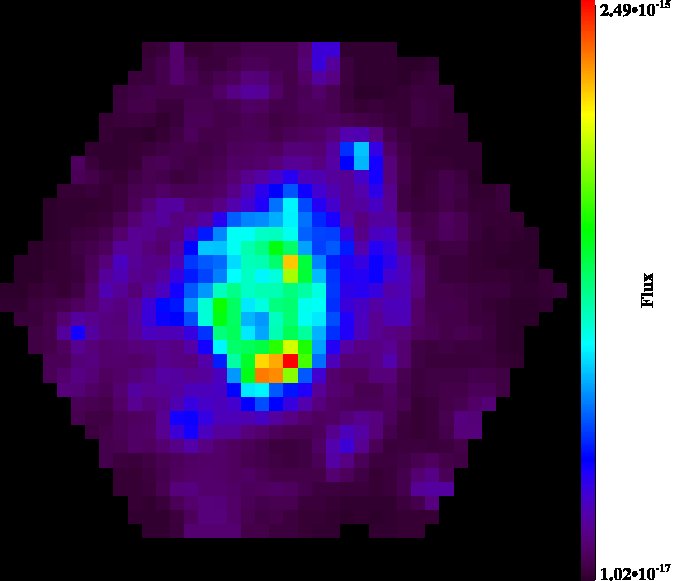}
\includegraphics[width=6cm]{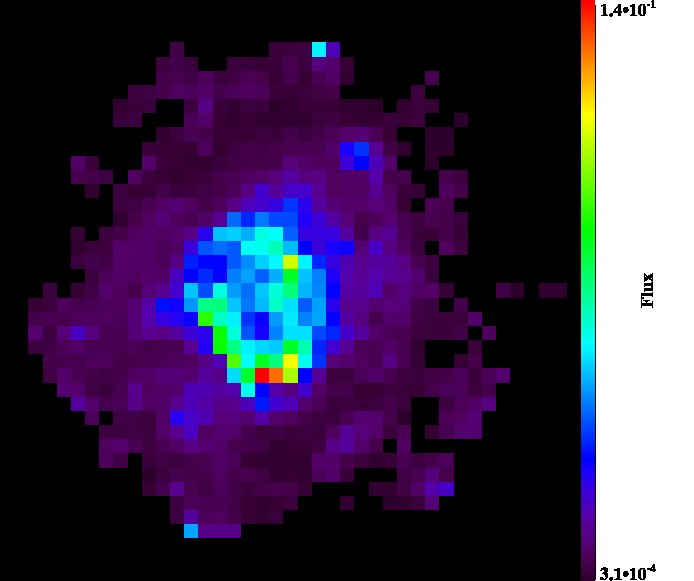}
\caption{ {\small  Left: $H\alpha$ flux map. Right: SFR from $H\alpha$ corrected from extinction. The SN is located in the star-bursting ring.}}
\label{sn1999dq_sfr}
\end{figure}

 \subsubsection{Metal content}
\begin{figure}[htb]
\centering
\includegraphics[width=6cm]{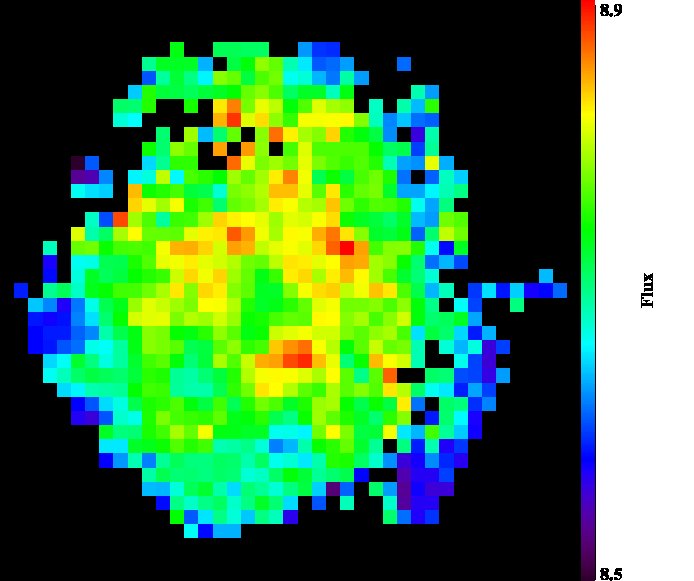}
\includegraphics[width=6cm]{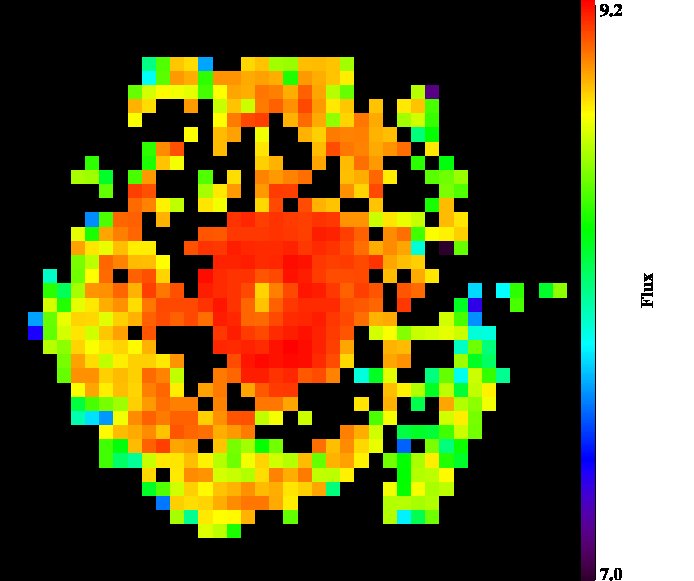}
\caption{ {\small  Left: 12+log(O/H) map derived from the O3N2 parameter and the calibration of \citet{2004MNRAS.348L..59P}. Right: 12+log(O/H) map derived from the R23 parameter and the  \citet{2004ApJ...613..898T} calibration. }}
\label{sn1999dq_Z}
\end{figure}

The metal content of a galaxy can be estimated by measuring the oxygen abundance. The several oxygen abundance calibrators that can be found in the literature have all their pros and cons. For this reason, I have derived metallicities maps using 3 different calibrations. In the left panel of figure \ref{sn1999dq_Z}, I present the 12+log(O/H) map derived from the O3N2 parameter and the calibration of \citet{2004MNRAS.348L..59P}. The regions with higher gas metallicities corresponds to the regions with higher star formation.  The ring and spiral arms are well visible. The high-metallicity region in the south of the ring correspond to the peak of the $H\alpha$ emission. The metallicities range from 8.9 in the inner ring to 8.6 in the outskirts disk. The right panel of figure \ref{sn1999dq_Z} is the 12+log(O/H) map derived from the R23 parameter and the \citet{2004ApJ...613..898T} calibration. The R23 map is less complete than the O3N2 due to the difficulties on detecting the [OII] lines. The reasons of the none detection in some spaxels may be physical (intrinsic luminosity of the line in these regions), but also due to instrumental limitations. Indeed, the [OII] line falls in the borders of the detector and therefore it is affected by vigneting effects and by the low efficiency of the instrument at blue wavelength. However, the R23 and O3N2 maps show similar trends: the region with higher SFR have higher gas metallicities. The R23 have a smaller range on metallicities variations. The difference between the two maps can be due to the dependence of the $R23$ parameter on the ionization parameter. 
   
 \subsubsection{The SN Ia environment}
 I have measured the properties of the host in the location of the SN 1999dq and compared it with the median values in the host. The deviation from the mean has been calculated following the expression proposed by \citet{2008A&A...490...45C}, written here for the SFR: 
 \begin{equation}
Deviation=\frac{SFR(SN\,Ia) - <SFR>}{\sigma(SFR)}
\end{equation}
where <SFR> is the mean value for the SFR, and $\sigma$(SFR) is the standard deviation of the regions with a S/N>10. The results are shown in Table \ref{table_SFR_dev}. The properties of the SN 1999dq environment do not deviate strongly from the mean properties of the host galaxy. The only significant deviation is the one of the SFR: The SN\,Ia is located in the region with more star-formation of the all galaxy.  
 
 \begin{table}[htb]
\label{table_SFR_dev} 
\centering
\begin{tabular}{l | c c c c c c}
\hline
\hline
&SFR&log(O/H)+12&log(O/H)+12&$A_V$&$EQW(H\alpha)$&$SII ratio$\\
&&O3N2&R23 T04&&&\\
\hline
SN position&0.14&8.78&9.05&2.20&&1.15\\
&$M_\odot \,yr^{-1}$&$dex$&$dex$&$mag$&$\lambda$&\\
\hline
Deviation&2.95&1.41&0.81&1.48&&0.91\\
& \multicolumn{6}{c}{$in\,sigma$} \\
\hline 
\end{tabular} 
\caption{{\small Main properties of the host in the region of the SN and their deviations to the mean values in the galaxy (with S/N > 10). }}
\end{table}

\subsection{Conclusions}
The pilot observational run of November 2009 has allowed us to collect IFS data for 6 SN Ia hosts. The interpretation of the data is still on progress and I therefore present here a preliminary study of one SN Ia host. This pilot run has allowed us to define a strategy to reduce and interpret the 3D data, as well as to investigate the potential of the study of SN Ia hosts with IFU spectroscopy. For instance, it has reveled some unsurprising limitations such as the low S/N in the external regions of the galaxy. We are now investigating the use of the Voronoi binning method to bin the cube with spaxel of equal S/N. We expect to increase considerably the S/N of the outskirt regions and thus derived more reliable parameters from their spectra. We are also analyzing the properties of the several regions of the galaxy - arms, ring, bulge - and comparing it with those of the SN localization in order to search for  systematic deviation of the SN environment. Preliminary results on the limited sample of six late-type galaxies seem to point out that SNe Ia occur systematically in region with the higher metal content and star formation rate. We are also comparing the properties near the SN Ia measured from our 3D spectroscopy with those that have been derived from integrated spectroscopy (SDSS, Pietro, long-slit Gallagher  2006,2008) and SED fitting of aperture photometry.

        \FrameThisInToc
\part*{Conclusion and perspective}
\begin{flushright}
\textit{ "The most exciting phrase to hear in science,\\
 the one that heralds new discoveries is not\\ 'Eureka!' (I've found it!),\\ but 'that's funny...'"\\} 
Isaac Asimov
 \end{flushright}

 \EmptyNewPage
\section*{Synthesis of the results}

\subsubsection{The crucial importance of methodology}
Along my research on the topic of the physical properties of distant galaxies, I have been confronted to the critical role of the methodology chosen to derived them.  
The study of intermediate redshift galaxies is frequently conducted in the framework of large surveys  such as the CFRS, GOODS, zCOSMOS, VVDS or DEEP. They aim to collect a large number of galaxies in order to minimize the statistical uncertainty. In return, they tend to rely on low time-consuming observations like photometry or low spatial resolution spectroscopy which can be important sources of uncontrolled systematic errors. Indeed, reliable estimations of the physical properties of distant galaxies can hardly be reached using these kind of data. \citet{2004A&A...417..905L} have shown that a reliable estimation of SFR, extinction and metallicity requires at least moderate spectral resolution (R>1000). I have confirmed this result with a new set of observations: $R_{23}$ metallicities derived from low resolution spectra (R=250) are overestimated by $+0.27\,dex$. Distant galaxies are far more complex than local galaxies, i.e. in terms of morphology and kinematics. Their study is at least equally demanding of precise observation strategies than what has been used for local galaxies. To better control the biases on the derivation of the physical properties of faint and distant galaxies, I have evaluated them with at least two independent measurements. Indeed, each measurement can be affected by different errors which can be of physical nature, such as the sensitivity of the method to the star formation history, or due to the measurement themselves, but a systematic double estimation allowed us to get a better control on the systematics.
 
\subsubsection{Galaxies evolved as closed-box systems during the past 8 Gyr}
In the framework of the IMAGES survey, I have derived the properties of the ISM from integrated spectroscopy data of a representative sample of z=0.6 intermediate-mass galaxies. The galaxy spectra have been acquired with FORS2 at moderate spatial resolution (R=1000) and have S/N comparable to those of local SDSS. The main results of this study are: 
\begin{itemize}
\item The M-Z relation at $z\sim0.6$ puts on evidence the strong evolution ($\sim0.3\,dex$) of metal content in the gaseous phase of galaxies. 
\item The M-Z relation at $z\sim0.6$ had a larger dispersion compared to the local relation. 
\item The gas fraction has doubled from z=0 to $z=0.6$. 
\item The effective yield of $z\sim0.6$ galaxies is concordant with a closed-box model. However, small contributions of outflows and infalls cannot be excluded.
\item There are no evidence of outflows in $z\sim0.6$ galaxy spectra and therefore, which implies that large-scale outflows do not play an important role in intermediate mass galaxies at this redshift. 
\end{itemize}
These observations suggest that the evolution of galaxies is consistent with a closed-box scenario in the past 8\,Gyr. In other words galaxies evolve simply by transforming their gaseous content into stars. Such a result does not exclude external sources of gas but it simply show that they are not required to explain the observed evolution.
On the other hand, IMAGES has put on evidence merger events as a central process of the recent evolution of intermediate-mass galaxies. For instance, the large dispersion of the M-Z relation at $z\sim0.6$ can be explain by the diversity of galaxy building histories within the framework of a merger scenario. These two apparently contradictory observations suggest a disk formation scenario in which local spiral are remnant of the major merger between two gas-rich galaxies progenitor at $z\sim1-2$. At z=0.6, we observed the final stages of the merger and the relaxation of the remanent which will then evolved secularly until today.   


\subsubsection{The SFR: a fundamental parameter to constrain the stellar mass}
Stellar mass is one of the most difficult quantity to constrain in distant galaxies. I have implemented a new method to derive the stellar population and stellar mass by fitting a combination of stellar population templates to broad-band photometry. I aimed to break the degeneracies inherent to this problem by adding the SFR as a new constrain. The SFR allows us to constrain the number of young stars in the system. However, adding the SFR constraint introduces a plethora of local minimum to the minimization problem turning the localization of the best solution difficult with conventional  code based on the gradient or Hessian function. Therefore, I have used a metaheuristic approach, the Swarm intelligence algorithm, to find the global solution of the optimization problem. The preliminary tests on fakes galaxies are promising: the stellar masses are recover within a random uncertainty of $\pm0.1 dex$ independently of the SFH. I have applied this method to two IMAGES galaxies which have been modeled in detail with hydrodynamics simulations. I show that the observational constraint given by the composition of stellar population is a useful tool to reconstruct the past history of these galaxies.

\subsubsection{High quality sky extraction in fiber-fed instruments}
The study of the evolution of galaxies at extreme look-back time will require observations from large surveys of high-z galaxies with a new generation of instrumental facilities, such as the European-Extremely Large Telescope (E-ELT). In this framework, I had the opportunity of working on the phase-A of OPTIMOS-EVE, a fiber-fed visible-to-near infrared multi-object spectrograph design for the E-ELT. My work in the OPTIMOS-EVE consortium was to define a strategy to resolve the critical issue of sky subtraction. The currently existing algorithm and observation strategy of sky subtraction in fiber-fed instrument can hardly reach 1\% of accuracy for estimating the sky background, while the EVE science case requires an accuracy < 1\%. I have developed a new technique of sky extraction based on the reconstruction of spatial fluctuations of the sky background (both background and emission). Using Monte-Carlo simulations, I have demonstrated that this algorithm can reach an accuracy similar that provided by slit spectrographs. This is due to a careful reconstruction of a sky continuum surface as a function of wavelength using dedicated sky fibers distributed over the whole instrument field-of-view. This makes it possible to interpolate the sky contribution at the location of the science channels with good accuracy. Objects as faint as $f_{Ly\alpha} =1\times10^{-19}\,ergs/cm^2/s$ and $m_{cont}=30$, which fairly represent z=8.8 Lyman-alpha emitters, should be detectable by the IFU mode of OPTIMOS-EVE within 40 hours of integration time.

\subsubsection{Host galaxies of SNe 1a}
 During the last decade, SNe Ia were used as cosmological standard candles.  However, subtle systematic uncertainties stemming from our limited physical knowledge of SNe Ia progenitor stars are currently the major obstacles to fully exploit the potential of SNe Ia in cosmology. One approach to get new insight into the properties of SN Ia progenitors is to focus on their host galaxies. We have started a project to study SNe Ia host galaxies in the local Universe using integrate field spectroscopy, in order to derive the properties of the gas and stellar populations in the immediate vicinity of the SN Ia. I describe the results of the pilot observation with 6 SNe 1a host carried out with the PPAK/PMAS integral field spectrograph at the Calar Alto observatory in November 2009.  This data have allow us to produce, for the first time, maps of the principal properties of SNe Ia host, such as the metallicity, extinction, SFR and electronic density. I present the preliminary study of SN 1999dq host and  show the potential of the study of Sn Ia host with IFU spectroscopy.

\section*{Perspectives}
\subsubsection{Investigate the redistribution of gas after a major merger }
The results presented in Part II chapter 4 are in majority not published yet. I am currently writing an article on this topic to be summited soon. I will mainly focus on the observed evolution of gas fraction and yield.  

In the future, I aim to develop several points of this study. In particular, I am interested in the motions of the ionized gas and metal distribution in the remnants of major mergers. Merger events induce redistributions of gas \& star, and can power strong star-formation rates. During the past years, several authors have investigated the effects of interactions and mergers on the chemical content of galaxies. For instance, Rupke et al. (2010) have argued that the low metal content found in the nucleus of interacting galaxies is due to radial inflows of low-metallicity gas from the outskirts of the two merging galaxies. I have found that the two most under-metallic galaxies (compared to the median metallicity at given mass) are probably ongoing mergers. I am projecting to investigate more in detail this topic by doing a systematic study of morphologies of over- and under- metallic galaxies in the IMAGES sample. Maps of the metal distribution and those of other properties of the ISM of distant galaxies will be useful tools to understand the redistribution of the matter after a merger in gas-rich systems. Such observations can be done using X-shooter/VLT in IFU mode. I am co-I of a project on this topic (Principal investigator M. Puech): three z=0.6 galaxies have been observed with X-shooter IFU in november. In addition, the motions of the ionized gas kinematics on large scale around ongoing mergers or recent remnants can be investigated using the large IFU of VLT/GIRAFFE spectrograph. I have already submited several observational proposals (until now accepted but not scheduled) to observe the external disks and internal halos up to radii of 50 kpc in actively star-forming galaxies at z=0.6-0.7 with this instrument, see Fig. \ref{GIRAFFE_IFU}. 
\begin{figure}[!h]
\centering
\includegraphics[width=0.6\textwidth]{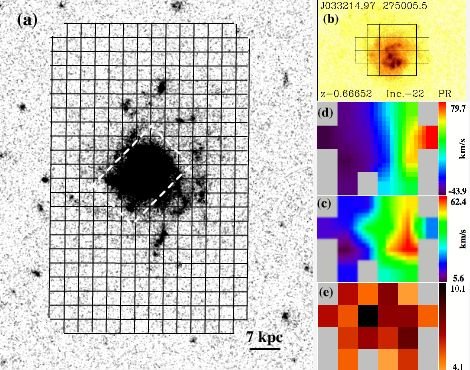}
\caption{{\small (a) Composite HST/ACS image in V-B-I-Z bands of The Firework galaxy. The white dashed square represents the region previously observed with the GIRAFFE IFU. The grid delimit the region planned to be observed with the ARGUS IFU. (b) Central part of the galaxy with the GIRAFFE/IFU region overplotted. (c) Velocity field (d) $\sigma$-map , the dispersion peak is offset from the center of the galaxy. (e) S/N map for the GIRAFFE pixels. }}
\label{GIRAFFE_IFU}
\end{figure}

At longer terms, the next observational facilities will allow us to study the physics of LSB galaxies. Their high gas fractions and the probable evolution in number density suggest that LSB galaxies may play a crucial role in the evolution of galaxies during the past 8 Gyr and even more at higher look-backtimes.

\subsubsection{Tracing back in time the star formation history from 7 Gyr to the early Universe} 
The preliminary results show that the constrain on the SFR improves the accuracy on the derived stellar mass and stellar population in star-forming galaxies.
In the next months, I will optimized and test the method. I am actually optimizing the parameter of the swarm intelligence algorithm to improve its convergence and decrease the execution time.  I  am planning to realize more test using fake galaxies with more complex SFHs, such as model galaxies taken from hydrodynamical merger simulations (Wuyts et al., 2009) and to evaluate the uncertainties due to the models and observations (Conroy et al., 2009, 2010; Conroy and Gunn, 2010). After the validation of the method, I will applied it to the IMAGES galaxies in order to derivate their star-formation history. It will allow us to constrain the evolution of the star formation rate density from 7 Gyrs to the early universe. Complementary studies with other photometric galaxy samples are also planned.

 \subsubsection{A new sky subtraction algorithm for fiber-fed spectrographs}
The tests realized on the $\lambda-\lambda$ sky reconstruction method to subtract sky light in fiber-fed spectrographs give very promising results. They show that this new technique can overcome the apparent limitation of fiber-fed instruments in recovering with high accuracy the sky contribution. If OPTIMOS-EVE was accepted for a phase B, my work on the next months on this topic would be to implement and optimized the algorithm in the data-reduction pipeline of OPTIMOS-EVE. Tests on real data with existing fiber-fed instrument, such as VLT/GIRAFFE and Caha/PKAS, are also planned. Independently of the acceptance of OPTIMOS-EVE for a phase B, I want to create a user-friendly interface for the sky-subtraction algorithm and adapt it to reduce data from several fiber-fed spectrographs. This software will be make public.

\subsubsection{Define strategy and tools for 3D surveys of SN Ia host}
As commented previously, the CENTRA-GEPI project on study of hosts of type Ia supernovae is in the first stages of realization. The data from the pilot observations in November 2009  has been crucial to establish a strategy to reduce and interpret the 3D data in view of the arriving data from the CALIFA survey. In the next 3 years, the CALIFA large survey at Calar Alto observatory will provides us with IFS data for approximatively 50 SNe Ia hosts and thus provide a sufficient number of hosts to deliver confident conclusions on the systematics induced by the SN Ia progenitors on the derivation of cosmological parameters.

\Annexes
\Annex{List of publications}
\subsubsection{Publications in journal with referees}
11 accepted papers:  
\begin{enumerate}
\item Puech, M.; Hammer, F.; Flores, H.; Neichel, B.; Yang, Y.; Rodrigues, M. , 2007A\&A, 476L, 21P : \textit{"First detection of a minor merger at z ~ 0.6}" 
\item Liang, Y. C.; Hammer, F.; Yin, S. Y.; Flores, H.; Rodrigues, M.; Yang, Y. B., 2007, A\&A, 473, 411L, "\textit{The direct oxygen abundances of metal-rich galaxies derived from electron temperature}" 
\item Rodrigues et al., "\textit{IMAGES IV: strong evolution of the oxygen abundance in gaseous phases of intermediate  mass galaxies from z$\sim$0.8}" 2008, A\&A,492, 371R 
\item Hammer, F.; Flores, H.; Yang, Y. B.; Athanassoula, E.; Puech, M.; Rodrigues, M.; Peirani, S. "\textit{A forming, dust  enshrouded disk at z=0.43: the first example of a late type disk rebuilt after a major merger?}" 2009, A\&A, 496, 381H
\item Yang, Y.; Hammer, F.; Flores, H.; Puech, M.; Rodrigues, M ; "\textit{A surviving disk from a galaxy collision at  z=0.4}", 2009, A\&A, 501, 437Y 
\item Hammer, F.; Flores, H.; Puech, M.; Athanassoula, E.; Rodrigues, M.; Yang, Y.; Delgado-Serrano, R. ; "\textit{The Hubble sequence: a vestige of merger events}",  2009, A\&A, 507, 1313H
\item Delgado-Serrano, R.; Hammer, F.; Yang, Y.ÊB.; Puech, M.; Flores, H.; Rodrigues, M.; "\textit{How was the Hubble sequence 6 Gyrs ago?}",  2010, A\&A, 509A, 78D
\item Puech, M.; Hammer, F.; Flores, H.; Delgado-Serrano, R.; Rodrigues, M.; Yang, Y; "\textit{The baryonic content and Tully-Fisher relation at z$\sim$0.6}", 2010, A\&A, 510A, 68P
\item Fuentes-Carrera, I.; Flores H.; Peirani S.; Yang Y.; Hammer F.; Rodrigues M.; Balkowski C., 2009, "\textit{Disentangling the morpho-kinematic properties of a face-on merger at z$\sim$0.7}", 2010, A\&A, 513A, 43F
\item Han, X. H.; Hammer, F.; Liang, Y. C.; Flores, H.; Rodrigues, M.; Hou, J. L.; Wei, J. Y.; " \textit{The Wolf-Rayet features and mass-metallicity relation of long-duration gamma-ray burst host galaxies}", 2010, A\&A, 514A, 24H
\item Chen, X. Y.; Liang, Y. C.; Hammer, F.; Prugniel, Ph.; Zhong, G. H.; Rodrigues, M.; Zhao, Y. H.; Flores, H.;  "\textit{Comparing six evolutionary population synthesis models through spectral synthesis on galaxies}", accepted for publication in A\&A
\end{enumerate}

2 in preparation and 1 submitted:
\begin{enumerate}
\item Rodrigues M.; Flores H.; Puech; M.; Yang Y.; Royer. F; in preparation ; "\textit{A method to subtract the skylight for the multi-fiber instrument E-ELT/OPTIMOS-EVE}" , 2010 SPIE
\item Rodrigues et al. 2010, in preparation ; " \textit{Extracting stellar population in star-forming galaxies}"
 \item Rodrigues et al. 2010, in preparation ; "\textit{Evolution of the gas fraction over the last 8 Gyrs}"
\end{enumerate}

\subsubsection{Press releases}
\begin{enumerate}
\item Press release Observatoire de Paris, 'No galaxy is an island'  \\ 
http://www.obspm.fr/actual/nouvelle/dec08/oxy.en.shtm
\item Press release Observatoire de Paris and ESO, "Hubble and ESO's VLT provide unique 3D views of remote galaxies", \\
http://www.eso.org/public/outreach/press-rel/pr-2009/pr-10-09.htm
\end{enumerate}

\Annex{Rodrigues et al. 2008, A\&A,492, 371R }
 \EmptyNewPage
\Annex{Hammer et al. 2009, A\&A, 496, 381H}
 \EmptyNewPage
\Annex{Fuentes-Carrera et al. 2010, A\&A, 513A, 43F}
 \EmptyNewPage
\Annex{Puech et al. 2010, A\&A, 510A, 68P}
 \EmptyNewPage
\Annex{Hammer et al. 2009, A\&A, 507, 1313H}
 \EmptyNewPage
\Annex{Liang et al. 2007, A\&A, 473, 411L }
 \EmptyNewPage
\Annex{Han et al. 2010, A\&A, 514A, 24H}
 \EmptyNewPage
\Annex{Rodrigues et al. 2010, SPIE submitted}
 \EmptyNewPage


\bibliographystyle{aa}
\DontWriteThisInToc
\bibliography{Main}
\DontWriteThisInToc

\end{document}